\documentclass[12pt,twoside]{documentclassmitthesis}
\usepackage{lgrind}
\usepackage{cmap}
\usepackage[T1]{fontenc}
\usepackage{pdfpages}
\usepackage{graphicx, subcaption}
\usepackage{amsmath}
\usepackage{amssymb}
\usepackage{mathtools}
\usepackage{float}
\usepackage[section]{placeins}
\usepackage{indentfirst}
\usepackage{hyperref}
\hypersetup{
    colorlinks,
    citecolor=black,
    filecolor=black,
    linkcolor=black,
    urlcolor=black
}
\extrafloats{100}
\usepackage{cleveref}
\usepackage{cite}
\begin{document}


\title{Light, Unstable Sterile Neutrinos:\\Phenomenology, a Search in the IceCube Experiment, and a Global Picture}
\author{Marjon H. Moulai}
\prevdegrees{B.A., University of California, Berkeley (2009)}
\department{Department of Physics}
\degree{Doctor of Philosophy }
\degreemonth{June}
\degreeyear{2021}
\thesisdate{April 23, 2021}
\supervisor{Janet M. Conrad}{Professor}
\chairman{Deepto Chakrabarty}{Associate Department Head of Physics}
\maketitle
\cleardoublepage
\setcounter{savepage}{\thepage}
\begin{abstractpage}
Longstanding anomalies in neutrino oscillation experiments point to the existence of a fourth, hypothetical neutrino: the sterile neutrino. Global fits to a sterile neutrino model find a strong preference for such a model over the massive neutrino Standard Model. However, the fit results suffer from inconsistencies between datasets, referred to as tension. This motivates more complicated models for new physics. This thesis considers a model of unstable sterile neutrinos, where the heaviest mass state can decay. First, the phenomenology of unstable sterile neutrinos is explored in the IceCube experiment, a gigaton neutrino detector located at the South Pole. Second, global fits to traditional and unstable sterile neutrino models are combined with one year of data from IceCube. A preference for the unstable sterile neutrino model is found, as well as a reduction in tension. Lastly, a high statistics search for unstable sterile neutrinos is performed in IceCube. The Standard Model is rejected with a $p$-value of 2.8\% and the traditional sterile neutrino model is rejected with a $p$-value of 4.9\%. The best-fit point is $\Delta m_{41}^2 = 6.7^{+3.9}_{-2.5}$ eV$^{2}$, $\sin^2 2 \theta_{24} = 0.33^{+0.20}_{-0.17}$, and $g^2 = 2.5 \pi \pm 1.5 \pi$, where $g$ is the coupling that mediates the neutrino decay. The best-fit corresponds to a lifetime of the heaviest neutrino of $\tau_4/m_4 = 6\times 10^{-16}$~s/eV. A Bayesian analysis finds a best model with similar sterile parameters.
\end{abstractpage}
\cleardoublepage
\section*{Acknowledgments}

First and foremost, I would like to thank my advisor, Janet Conrad. After several years away from academia, Janet gave me the opportunity to return, which had felt unlikely, if not impossible. She offered me a position in her research group and guided me into graduate school. Throughout graduate school, she believed in me. She shared interesting research projects with me and always had ideas for solving problems. I may not have gotten a graduate degree without her.

This thesis was made possible by collaborative work done by many people. In the IceCube Collaboration, I especially would like to thank Carlos Arg{\"u}elles, Spencer Axani, Alex Diaz, Ben Jones, Grant Parker, Austin Schneider, Ben Smithers, David Vannerom, Blake Watson, and Philip Weigel for their efforts and assistance. I thank Jason Koskinen and Klaus Helbing for their review of the analysis, Anna Pollman, Juanan Aguilar, and Ignacio Taboada for their help in getting the box opened, and Tom Gaisser for useful discussion.

For the phenomenology work in this thesis, I extend my thanks to my collaborators Zander Moss, Gabriel Collin, and Mike Shaevitz. I also thank my DCTPC/MITPC collaborators: Josh Spitz, Allie Hexley, Anna Jungbluth, Adrien Hourlier, and Jaime Dawson.

I would not have made it through graduate school without the friends I made at MIT, including Christina Ignarra, John Hardin, Alex Leder, Cheko Cantu, Axel Schmidt, and Lauren Yates, to name a few. Similarly, I am grateful for the friends I made at WIPAC, which include Logan Wille, Josh Wood, Ali Kheirandish, Sarah Mancina, Matt Meehan, Jeff Lazar, Ibrahim Safa, and Donglian Xu.

I would like to acknowledge Cathy Modica, Sydney Miller, Nicole Laraia, Kim Krieger, and Tina Chorlton for their administrative assistance. I thank the rest of my thesis committee: Joe Formaggio and Claude Canizares.

Lastly, I am grateful for my parents, Ellie Golestani and Javad Moulai, and my sister, Maryam Moulai, for their love and support.


\pagestyle{plain}

\tableofcontents
\newpage
\listoffigures
\newpage
\listoftables

\chapter{Introduction}
\section{Neutrino oscillations in vacuum}

\begin{figure}
  \centering
	\includegraphics[width=\textwidth]{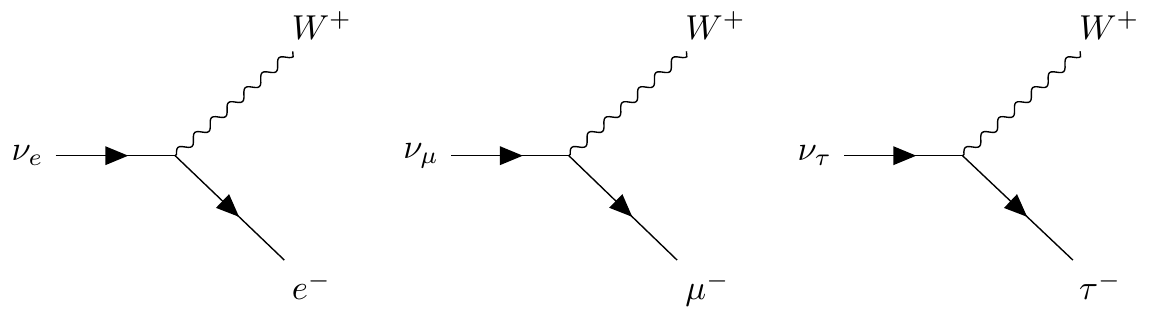}
  \caption[Feynman vertices for charged current neutrino interactions]{Feynman vertices for charged current neutrino interactions. Figure produced with TikZ-Feynman~\cite{Ellis:2016jkw}.}
  \label{fig:nu_cc_flavors}
\end{figure}

In the Standard Model of particle physics, there exist three neutrino flavor states: electron neutrino ($\nu_e$), muon neutrino ($\nu_\mu$), and tau neutrino ($\nu_\tau$)~\cite{Griffiths:2008zz, Thomson:2013zua, Giunti:2007ry, Zyla:2020zbs}. The flavor of a neutrino can only be determined in charged current interactions, which involve the exchange of a $W^\pm$ boson, and when the charged lepton that is produced can be identified. In these interactions, an incoming electron neutrino will produce an outgoing electron ($e^-$); a muon neutrino, a muon ($\mu^-$); and a tau neutrino, a tau ($\tau^-$), as shown in Fig~\ref{fig:nu_cc_flavors}. In the charged current interactions of antineutrinos, the corresponding positively charged lepton will be produced.

\begin{figure}
  \centering
	\includegraphics[width=0.5\textwidth]{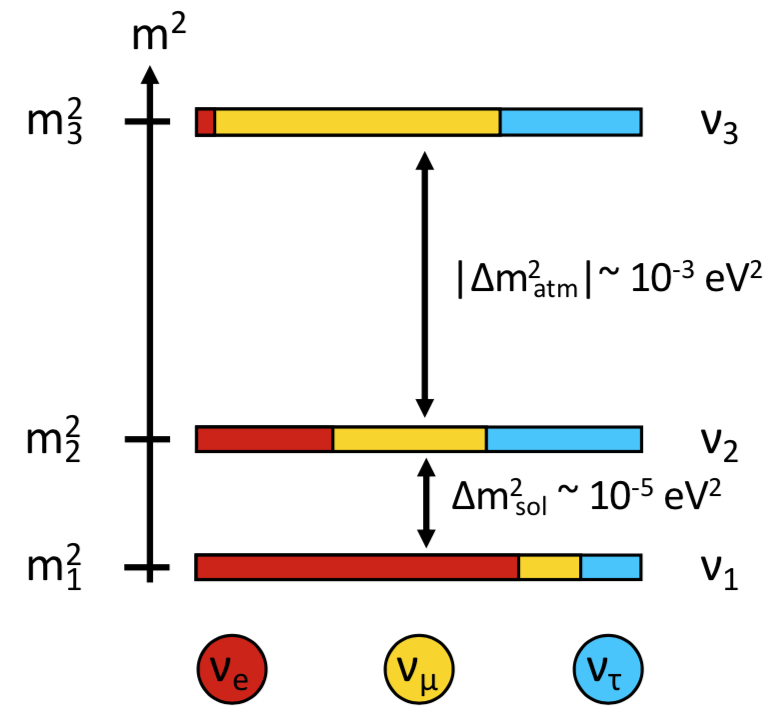}
  \caption[Schematic illustration of the Normal Ordering]{Schematic of the Normal Ordering, with approximate neutrino flavor contributions to each neutrino mass state.}
  \label{fig:3nu}
\end{figure}

\begin{figure}[htb]
  \centering
	\includegraphics[width=0.75\textwidth]{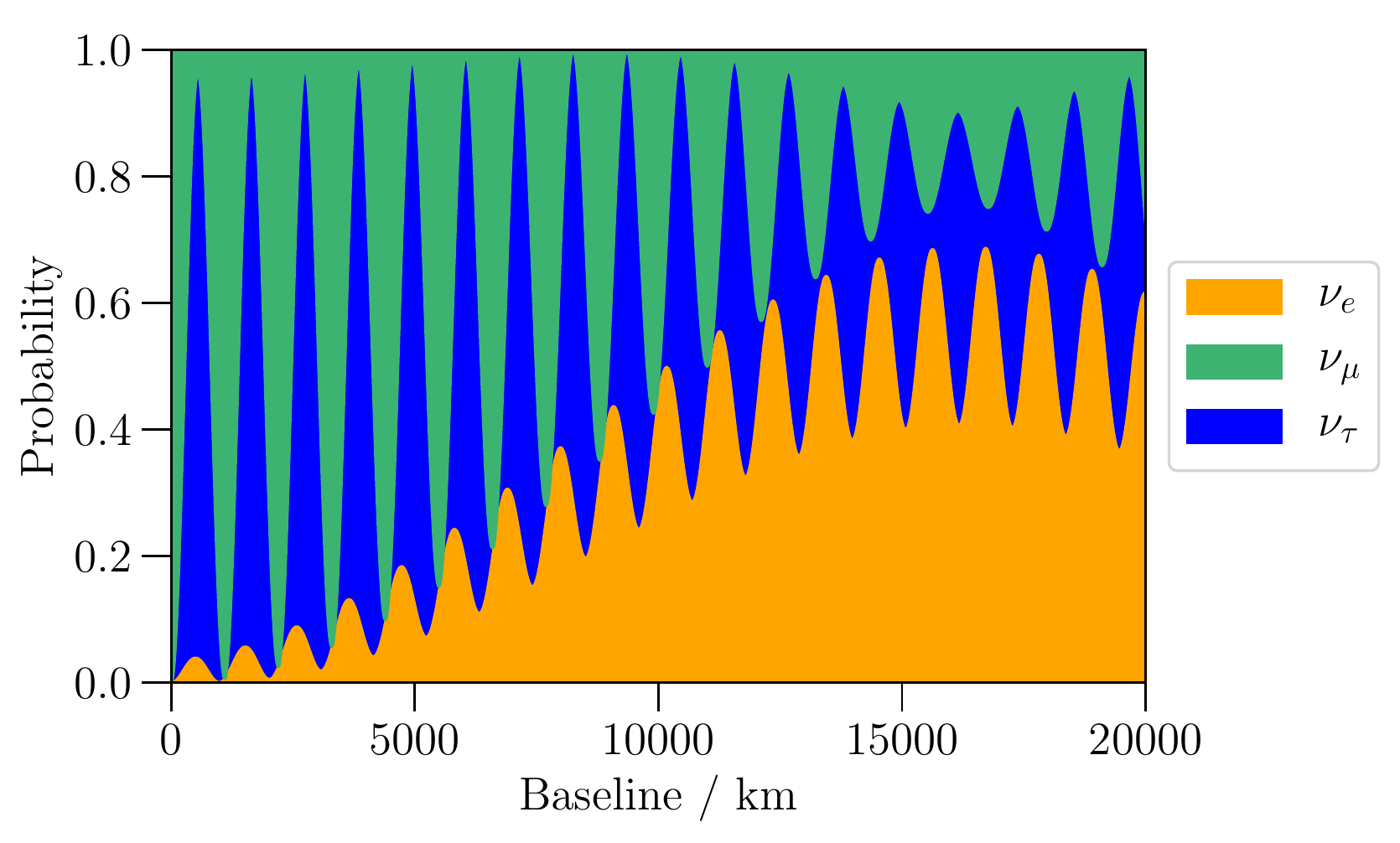}
  \caption[Neutrino oscillation probabilities for 1 GeV neutrinos]{Neutrino oscillation probabilities for 1 GeV neutrinos born in the muon flavor, shown as a function of distance from the point of creation, or baseline. The oscillation probability of $\nu_\mu \rightarrow \nu_e$ is shown in yellow. The survival probability of $\nu_\mu \rightarrow \nu_\mu$ is shown in green. The oscillation probability of $\nu_\mu \rightarrow \nu_\tau$ is shown in blue. Probabilities were calculated with \texttt{nuSQuIDS}~\cite{ARGUELLESDELGADO2015569, nusquids}.}
  \label{fig:std_oscillations}
\end{figure}

In the massive neutrino Standard Model, there exist three neutrino mass states ($\nu_1$, $\nu_2$, and $\nu_3$) that are distinct from the flavor states. The flavor and mass states mix via a unitary mixing matrix, the Pontecorvo-Maki-Nakagawa-Sakata (PMNS) matrix, $U$:
\begin{equation}
\label{eq:nu_mixing}
\begin{pmatrix}
\nu_e \\
\nu_\mu \\
\nu_\tau
\end{pmatrix}
=
\begin{pmatrix}
U_{e1} & U_{e2} & U_{e3} \\
U_{\mu 1} & U_{\mu 2} & U_{\mu 3} \\
U_{\tau1} & U_{\tau2} & U_{\tau3} \\
\end{pmatrix}
\begin{pmatrix}
\nu_1 \\
\nu_2 \\
\nu_3
\end{pmatrix}.
\end{equation} 
This is equivalent to
\begin{equation}
| \nu_\alpha \rangle = \sum_{i=1}^3 U_{\alpha i}^* |\nu_i \rangle,
\end{equation}
which follows the notation in \cite{Giunti:2007ry,Zyla:2020zbs}. This is illustrated schematically in Fig.~\ref{fig:3nu}. The mass states, $|\nu_i\rangle$, are eigenstates of the Hamiltonian. Assuming that the neutrino is a plane wave, the time evolution of the mass states is
\begin{equation}
|\nu_i(t) \rangle = e^{-i E_i t} | \nu_i \rangle,
\end{equation}
and therefore, the time evolution of the flavor states is
\begin{equation}
| \nu_\alpha (t) \rangle = \sum_{i=1}^3 U_{\alpha i}^* e^{-i E_i t} | \nu_i \rangle,
\end{equation}
where $E_i$ is the energy of the $i^{th}$ mass state and $t$ is time. The neutrino mass states have three distinct masses, $m_i$. The mass-squared splittings are
\begin{equation}
\Delta m^2_{ij} \equiv m_i^2 - m_j^2.
\end{equation}
Neutrinos are highly relativistic particles, which means
\begin{equation}
\begin{aligned}
E_i &= \sqrt{p^2 + m_i^2} \simeq p + \frac{m_i^2}{2E_i} \\
t &= L.
\end{aligned}
\end{equation}
Here it is assumed that all mass states have the same momentum, $p$, and $L$ is the baseline the neutrino has traveled, using natural units. All this together gives rise to the phenomenon of neutrino oscillation, where a neutrino born in one flavor state, $\nu_\alpha$, has an oscillatory probability of transforming into another state $\nu_\beta$, as it travels through space. This oscillation probability is given by
\begin{equation}
P(\nu_\alpha \rightarrow \nu_\beta) = |\langle \nu_\beta | \nu_\alpha (t) \rangle |^2 
\simeq \sum\limits_{\substack{k,j \in [1,2,3] \\ k \neq j}} U_{\alpha k}^* U_{\beta k} U_{\alpha j} U_{\beta j}^* \exp \bigg( \frac{-i \Delta m^2_{kj} L} {2E} \bigg)
\end{equation}
where $E$ is the energy of the neutrino. This formula is applicable if one can assume that the difference in energies of the various mass eigenstates can be neglected. A review on neutrino oscillation that explores the assumptions that were made here is in~\cite{Akhmedov:2019iyt}. The oscillation of 1 GeV neutrinos born in the muon flavor is shown in Fig.~\ref{fig:std_oscillations}. The observation of neutrino oscillation is the reason neutrinos are known to have mass, since nonzero $\Delta m^2$ requires at least one nonzero mass state~\cite{Esteban:2020cvm, Zyla:2020zbs, deSalas:2020pgw, Capozzi:2018ubv}.

Like any unitary matrix, the PMNS matrix may be decomposed as the product of rotation matrices. A convenient representation is:
\begin{equation}
\label{eq:pmns_rotation}
U = 
\begin{pmatrix}
1 & 0 & 0 \\
0  & \cos \theta_{23}  & \sin \theta_{23} \\
0  & -\sin \theta_{23}  & \cos \theta_{23} \\
\end{pmatrix}
\begin{pmatrix}
\cos \theta_{13} & 0 & \sin \theta_{13} e^{i\delta_{\rm{CP}}} \\
0 & 1 & 0 \\
-\sin \theta_{13}e^{-i \delta_{\rm{CP}}}  & 0 & \cos \theta_{13} \\
\end{pmatrix}
\begin{pmatrix}
\cos \theta_{12}  & \sin \theta_{12} & 0\\
-\sin \theta_{12}  & \cos \theta_{12} & 0\\
0 & 0 & 1
\end{pmatrix}.
\end{equation}
In Eq.~\ref{eq:pmns_rotation}, it is assumed that neutrinos are Dirac and not Majorana, which is equivalent to assuming that neutrinos are not their own antiparticles. If neutrinos are Majorana, the representation of the PMNS matrix in Eq.~\ref{eq:pmns_rotation} would need to be modified by a fourth factor, a complex matrix, at the far right-hand-side. The Majorana nature of neutrinos is largely irrelevant to this thesis.

The phase angle $\delta_{\rm{CP}}$ in Eq.~\ref{eq:pmns_rotation} represents violation of charge conjugation parity (CP) symmetry. A value of $\delta_{\rm{CP}}$ that is neither 0$^\circ$ nor 180$^\circ$ would mean that neutrinos and antineutrinos have different oscillation probabilities. The global best-fit value of $\delta_{\rm{CP}}$ is consistent with 180$^\circ$~\cite{Esteban:2020cvm, deSalas:2020pgw}. The work in this thesis is largely insensitive to CP violation. For simplicity, $\delta_{\rm{CP}}$ is assumed to be 0 in this work; equivalently, neutrinos and antineutrinos are assumed to have identical oscillation probabilities.

Another unknown is the neutrino mass ordering, that is, the order of $m_1$, $m_2$, and $m_3$. Solar neutrino experiments have determined 
that $m_2$ is greater than $m_1$, with $\Delta m_{21}^2 \approx 7 \times 10^{-5}$~eV$^2$~\cite{Zyla:2020zbs}. Atmospheric neutrino experiments have determined that $|\Delta m_{31}|^2$ is much greater than $|\Delta m^2_{21}|$, with $|\Delta m_{31}|^2 \approx 2 \times 10^{-3}$~eV$^2$. The Normal Ordering refers to the scenario where $m_3$ is larger than $m_1$, as depicted in Fig.~\ref{fig:3nu}, and the Inverted Ordering is the alternative. The work in this thesis is largely insensitive to the ordering of the masses for the three known neutrinos. The Normal Ordering is assumed.

In the approximation of two neutrino flavors and two neutrino mass states, there is only one rotation angle, $\theta$, and one mass-squared splitting, $\Delta m^2$. In this scenario, the probability for a $\nu_\alpha$ to oscillate to $\nu_\beta$ is given by Eq.~\ref{eq:twoflavorosc} and the survival probability is for $\nu_\alpha$ to remain $\nu_\alpha$ is given by Eq.~\ref{eq:twoflavorsurv}. The amplitude of oscillation is set by the mixing angle, $\theta$, and the frequency is set by $\Delta m^2$:
\begin{equation}
\label{eq:twoflavorosc}
P(\nu_\alpha \rightarrow \nu_\beta) =\sin^2 (2 \theta) \sin^2 \bigg( \frac{\Delta m^2 L}{4 E} \bigg)
\end{equation}

\begin{equation}
\label{eq:twoflavorsurv}
P(\nu_\alpha \rightarrow \nu_\alpha) = 1 - \sin^2 (2 \theta) \sin^2 \bigg( \frac{\Delta m^2 L}{4 E} \bigg)
\end{equation}

\section{Anomalies}

Some long-standing neutrino experimental results do not fit within the three-neutrino oscillation framework~\cite{Conrad:2013mka}. Two of the most significant of these are the Liquid Scintillator Neutrino Experiment (LSND) and MiniBooNE. While inconsistent with the massive neutrino Standard Model, these results are consistent with each other, motivating searches for new physics, including the work in this thesis. These experiments are reviewed below and a more detailed review can be found in~\cite{Conrad:2013mka}.

LSND ran at the Los Alamos National Laboratory LAMPF/LANSCE accelerator during 1993-1998~\cite{Athanassopoulos:1996ds,Aguilar:2001ty}. A 798 MeV proton beam struck a target, producing charged pions ($\pi^\pm$) and muons that came to rest in the beam dump. Stopped negatively charged pions would undergo nuclear capture, while stopped positively charged pions would undergo the decay chain
\begin{equation}
\begin{aligned}
\pi^+ \rightarrow \nu_\mu \: &\mu^+ \\
&\mu^+ \rightarrow \bar{\nu}_\mu \: \nu_e \: e^+,
\end{aligned}
\end{equation} 
where the muon also comes to rest before it decays. This produced a beam containing muon antineutrinos, muon neutrinos and electron neutrinos, with electron antineutrino contamination at the level of approximately $10^{-4}$ of the muon antineutrino flux in the relevant energy range~\cite{BURMAN1990621,BURMAN1996416,Conrad:2013mka}. The detector was a tank filled with mineral oil doped with scintillator and  was instrumented with photomultiplier tubes. The center of the detector was 30~m away from the beam stop. In the detector, electron antineutrinos would interact via inverse beta decay,
\begin{equation}
\bar{\nu}_e \: p \rightarrow e^+ \: n,
\end{equation}
while electron neutrinos would interact by scattering off carbon,
\begin{equation}
\nu_e + \:^{12}\rm{C} \rightarrow  ^{12}\rm{N}_{\rm{ground\:state}} + e^-.
\end{equation}
These two interactions were distinguishable, as the electron antineutrino signal had two features, annihilation of the positron ($e^+$) and a 2.2 MeV gamma from neutron capture, while the electron neutrino signal had one, the electron.

\begin{figure}[h]
  \centering
	\includegraphics[width=0.5\textwidth]{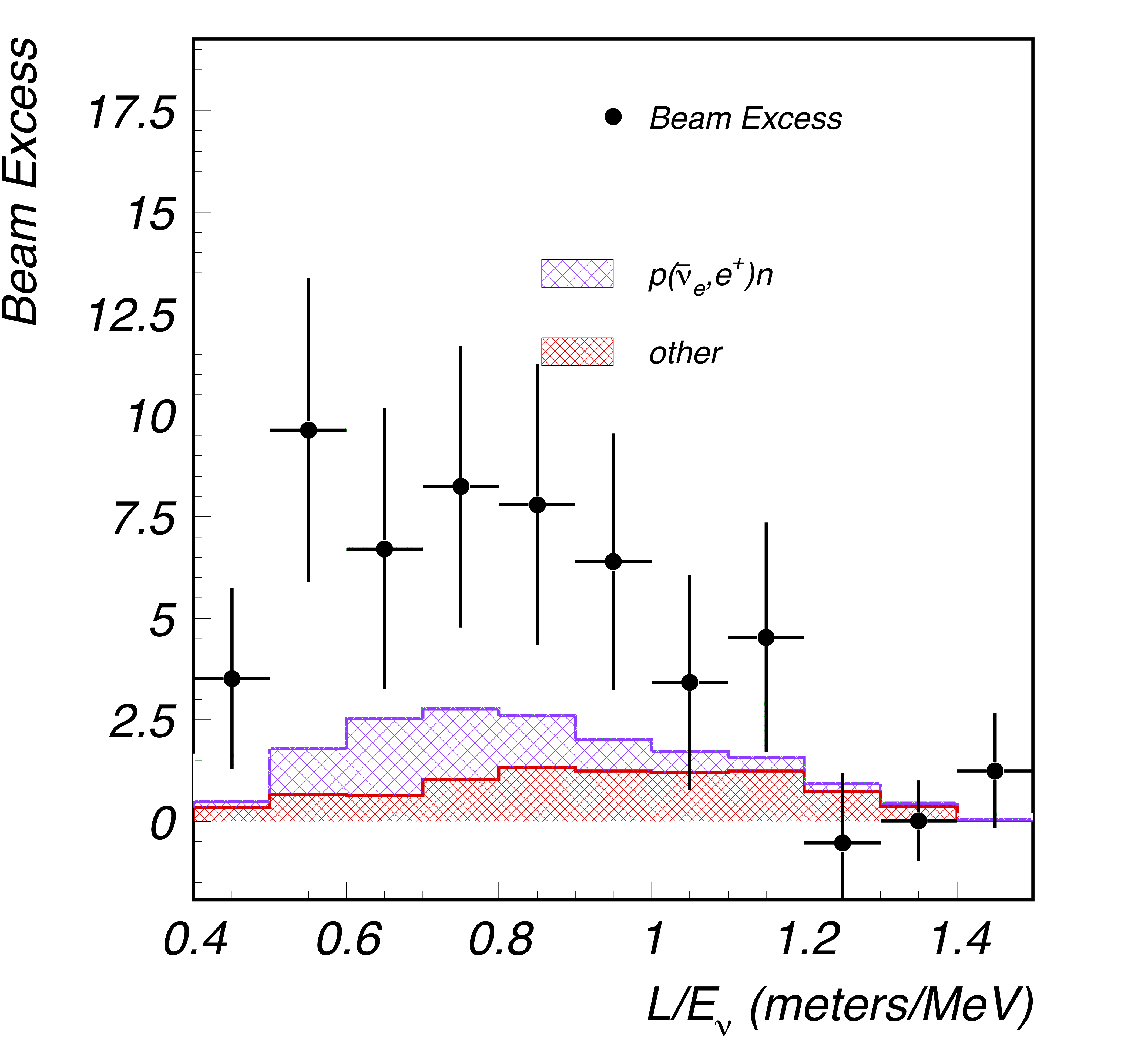}
  \caption[LSND anomaly]{Distribution of LSND beam-on events as a function of baseline/energy. Figure adapted from \cite{Aguilar:2001ty}.}
	\label{fig:lsnd}
\end{figure}

LSND observed an excess of electron antineutrino events~\cite{Aguilar:2001ty}. The LSND result is shown in Fig.~\ref{fig:lsnd}. Data are shown as black points. The purple and red stacked histograms are the expected background contributions. The beam-off backgrounds have been subtracted, which can explain the negative data point. The significance of the LSND anomaly is 3.8$\sigma$.

The MiniBooNE experiment was designed to probe the LSND anomaly~\cite{Aguilar-Arevalo:2013pmq, Aguilar-Arevalo:2020nvw}. MiniBooNE ran at Fermi National Accelerator Laboratory during 2002 - 2019. MiniBooNE used the Booster Neutrino Beamline, created from 8~GeV protons hitting a target. Charged mesons, namely, pions and kaons were produced, and the polarity of the magnetic focusing horn would select for either positively or negatively charged mesons. These mesons would decay in flight to produce a muon neutrino or muon antineutrino beam. The MiniBooNE detector sat 541~m downstream of the target and was composed of a large tank filled with mineral oil and instrumented with 
photomultiplier tubes.

\begin{figure}
 \centering
\includegraphics[width=0.7\textwidth]{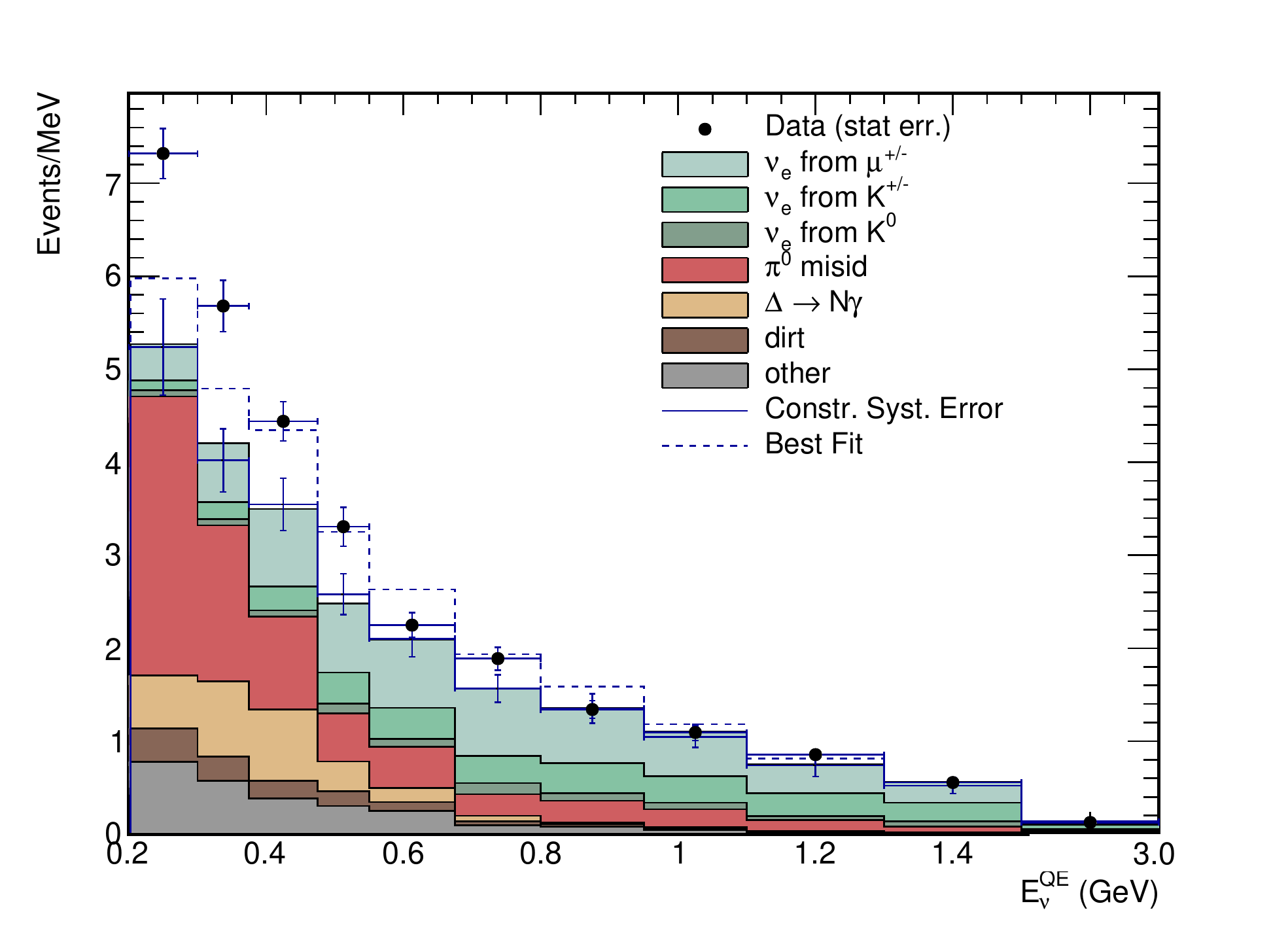}

	\includegraphics[clip, trim = 0cm 12.2cm 0cm 0cm, width=0.65\textwidth]{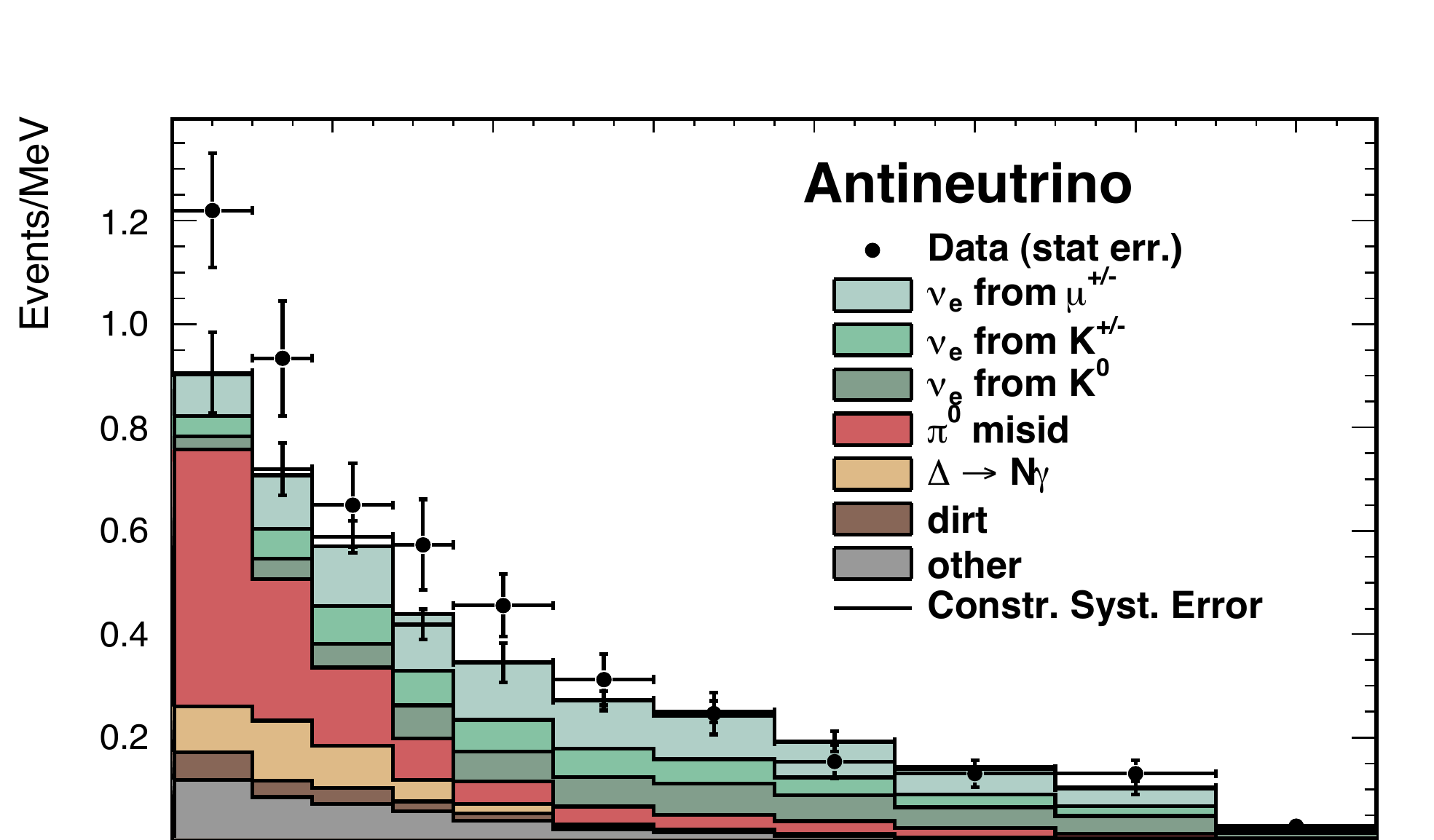}
	\includegraphics[clip, trim = 0cm 0cm 0cm 21.9cm, width=0.65\textwidth]{figs_intro/mb-fig1.pdf}
  \caption[MiniBooNE anomaly]{(Top) MiniBooNE anomaly in neutrino mode. Figure from \cite{Aguilar-Arevalo:2020nvw}. (Bottom) MiniBooNE anomaly in antineutrino mode. Figure adapted from \cite{Aguilar-Arevalo:2013pmq}.}
  \label{fig:miniboone_anomaly}
\end{figure}

MiniBooNE observed an excess of electron-like events in both the muon neutrino beam and muon antineutrino beam, shown in Fig.~\ref{fig:miniboone_anomaly}. The data are shown as black points and the expected backgrounds are shown as stacked histograms. At lower energies, the data sits above the background expectation. This is known as the MiniBooNE anomaly, or MiniBooNE low-energy excess. The significance of the MiniBooNE anomaly is 4.8$\sigma$.

If the appearance of electron-like events in MiniBooNE and LSND is interpreted as a $\nu_\mu \rightarrow \nu_e$ or $\bar{\nu}_\mu \rightarrow \bar{\nu}_e$ oscillation within a two-flavor model, then using Eq.~\ref{eq:twoflavorosc} the data point to value a $\Delta m^2$ larger than $10^{-2}$~eV$^{-2}$.  This cannot be accommodated with the three-neutrino framework, because the two known $|\Delta m_{ij}^2|$ do not add to a large enough value. Curiously, while the LSND and MiniBooNE experiments had significant systematic differences, such as a factor of ten in energy and different types of beams (decay-at-rest vs. decay-in-flight), the allowed regions of [$\Delta m^2$, $\sin^2 2 \theta$] in the oscillation interpretations of LSND and MiniBooNE are highly compatible~\cite{Aguilar-Arevalo:2020nvw}. The combined LSND and MiniBooNE excesses have a significance of $6.1\sigma$.

\clearpage

\begin{figure}
 \centering
\includegraphics[width=0.6\textwidth]{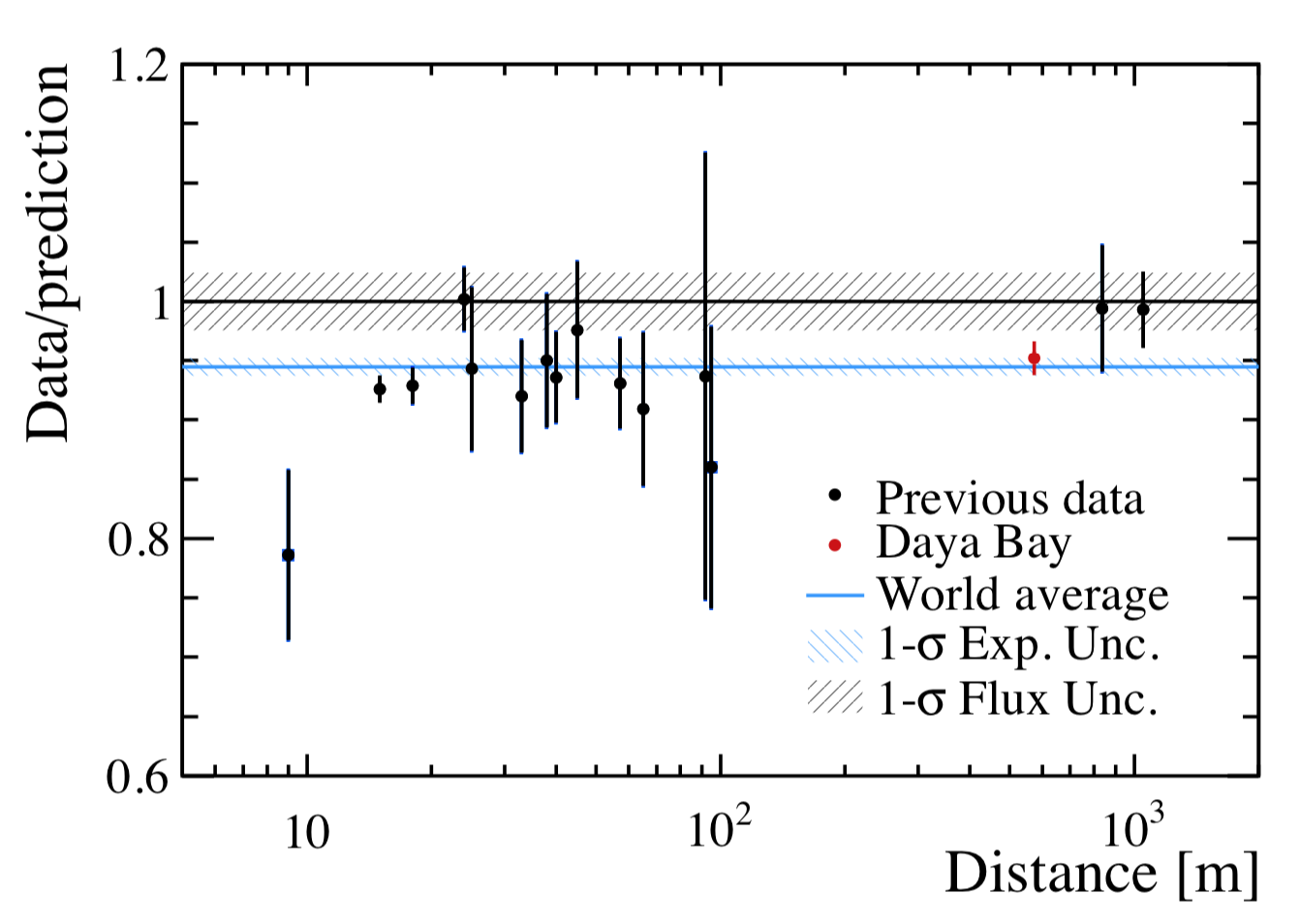}
 \caption[Reactor Antineutrino Anomaly]{Ratio of reactor antineutrino flux measurement to prediction as a function of distance from the reactor core for Daya Bay and prior experiments. Figure from~\cite{Adey:2018qct}.}
\label{fig:daya_bay}
\end{figure} 

While LSND and MiniBooNE both observed anomalies in muon (anti)neutrino beams, anomalous results have also been observed in electron (anti)neutrino beams. These include the reactor anomalies and the gallium anomaly. The reactor antineutrino anomaly is a deficit in the flux of electron antineutrinos emerging from nuclear reactors as compared to theoretical predictions~\cite{Mention_PhysRevD.83.073006, Mueller:2011nm, Huber:2011wv, Giunti:2019qlt}. This anomaly has been observed in a number of experiments at various baselines from the reactor cores. A recent precision measurement from the experiment Daya Bay, as well as measurements from prior experiments, is shown in Fig.~\ref{fig:daya_bay}~\cite{Adey:2018qct}. The measurements are normalized to theoretical predictions, accounting for the effects of oscillations. The world average measurement sits significantly below the prediction. Ratios of measurements at different baselines from reactors have found small signals, at the level of $1-2\sigma$~\cite{Atif:2020glb, Diaz:2019fwt}. Finally, several experiments have observed a ``5~MeV bump,'' an unexpected feature in the antineutrino spectra with a magnitude of about 10\%~\cite{Huber:2016xis}.

Lastly, the gallium anomaly is a deficit of electron neutrinos observed in the electron-capture decays of $^{37}$Ar and $^{51}$Cr by the GALLEX and SAGE experiments~\cite{Anselmann:1994ar, Hampel:1997fc, Kaether:2010ag, Abdurashitov:1996dp, Abdurashitov:1998ne, Abdurashitov:2005tb, Abdurashitov:2009tn}. These were gallium-based experiments designed to measure the solar neutrino flux and which used $^{37}$Ar and $^{51}$Cr for calibration. These are electron neutrino emitting radioactive sources. These experiments observed a $2.3 \sigma$ deficit in electron neutrinos from these radioactive sources~\cite{Kostensalo:2019vmv}.

\section{Traditional 3+1 model}
\begin{figure}
 \centering
\includegraphics[width=0.6\textwidth]{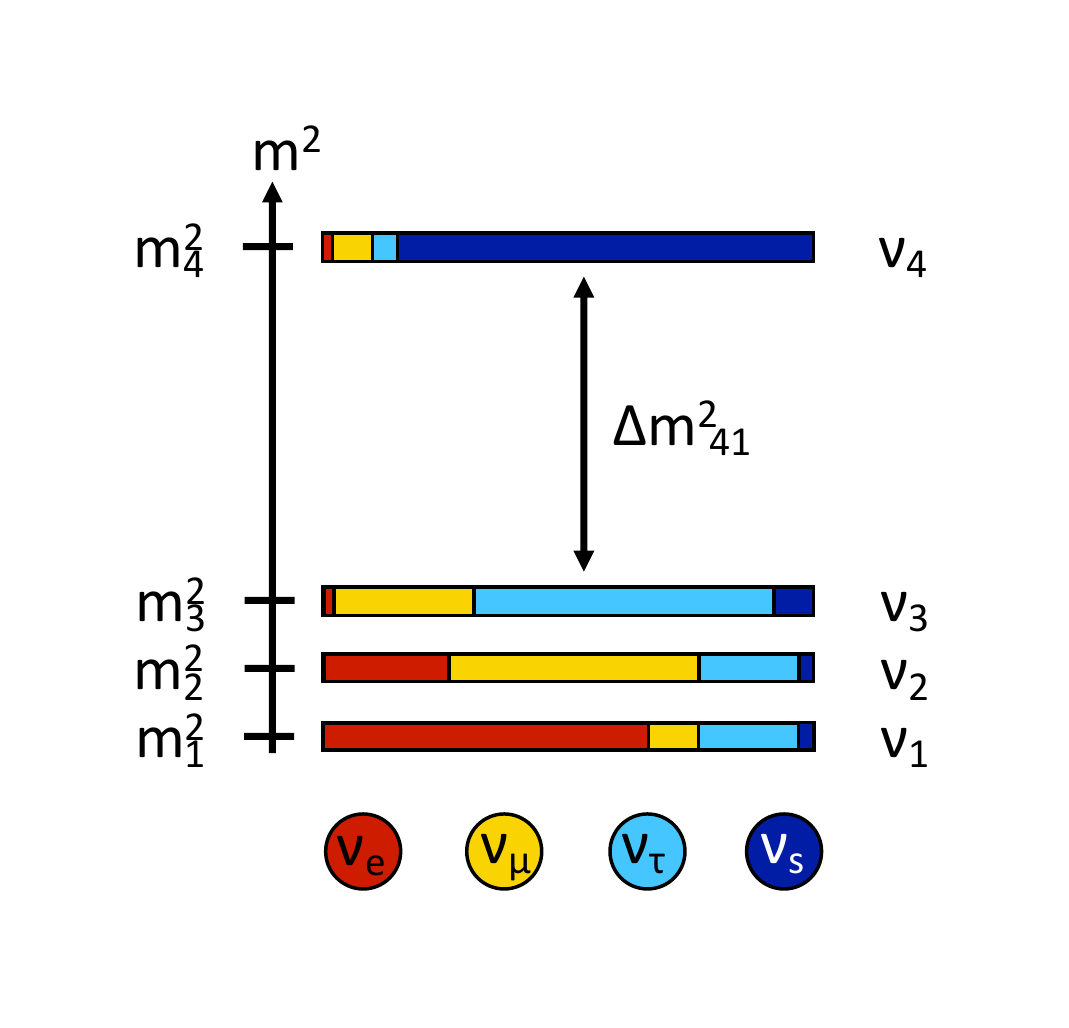}
 \caption[Schematic illustration of a 3+1 sterile neutrino model]{Schematic illustration of a 3+1 sterile neutrino model, assuming the normal neutrino ordering. The various components of the mass states, as well as the differences in the masses, are not to scale. This figure is for illustrative purposes only.}
\label{fig:31schematic}
\end{figure}

One model that has been put forth to account for the anomalies previously described is a 3+1 sterile neutrino model~\cite{Diaz:2019fwt, Kopp:2013vaa, Abazajian:2012ys}. A sterile neutrino, $\nu_s$, is a hypothetical flavor of neutrino which does not interact via the weak nuclear force. A 3+1 model is schematically depicted in Fig.~\ref{fig:31schematic}, for the case of the normal ordering $(m_3 > m_2)$. In a 3+1 model, a fourth, heavier mass state, $\nu_4$ is introduced. This mass state is mostly comprised of the sterile flavor, $\nu_s$. The three lighter mass states include a small sterile component. If the fourth mass state is heavy enough, the three lighter masses become degenerate.

In a 3+1 sterile neutrino model, the three-flavor mixing matrix, that is, the PMNS matrix, given in Eq.~\ref{eq:nu_mixing}, is expanded by one column, to account for the fourth mass state, and one row, to account for the sterile flavor. This expanded, four-flavor mixing matrix is shown in Eq.~\ref{eq:nu_mixing_31}:
\begin{equation}
\label{eq:nu_mixing_31}
\begin{pmatrix}
\nu_e \\
\nu_\mu \\
\nu_\tau \\
\nu_s
\end{pmatrix}
=
\begin{pmatrix}
U_{e1} & U_{e2} & U_{e3} & U_{e4}\\
U_{\mu 1} & U_{\mu 2} & U_{\mu 3} & U_{\mu 4} \\
U_{\tau1} & U_{\tau2} & U_{\tau3} & U_{\tau 4}\\
U_{s 1} & U_{s2} & U_{s3} & U_{s4} \\
\end{pmatrix}
\begin{pmatrix}
\nu_1 \\
\nu_2 \\
\nu_3 \\
\nu_4
\end{pmatrix}
\end{equation} 
The representation of the mixing matrix as the product of rotation matrices, given for three-flavor case in Eq.~\ref{eq:pmns_rotation}, requires equivalent modification. Following \cite{Barry:2011wb}, the four-flavor mixing matrix can be written as
\begin{equation}
\label{eq:mixing_matrix_rotation_matrices_4_flavors}
U = R_{34} \tilde{R}_{24} \tilde{R}_{14} R_{23} \tilde{R}_{13} R_{12} P,
\end{equation}
where $R_{ij}$ and $\tilde{R}_{ij}$ are, respectively, real and complex rotation matrices in $ij$ space. For example,
\begin{equation}
R_{34} \equiv \begin{pmatrix}
1 & 0 & 0 & 0 \\
0 & 1 & 0 & 0 \\
0 & 0 & \cos \theta_{34} & \sin \theta_{34} \\
0 & 0 & - \sin \theta_{34} & \cos \theta_{34} 
\end{pmatrix}
\textrm{~and~}
\tilde{R}_{14} \equiv \begin{pmatrix}
\cos \theta_{14} & 0 & 0 & \sin \theta_{14} e^{-i\delta_{14}} \\
0 & 1 & 0 & 0 \\
0 & 0 & 1 & 0 \\
- \sin \theta_{14} e^{i \delta_{14}}& 0 & 0 & \cos \theta_{14} 
\end{pmatrix}.
\end{equation}
The factor $P$ is the diagonal Matrix:
\begin{equation}
P = \textrm{diag}(1, e^{i \alpha /2}, e^{i(\beta/2 + \delta_{13})}, e^{(\gamma/2+\delta_{14})}),
\end{equation}
where $\alpha$, $\beta$ and $\gamma$ are the Majorana phases. These phases are only nonzero if neutrinos are Majorana particles, i.e. they are their own antiparticles. 

In a 3+1 model, neutrinos oscillate between all four flavors. This could account for electron neutrinos appearing in a muon neutrino beam, as seen by LSND and MiniBooNE, as well as a disappearance of electron neutrinos from an electron neutrino beam, as seen in the reactor and gallium experiments. The oscillation amplitudes for oscillation channels relevant to this thesis, and assuming a two-flavor approximation, are given in Table~\ref{tab:osc_amplitudes}. These amplitudes are parameterized in three ways: in terms of the mixing matrix elements, $U_{\alpha i}$, which are the physical parameters; in terms of effective mixing angles, $\theta_{\alpha \beta}$; and in terms of rotation angles, $\theta_{ij}$. In this thesis, plots and discussion related to global fits will use oscillation amplitudes given in terms the effective angles. Plots and discussion related only to the experiment IceCube will use oscillation amplitudes given in terms of rotation angles.

\begin{table}[h]
\label{tab:osc_amplitudes}
\begin{tabular}{c|c|c|c}
Channel &
\begin{tabular}{c} Mixing \\ matrix \\ elements \end{tabular} & 
\begin{tabular}{c} Effective \\ mixing \\ angles \end{tabular} & 
\begin{tabular}{c} Rotation \\ angles \end{tabular} \\

\hline
$\nu_e \rightarrow \nu_e$ &
$4(1 - |U_{e4}|^2)|U_{e4}|^2 $& 
$\sin^2 2\theta_{ee}$&
$\sin^2 2\theta_{14}$ \\
$\nu_\mu \rightarrow \nu_\mu$ &
$4(1 - |U_{\mu4}|^2)|U_{\mu4}|^2 $& 
$\sin^2 2\theta_{\mu\mu}$&
$4\cos^2 \theta_{14}\sin^2 \theta_{24}(1 - \cos^2 \theta_{14}\sin^2 \theta_{24})$ \\
$\nu_\mu \rightarrow \nu_e$ &
$4|U_{e4}|^2|U_{\mu 4}|^2 $& 
$\sin^2 2\theta_{\mu e}$&
$\sin^2 2\theta_{14}\sin^2 \theta_{24}$ \\
\end{tabular}
\caption[Oscillation amplitudes for 3+1 model]{Oscillation amplitudes for three neutrino oscillation channels given in terms of the mixing matrix elements, the effective mixing angles, and rotation angles. Note that $\sin^2 2\theta_{24} = \sin^2 2\theta_{\mu\mu}$ for $\theta_{14} = 0$.}
\label{tab:osc_amplitudes}
\end{table}

\section{Tension in the 3+1 fits}

The oscillation amplitudes for the three channels shown in Table~\ref{tab:osc_amplitudes}, which are $\nu_e \rightarrow \nu_e$, $\nu_\mu \rightarrow \nu_\mu$  and $\nu_\mu \rightarrow \nu_e$, depend on only two mixing matrix elements: $U_{e 4}$ and $U_{\mu4}$. This means that these two parameters are over-constrained by experimental data spanning the three distinct channels. Global fits spanning these channels have been performed by several groups, and they find similar allowed regions, with $\Delta m_{41}^2$ around 1 eV$^2$~\cite{Diaz:2019fwt, Dentler:2018sju, Gariazzo:2017fdh}. The most recent global fit finds a $5\sigma$ improvement over the null hypothesis, i.e. only three neutrinos~\cite{Diaz:2019fwt}. Nevertheless, all three fitting groups find a similar problem in the fits: inconsistency between subgroups of the data.

One natural way to split up the data is into ``appearance'' and ``disappearance'' experiments. Appearance experiments are those in the $\nu_\mu \rightarrow \nu_e$ channel, while disappearance experiments are those in the $\nu_e \rightarrow \nu_e$ and $\nu_\mu \rightarrow \nu_\mu$ channels. One method of characterizing consistency between subgroups of data is the \textit{parameter goodness-of-fit} (PG)\cite{Maltoni:2002xd,Maltoni:2003cu}. 
One performs three separate fits: one on the appearance experiments, one on the disappearance experiments,  and one on all the datasets. An effective $\chi^2$ is defined~\cite{Diaz:2019fwt}:
\begin{equation}
\chi^2_{\textrm{PG}} = \chi^2_{\textrm{global}} - (\chi^2_{\textrm{appearance}} + \chi^2_{\textrm{disappearance}}).
\end{equation}
The effective number of degrees of freedom is:
\begin{equation}
N_{\textrm{PG}} = (N_{\textrm{appearance}} + N_{\textrm{disappearance}}) - N_{\textrm{global}}.
\end{equation}
The $\chi^2_{\textrm{PG}}$ is then interpreted as a $\chi^2$ with $N_{\textrm{PG}}$ degrees of freedom and a probability ($p$-value) is calculated. A recent fit to short-baseline experiments found the $p$-value for the PG test to be 4E-6, which has a significance of 4.5$\sigma$~\cite{Diaz:2019fwt}. This means that the appearance datasets and the disappearance datasets are highly inconsistent. This is referred to as tension in the fits. This motivates the consideration of models beyond a traditional 3+1 sterile neutrino model that could relieve this tension.

\chapter{Status of 3+1 sterile neutrinos in IceCube}

\section{Neutrino oscillations in matter}

The experiment described in this thesis, IceCube, is uniquely capable of searching for a signature of sterile neutrinos that involves matter effects in oscillations. Neutrinos traversing matter experience different oscillation probabilities than those traversing vacuum~\cite{wolfenstein1978neutrino, mikheyev1986resonant, liu1998parametric, akhmedov2000parametric}. This is because the different neutrino flavors undergo different interactions with matter, therefore experiencing different potentials, modifying the Hamiltonian. In fact, oscillations in matter may have a larger amplitude than oscillations in vacuum.

Following~\cite{Akhmedov:1999uz}, the Schr\"{o}dinger equation for a two-flavor approximation in vacuum can be written as
\begin{equation}
i \frac{d}{dt} \begin{pmatrix}\nu_\mu \\ \nu_s \end{pmatrix}
=
\begin{pmatrix}
-\frac{\Delta m^2}{4E} \cos 2 \theta_0 & \frac{\Delta m^2}{4E} \sin 2 \theta_0 \\ 
\frac{\Delta m^2}{4E} \sin 2 \theta_0 & \frac{\Delta m^2}{4E} \cos 2 \theta_0 
\end{pmatrix}
=
\begin{pmatrix}\nu_\mu \\ \nu_s \end{pmatrix},
\end{equation}
where $\sin^2 2\theta_0$ is the oscillation amplitude in vacuum. In the case of oscillations in matter, the $\nu_\mu$ state experiences a potential due to neutral current (NC) interactions off quarks, while the $\nu_s$ state does not. This potential is
\begin{equation}
V_{\textrm{NC}} = \mp \frac{\sqrt{2}}{2} G_F N_n,
\end{equation}
where $-(+)$ corresponds to $\nu_\mu$($\bar{\nu}_\mu$), $G_F$ is the Fermi constant and $N_n$ is the neutron number density. Adding this potential to the effective Hamiltonian,
\begin{equation}
i \frac{d}{dt} \begin{pmatrix}\nu_\mu \\ \nu_s \end{pmatrix}
=
\begin{pmatrix}
-\frac{\Delta m^2}{4E} \cos 2 \theta_0 \mp \frac{\sqrt{2}}{2} G_F N_n & \frac{\Delta m^2}{4E} \sin 2 \theta_0 \\ 
\frac{\Delta m^2}{4E} \sin 2 \theta_0 & \frac{\Delta m^2}{4E} \cos 2 \theta_0 
\end{pmatrix}
=
\begin{pmatrix}\nu_\mu \\ \nu_s \end{pmatrix},
\end{equation}
and assuming constant matter density, one can diagonalize the effective Hamiltonian to find the oscillation amplitude in matter:
\begin{equation}
\label{eq:msw_amp}
\sin^2 2 \theta_{\rm{matter}} = \frac{\big(\frac{\Delta m^2}{2E}\big)^2 \sin^2 2 \theta_0}{ \big(\frac{\Delta m^2}{2E} \cos 2 \theta_0 \pm \frac{\sqrt{2}}{2} G_F N_n \big)^2   + 
\big(\frac{\Delta m^2}{2E}\big)^2  \sin^2 2 \theta_0}.
\end{equation}
The eigenstates in matter are a linear combination of the flavor eigenstates. A critical energy exists where regardless of the magnitude of the oscillation amplitude in vacuum, the oscillation amplitude in matter becomes unity either for neutrinos or antineutrinos. This critical energy is
\begin{equation}
E_{\rm{critical}} = \mp \frac{\Delta m^2 \cos 2\theta_0}{\sqrt{2} G_F N_n}.
\end{equation}
If $\Delta m^2$ is positive and the mixing angle is relatively small, that is, $|\theta_0|< \frac{\pi}{4}$, the resonance occurs for antineutrinos rather than neutrinos. For the following values, 
\begin{equation}
\begin{split}
\begin{aligned}
\Delta m_{41}^2 &\approx 1~\textrm{eV}^2 \\
\cos 2\theta_0 &\approx 1 \\
\rho_{\textrm{Earth}} &\approx 6~\textrm{g/cm}^3,
\end{aligned}
\end{split}
\end{equation}
where $\rho_{\textrm{Earth}}$ is the average density of the Earth, and with the approximation that half of the Earth's mass is from neutrons, one finds the critical energy to be about 3~TeV. This suggests that for sterile parameter values consistent with global fit findings, a resonant transition would occur for antineutrinos traversing the Earth at TeV energies~\cite{Nunokawa:2003ep, Choubey:2007ji, Razzaque:2011ab}.

\begin{figure}[ht]
  \centering
  \includegraphics[width=\textwidth]{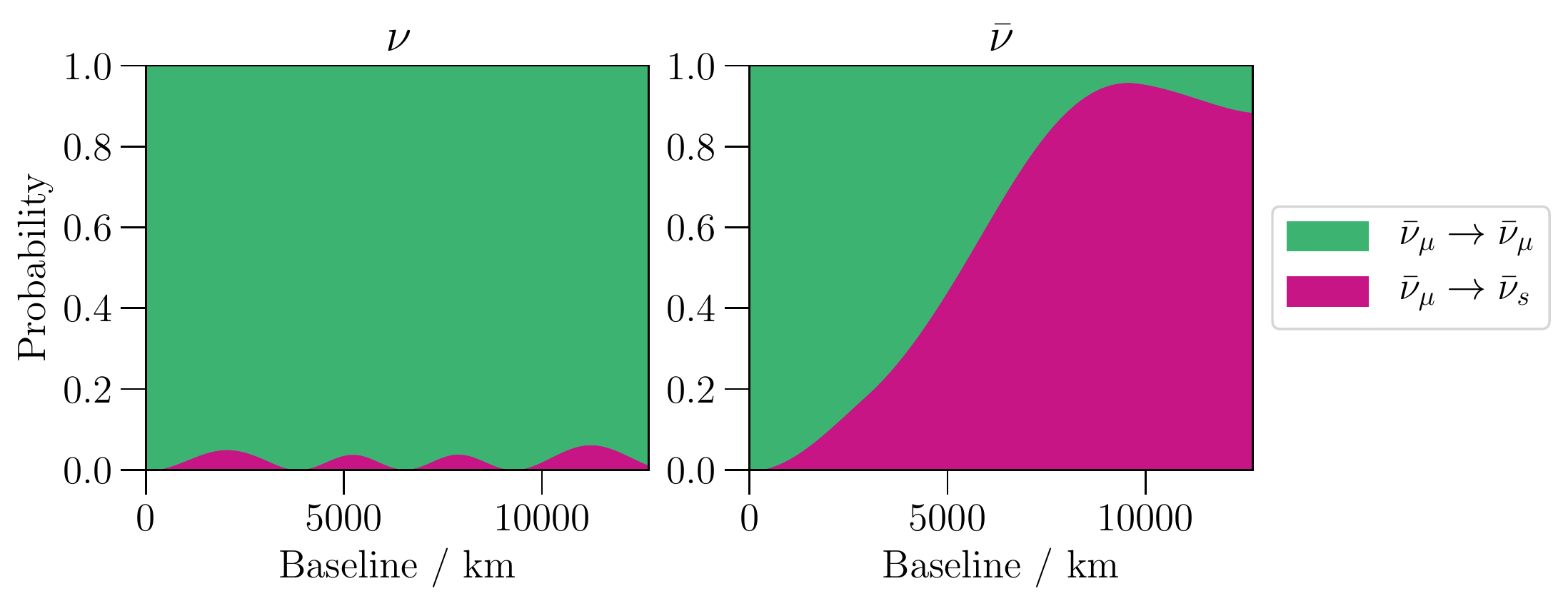}
  \caption[Traditional 3+1 oscillation probabilities versus baseline along Earth's diameter]{Oscillation probabilities as a function of baseline for a muon neutrino (left) and a muon antineutrino (right) traversing the diameter of the Earth for $\Delta m_{41}^2 = 1~\textrm{eV}^2$, $\sin^2 2 \theta_{24} = 0.1$, and 2.3~TeV of energy. Probabilities are calculated with \texttt{nuSQuIDS}~\cite{nusquids}.}
  \label{fig:matter_oscillation_prob_vs_baseline}
\end{figure}

\begin{figure}[hb]
  \centering
  \includegraphics[width=0.75\textwidth]{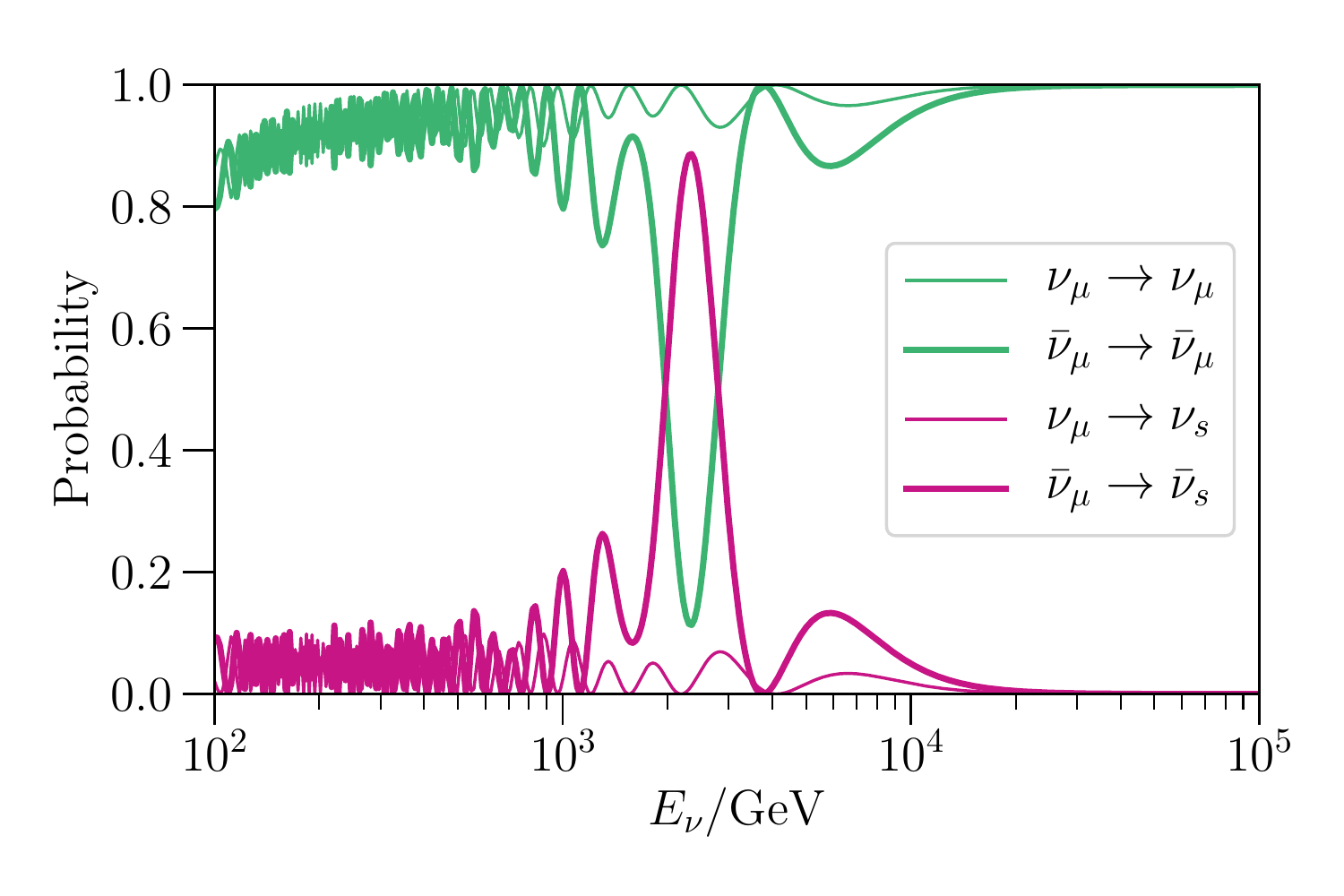}
  \caption[Traditional 3+1 oscillation probabilities versus energy for neutrinos traversing the Earth]{Oscillation probabilities as a function of energy for muon neutrinos (thin lines) and muon antineutrinos (thick lines) after having traversed the diameter of the Earth, for $\Delta m_{41}^2 = 1~\textrm{eV}^2$ and $\sin^2 2 \theta_{24} = 0.1$. The probability of remaining in the muon flavor is shown in green. The probability of oscillating into the sterile flavor is shown as magenta.}
  \label{fig:matter_oscillation_prob_vs_energy}
\end{figure}

This derivation made two simplifying assumptions: constant density and two-flavor oscillations. Reality is more complicated. Oscillation probabilities for neutrinos traversing the diameter of the Earth, accounting for both the radially varying Earth density and a full four-neutrino oscillation framework, are calculated with \texttt{nuSQuIDS} and shown in the Figs.~\ref{fig:matter_oscillation_prob_vs_baseline} and \ref{fig:matter_oscillation_prob_vs_energy}. In these calculations, the sterile parameters are
\begin{equation}
\begin{split}
\begin{aligned}
\Delta m^2 &= 1~\textrm{eV}^2 \\
\sin^2 2 \theta_{24} &= 0.1, \\
\end{aligned}
\end{split}
\end{equation}
and all other sterile mixing angles and CP-violating angles are zero. The neutrinos are born in the muon flavor. In Fig.~\ref{fig:matter_oscillation_prob_vs_baseline}, the oscillation probabilities for neutrinos and antineutrinos are shown as a function of baseline across the diameter of the Earth, for the energy 2.3~TeV. At this energy, the probability of oscillating from the muon flavor into either the electron or tau flavors is minuscule. The left panel corresponds to neutrinos; only small oscillations from the muon to sterile flavor are observed, shown in magenta. The right panel corresponds to antineutrinos; a resonant oscillation into the sterile flavor is observed. The magnitude of this conversion is much greater than 0.1 which is the value of $\sin^2 \theta_{\mu \mu}$, and would be the amplitude of the oscillation in vacuum.

Figure~\ref{fig:matter_oscillation_prob_vs_energy} shows the same phenomenon, over the energy range between 100~GeV and 100~TeV. This figure only shows the final oscillation probabilities across the diameter of the Earth. The thick magenta line shows the probability of a muon antineutrino oscillating into the sterile flavor, and the peak of the resonance is found at 2.3~TeV, corresponding to the scenario shown in Fig.~\ref{fig:matter_oscillation_prob_vs_baseline}. The thick green line shows the probability of such a muon antineutrino remaining in the muon flavor. The thin lines show the corresponding probabilities for neutrinos, rather than antineutrinos. At the lowest energies, oscillation into the electron and tau flavors is non-negligible, which accounts for the shown probabilities not adding up to unity.

\section{Brief description of IceCube}
\begin{figure}[h]
  \centering
  \includegraphics[width=0.65\textwidth]{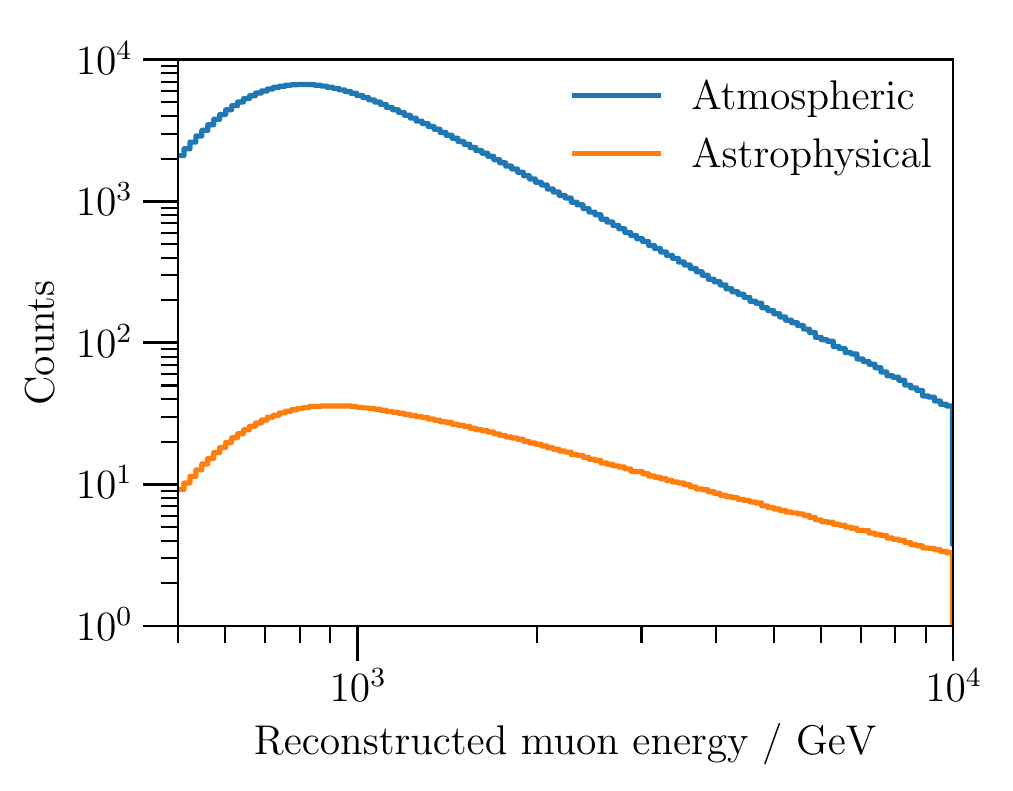}
  \caption[Predicted atmospheric and astrophysical event distribution]{Predicted event distribution of muons associated with atmospheric and astrophysical neutrinos. This prediction corresponds to a livetime of 7.6 years. These events originate below the horizon.}
  \label{fig:event_dist_atmo_astro}
\end{figure}

IceCube is a neutrino detector that can observe TeV neutrinos originating from across the Earth, allowing for a unique search for sterile neutrinos that makes use of the matter effect parametric resonance discussed previously. A brief description of IceCube is given here to facilitate understanding of the rest of this chapter and the following one. A longer description is given in Chapter~\ref{chap:icecube}. IceCube is a gigaton ice-Cherenkov neutrino detector located deep in the ice at the South Pole. Over 5000 optical sensors are deployed in the ice to observe light produced in neutrino interactions. IceCube observes atmospheric and astrophysical neutrinos coming from all directions, although the work in this thesis uses upward-going neutrinos, that is, those originating from below the horizon. IceCube detects neutrinos with energies above 100~GeV, and as high as several PeV. The atmospheric and astrophysical neutrino fluxes fall steeply with energy, while the efficiency of the detector increases with energy. This results in a peak of events near one TeV~\cite{Aartsen:2015rwa}, which is shown in Fig.~\ref{fig:event_dist_atmo_astro}.

\section{Eight-year search results}
\begin{figure}[h]
  \centering
  \includegraphics[width=0.65\textwidth]{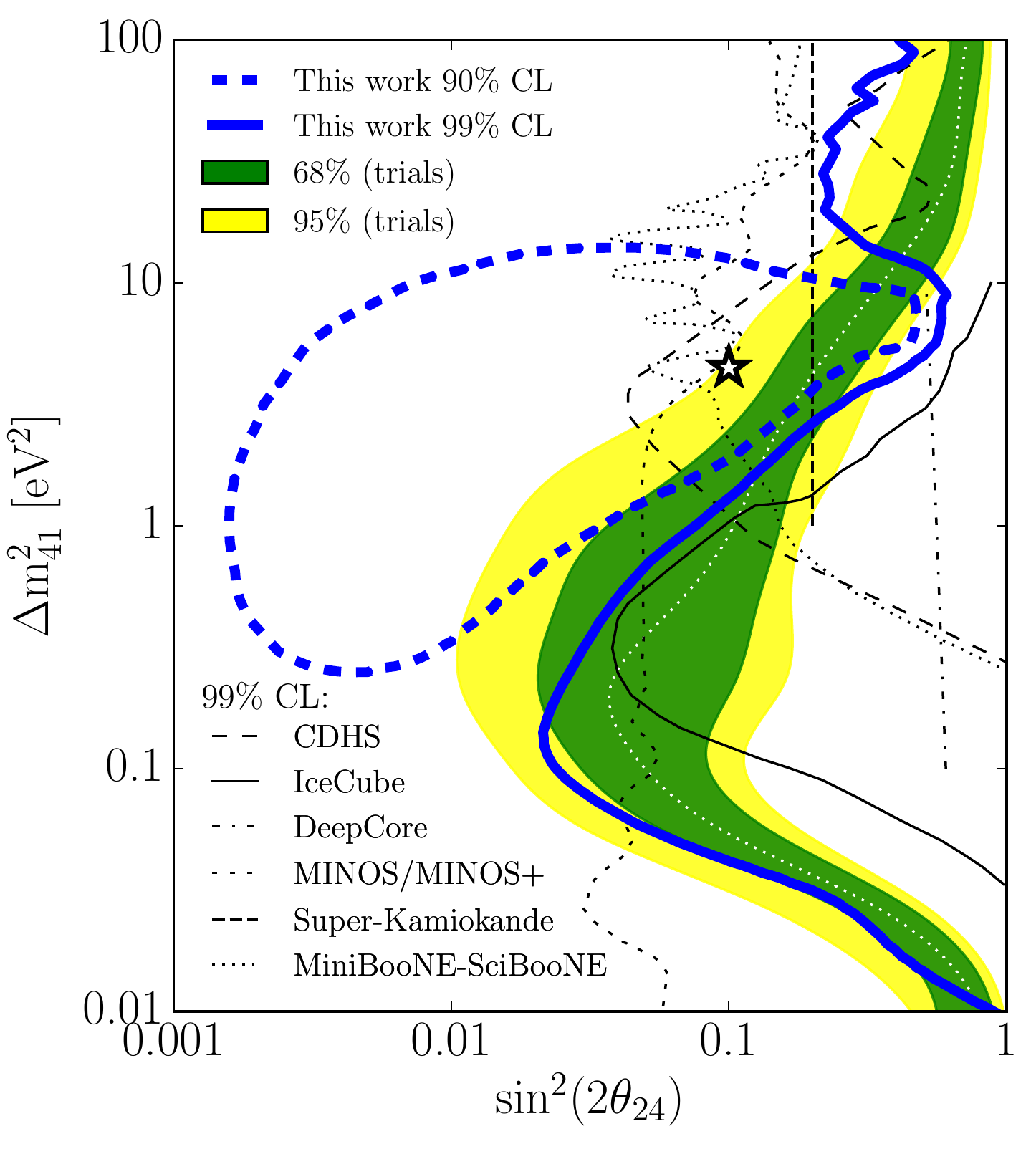}
  \caption[Frequentist, traditional 3+1, eight-year result]{Frequentist result from an eight-year search for sterile neutrinos in IceCube. The best-fit point is marked with a star. The confidence level contours are found assuming Wilks' theorem, and are shown at 90\% and 99\% in dashed and solid blue curves, respectively. The median sensitivity at 99\% confidence level is found from pseudoexperiments, and is shown as a thin dashed white curve. The $1\sigma$ and $2\sigma$ ranges of the sensitivity are shown in green and yellow bands, respectively. Results from other experiments are shown as black curves, and they reject the region to their right. Figure from~\cite{Aartsen:2020iky}.}
  \label{fig:meows_freq}
\end{figure}

\begin{figure}[h]
  \centering
  \includegraphics[width=0.75\textwidth]{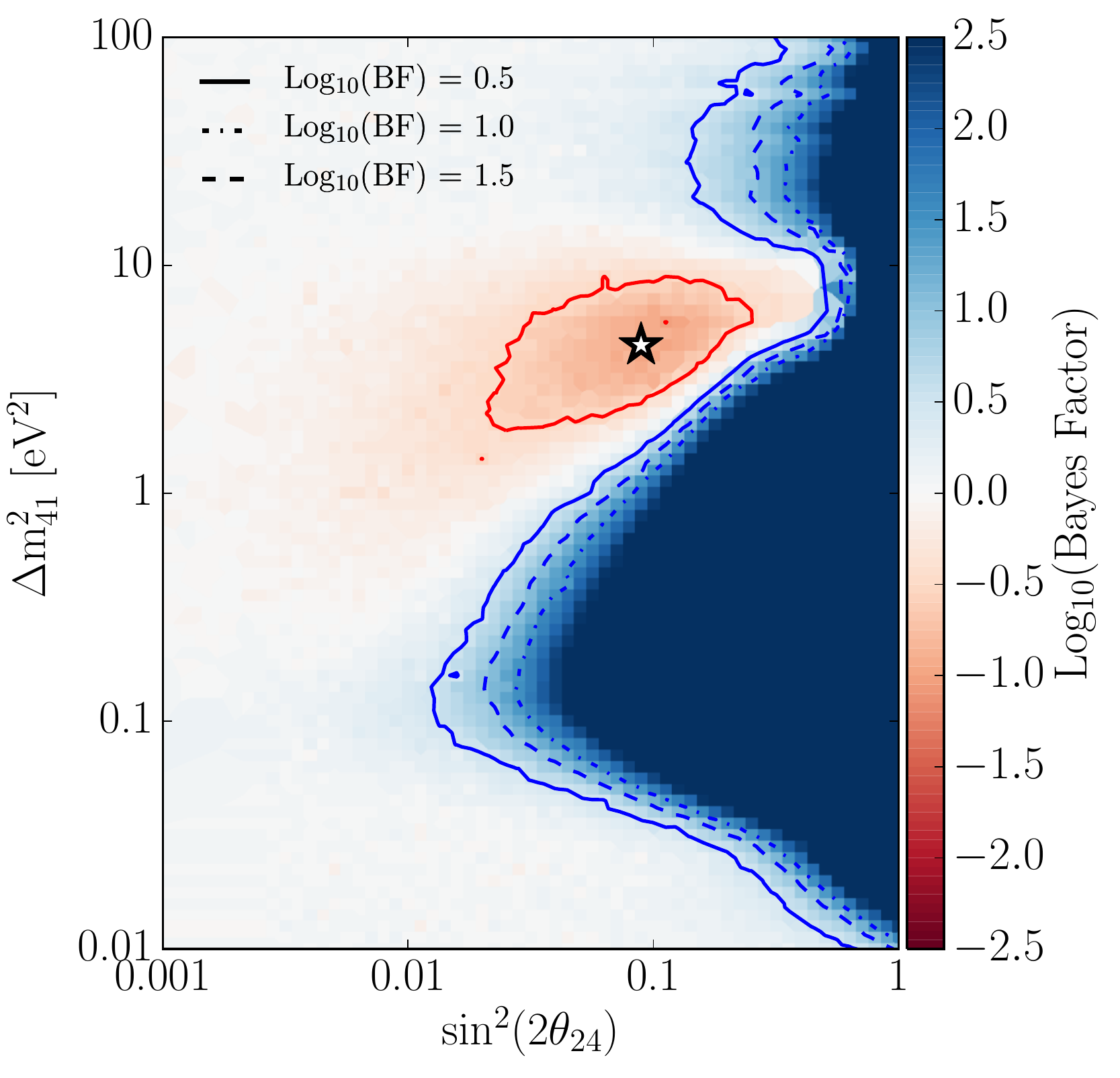}
  \caption[Bayesian, traditional 3+1, eight-year result]{Bayesian result from an eight-year search for sterile neutrinos in IceCube. The logarithm of the Bayes factor~\cite{Jeffreys:1939xee} for a particular sterile model relative to the null hypothesis is shown. The best-model point is marked with a star. Figure from~\cite{Aartsen:2020iky}.}
  \label{fig:meows_bayes}
\end{figure}

A one-year search for an eV-scale sterile neutrino at TeV-energies was performed in 2015 in IceCube and found no evidence for them~\cite{Jones:2015bya, delgado2015new, TheIceCube:2016oqi}. A subsequent eight-year search, with a fifteen-fold increase in statistics, was performed~\cite{Axani:2020zdv, Aartsen:2020iky, Aartsen:2020fwb}. This search made use of the atmospheric neutrino flux systematic treatment described in Chapter~\ref{chap:flux}. Both a frequentist analysis and a Bayesian analysis were performed. The result of the frequentist analysis is shown in Fig.~\ref{fig:meows_freq} and the result of the Bayesian analysis is shown in Fig.~\ref{fig:meows_bayes}. The frequentist analysis found a best-fit point at $\Delta m^2_{41} = 4.5~\textrm{eV}^2$ and $\sin^2 2\theta_{24} = 0.10$, and found that the null hypothesis was rejected with a $p$-value of 8\%. The Bayesian analysis found a best-model location with approximately the same sterile parameter values, and found that model to be strongly preferred, by a factor of about 10, to the null hypothesis, that is, no sterile neutrino.

\section{Jumping off point for the work in this thesis}

The result of the eight-year search for sterile neutrinos in IceCube is intriguing. The best-fit point and the point corresponding to the best-model are in the region of interest from global fits. This result is likely to modify the global fit results. Nevertheless, the result is not very statistically significant and tension is likely to remain in the global fits. An alternative model that could resolve the tension is motivated. The rest of this thesis describes such a sterile neutrino model and a search for it in the IceCube experiment.
\chapter{Phenomenology of unstable sterile neutrinos in IceCube}

\section{Unstable sterile neutrino model in IceCube}
\label{sec:prd_decay}

In the Standard Model of Particle Physics, stable particles are those that are protected by a fundamental symmetry. Others decay. There is no fundamental symmetry that protects neutrinos. In the massive neutrino Standard Model, the heavier two of the three mass states can decay radiatively. Two processes for a heavier mass state, $\nu_i$, to decay to a lighter mass state, $\nu_j$, are
\begin{equation}
\begin{split}
\begin{aligned}
\nu_i \rightarrow \nu_j + \gamma \hspace{2.5cm}&\textrm{~with lifetime~} \tau \simeq 10^{36} (m_i/\textrm{eV})^{-5} ~\textrm{years} \\
\nu_i \rightarrow \nu_j + \gamma +\gamma \hspace{2.5cm}&\textrm{~with lifetime~} \tau \simeq 10^{67} (m_i/\textrm{eV})^{-9} ~\textrm{years},
\end{aligned}
\end{split}
\end{equation}
where $\gamma$ is a photon~\cite{Diaz:2019fwt, PhysRevD.25.766, PhysRevD.28.1664}. These decays are slow; for the neutrinos of the massive neutrino Standard Model, the lifetimes are many times longer the age of the Universe.

If there is a fourth mass state, it may decay. The decay of a heavy, fourth mass state, in the range 1~keV - 1~MeV was previously considered  to explain the LSND anomaly~\cite{PalomaresRuiz:2005vf, PalomaresRuiz:2006ue}. In the following publication~\cite{Moss:2017pur}, a model involving oscillations and decay for eV-scale neutrinos was developed and applied to the IceCube detector. This used the open dataset from the one-year high-energy sterile neutrino search~\cite{TheIceCube:2016oqi}.

\label{sec:decay_prd}
\addcontentsline{toc}{subsection}{\textit{Publication: Exploring a nonminimal sterile neutrino model involving decay at IceCube}}
\includepdf[pages={-}]{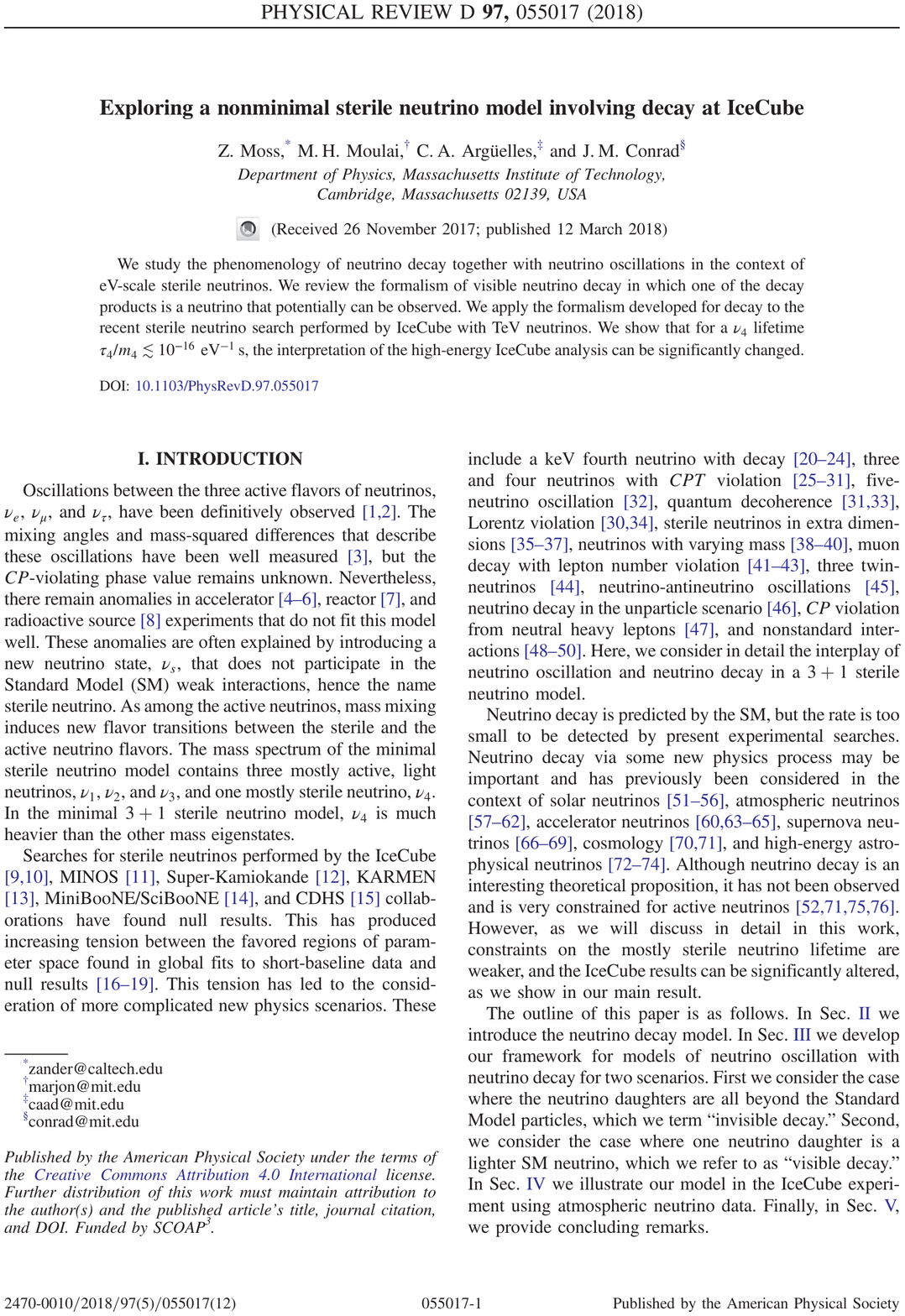}

\section{Incorporating IceCube data into global fits}

Following~\cite{Moss:2017pur}, global fits were performed on short-baseline data using both a traditional $3+1$ model and the invisible decay version of the unstable sterile neutrino model, referred to as $3+1+\textrm{decay}$~\cite{Diaz:2019fwt}. These fits used data from fourteen different neutrino experiments, excluding IceCube. Invisible decay refers to having no detectable daughter particles. It was found that the best-fit point of the unstable sterile neutrino model is preferred to that of the traditional $3+1$ model. The significance of the improvement is $2.6\sigma$. Moreover, the observed tension between appearance and disappearance experiments is reduced, but not resolved. In the following publication, the one-year dataset from IceCube was incorporated into the global fit results~\cite{Moulai:2019gpi}. 

\label{sec:fits_prd}
\addcontentsline{toc}{subsection}{\textit{Publication: Combining sterile neutrino fits to short-baseline data with IceCube data}}
\includepdf[pages={-}]{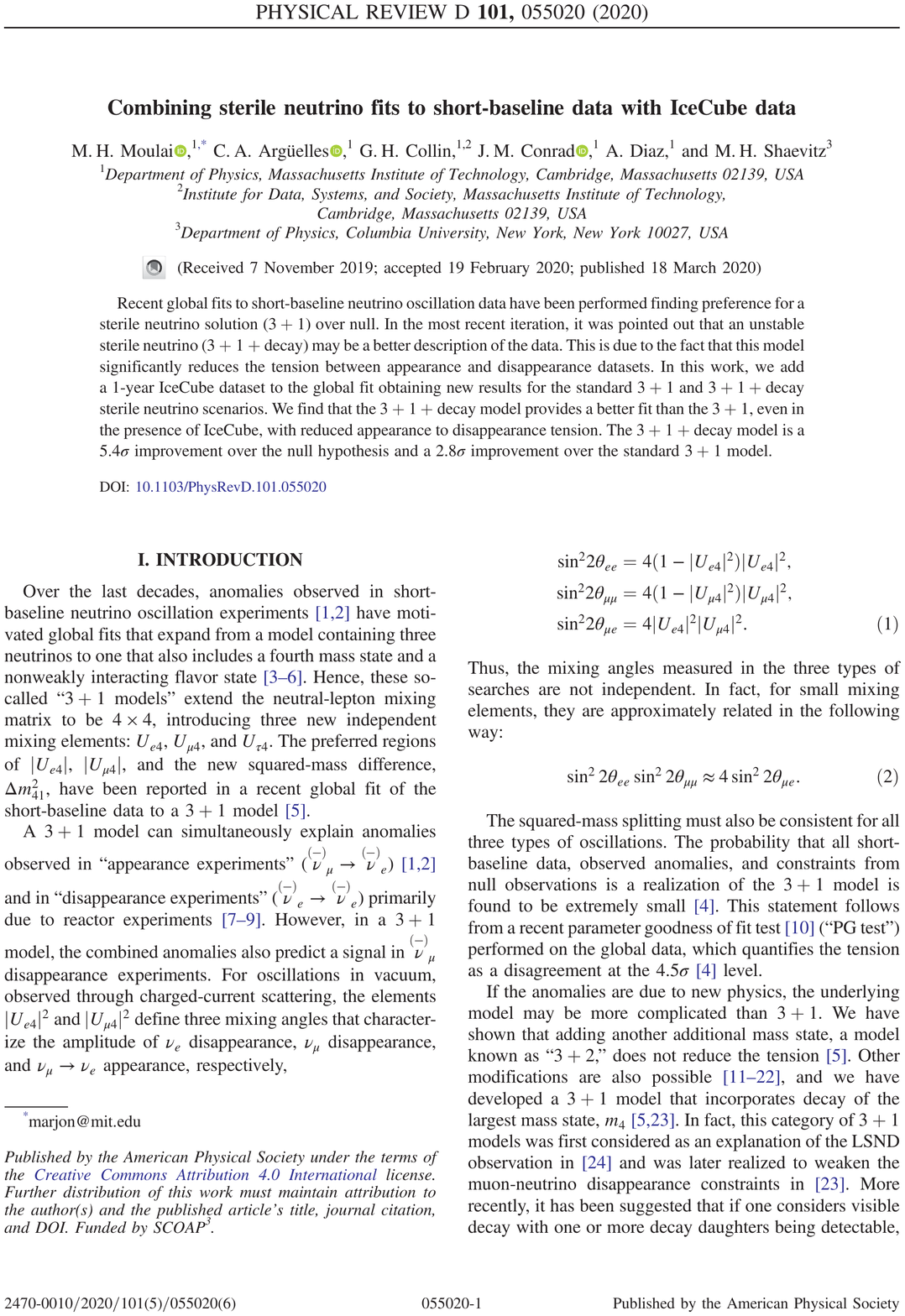}
\includepdf[pages={-}]{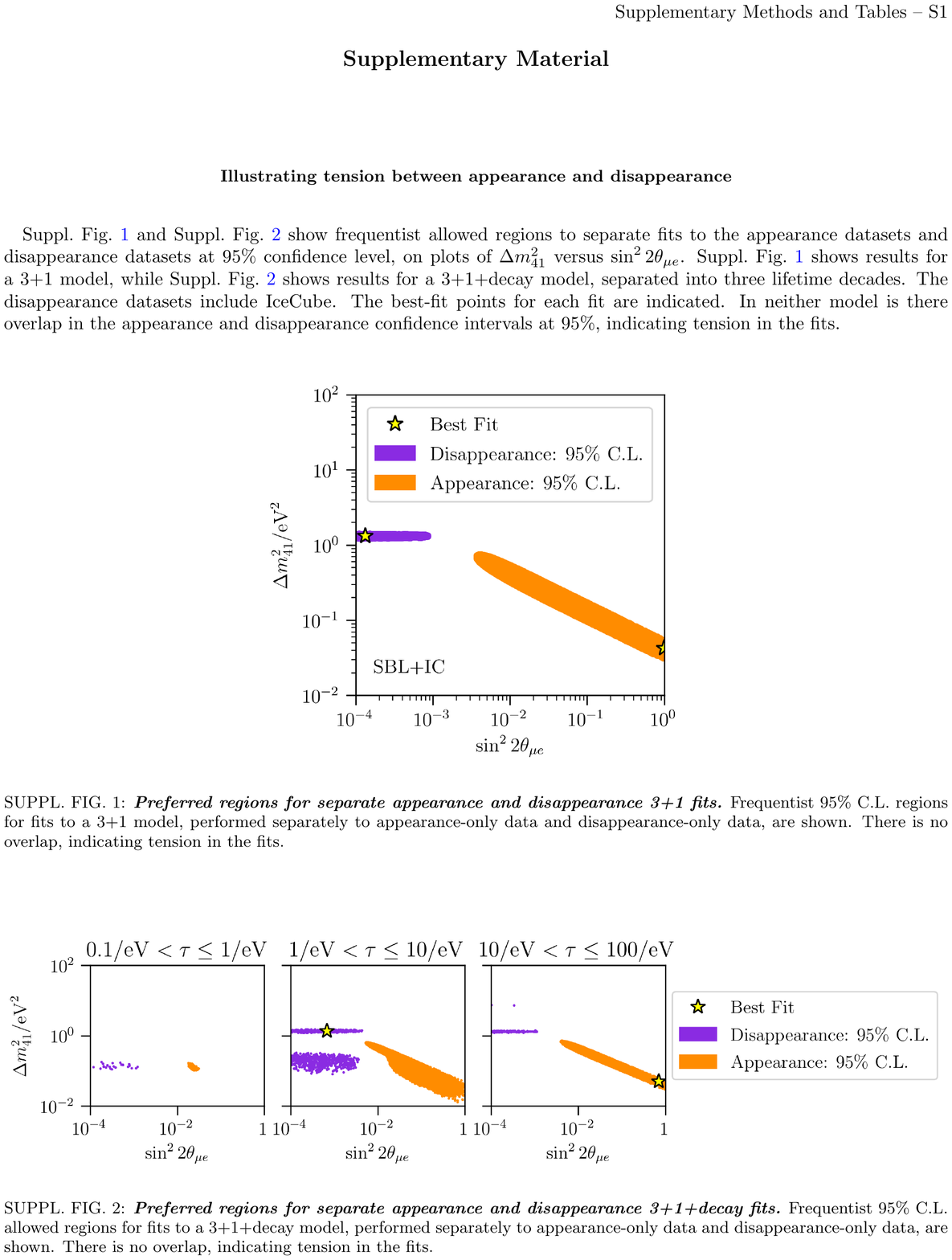}

\section{More on the phenomenology of unstable sterile neutrinos in IceCube}
\begin{figure}[h!]
  \centering
  \includegraphics[width=\textwidth]{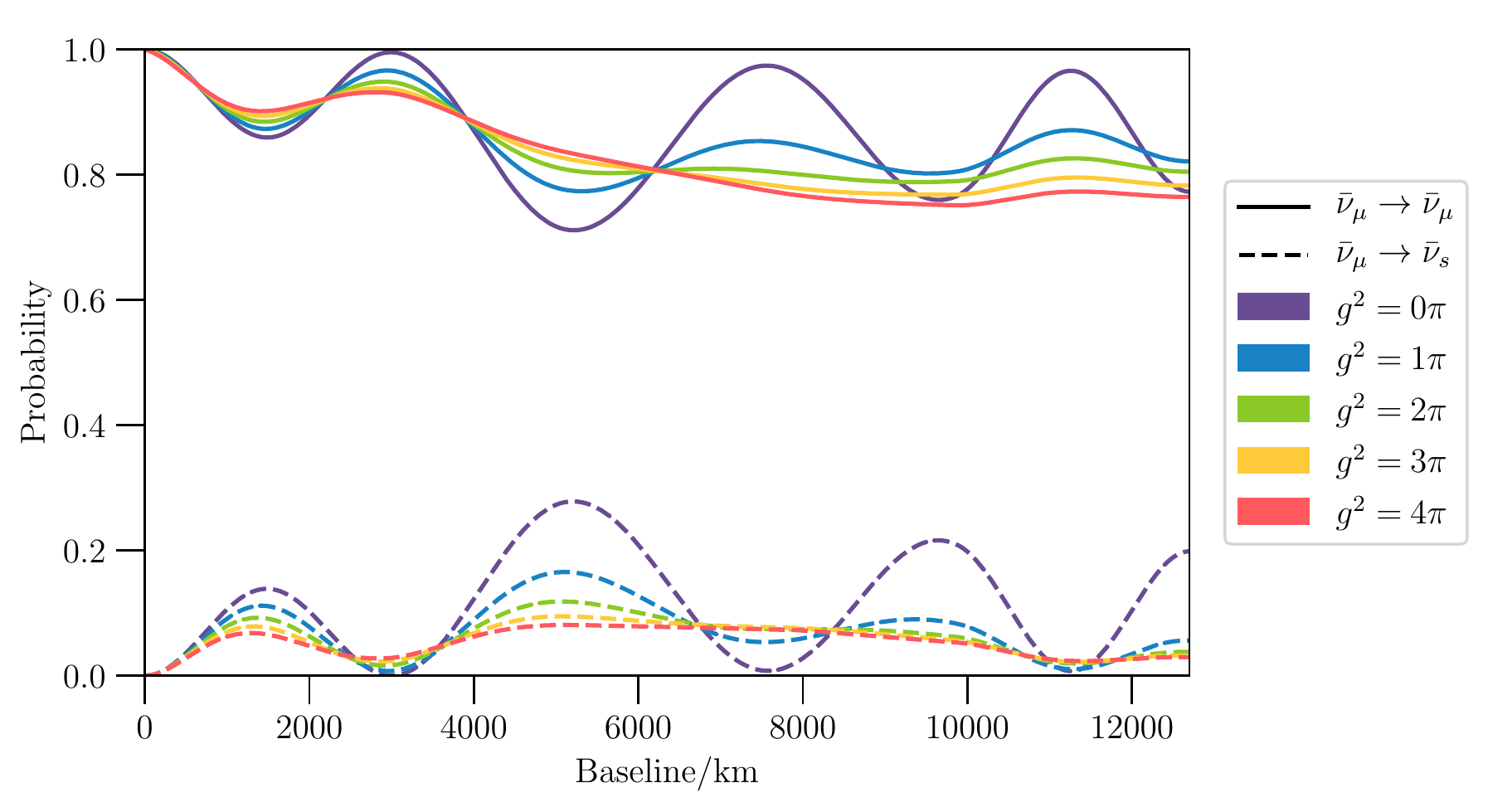}
  
  \includegraphics[width=\textwidth]{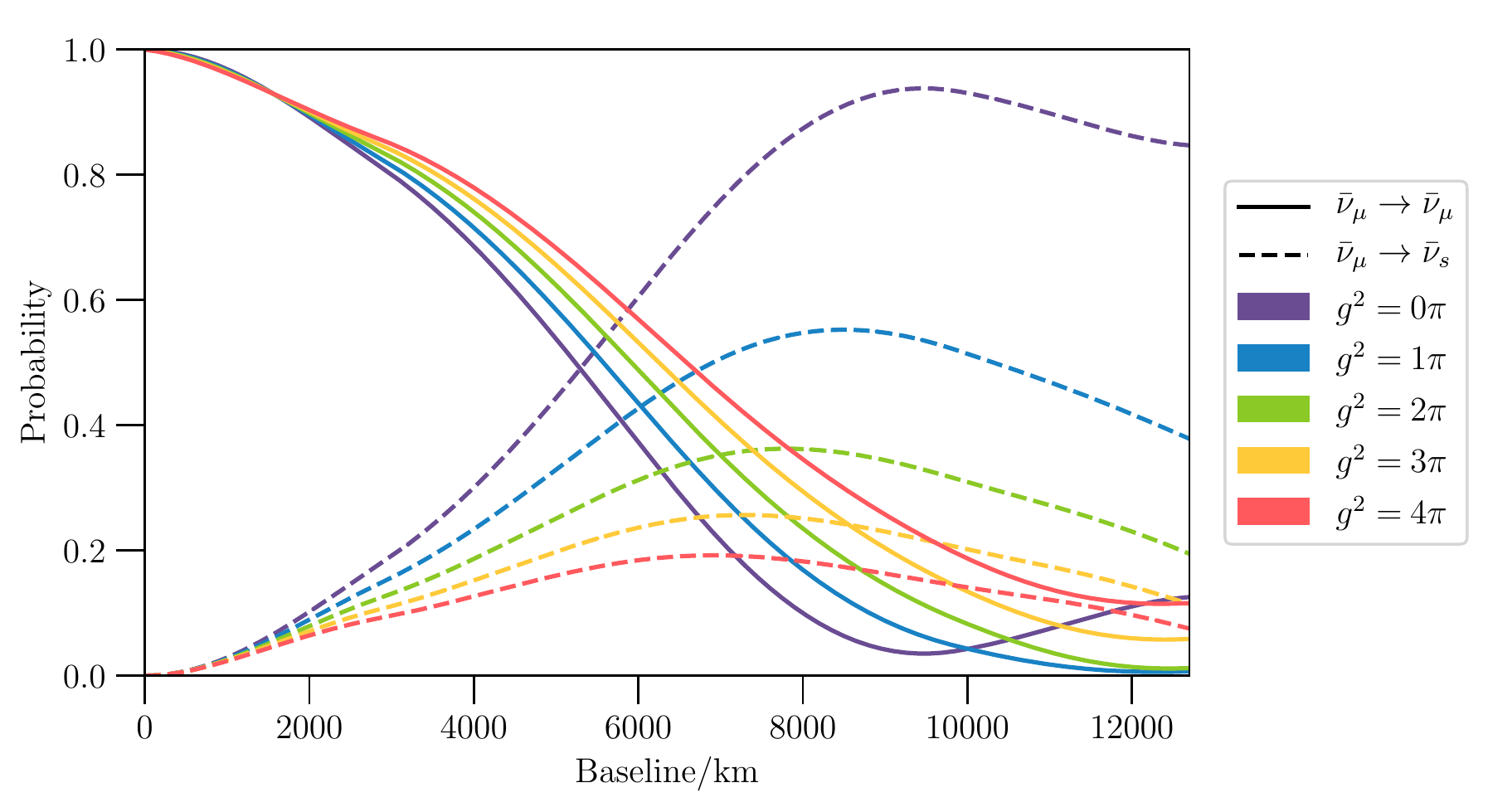}
\caption[Transition probabilities versus baseline along Earth's diameter for 3+1+decay]{Transition probabilities of a 1~TeV (top) and 2.3~TeV (bottom) muon antineutrino traversing the diameter of the Earth for an unstable sterile neutrino model with invisible decay and parameters $\Delta m^2_{41} = 1.0 \: \textrm{eV}^2$ and $\sin^2 2\theta_{24} = 0.1$. The four different colors correspond to four values of the decay-mediating coupling, $g^2$, where $g^2 = 0$ corresponds to a traditional 3+1 model, i.e. no neutrino decay. The solid curves correspond to muon antineutrino survival probabilities. The dashed curves correspond to the probability of antineutrino oscillating into the sterile flavor.}
 \label{fig:osc_decay_baseline}
\end{figure}

\begin{figure}[h!]
  \centering
  \includegraphics[width=\textwidth]{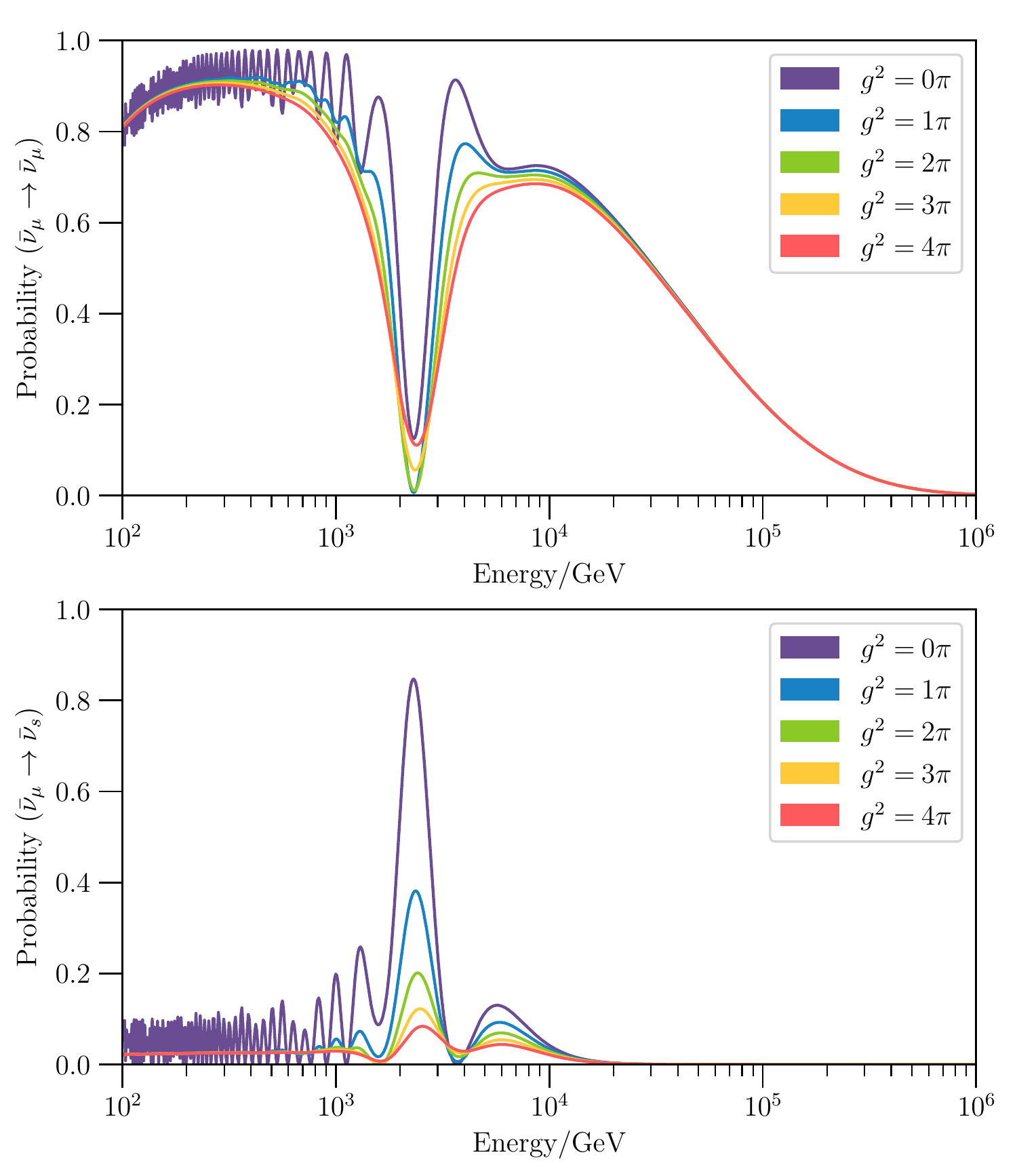}
  \caption[Transition probabilities versus energy for antineutrinos traversing the Earth for 3+1+decay]{Probabilities, shown as a function of energy, of muon antineutrinos traversing the diameter of the Earth to (top) survive in the muon flavor and (bottom) transition to the sterile flavor, assuming an unstable sterile neutrino model with invisible decay and parameters $\Delta m^2_{41} = 1.0 \: \textrm{eV}^2$ and $\sin^2 2\theta_{24} = 0.1$. The four different colors correspond to four values of the decay-mediating coupling, $g^2$, where $g^2 = 0$ corresponds to a traditional 3+1 model, i.e. no neutrino decay. At the highest energies, the survival probability of muon antineutrinos goes to zero because of the likelihood of interactions within the Earth.}
  \label{fig:osc_decay_escan_antinu}
\end{figure}

\begin{figure}[h!]
  \centering
  \includegraphics[width=\textwidth]{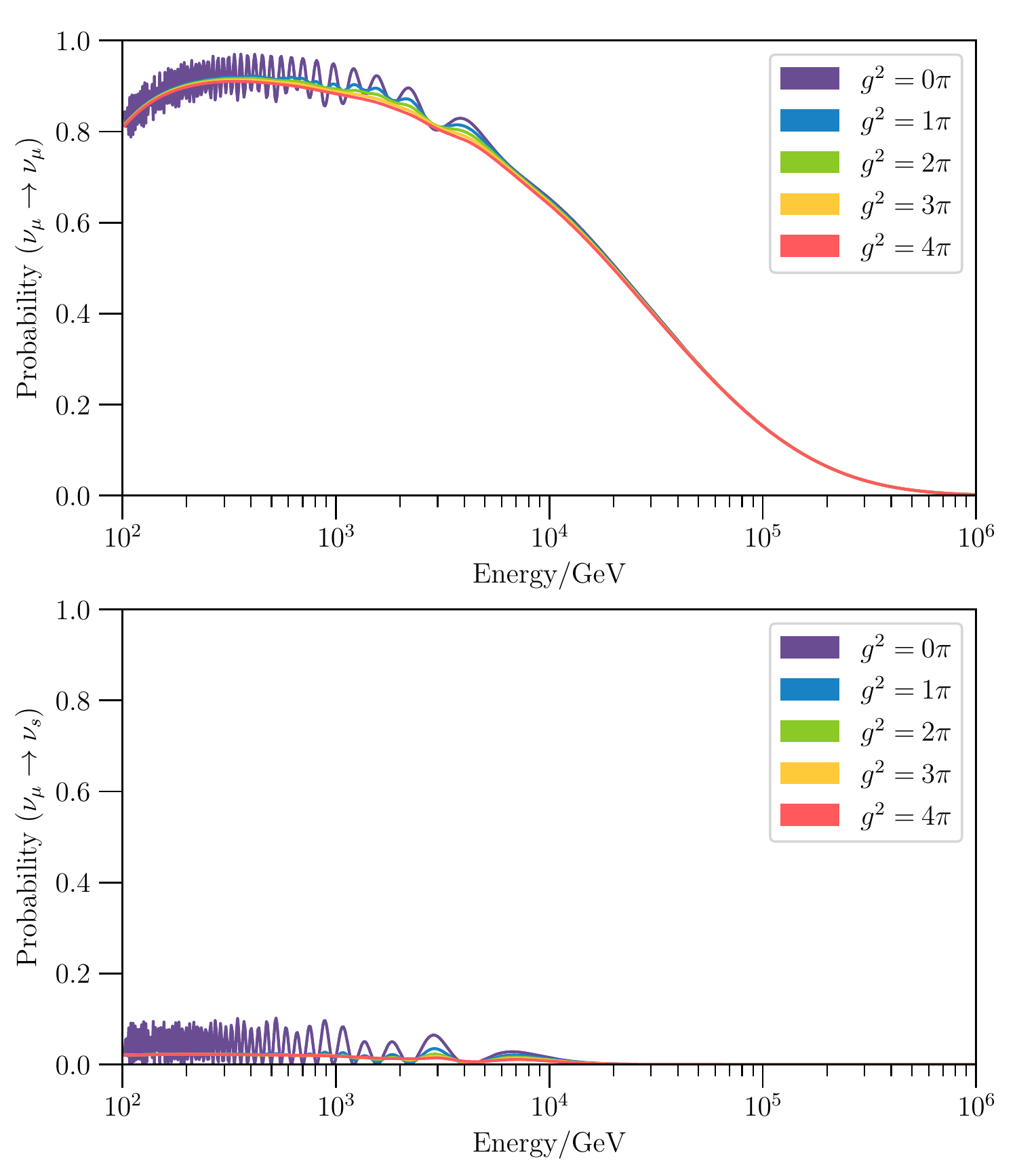}
  \caption[Transition probabilities versus energy for neutrinos traversing the Earth for 3+1+decay]{Probabilities, shown as a function of energy, of muon neutrinos traversing the diameter of the Earth to (top) survive in the muon flavor and (bottom) transition to the sterile flavor, assuming an unstable sterile neutrino model with invisible decay and parameters $\Delta m^2_{41} = 1.0 \: \textrm{eV}^2$ and $\sin^2 2\theta_{24} = 0.1$. The four different colors correspond to four values of the decay-mediating coupling, $g^2$, where $g^2 = 0$ corresponds to a traditional 3+1 model, i.e. no neutrino decay. At the highest energies, the survival probability of muon neutrinos goes to zero because of the likelihood of interactions within the Earth.}
  \label{fig:osc_decay_escan_nu}
\end{figure}

The results of combining the one-year IceCube data set with the global fit results from short-baseline measurements motivated further study of the unstable sterile neutrino model in IceCube. Decay of the fourth mass state causes two effects. One effect is a dampening of oscillation. The second effect is an overall disappearance of these particles. Transition probabilities as a function of baseline for muon antineutrinos traversing the diameter of the Earth are shown in Fig.~\ref{fig:osc_decay_baseline}. Transition probabilities as a function of energy for muon antineutrinos of energies 1~TeV and 2.3~TeV are shown in Fig.~\ref{fig:osc_decay_escan_antinu} for muon antineutrinos and Fig.~\ref{fig:osc_decay_escan_nu} for muon neutrinos. These probabilities are calculated with \texttt{nuSQUIDSDecay}\cite{Moss:2017pur}. In all of these figures, the additional mass-squared splitting and mixing amplitude are $\Delta m^2_{41} = 1.0~\textrm{eV}^2$ and $\sin^2 2\theta_{24} = 0.1$. Furthermore, all these figures show transition probabilities for four values of the square of the coupling that mediates the decay, $g^2$. The relationship between this coupling, $g$, the mass of the fourth neutrino mass state, $m_4$, and its lifetime, $\tau$ is~\cite{Moss:2017pur}
\begin{equation}
{\tau} = \frac{16 \pi}{g^2 m_4},
\label{eq:coupling_mass_lifetime}
\end{equation}
where it has been approximated that the lightest mass state is $\nu_1$ and it is massless. The traditional $3+1$ model corresponds to $g^2 = 0$.

Decay of the fourth mass state causes a dampening of the oscillation between the muon and sterile flavors, shown in Fig.~\ref{fig:osc_decay_baseline}. In the top panel, the scenario with $g^2 = 0$ causes oscillations that appear in the true neutrino cosine zenith distribution, shown in the oscillograms in the publication in~\Cref{sec:decay_prd} and later in~\Cref{sec:oscillgorams_and_expected_event_distributions}. This effect disappears very quickly with non-zero decay-mediating coupling, $g^2$. While IceCube has good angular resolution, the angular binning and the energy resolution wash out these fine features in the reconstructed event distribution, as will be shown in~\Cref{sec:oscillgorams_and_expected_event_distributions}.

Decay of the fourth mass state shifts the disappearance maximum to a longer baseline, shown in Fig.~\ref{fig:osc_decay_baseline}. As $g^2$ goes from zero to $4\pi$, there is a crossover in disappearance probabilities at fixed baseline. The shift in the position of the disappearance maximum moves the resonance in cosine zenith. This will be observable in the expected event distributions.

Decay of the fourth mass state widens the resonance seen in Fig.~\ref{fig:osc_decay_escan_antinu} (top). Below 200~GeV, oscillation between the muon and tau flavors becomes significant. This appears as disappearance of the muon flavor in Figs~\ref{fig:osc_decay_escan_antinu} and~\ref{fig:osc_decay_escan_nu}, without associated increase of the sterile flavor. At the highest energies, the Earth becomes opaque to neutrinos: a muon neutrino is likely to interact before crossing the diameter of the Earth. This causes the dramatic disappearance of muon flavor without associated increase in the sterile flavor.

\chapter{The IceCube neutrino detector}
\label{chap:icecube}

\section{The IceCube Neutrino Observatory}
\label{sec:icno}
\begin{figure}[h]
  \centering
  \includegraphics[width=\textwidth]{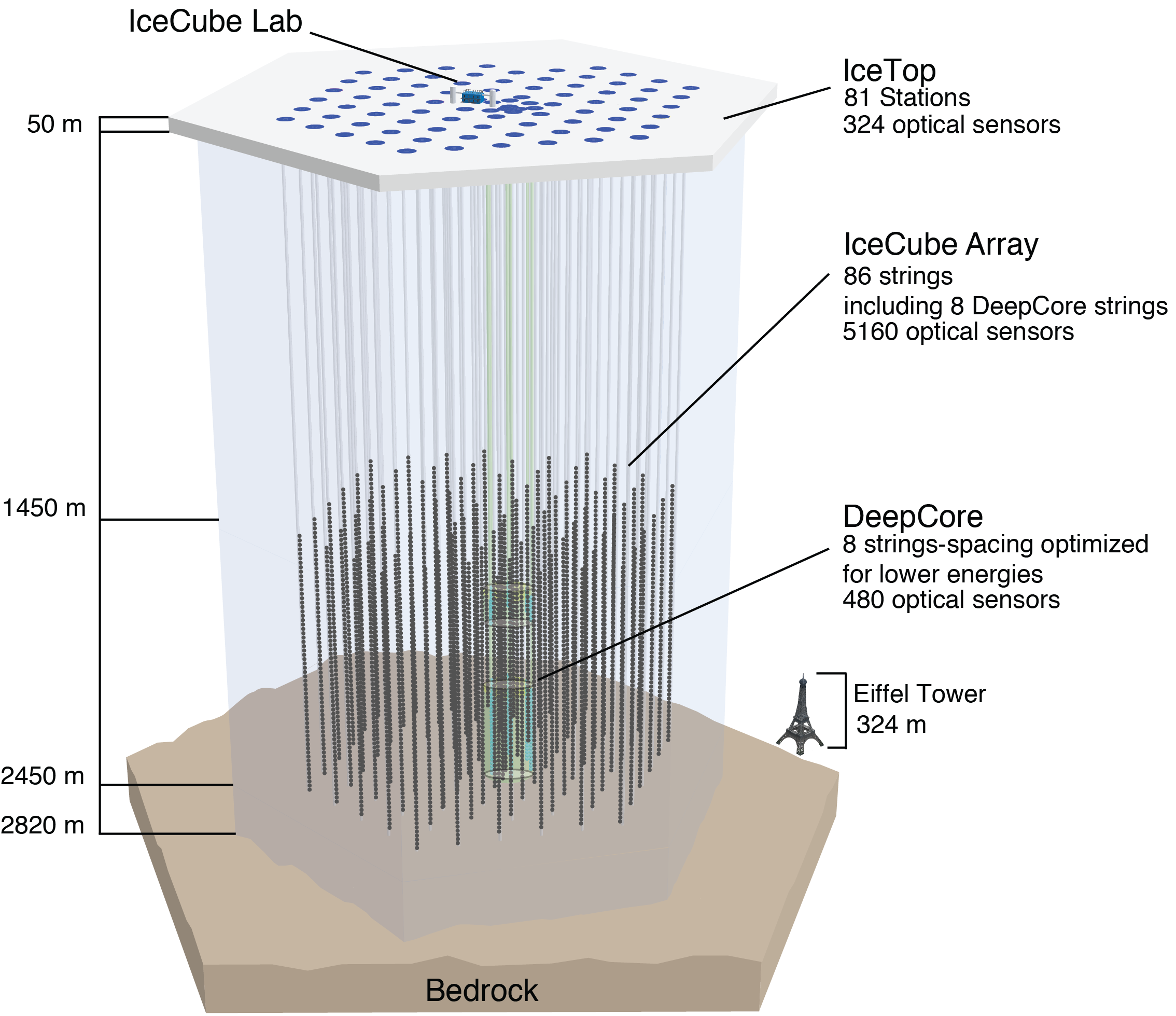}
  \caption[Schematic illustration of IceCube Neutrino Observatory]{Schematic illustration of the IceCube Neutrino Observatory. Credit: IceCube Collaboration.}
  \label{fig:ICNO}
\end{figure}

The IceCube Neutrino Observatory (ICNO) is located at the Amundsen-Scott South Pole Station in Antarctica~\cite{Aartsen:2016nxy}. The ICNO is depicted in Fig.~\ref{fig:ICNO}, with the Eiffel Tower shown for scale. The ICNO has two detector components. One component is the in-ice array, IceCube, which uses a cubic kilometer of clear, Antarctic ice to detect high-energy neutrinos via the emission of Cherenkov radiation. IceCube is located deep in the ice and is described in section~\ref{sec:inice}. The second component, IceTop, is a surface array that detects cosmic ray air showers and is only mentioned for completeness. As depicted in Fig.~\ref{fig:ICNO}, the in-ice array has a more-densely instrumented sub-array, named DeepCore, which detects neutrinos at lower energies than the main array~\cite{Collaboration:2011ym}. The IceCube Laboratory, housing a farm of computer servers, is at the surface~\cite{Aartsen:2016nxy}. Data collected in the array are sent to the IceCube Laboratory, where they are processed and filtered. Interesting data are sent to the north over satellite.
 
\section{IceCube in-ice array}
\label{sec:inice}
IceCube instruments a cubic kilometer of ice with optical sensors~\cite{Aartsen:2016nxy}. This large volume is desirable because the atmospheric and astrophysical neutrino fluxes fall steeply with energy. The South Pole ice cap is about 3~km deep, and IceCube instruments depths between 1450~m and 2850~m below the surface. The optical sensors, called Digital Optical Modules (DOMs), are described in~\Cref{sec:dom}. Altogether, 5160 DOMs hang on 86 bundled cables called strings. Of the 86 strings, 78 are spaced 125~m apart horizontally in a hexagonal array. On these strings, the 60 DOMs per string are separated by 17~m vertically. There are 8 additional strings that are part of DeepCore. These DeepCore-specific strings are positioned closer together horizontally, and below 1750~m below the surface, the DOMs are positioned closer together vertically~\cite{Collaboration:2011ym}. 

\begin{figure}[t!]
  \centering
  \includegraphics[width=0.75\textwidth]{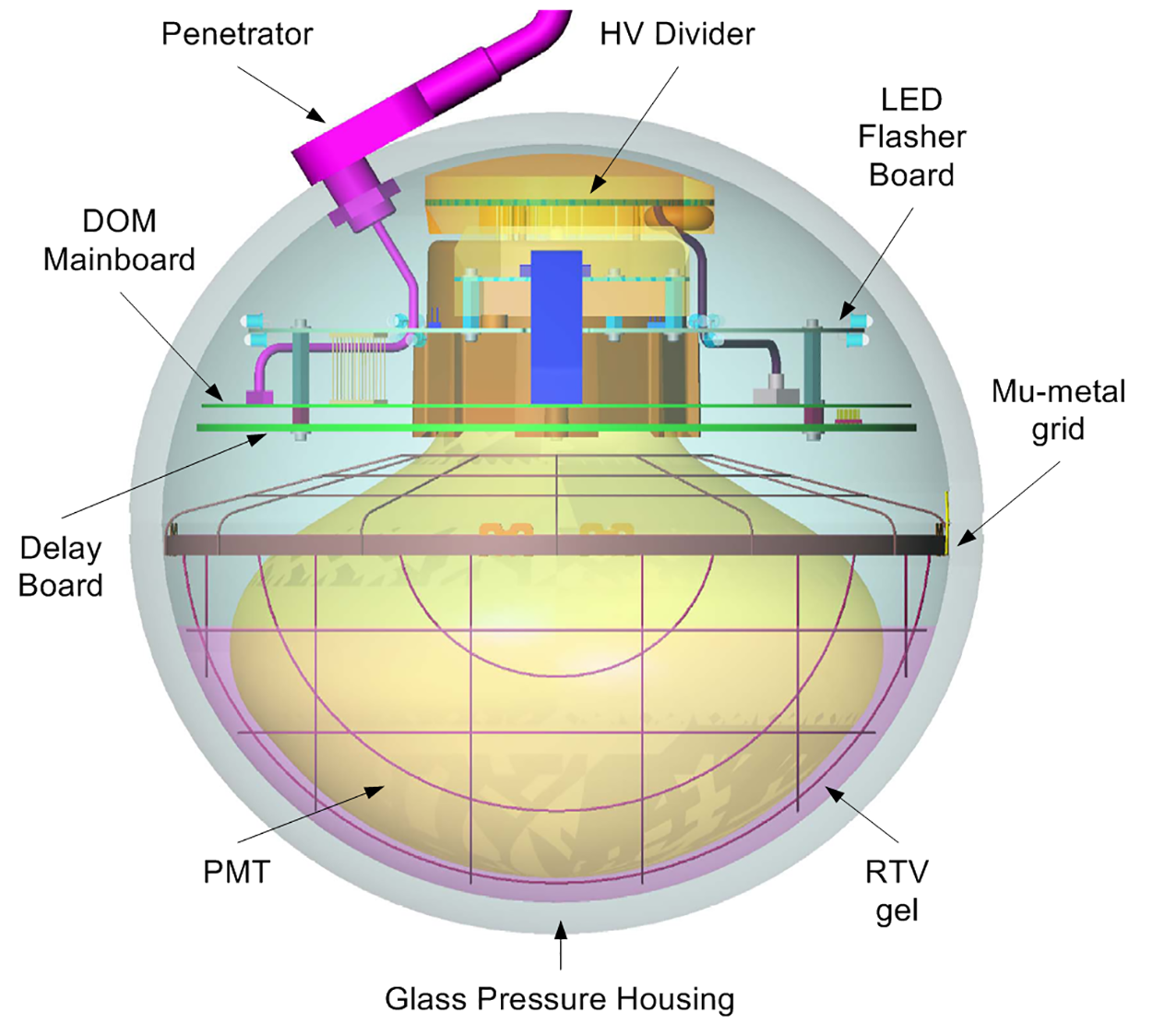}
  \caption[IceCube digital optical module]{Schematic illustration of an IceCube digital optical module (DOM). Credit: IceCube Collaboration.}
  \label{fig:dom}
\end{figure}

\section{Digital Optical Module}
\label{sec:dom}
The DOM is the basic unit of both detection and data acquisition (DAQ) in IceCube. DOMs are capable of bidirectional  communication with adjacent DOMs and the central DAQ at the IceCube Laboratory and executing calibration tasks. A schematic of a digital optical module (DOM) is shown in Fig.~\ref{fig:dom}. Each DOM houses a 10'' Hamamatsu photo-multiplier tube (PMT). The regular DOMs use Hamamatsu R7081-02 PMTs while some DOMs in DeepCore use a higher quantum efficiency version, Hamamatsu 47081-02MOD~\cite{Abbasi_2010}. The PMT is positioned downwards-facing in a glass pressure sphere. Approximately 1~cm of clear silicone gel between the PMT window and the glass housing both mechanically supports the PMT and optically couples it to the glass. The glass housing is made of low-radioactivity borosilicate glass and can withstand 250~bar~\cite{Aartsen:2016nxy}. A mu-metal grid surrounds the bulb of the PMT to shield it from the ambient South Pole magnetic field, which is 550~mG at an angle of 17$^{\circ}$ from vertical.

DOMs have several circuit boards. The Main Board digitizes PMT signals, communicates with adjacent DOMS and the central DAQ, and powers and directs all DOM subsystems as needed. The LED Flasher Board produces light within the array for the purposes of calibration~\cite{Aartsen:2013rt,Aartsen:2016nxy}. The standard Flasher Board has 12 light-emitting diodes, with six pointing out horizontally and six tilted up at 51.6$^{\circ}$. The Flasher Board is used to calibrate the position of DOMs in the ice and to measure the optical properties of the ice. The remaining circuit boards produce the high voltage for the PMT, set and monitor the PMT high voltage, and delay PMT signals.

\section{The ice}
\label{sec:ice}

The deep glacial ice at the South Pole is a good medium for detection of Cherenkov radiation because it is exceptionally clear: the effective scattering length and absorption length are order 10~m and 100~m, respectively, for light at 400~nm. At depths below 1450~m, there are no bubbles naturally present~\cite{Lundberg:2007mf}. There are, however, impurities such as dust. The concentration of dust varies throughout the ice in a tilted, depth-dependent manner~\cite{Ackermann:2006pva}. Most notable is the presence of an approximately 100~m thick dust layer at a depth of about 2000~m; in this dust layer, scattering and absorption of light is much worse~\cite{Aartsen:2013rt}. The temperature and pressure also vary in the ice, modifying photon propagation. As well as inhomogeneous, the optical properties of the ice have been found to be anisotropous~\cite{Aartsen:2013ola}. Light birefringence in ice polycrystals has been proposed as an explanation of the anisotropy~\cite{Chirkin:2019vyq}. The optical properties thus requires calibration and modeling~\cite{Aartsen:2016nxy, Lundberg:2007mf, Aartsen:2013ola}.  

The dust concentration and tilt of the ice have been measured with laser dust loggers. After six IceCube boreholes were water drilled for the installation of strings, laser dust loggers were deployed to measure the amount of dust. Optical profiles of the boreholes were produced, with a resolution of approximately 2~mm~\cite{doi:10.1029/2009JD013741}. A tilt map of the ice was produced from comparison of measurements as a function of depth for the different boreholes.

\begin{figure}[htb!]
  \centering
  \includegraphics[width=\textwidth]{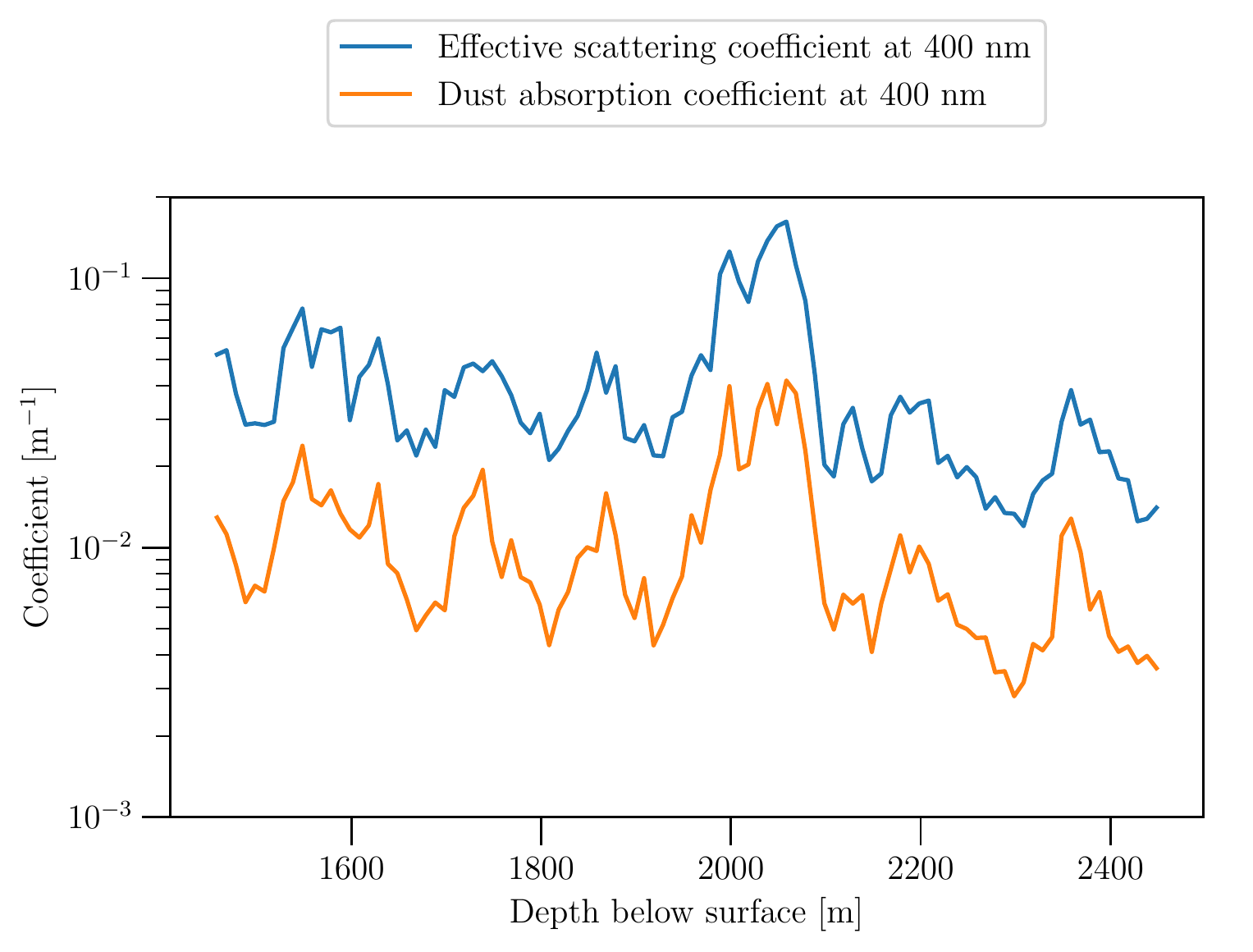}
  \caption[Ice model scattering and absorption coefficients]{Effective scattering and dust absorption coefficients at 400~nm from the Spice3.2 ice model.}
  \label{fig:ice_coefficients}
\end{figure}

The optical properties of the glacial ice can be described with a six-parameter ice model and a table of depth-dependent coefficients~\cite{Ackermann:2006pva, Aartsen:2013rt}. The scattering component of the model makes use of Mie scattering theory, which describes the scattering of photons off targets of general sizes and predicts the scattering angle distribution for any wavelength~\cite{Lundberg:2007mf, doi:10.1002/andp.19083300302}. The depth-dependent coefficients correspond to scattering and absorption lengths.  The absorption coefficient is the inverse of the average distance a photon travels before absorption. The scattering coefficient is the inverse of the average distance between consecutive scatters; an effective scattering coefficient is more useful for modeling the ice, as most scatters are in the forward direction. The coefficients are estimated for the wavelength 400~nm. The wavelength-dependency of these coefficients is modelled. The coefficients can be considered average values for 10~m thick layers. Calibration data, using light from the LEDs in the DOMS, is fit to this model to determine these coefficients~\cite{Aartsen:2013rt}. There is excellent agreement between the effective scattering coefficient extracted from flasher data and the average dust log~\cite{Aartsen:2013rt}. The values of the effective scattering coefficient and the dust absorption coefficient at 400~nm for the ice model used in this work are shown in Fig.~\ref{fig:ice_coefficients}.

The ``hole ice'' is the refrozen ice in the boreholes that had been hot-water drilled for installation of the detector~\cite{Aartsen:2013rt, Aartsen:2016nxy}. The hole ice is about 60~cm in diameter and froze naturally. It has worse optical properties than the undisturbed glacial ice. The hole ice has two components. The outer component, at radii about 8-30~cm, is clear, but has an effective scattering length of about 50~cm, greatly diminished compared to that of the glacial ice. The inner component, about 16~cm in diameter, is called the bubble column. It is hypothesized that air introduced during drilling is pushed to the center of the borehole during refreezing. The bubble column has an even shorter scattering length. Two video cameras deployed on one of the strings recorded the freezing process and confirm the presence of the bubble column, which appears opaque, as well as fits to calibration flasher data.

\begin{figure}[!tb]
  \centering
  \includegraphics[width=0.75\textwidth]{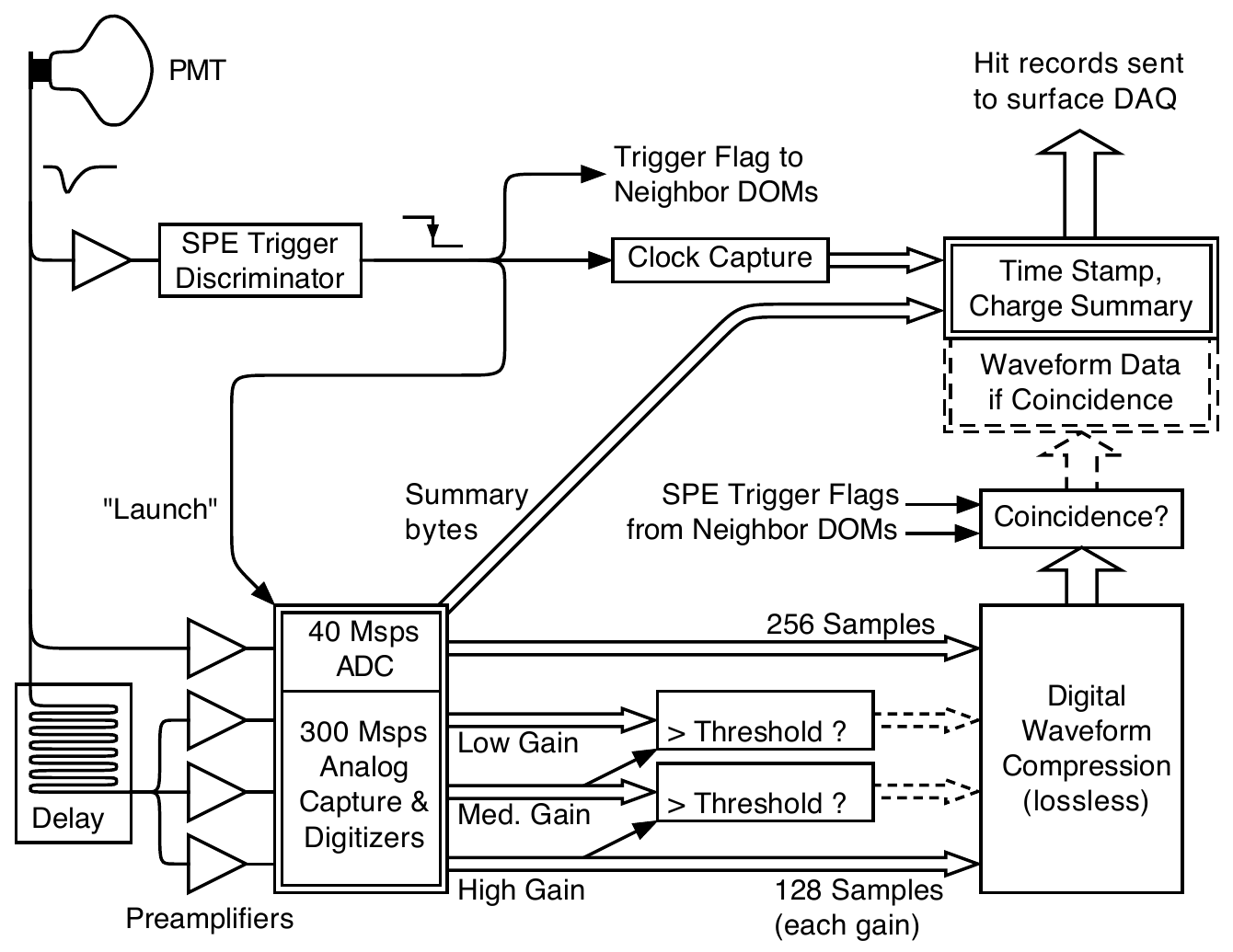}
  \caption[Data flow from PMT to central DAQ]{Data flow from PMT to central DAQ at the surface. Figure from~\cite{Aartsen:2016nxy}.}
  \label{fig:dom_dataflow}
\end{figure}

\section{Data acquisition}
\label{sec:dom_daq}

Data flow from the PMT to the central DAQ at the IceCube Laboratory is depicted schematically in Fig.~\ref{fig:dom_dataflow}. The PMT gain is about $10^7$~\cite{Abbasi_2010} for a single photon. The output signal of the PMT is AC-coupled with a toroidal transformal~\cite{Aartsen:2016nxy}. This output is then routed to a discriminator, which is set to fire when the voltage corresponds to 0.25 single photoelectrons (SPE). This is called a ``hit'' and begins a ``launch''. The time is captured. Four digitizers, one continuously sampling fast Analog-to-Digital Converter (fADC) and three custom Analog Transient Waveform Digitizers (ATWD), are used to capture the PMT signal. The signal has been routed through a delay board before reaching the three ATWDs. The three ATWDs have low, medium, and high gain, to capture a wide range of PMT signals. Using dedicated wiring, the DOM communicates with the nearest and next-to-nearest DOMS up and down on the string, to determine whether there is a local coincidence of hits. This will determine how much information is transmitted to the surface. In the case of a local coincidence, the hit is called called a Hard Local Coincidence (HLC) hit, and a timestamp, charge summary and waveform data are sent. In the case of no local coincidence, the hit is called Soft Local Coincidence (SLC) hit, and only a timestamp and charge summary are sent. 

At the IceCube Laboratory, a filter searches for events likely to correspond to muons~\cite{Aartsen:2016nxy}. Due to the morphology, these events are called tracks. All up-going tracks are selected. Data are sent to the north over satellite.

\clearpage
\section{Description of neutrino interaction}

\begin{figure}[!htb]
  \centering
  \includegraphics[width=0.75\textwidth]{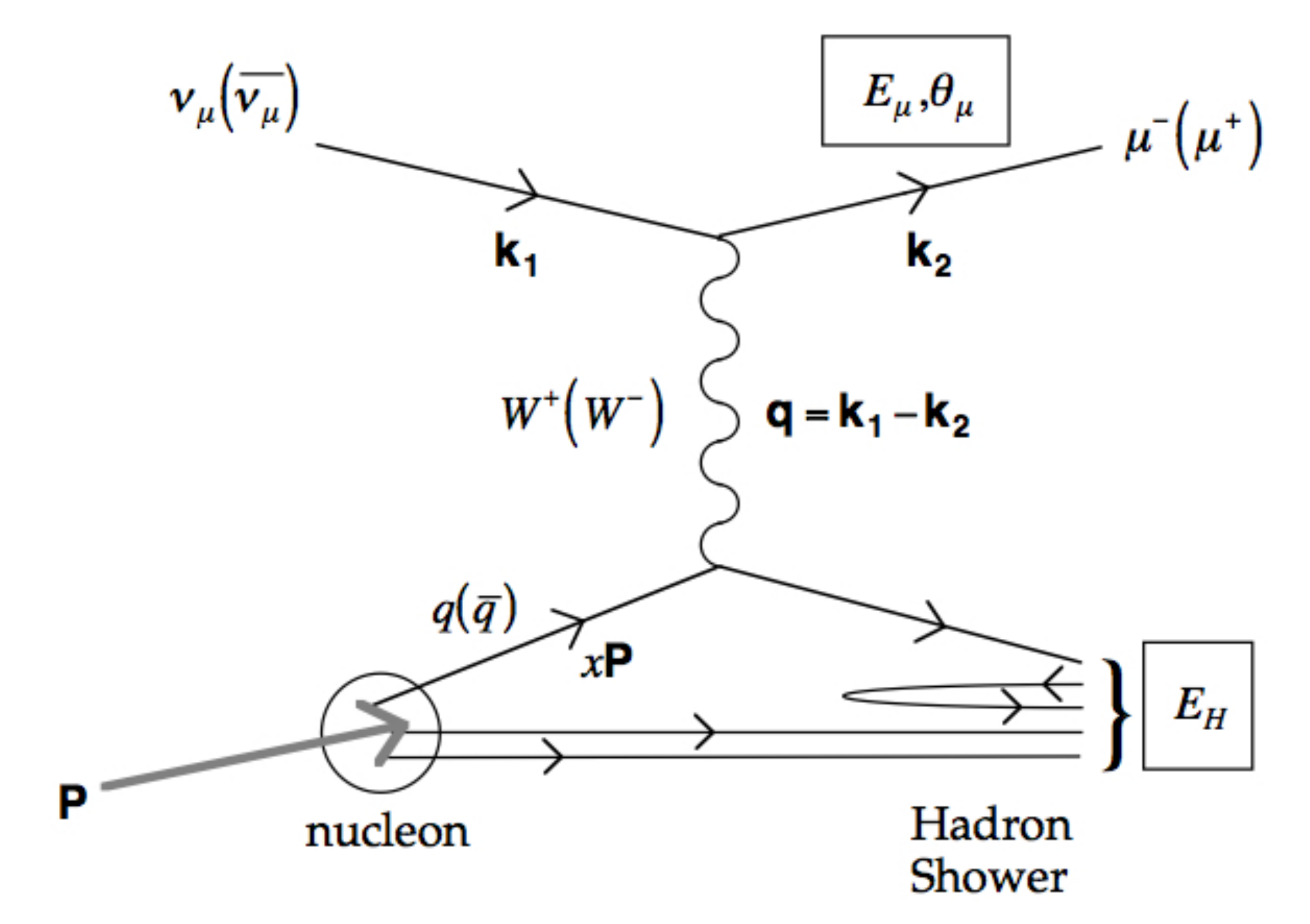}
  \caption[Feynman diagram for deep inelastic scattering]{Feynman diagram for the first-order contribution to muon-neutrino, charged-current DIS. Figure from~\cite{Conrad:1997ne}.}
  \label{fig:dis}	
\end{figure}

\begin{figure}[htb!]
  \centering
  \includegraphics[clip, trim =0cm 0cm 10cm 0cm , width=0.75\textwidth]{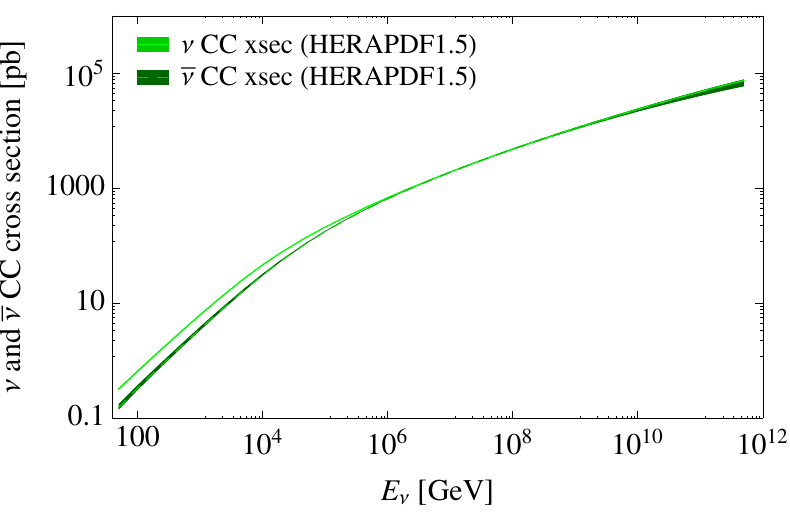}
  \caption[Charged-current cross sections]{Neutrino and antineutrino charged-current cross sections. Figure from~\cite{CooperSarkar:2011pa}.}
  \label{fig:xs}	
\end{figure}

At energies above 100~GeV, neutrinos interact via Deep Inelastic Scattering (DIS), where a neutrino scatters off one quark in a nucleon~\cite{Formaggio:2013kya}. The products of this interaction are a hadronic shower and an outgoing lepton. In muon neutrino charged-current interactions, the outgoing lepton is a muon. The leading-order Feynman diagram for charged-current muon neutrino DIS is shown in Fig.~\ref{fig:dis}. The cross sections for charged-current interactions above 100~GeV are given in Fig.~\ref{fig:xs}, for both neutrinos, in light green, and antineutrinos, in dark green.

\begin{figure}[htb!]
  \centering
  \includegraphics[width=0.75\textwidth]{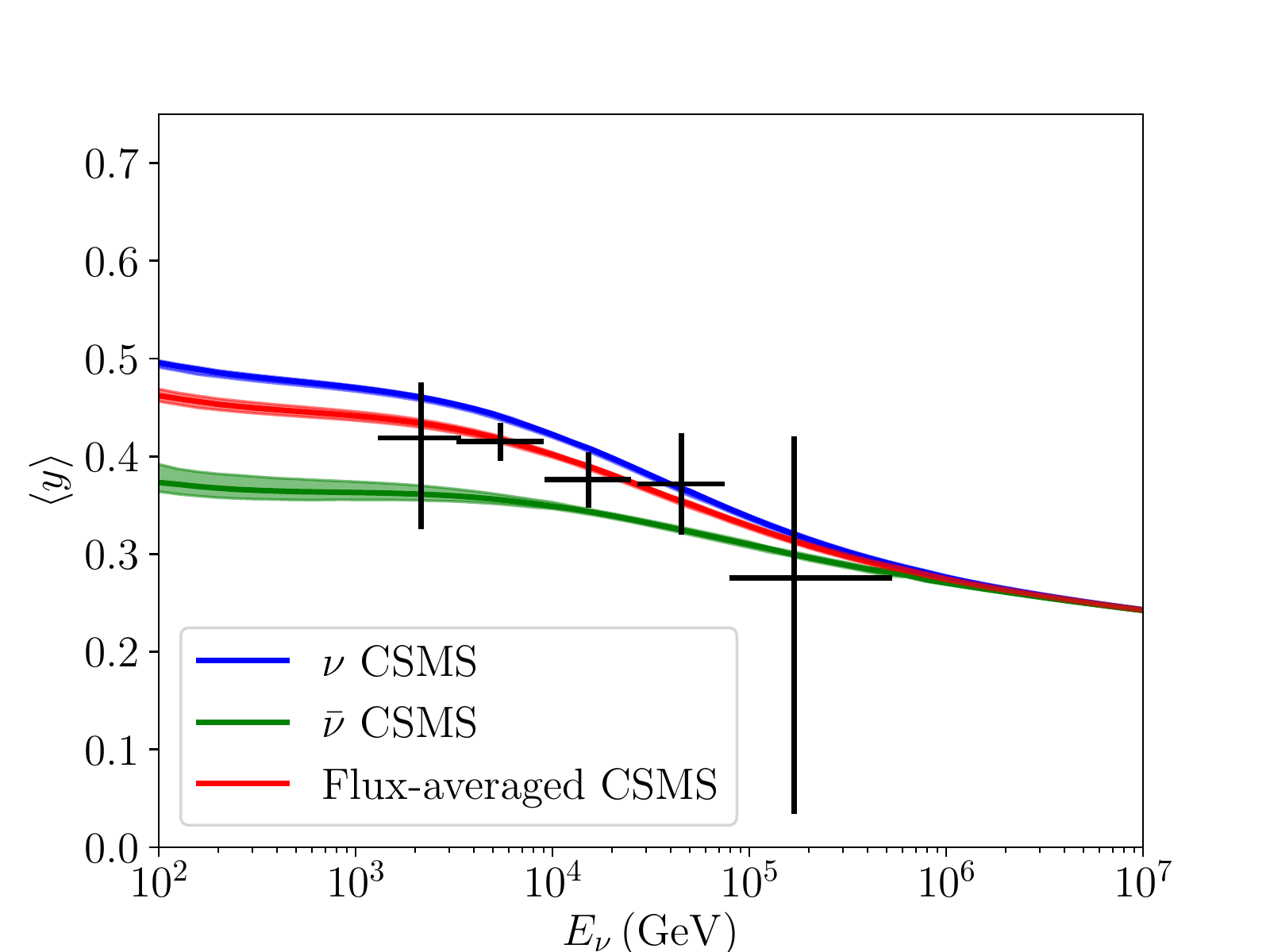}
  \caption[Mean inelasticity versus neutrino energy]{Mean inelasticity as a function of neutrino energy. Figure from~\cite{Aartsen:2018vez}.}
  \label{fig:inelasticity}	
\end{figure}

The fraction of neutrino energy, $E_\nu$, that goes into the hadronic shower is the in-elasticity, $y$. The muon energy, $E_\mu$, is then $(1-y)E_\nu$. For energies in the range 100~GeV -- 20~TeV, the mean inelasticity, $\langle y \rangle$, is approximately 0.4 -- 0.5~\cite{Gandhi:1995tf, Aartsen:2018vez}. The mean inelasticity as a function of neutrino energy is shown in Fig.~\ref{fig:inelasticity}. The curves are mean inelasticity predictions for neutrinos (blue), antineutrinos (green), and the flux-averaged combination of neutrinos and antineutrinos (red). The black crosses show measurements from IceCube.
 
\clearpage

\begin{figure}[h]
\centering
\includegraphics[width=4in]{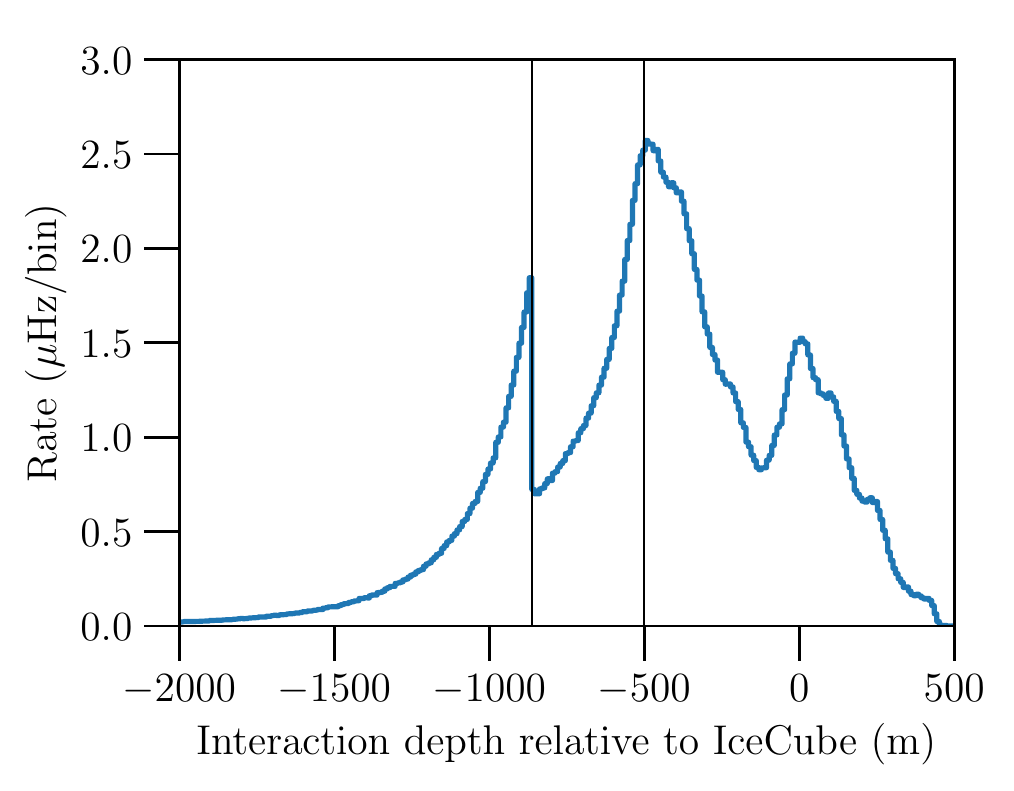}
\caption[Depth distribution of interactions from simulation]{Distribution of neutrino interaction depth relative to IceCube, from simulation. The vertical line at -862~m denotes the transition between bedrock and ice. The vertical line at -500~m denotes the bottom of the detector.}
\label{fig:bedrock_position}
\end{figure}

\begin{figure}
  \centering
  \includegraphics[width=0.75\textwidth]{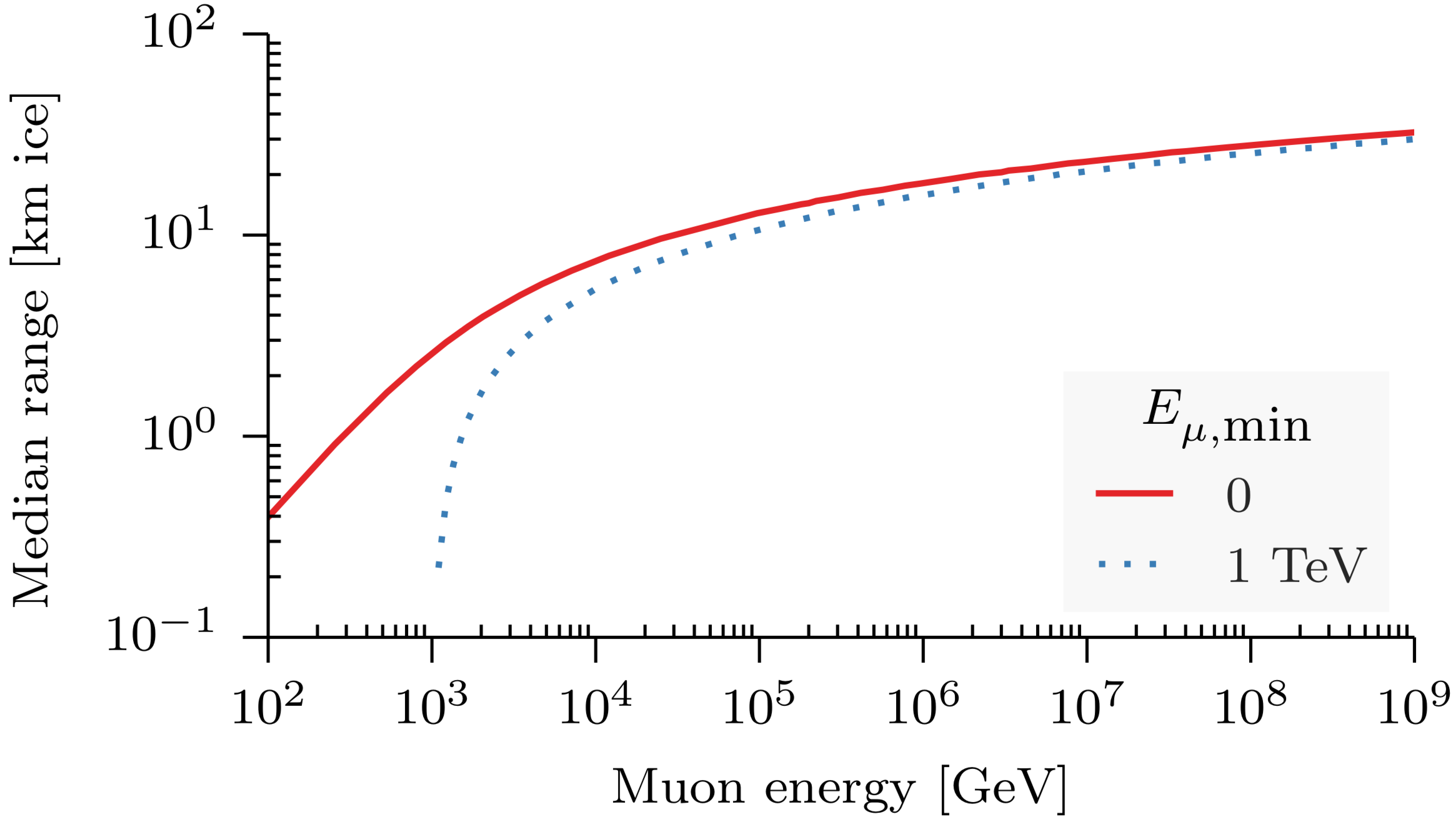}
  \caption[Muon range in ice]{Muon range in ice as a function of initial muon energy, calculated with \texttt{PROPOSAL}~\cite{KOEHNE20132070}. The solid, red curve shows the median muon range. The dotted blue curve shows the range at which 50\% of muons will have kinetic energy larger than 1~TeV. Figure from \cite{vanSanten:2014kqa}.}
  \label{fig:jvs_muon_range}	
\end{figure}

\begin{figure}[htb!]
  \centering
  \includegraphics[width=0.6\textwidth]{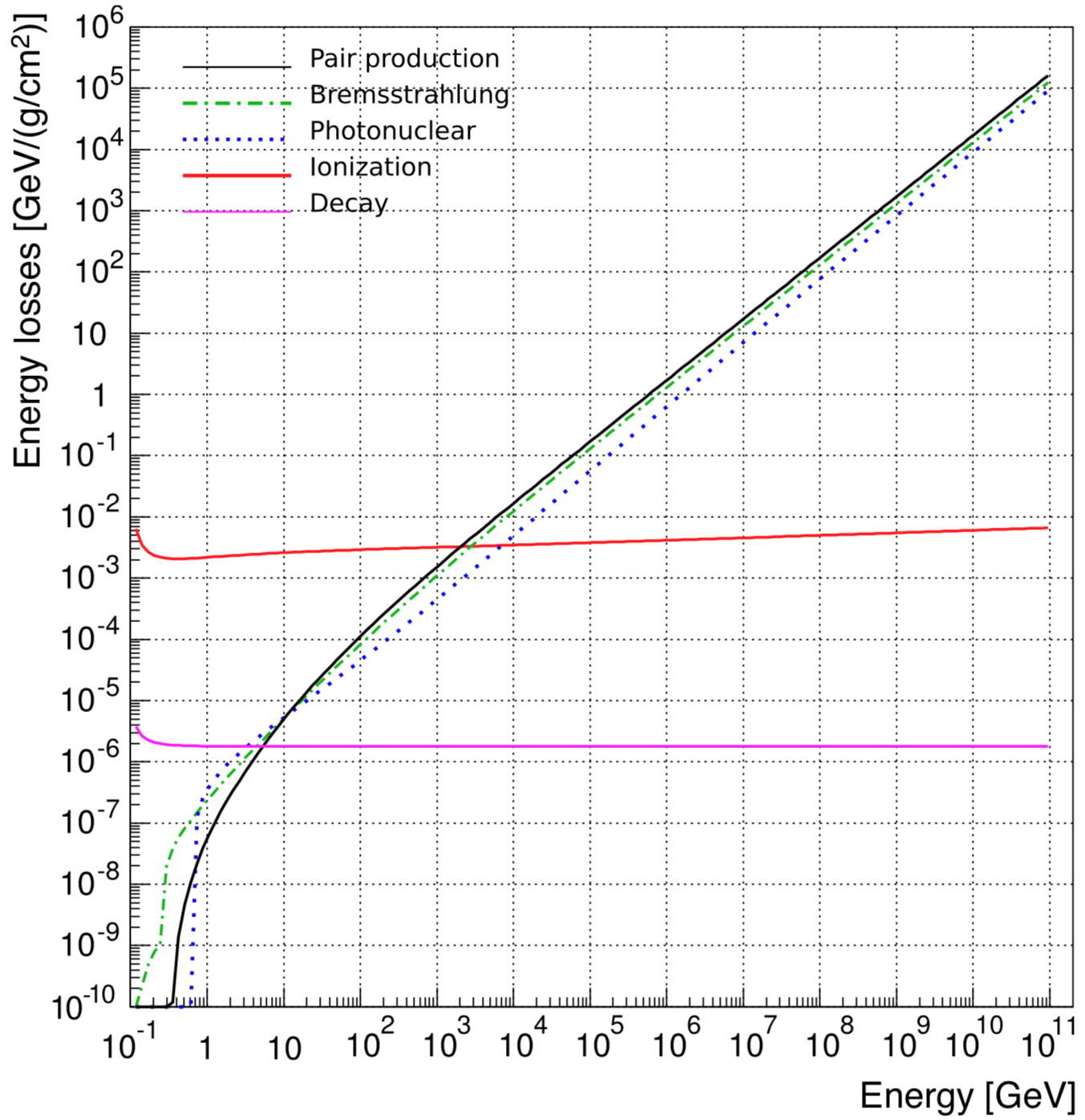}
  \includegraphics[width=0.6\textwidth]{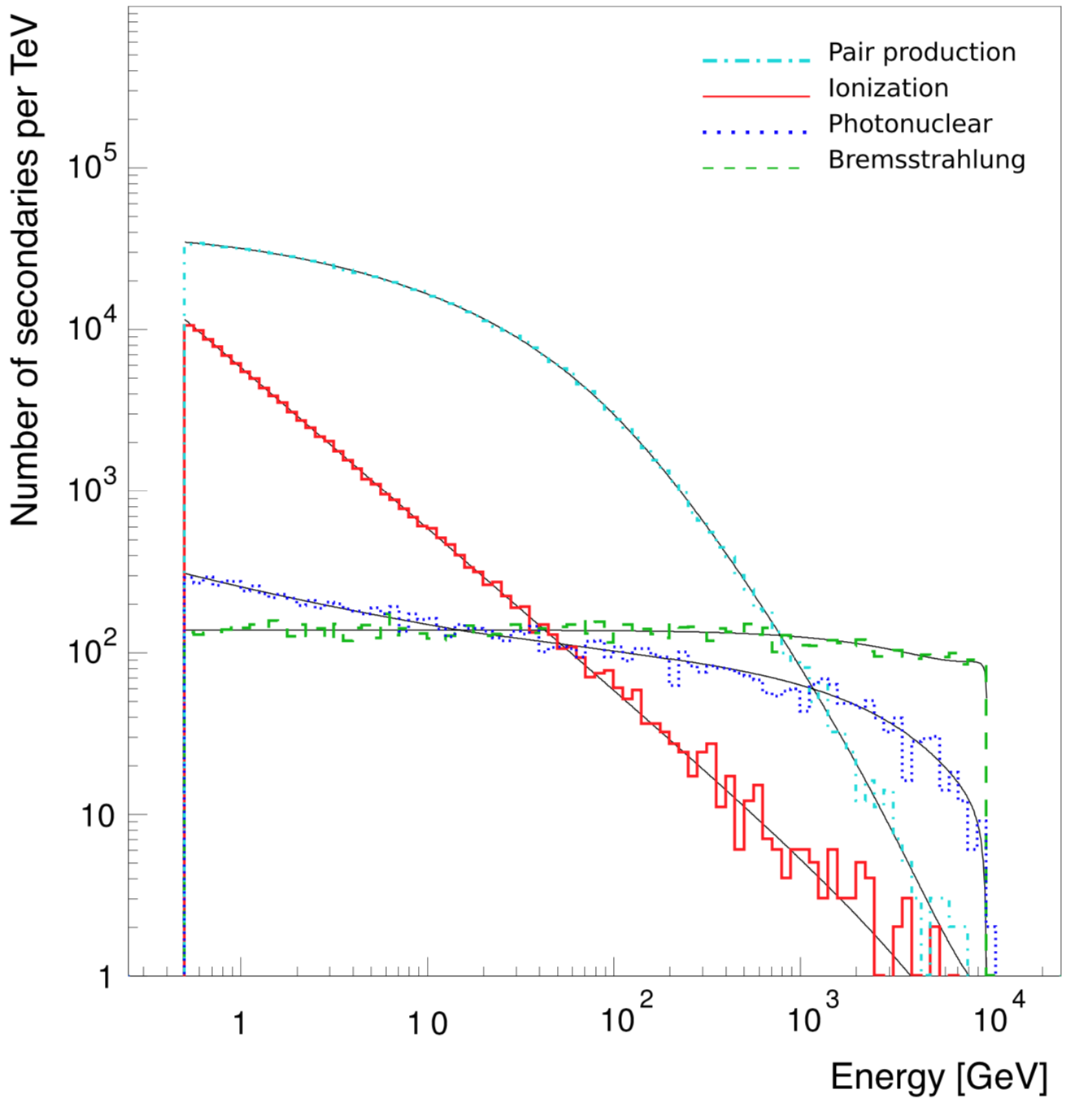}
  \caption[Muon energy losses in ice and secondary spectra]{(Top) Energy losses for muons in ice. (Bottom) Energy spectra of secondary particles for a 10~TeV muon traveling through rock. All calculations performed with \texttt{PROPOSAL}~\cite{KOEHNE20132070}. Figures from \cite{KOEHNE20132070}. }
  \label{fig:PROPOSAL}	
\end{figure}

\begin{figure}[h]
 \centering
 \includegraphics[clip, trim = 5cm 0cm 5cm 0, width=0.9\textwidth]{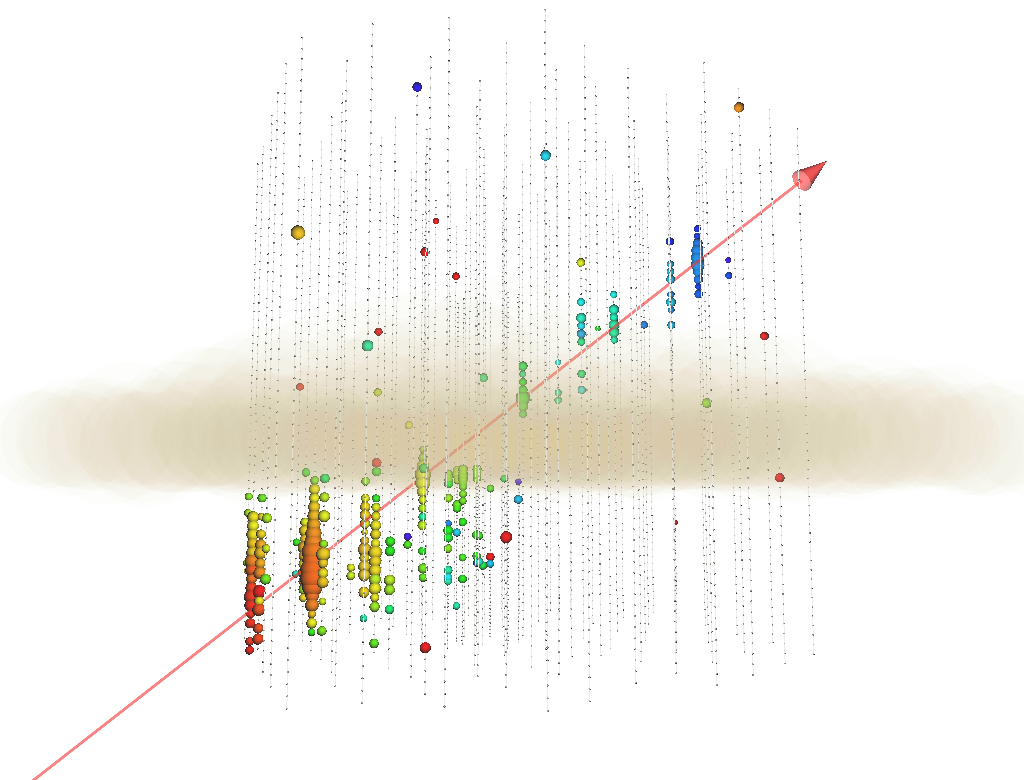}
 \caption[Event display for a 5.2 TeV muon]{Event display for an up-going muon. The thin, grey, vertical lines correspond to the 86 IceCube strings. The small grey dots on the strings are the DOMs that did not see light. The DOMs that observed light are shown as colored spheres. The color of the sphere indicates timing, where red DOMs saw light at the earliest times and blue DOMs at the latest times. The larger the size of the sphere, the more light was observed. The reconstructed muon trajectory is shown a red line. The reconstructed muon energy is 5.2~TeV. Figure courtesy of Sarah Mancina.}
 \label{fig:event_display0}	
\end{figure}

\begin{figure}[h]
 \centering
 \includegraphics[clip, trim = 5cm 0cm 5cm 0, width=0.9\textwidth]{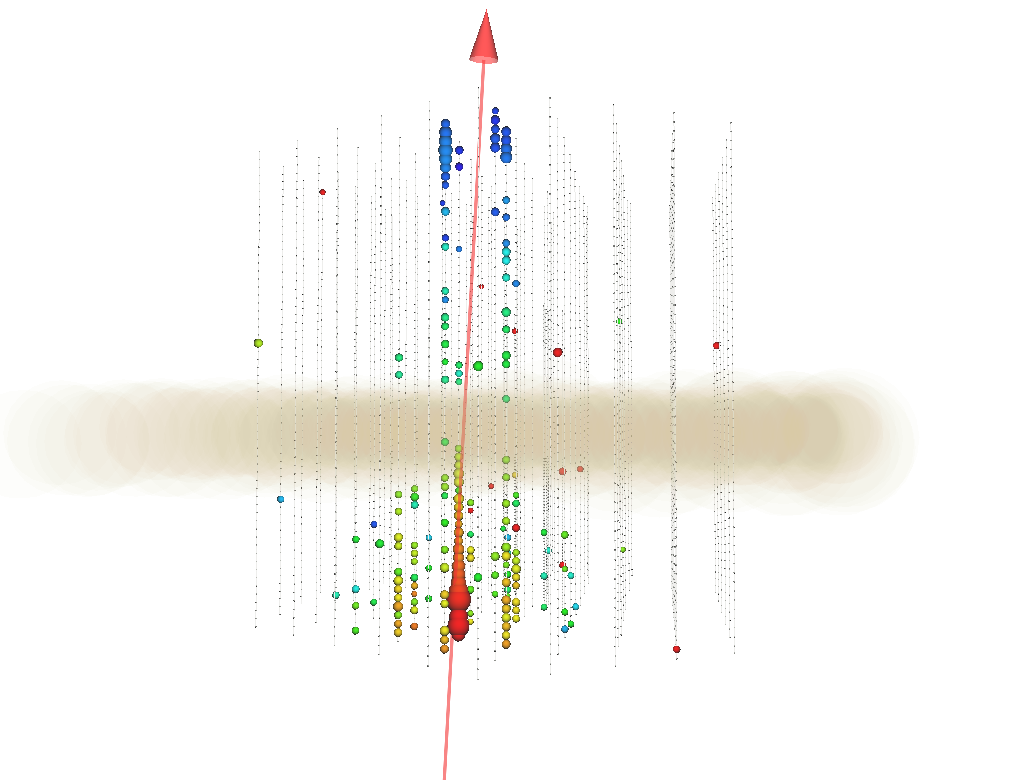}
 \caption[Event display for a 7.0 TeV muon]{Event display for an up-going muon. The thin, grey, vertical lines correspond to the 86 IceCube strings. The small grey dots on the strings are the DOMs that did not see light. The DOMs that observed light are shown as colored spheres. The color of the sphere indicates timing, where red DOMs saw light at the earliest times and blue DOMs at the latest times. The larger the size of the sphere, the more light was observed. The reconstructed muon trajectory is shown a red line. The reconstructed muon energy is 7.0~TeV. Figure courtesy of Sarah Mancina.}
 \label{fig:event_display1}	
\end{figure}

In the events relevant for this thesis, neutrino interaction occurs either in the bedrock or ice below IceCube. About 20\% of the interactions occur in the bedrock, shown in Fig.~\ref{fig:bedrock_position}. The hadronic shower is contained outside the detector. The muon enters the detector and is observed. The muon, traveling faster than the speed of light in ice, will radiate Cherenkov light, which may be detected by DOMs. The muon will lose energy as it traverses the ice. Ionization, pair production, bremsstrahlung and photonuclear interactions all contribute to muon energy losses~\cite{Groom:2001kq}. The average stopping power of each process as a function of muon energy is shown in Fig.~\ref{fig:PROPOSAL} (top). Below approximately 2~TeV, ionization dominates. Above this energy, the other, radiative processes dominate. Secondary particles produced in these interactions include delta electrons, bremmsstrahlung photons, excited nuclei and pairs of electrons~\cite{KOEHNE20132070}. The energy spectra for these secondary particles, for the case of a 10~TeV muon traveling through rock, is given in Fig.~\ref{fig:PROPOSAL} (bottom). Charged secondaries of sufficiently high energy will themselves produce Cherenkov radiation, which may be detected by DOMs. These processes are stochastic, resulting in bursts of energy loss along the muon's trajectory. The muon is likely to exit the detector, as muons in the energy range 500 GeV -- 10 TeV can travel kilometers of ice. The median range of muons is shown in Fig.~\ref{fig:jvs_muon_range}. Event displays for two muons are shown in Figs.~\ref{fig:event_display0} and \ref{fig:event_display1} .

Muons and antimuons have opposite electric charges. Other experiments may distinguish them by imposing a magnetic field and observing the direction in which their trajectories curve. IceCube has no such magnetic field, and therefore no ability to distinguish between muons and antimuons. This means that IceCube cannot distinguish between muon neutrinos, which would be tagged by the presence of a muon, and muon antineutrinos, which would be tagged by the presence of an antimuon.

\section{Muon reconstruction}
The analysis is performed using reconstructed muon properties: the reconstructed muon energy -- also referred to as the muon energy proxy -- and the reconstructed cosine of the muon's zenith angle, $\cos \theta_{\rm{z}}$, where the zenith angle is the angle with respect to the North Pole-South Pole axis. Events from the horizon have $\cos \theta_{\rm{z}} = 0$, while events coming from the North Pole have  $\cos \theta_{\rm{z}} = -1$. The direction of the muon is reconstructed based off the arrival time of the first photon to reach each DOM, as well as the charge of all pulses from each DOM~\cite{Weaver:2015bja, Axani:2020zdv}. The reconstruction algorithm considers the shape of Cherenkov cone, as well as scattering and absorption of photons within the ice. The resolution of the angular reconstruction is 0.005 -- 0.015 in $\cos \theta_{\rm{z}}$. The angular reconstruction is described further in~\cite{Axani:2020zdv}.

\begin{figure}[h]
  \centering
  \includegraphics[width=0.9\textwidth]{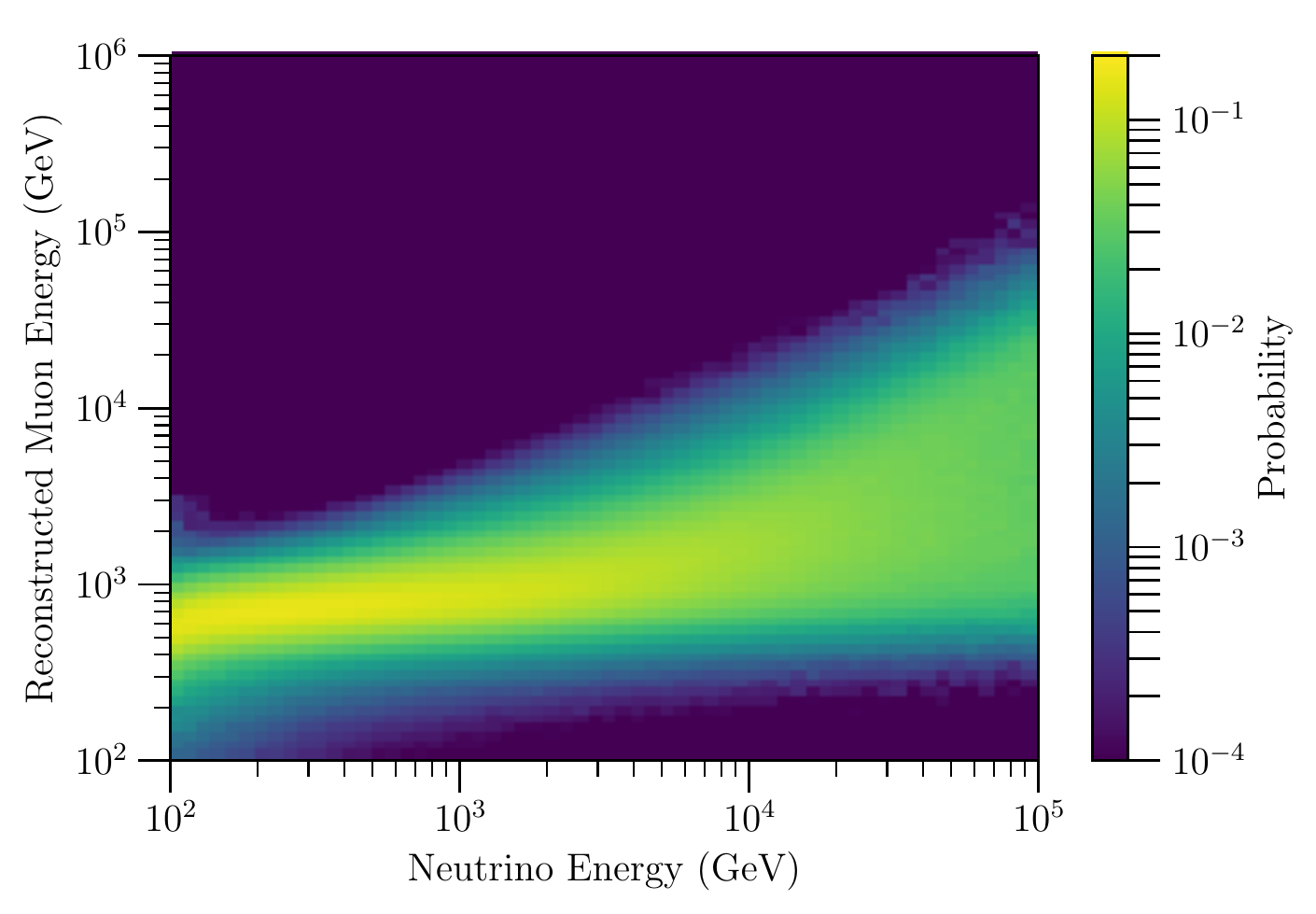}
  \caption[Reconstructed muon energy versus true neutrino energy]{The probability of obtaining a given reconstructed muon energy as a function of true neutrino energy.}
  \label{fig:muex_vs_nu_true}	
\end{figure}

\begin{figure}[h]
  \centering
  \includegraphics[width=0.9\textwidth]{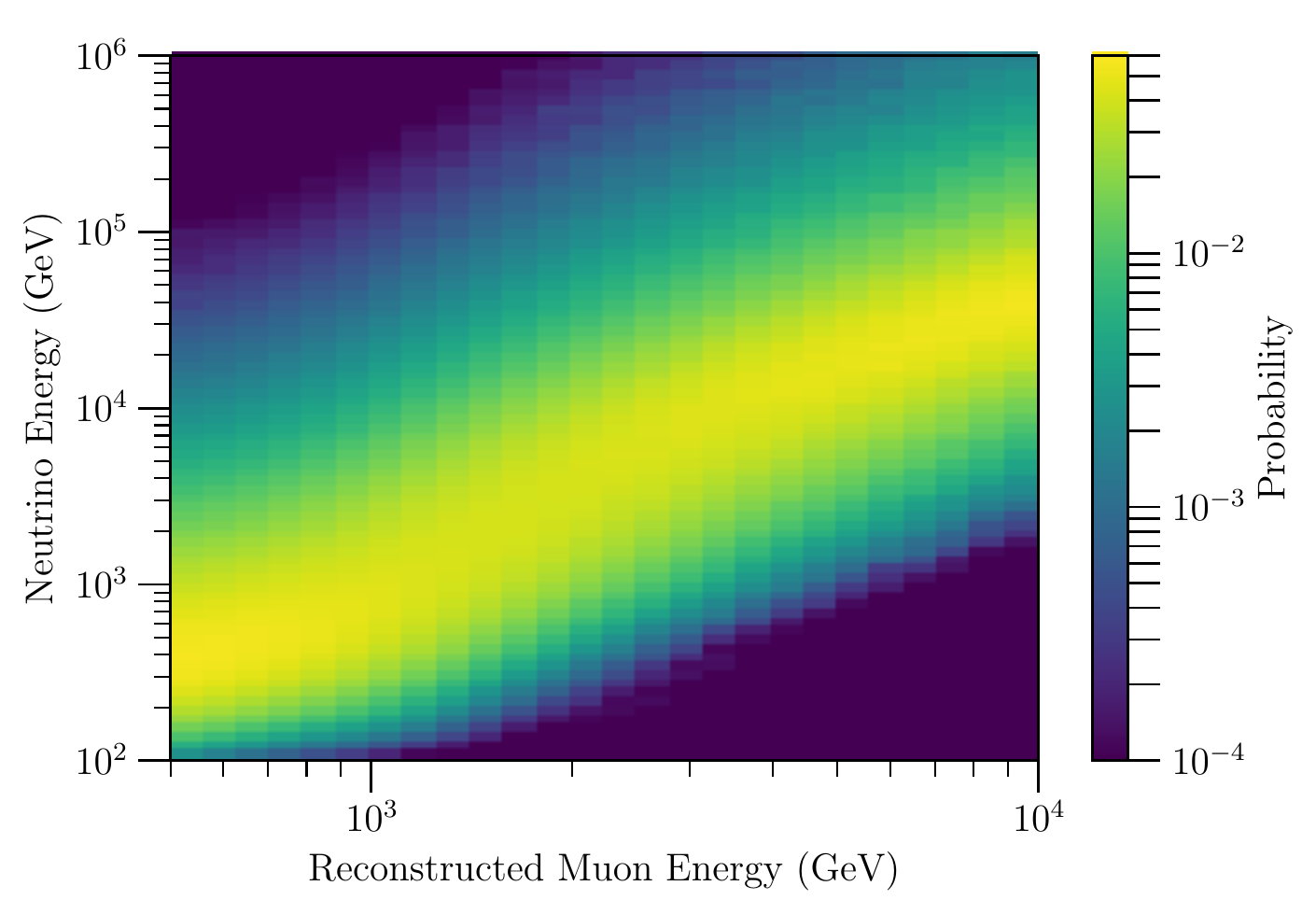}
  \caption[True neutrino energy versus reconstructed muon energy]{The probability distribution of neutrino energy as a function of reconstructed muon energy. This probability distribution assumes a neutrino flux proportional to $E^{-2}$, where $E$ is the neutrino energy.}
  \label{fig:nu_true_vs_muex}	
\end{figure}

The energy resolution is much worse than the angular resolution. It is so large it is described in terms of the $\log_{10}$ of the energy:  $\sigma_{\log_{10}(E_\mu)} \sim 0.4$. This is due to the stochastic nature of the muon's energy loss and the fact that the muon is not contained within the detector. Both of these are evident in Figs~\ref{fig:event_display0} and \ref{fig:event_display1}. The muon energy is related to its energy loss per unit length, which is proportional to the light produced and thus the light observed by the detector. The muon energy reconstruction algorithm used here, \texttt{MuEx}, uses an analytical expression for the observed light, taking into consideration the depth-dependent optical properties of the ice~\cite{Weaver:2015bja, Aartsen:2013vja}. The relationship between neutrino energy and reconstructed muon energy is shown in Figs.~\ref{fig:muex_vs_nu_true} and \ref{fig:nu_true_vs_muex}.

\chapter{Atmospheric neutrino fluxes}
\label{chap:flux}

\section{The coupled cascade equations}
\begin{figure}[H]
  \centering
	\includegraphics[width=0.75\textwidth]{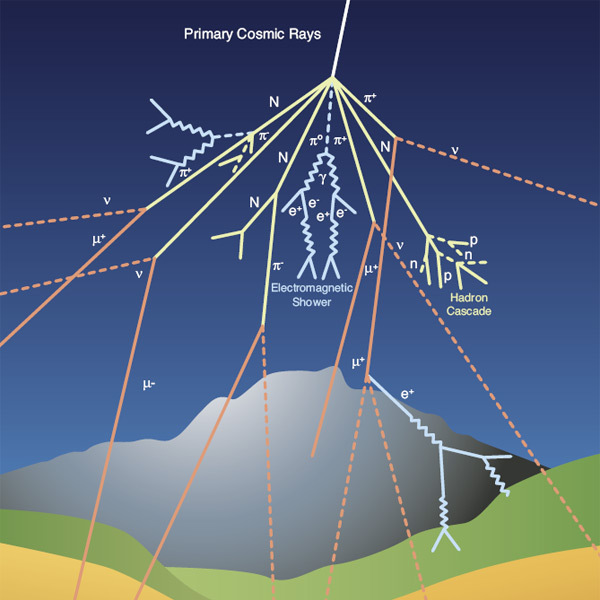}
  \caption[Schematic illustration of a cosmic ray air shower]{Schematic illustration of a cosmic ray air shower. Figure from CERN~\cite{CERN_cosmicrays}.}
  \label{fig:cosmic_ray_air_shower}
\end{figure}

Atmospheric neutrinos comprise the vast majority of neutrinos observed by the IceCube detector. They arise from the interactions of primary cosmic rays with nuclei in the Earth's atmosphere~\cite{gaisser2016cosmic}. These collisions produce hadronic showers. Secondary particles decay to atmospheric neutrinos. A cosmic ray air shower is depicted schematically in Fig.~\ref{fig:cosmic_ray_air_shower}. The evolution of particle populations during the hadronic cascade is described by the coupled cascade equations~\cite{gaisser2016cosmic}. These are coupled, differential transport equations that describe the propagation, interaction, and decay of particles. The coupled cascade equation for the flux, $\Phi_{E_i}^h$, of particle $h$ in discrete energy bin $E_i$ and at atmospheric slant depth, $X$, is given by Eq.~\ref{eq:cascade}, 

\begin{equation}
\begin{split}
\frac{d \Phi_{E_i}^h}{d X} = 
&- \frac{\Phi_{E_i}^h}{\lambda^h_{\textrm{int},\:E_i}} \\
&- \frac{\Phi_{E_i}^h}{\lambda^h_{\textrm{dec},\:E_i}(X)} \\
&+ \sum_{E_k \geq E_i} \sum_l \frac{c_{l(E_k) \rightarrow h(E_i)}}{\lambda^l_{\textrm{int},\:E_k}} \Phi^l_{E_k} \\
&+ \sum_{E_k \geq E_i} \sum_l \frac{d_{l(E_k) \rightarrow h(E_i)}}{\lambda^l_{\textrm{dec},\:E_k}(X)} \Phi^l_{E_k}.
\end{split}
\label{eq:cascade}
\end{equation}
The first term accounts for the loss of particle $h$ due to interactions, with interaction length
\begin{equation}
\lambda^h_{\textrm{int},\:E_i} = \frac{m_{\textrm{air}}}{\sigma^{\textrm{inel}}_{p-\textrm{air}}},
\end{equation}
where $m_{\textrm{air}}$ is the mean mass of nuclei in the air, and $\sigma^{\textrm{inel}}_{p-\textrm{air}}$ is the inelastic cross section between a proton and an air nucleus. The second term accounts for loss of particle $h$ due to decay, with decay length
\begin{equation}
\lambda^h_{\textrm{dec},\:E_i}(X) = \frac{c \tau_h E_i \rho_{\textrm{air}}(X)}{m_h},
\end{equation}
where $\tau_h$ and $m_h$ are the lifetime and mass of particle $h$, and $\rho_{\textrm{air}}(X)$ is the density of the air at slant depth $X$. The third term accounts for gain of particle $h$ due to interactions of other particle species; the coefficients representing this, $c_{l(E_k) \rightarrow h(E_i)}$, are determined from hadronic interaction models. The fourth term accounts for gain of particule $h$ due to the decay of other particle species. The atmospheric slant depth, $X$ is given by
\begin{equation}
\label{eq:slantdepth}
X(h_0) = \int_0^{h_0} dl \rho_{\rm{air}}(h_{\rm{atm}}(l)),
\end{equation}
where $\rho_{\rm{air}}$ is the mass density of the air as a function of the atmospheric height, $h_{\rm{atm}}$, and the density is integrated along the path, $l$, of the cascade.

In this work, the atmospheric neutrino flux is calculated with the program \texttt{Matrix Cascade Equations} (\texttt{MCEq}). \texttt{MCEq} casts the coupled cascade equations as a matrix equation, which it solves numerically~\cite{Fedynitch:2015zma, fedynitch2012influence}. Over fifty baryons, mesons, and leptons are considered in the calculations. The calculation requires three inputs; these inputs are informed by a wide range of collider and fixed target experiments, as we will discuss below. One input is the primary cosmic ray model, which describes the composition and spectra of the incoming particles. Another is the hadronic interaction model, which describes the cross sections of interactions in the cascade. The last is the description of the temperature or density profile of the atmosphere. 

\begin{figure}
  \centering
	\includegraphics[clip, trim = 13cm 8.5cm 0cm 0cm, width=0.8\textwidth]{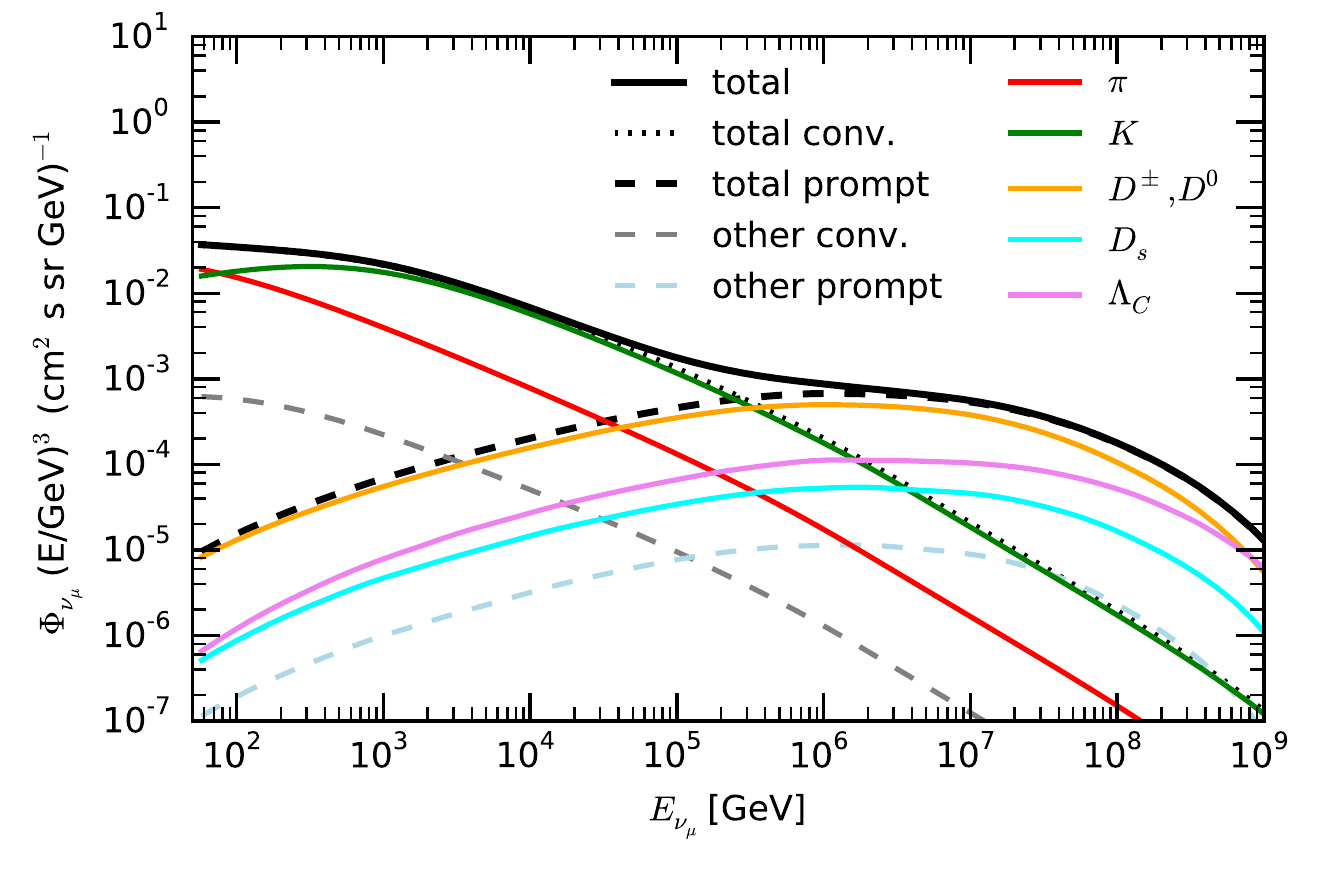}
  \caption[Energy spectra of atmospheric muon neutrinos and their various constituents]{Energy spectrum of atmospheric muon neutrinos and antineutrinos, as well as the partial contributions to the total flux, as calculated with \texttt{MCEq}. The spectrum is scaled by the energy cubed. Figure from \cite{Fedynitch:2015zma}.}
  \label{fig:mceq_anatoli_fluxes}
\end{figure}

The combined muon neutrino and antineutrino spectrum, and partial contributions from the various parent particles, are shown in Fig.~\ref{fig:mceq_anatoli_fluxes}. These spectra were calculated with \texttt{MCEq} and assume Thunman-Ingelman-Gondolo~\cite{Gondolo:1995fq} as the primary cosmic ray flux model and SIBYLL-2.3 RC1~\cite{Engel:2015} as the hadronic interaction model. The atmospheric muon neutrino flux is comprised of the \textit{conventional} component and the \textit{prompt} component, shown in Fig.~\ref{fig:mceq_anatoli_fluxes}. The conventional flux originates in the decays of charged pions, charged kaons, and muons. Muon neutrinos come from the decays:
\begin{equation}
\begin{split}
\begin{aligned}
\pi^+ &\rightarrow \mu^+ + \nu_\mu \\
K^+ &\rightarrow \mu^+ + \nu_\mu \\
\mu^- &\rightarrow e^- + \bar{\nu}_e + \nu_\mu
\end{aligned}
\end{split}
\label{eq:atm_flux_decays}
\end{equation}
while muon antineutrinos come from the decays of the oppositely charged particles. Above around 100~GeV, neutrinos from kaon decays are predominant to those from either pion or muon decays. The prompt flux comes from the decays of heavier particles, such as charmed mesons, although it has yet to be observed. It is expected to be predominant at energies above about $10^5$~GeV.

\section{Cosmic ray and hadronic interaction models}

\begin{figure}[H]
  \centering
  \includegraphics[width=0.85\textwidth]{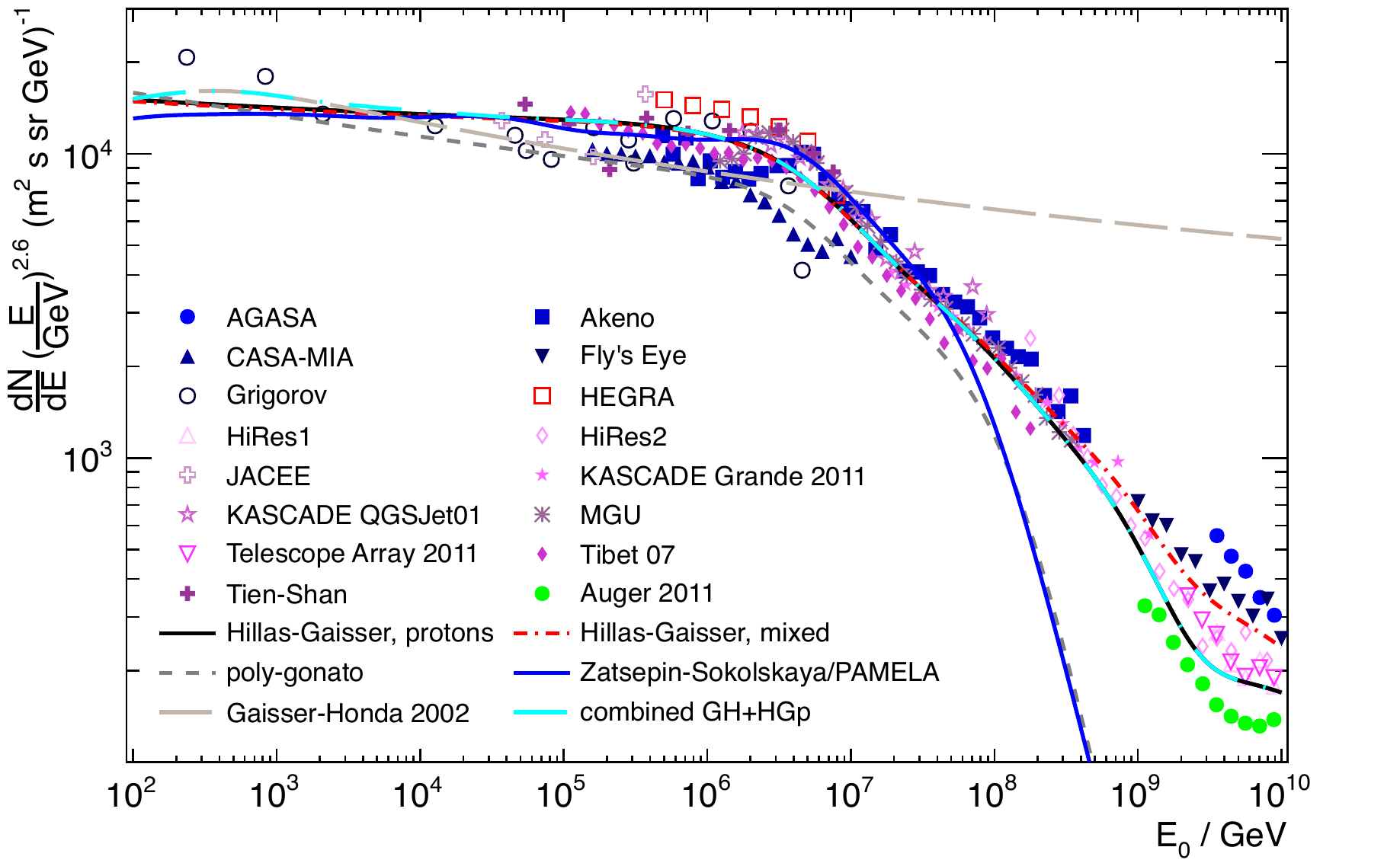}
  \caption[All-particle cosmic ray spectrum]{All-particle cosmic ray spectrum. Model predictions are shown as solid or dashed curves. Data from ground-based arrays are shown as points~\cite{takeda2003energy, nagano1984energy, Arqueros:1999uq, bird1994cosmic, Grigorov:1970xu, Glasmacher:1999id, Abbasi:2007sv, asakimori_93, asakimori_91, Apel:2011mi, ANTONI20051, danilova_77, Fomin_91, Tsunesada:2011mp, Amenomori:2008aa, Amenomori:1995nt}. Figure from \cite{fedynitch2012influence}.}
  \label{fig:fedynithc_cr_models}

\end{figure}

There is a range of available models that describe the flux and composition of primary cosmic rays. The various models make different assumptions about the sources of cosmic rays and also are informed by different datasets. The models considered in this work are PolyGonato~\cite{Hoerandel:2002yg}, Hillas-Gaisser 2012 H3a~\cite{Gaisser:2012zz}, and Zatsepin-Sokolskaya/PAMELA~\cite{Zatsepin:2006, Adriani69}. These models account for cosmic ray hydrogen, helium, and heavier nuclei. The Hillas-Gaisser 2012 H3a model, which is the one used in this work, includes five groups of nuclei: hydrogen, helium, carbon-nitrogen-oxygen, magnesium-aluminium-silicon, and iron. The all-particle cosmic ray energy spectrums predicted by these models are shown by the curves on Fig.~\ref{fig:fedynithc_cr_models}. Data, from various ground-based arrays, are shown as points~\cite{takeda2003energy, nagano1984energy, Arqueros:1999uq, bird1994cosmic, Grigorov:1970xu, Glasmacher:1999id, Abbasi:2007sv, asakimori_93, asakimori_91, Apel:2011mi, ANTONI20051, danilova_77, Fomin_91, Tsunesada:2011mp, Amenomori:2008aa, Amenomori:1995nt}. The abrupt change in the all-particle spectrum at about 10$^{6.5}$~GeV is called the ``knee'' of the cosmic ray spectrum. One can see that the data sets have a large spread in the knee region, and this will need to be addressed as a systematic uncertainty.

A selection of models are available to describe the hadronic interactions in cosmic ray air showers. The models considered in this work are SIBYLL-2.3c~\cite{Riehn:2017mfm}, SIBYLL-2.3~\cite{Riehn:2015oba}, EPOS-LHC~\cite{Pierog:2013ria}, and QGSJET-II-04~\cite{Ostapchenko:2010vb}. These models are tuned to data from accelerator experiments. Data from Large Hardon Collider, which has run at around 10~TeV, can constrain the interactions of cosmic ray particles at the knee~\cite{dEnterria:2011twh}. This is because of the Lorentz transformation between these two reference frames: detectors at the Large Hardon Collider operate in the center-of-mass frame, while the atmosphere, which is impinged by cosmic rays, is effectively a fixed target. The center-of-mass energy of cosmic ray interactions is
\begin{equation}
\sqrt{s} \simeq \sqrt{2 E m},
\end{equation}
where $E$ is the energy of the cosmic ray, $m$ is the mass of the target particle, and the mass of the cosmic ray particle is negligible compared to its energy. 

The hadronic interaction models do not come with uncertainties and there is significant discrepancy between the models~\cite{Pierog_2019}. Three of the properties that vary between the models are the inelastic cross sections, the shower multiplicity, and interaction elasticity. This causes different predictions of the atmospheric neutrino and antineutrino fluxes. In particular, the ratio of the muon neutrinos to muon antineutrinos varies largely. This is an important systematic uncertainty for the work in this thesis, as the matter resonance due to the presence of a sterile neutrino is expected to appear in the muon antineutrino flux alone.

\section{Barr scheme for hadronic interaction uncertainties}

The Barr scheme is a method for assigning hadronic interaction uncertainties for the purpose of calculating conventional atmospheric neutrino fluxes~\cite{Robbins:820971, Barr:2006it}. Uncertainties are assigned based on measurements of the production of charged pions and kaons from projectiles impinging on targets,
\begin{equation}
\begin{split}
\begin{aligned}
p N &\rightarrow \pi^\pm X \\
p N &\rightarrow K^\pm X. \\
\end{aligned}
\end{split}
\end{equation}
The projectile, $p$, is most often a proton. The target $N$, is a light element, for example, beryllium, carbon or aluminium. The other products, $X$, are unconstrained. Uncertainties are given as a function of $E_i$, the incident particle's energy, and $x_{\textrm{lab}}$, the fraction of the incident particle's energy that is carried by the secondary meson. 

Uncertainties are assigned based off the consistency between datasets, as well as the amount of extrapolation that is necessary between different nuclear targets, energies, and values of the transverse momentum. The assigned uncertainties for incident particle energies above 30~GeV are given in Table~\ref{tab:barr_uncertainties}.

\begin{table}
\centering
\begin{tabular}{c c c c c}
\hline
Label & Incident energy, $E_i$ & Meson & $x_{\textrm{lab}}$ & Uncertainty \\
\hline
\hline
G & >30 GeV & $\pi^{\pm}$ & $0-0.1$ & 30\% \\
H & >30 GeV & $\pi^{\pm}$ & $0.1-1$ & 15\% \\
I & >500 GeV & $\pi^{\pm}$ & $0.1-1$ & 12.2\% $\times \log_{10}(E_i/500~\textrm{GeV})$\\
\hline
W & >30 GeV & $K^{\pm}$ & $0-0.1$ & 40\% \\
Y & >30 GeV & $K^{\pm}$ & $0.1-1$ & 30\% \\
Z & >500 GeV & $K^{\pm}$ & $0.1-1$ & 12.2\% $\times \log_{10}(E_i/500~\textrm{GeV})$\\
\hline

\end{tabular}
\caption[Estimated uncertainties in the production of charged kaons and pions]{Estimated uncertainties in the production of charged kaons and pions. Values from \cite{Barr:2006it}.}
\label{tab:barr_uncertainties}
\end{table}

Since the conventional flux is predominantly composed of kaons, in the energy range relevant for this thesis, the kaon production uncertainties are more important than the pion production uncertainties. The G, H, and I uncertainties from Table~\ref{tab:barr_uncertainties} are therefore not explicitly incorporated into the analysis. The uncertainties on the production of oppositely charged kaons are considered separately. Six \textit{Barr parameters} -- WP, WM, YP, YM, ZP, ZM -- refer to the uncertainties in Table~\ref{tab:barr_uncertainties}, where the suffix P indicates it references the production of positively charged kaons, and the suffix M indicates it references that of negatively charged kaons.

The propagation of these uncertainties to the muon neutrino and antineutrino fluxes is calculated with \texttt{MCEq} as in~\cite{Fedynitch:2017M4, Yanez:20197d}. For each of the six Barr parameters, $\mathcal{B}_i$, muon neutrino and antineutrino flux derivatives are calculated:
\begin{equation}
\frac{\partial \Phi}{\partial \mathcal{B}_i} = \frac{\Phi(\mathcal{B}_i = \delta) - \Phi_(\mathcal{B}_i = -\delta)}{2 \delta}.
\end{equation} 
For example, $\Phi(\textrm{YP} = \delta)$, is the flux where $K^+$ production in the region correlating to Y has been increased by the relative amount $\delta$ compared to the nominal model. If one assumes the nominal flux prediction, $\Phi_0$, underestimates charged meson production by $\mathcal{B}_i$ for each of the respective regions in this parameterization, a corrected flux prediction is
\begin{equation}
\Phi(\mathcal{B}_1,\mathcal{B}_2,...) = \Phi_0 + \sum_i \mathcal{B}_i \frac{\partial \Phi}{\partial \mathcal{B}_i}.
\end{equation}

\section{Atmospheric flux uncertainty treatment}
The various combinations of cosmic ray and hadronic interaction models yield discrete predictions for the atmospheric muon neutrino and antineutrino fluxes, as well as their ratio. It is unknown which is the closest to the truth, and how close it is. A set of eight physically-motivated systematic uncertainties allows for a continuous range of flux predictions that cover the discrete ones. The normalization and composition of cosmic rays above the knee have a large uncertainty~\cite{fedynitch2012influence}. This leads to a large uncertainty in the normalization of the atmospheric neutrino flux. Here it is taken to be 40\%. Additionally, there is an uncertainty in the spectral index of the cosmic ray flux, leading to one in the neutrino flux. In this work, the effect of the uncertainty in the spectral index, $\Delta \gamma$, on the neutrino flux is parameterized as
\begin{equation}
\Phi(E, \: \Delta \gamma) = \Phi (E) \bigg(\frac{E}{2.2~\textrm{TeV}} \bigg)^{-\Delta \gamma}.
\end{equation}
The uncertainty on $\Delta \gamma$ is taken to be 0.03. As discussed previously, the six Barr parameters, WP, WM, YP, YM, ZP, and ZM, represent the uncertainty in the production of charged kaons.

\begin{figure}[H]
  \centering
  \includegraphics[clip, trim = 9.3cm 1.3cm 0.5cm 3.5cm, width=0.4\textwidth]{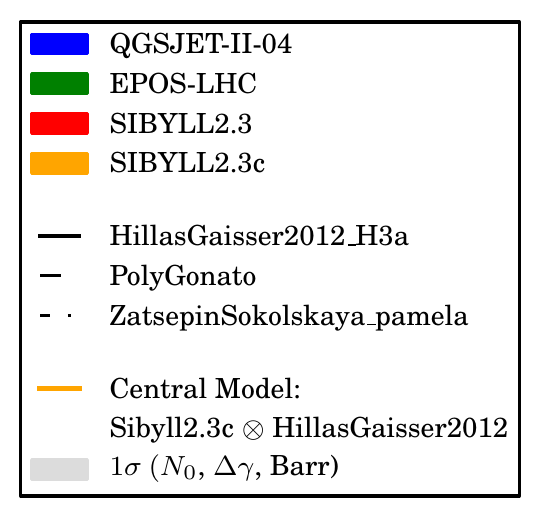}
	\includegraphics[width=0.6\textwidth]{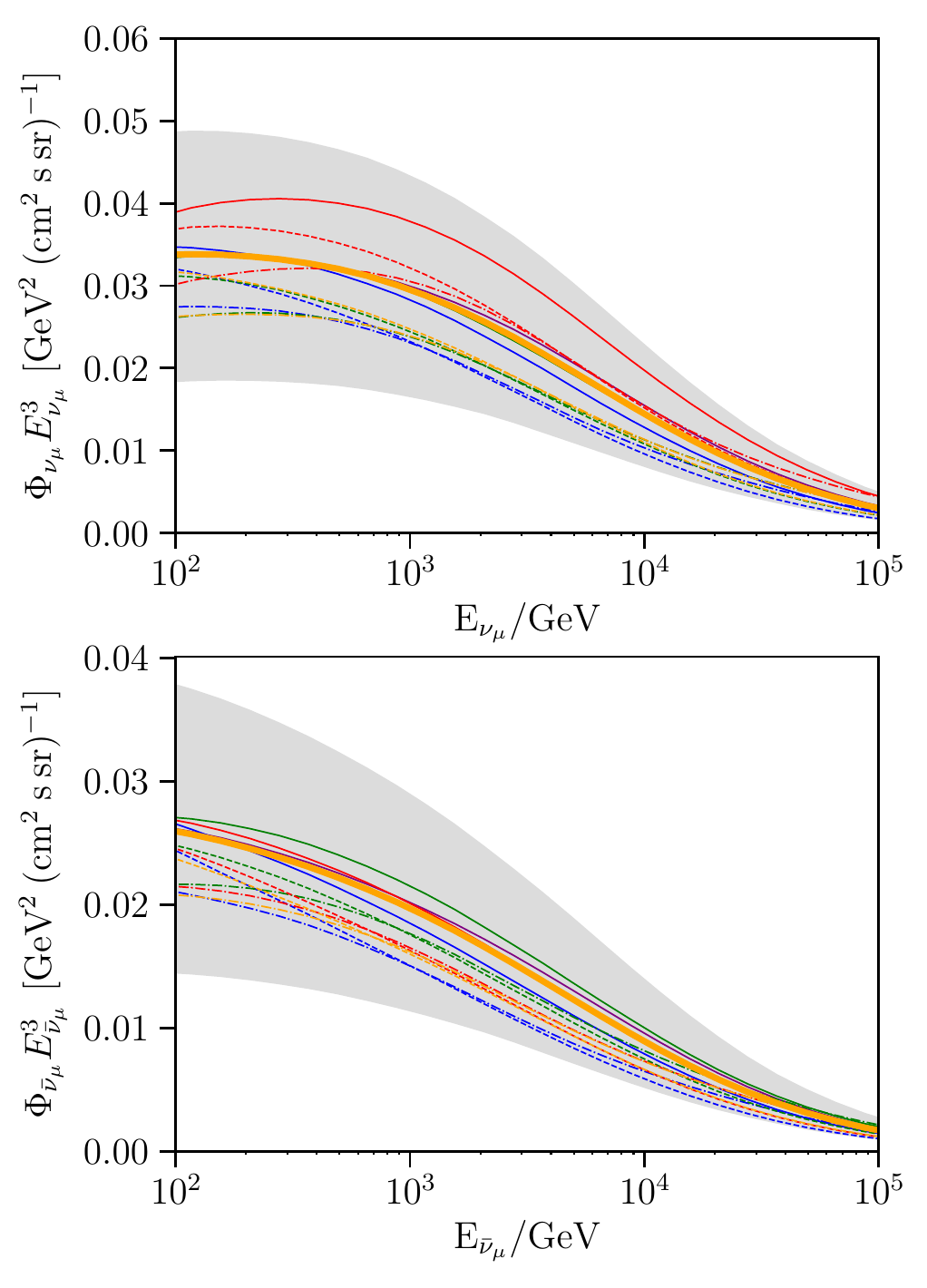}
  \caption[Energy spectrum band from Barr scheme]{The spectrum of atmospheric muon (top) neutrinos and (bottom) antineutrinos. The fluxes are scaled by energy cubed. The different colors correspond to different hadronic interaction models. The different linestyles correspond to different primary cosmic ray models. The grey band represents the $1\sigma$ uncertainty around the central model when considering eight relevant nuisance parameters. The fluxes shown here are averaged over the cosine zenith angles.}
  \label{fig:barr_envelope_edist}
\end{figure}

\begin{figure}[H]
  \centering
  \includegraphics[clip, trim = 9.3cm 1.3cm 0.5cm 3.5cm, width=0.4\textwidth]{figs_flux/plot_just_for_legend.pdf}
	\includegraphics[width=0.6\textwidth]{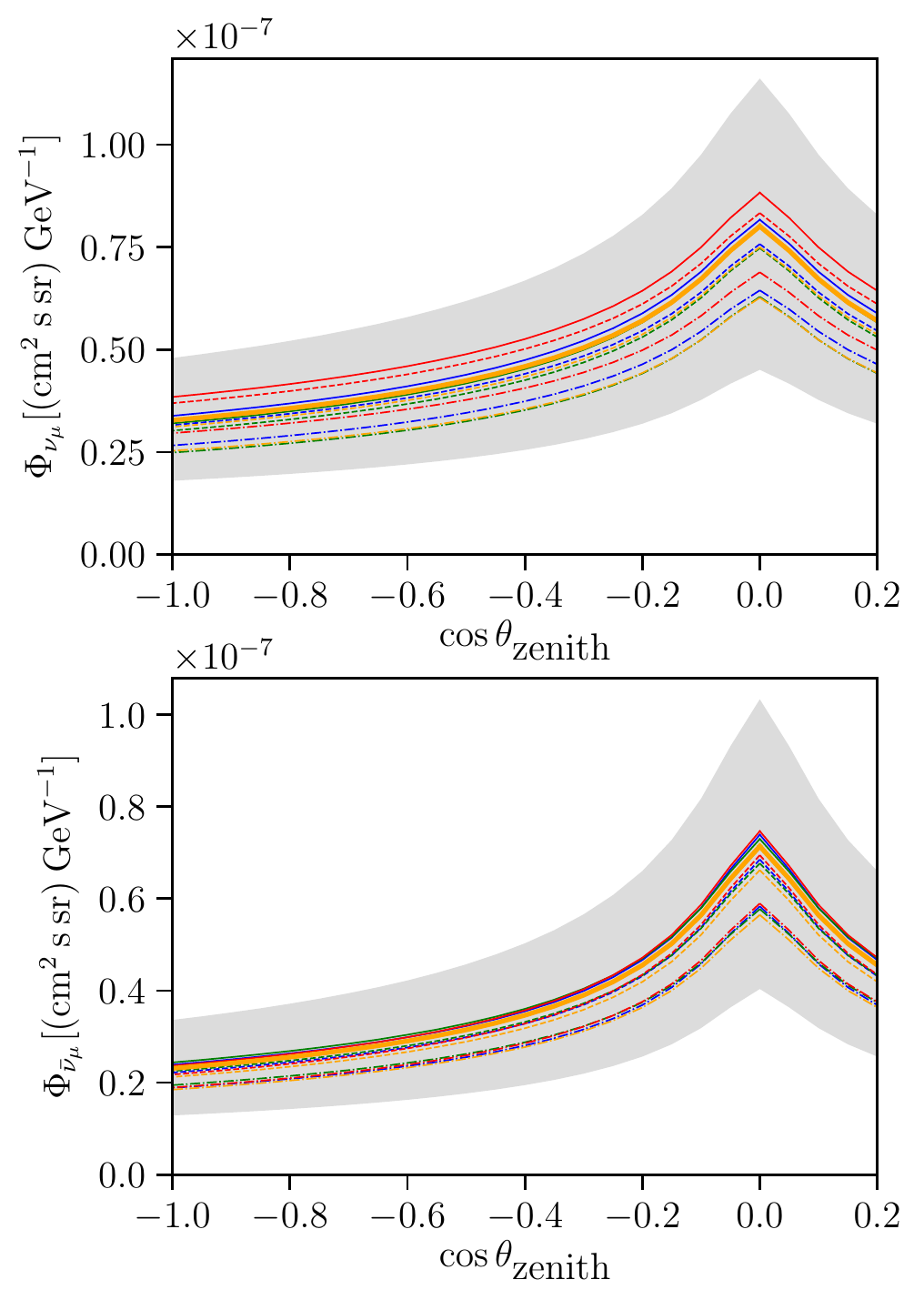}
  \caption[Angular distribution spanned by Barr scheme at 90~GeV]{The angular distribution of atmospheric muon (top) neutrinos and (bottom) antineutrinos at 90~GeV. The different colors correspond to different hadronic interaction models. The different linestyles correspond to different primary cosmic ray models. The grey band represents the $1\sigma$ uncertainty around the central model when considering eight relevant nuisance parameters. The fluxes shown here are averaged over the cosine of the zenith angles.}
  \label{fig:barr_envelope_edist_90GeV}
\end{figure}

\begin{figure}[H]
  \centering
  \includegraphics[clip, trim = 9.3cm 1.3cm 0.5cm 3.5cm, width=0.4\textwidth]{figs_flux/plot_just_for_legend.pdf}
	\includegraphics[width=0.6\textwidth]{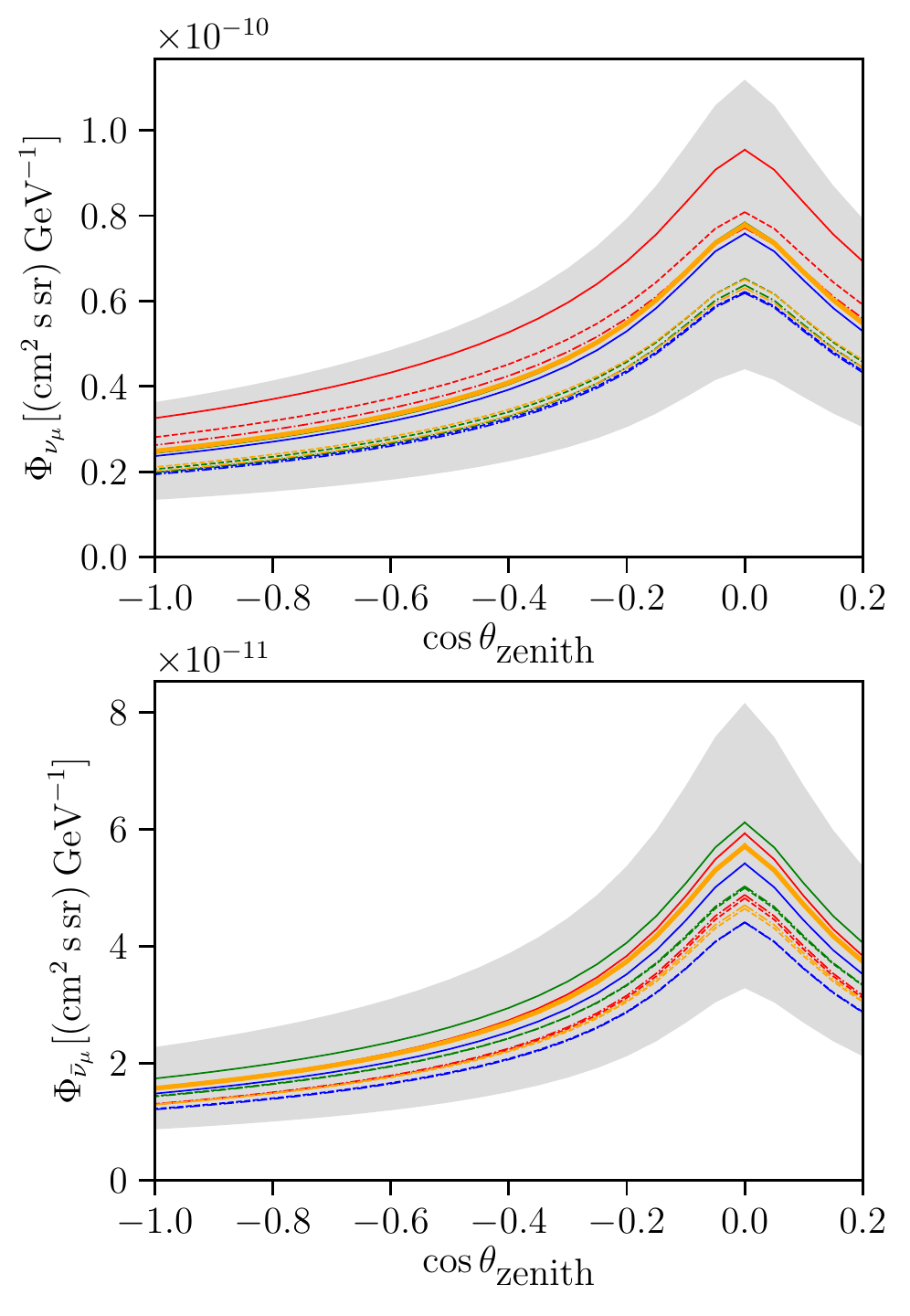}
  \caption[Angular distribution spanned by Barr scheme at 900~GeV]{The angular distribution of atmospheric muon (top) neutrinos and (bottom) antineutrinos at 900~GeV. The different colors correspond to different hadronic interaction models. The different linestyles correspond to different primary cosmic ray models. The grey band represents the $1\sigma$ uncertainty around the central model when considering eight relevant nuisance parameters. The fluxes shown here are averaged over the cosine of the zenith angles.}
  \label{fig:barr_envelope_edist_900GeV}
\end{figure}

\begin{figure}[H]
  \centering
  \includegraphics[clip, trim = 9.3cm 1.3cm 0.5cm 3.5cm, width=0.4\textwidth]{figs_flux/plot_just_for_legend.pdf}
	\includegraphics[width=0.6\textwidth]{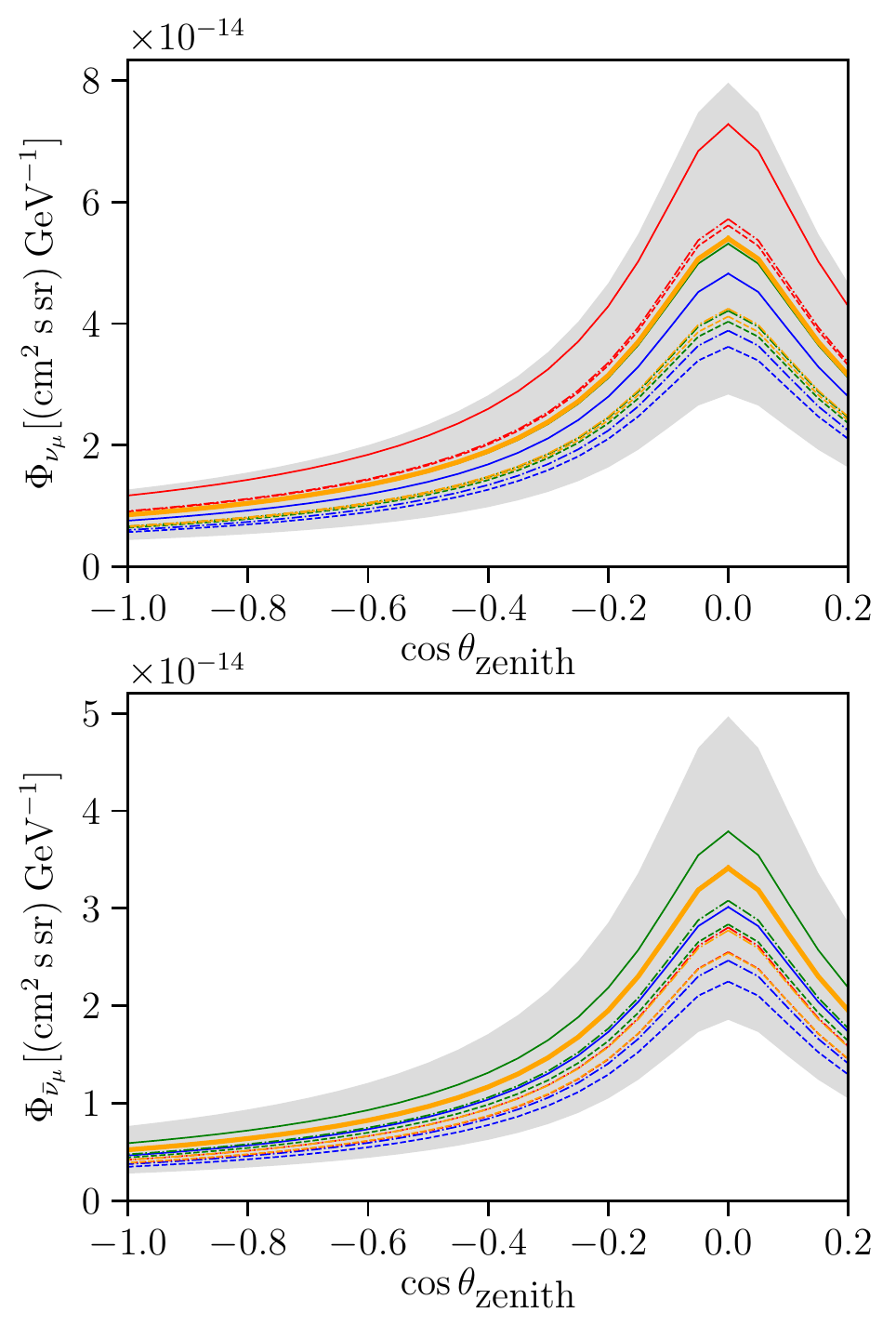}
  \caption[Angular distribution spanned by Barr scheme at 9~TeV]{The angular distribution of atmospheric muon (top) neutrinos and (bottom) antineutrinos at 9~TeV. The different colors correspond to different hadronic interaction models. The different linestyles correspond to different primary cosmic ray models. The grey band represents the $1\sigma$ uncertainty around the central model when considering eight relevant nuisance parameters. The fluxes shown here are averaged over the cosine of the zenith angles.}
  \label{fig:barr_envelope_edist_9TeV}
\end{figure}

\begin{figure}[H]
  \centering
  \includegraphics[clip, trim = 9.3cm 1.3cm 0.5cm 3.5cm, width=0.4\textwidth]{figs_flux/plot_just_for_legend.pdf}
	\includegraphics[width=0.6\textwidth]{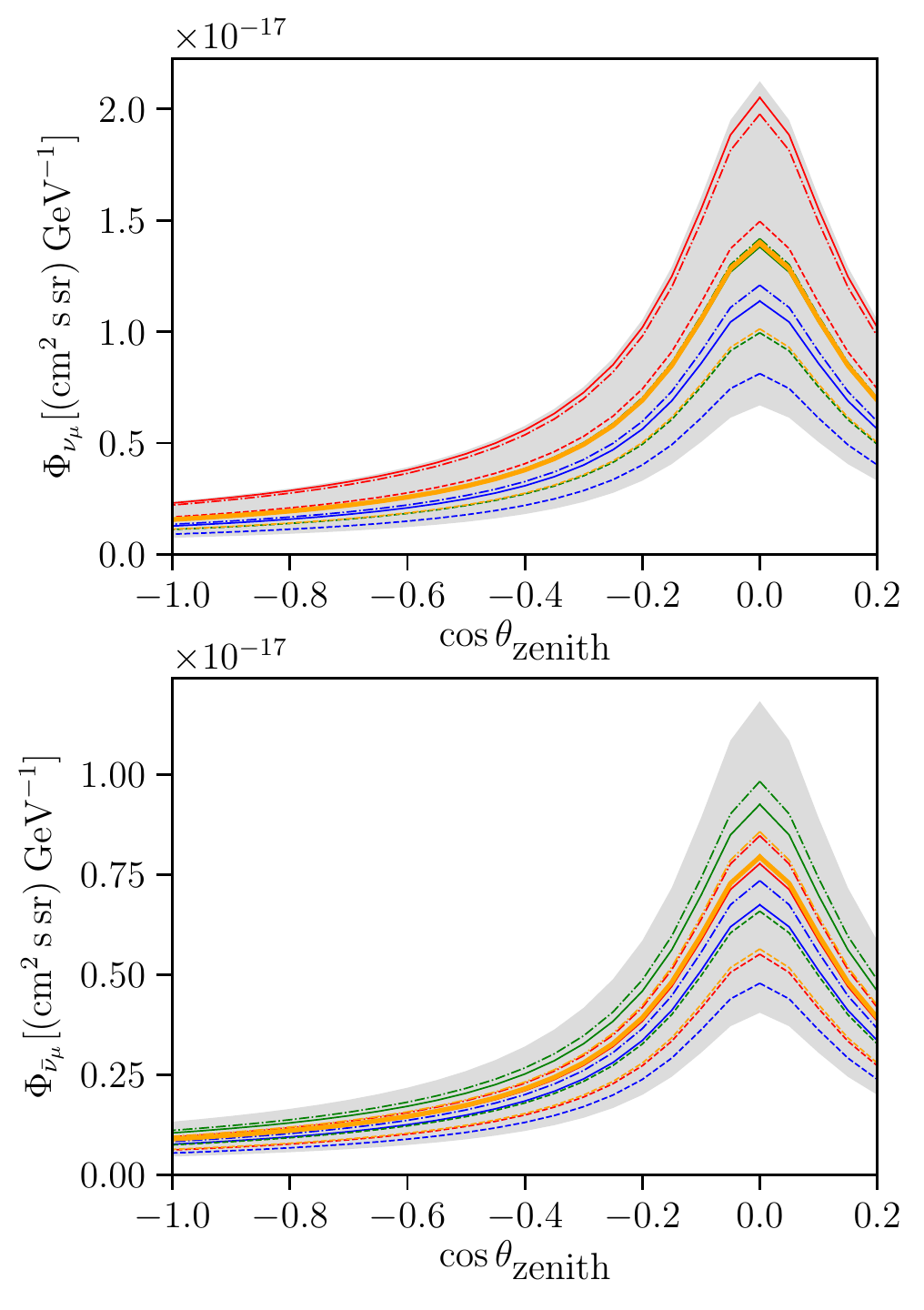}
  \caption[Angular distribution spanned by Barr scheme at 90~TeV]{The angular distribution of atmospheric muon (top) neutrinos and (bottom) antineutrinos at 90~TeV. The different colors correspond to different hadronic interaction models. The different linestyles correspond to different primary cosmic ray models. The grey band represents the $1\sigma$ uncertainty around the central model when considering eight relevant nuisance parameters. The fluxes shown here are averaged over the cosine of the zenith angles.}
  \label{fig:barr_envelope_edist_90TeV}
\end{figure}

\begin{figure}[H]
  \centering
  \includegraphics[clip, trim = 9.3cm 1.3cm 0.5cm 3.5cm, width=0.4\textwidth]{figs_flux/plot_just_for_legend.pdf}
 \includegraphics[width=0.75\textwidth]{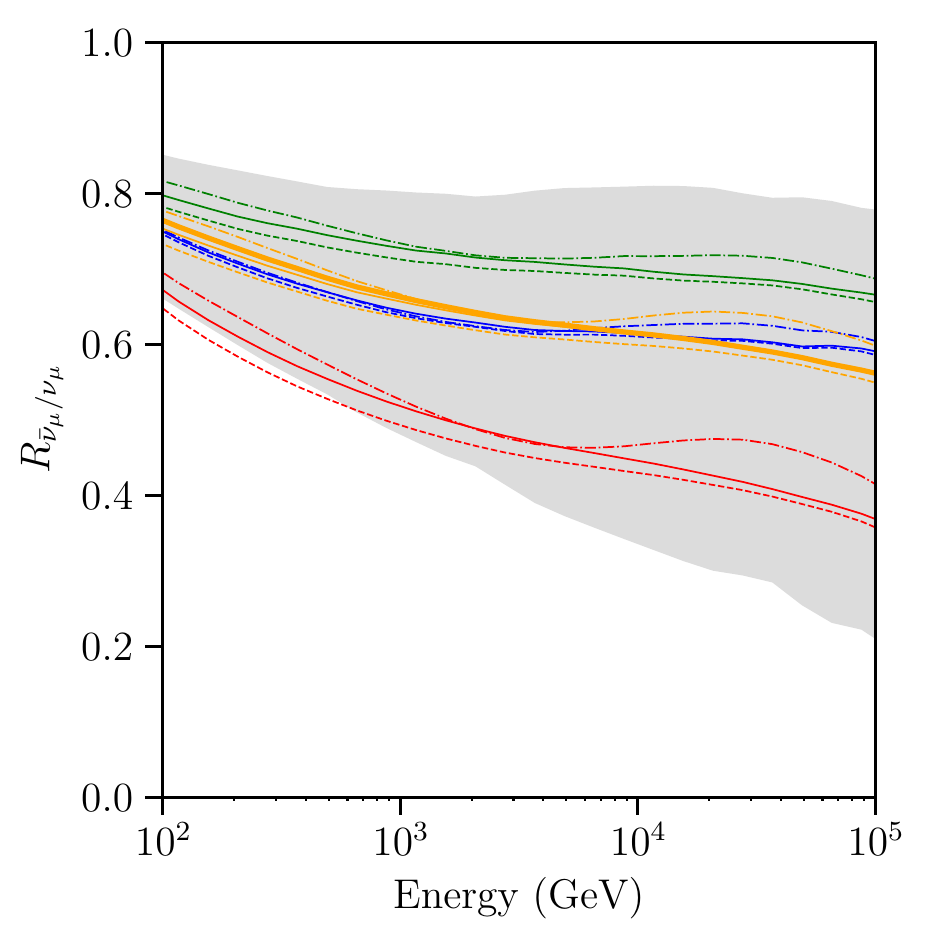}
  \caption[Ratio of antineutrinos to neutrinos spanned by Barr scheme]{Ratio of atmospheric muon antineutrinos to neutrinos, as a function of energy. The different colors correspond to different hadronic interaction models. The different linestyles correspond to different primary cosmic ray models. The grey band represents the $1\sigma$ uncertainty around the central model when considering eight relevant nuisance parameters. The fluxes shown here are averaged over the cosine of the zenith angles.}
  \label{fig:barr_nubar_nu_true}
\end{figure}

In this work, the central prediction of the neutrino flux is that from Hillas-Gaisser 2012 H3a as the cosmic ray model, and Sibyll2.3c as the hadronic interaction model. Figs.~\labelcref{fig:barr_envelope_edist,fig:barr_envelope_edist_90GeV,fig:barr_envelope_edist_900GeV,fig:barr_envelope_edist_9TeV,fig:barr_envelope_edist_90TeV,fig:barr_nubar_nu_true} show that the uncertainty on the flux prediction using this central model and considering this set of eight systematic uncertainties spans the predictions from the combinations of three cosmic ray models and four hadronic interaction models. In these plots, the curves represent the discrete flux predictions, the color indicates the hadronic interaction model, and the linestyle indicates the cosmic ray model. The grey band shows the $1\sigma$ range of predictions based off the central model and the eight systematic uncertainties considered here. These uncertainties are considered uncorrelated, and their effects are added in quadrature. Figure~\ref{fig:barr_envelope_edist} shows that these eight systematic uncertainties cover the discrete, predicted energy distributions of the muon neutrino and antineutrino fluxes. The distributions shown are averaged over the zenith angles. Figures~\labelcref{fig:barr_envelope_edist_90GeV,fig:barr_envelope_edist_900GeV,fig:barr_envelope_edist_9TeV,fig:barr_envelope_edist_90TeV} show these uncertainties cover the discrete, predicted cosine zenith distributions, at the energies 90~GeV, 900~GeV, 9~TeV and 90~TeV. Finally, Fig.~\ref{fig:barr_nubar_nu_true} shows that the discrete predictions of the muon antineutrino to neutrino ratio, as a function of energy, are spanned by these systematic uncertainties.

A benefit of this uncertainty parameterization is that the discrepancy in the atmospheric fluxes from the various models can be included in an analysis in a continuous manner. It is not necessary to 
perform the sterile search analysis multiple times, assuming a different, discrete model each time, as was done in the one-year search~\cite{Jones:2015bya,Collin:2018juc}. Instead one can incorporate these eight systematic uncertainties as continuous nuisance parameters that reflect the underlying physical uncertainties. Furthermore, the posterior distributions of these nuisance parameters are informative of the respective hadronic processes.

In addition to the eight atmospheric flux systematic uncertainties discussed here, the analysis in this thesis includes two more, mentioned here for completeness. One is the atmospheric density. The other is the cross section of kaons interacting with atmospheric nuclei. These will be discussed further in \Cref{ch:analysis} and are not included in the envelopes shown here.

\hfill

\section{Neutrino flux at detector}
For the work in this thesis, nine components to the total neutrino flux are calculated. Seven relate to the conventional atmospheric flux: the central model and the six Barr fluxes. The atmospheric density profile assumed in the calculation of these fluxes is based off data from AIRS satellite~\cite{AIRS}. The prompt component of the conventional flux is included based on the prediction from the BERSS model~\cite{Bhattacharya:2015jpa}. The final component is the astrophysical flux.

Astrophysical neutrinos were discovered by IceCube~\cite{aartsen2014observation}; their origin remains unknown. The model of the astrophysical flux used in this work is based on previous measurements from IceCube~\cite{Abbasi:2020jmh, aartsen2016observation, Schneider:2019ayi, Stachurska:2019}. The astrophysical flux is assumed to be isotropic and to follow a single power law,
\begin{equation}
\frac{dN_{\nu_\mu,\bar{\nu}_\mu}}{dE} = N_{\textrm{astro}} \bigg( \frac{E}{100~\textrm{TeV}} \bigg)^{-2.5},
\end{equation}
where the normalization is
\begin{equation}
N_{\textrm{astro}}  = 0.787 \times 10^{-18}~\textrm{GeV}^{-1}~\textrm{sr}^{-1}~\textrm{s}^{-1}~\textrm{cm}^{-2}.
\end{equation}
This assumes the astrophysical flux is equally composed of the electron, muon and tau flavors, and is equally composed of neutrinos and antineutrinos. The uncertainties on the normalization and spectral index are significant and correlated, and discussed further in \Cref{ch:analysis}.

The prediction of the neutrino flux incident on the Earth is converted into a prediction of the flux at the IceCube detector. This calculation is done with the neutrino evolution code \texttt{nuSQuIDS}~\cite{ARGUELLESDELGADO2015569}. This software accounts for processes that affects the propagation of neutrinos through the Earth. One process is neutrino oscillations between four flavors, including the effects due to matter. Another significant process is charged and neutral current interactions in the Earth, which lead to neutrino absorption at high energies and long baselines. A subdominant process is tau regeneration, which is the production of neutrinos from charged current interactions of tau neutrinos and subsequent tau decay~\cite{PhysRevLett.81.4305}. The Earth is assumed to be spherically symmetric. The density profile of the Earth is assumed to follow the Preliminary Reference Earth Model (PREM)~\cite{DZIEWONSKI1981297}.

Lastly, neutrino decay affects the propagation of neutrinos. This effect is handled by~\texttt{nuSQUIDSDecay}, a class specialization of \texttt{nuSQuIDS}~\cite{Moss:2017pur}. Neutrinos are assumed to be Dirac. The coupling that mediates the decay is assumed to be purely scalar. The fourth mass state may decay to a massless scalar and another neutrino. A right-handed neutrino would decay to a left-handed neutrino, while a right-handed antineutrino would decay to a left-handed antineutrino. These daughter neutrinos, as well as the daughter scalar particle, are not detectable. The value of the coupling is converted to a lifetime with Eq.~\ref{eq:coupling_mass_lifetime}. The \texttt{nuSQUIDSDecay} partial rate constructor is used, where the chirality-violating process rate is the inverse of the lifetime.

\chapter{Search for unstable sterile neutrinos}
\label{ch:analysis}

\section{Data selection}
This analysis uses a large set of well-reconstructed, upwards-going, high-energy, muon data events. This dataset is assembled using the event selection described briefly in this section. More details on the event selection can be found in~\cite{Axani:2020zdv, Aartsen:2020fwb, Aartsen:2015rwa, Weaver:2015bja}.

The event selection is a set of criteria designed to select signal events with high efficiency and to reject background events, yielding a high purity. The signal events are upwards-going muons. These are known to come from the charged-current interactions of atmospheric and astrophysical muon neutrinos and antineutrinos. The dominant background is atmospheric muons from cosmic ray air showers. IceCube detects these at a rate of about 3 kHz, while the rate of signal events is about 1~mHz~\cite{Aartsen:2015nss}. Simulation is used to tune the background rejection of the event selection. The software \texttt{CORSIKA} (COsmic Ray SImulations for KAscade) is used to simulate muon production in cosmic ray air showers in the atmosphere\cite{heck1998corsika, heck2000extensive}. \texttt{PROPOSAL} is then used to propagate the muons through the firn and ice to the detector~\cite{KOEHNE20132070}. The simulated cosmic ray muons are then weighted according to the Hillas-Gaisser H3a cosmic ray model~\cite{Gaisser:2013bla}. Other backgrounds include neutral current neutrino events and charged-current electron and tau neutrino events. These events are associated with cascades, which have a very different morphology than the signal track-like events, allowing them to be readily rejected. Furthermore, the flux of atmospheric electron and tau neutrinos is subdominant to that of muon neutrinos. The electron and tau neutrino backgrounds are also simulated.

The event selection criteria is the union of two sets of criteria, or filters, described further  in~\cite{Axani:2020zdv, Aartsen:2020fwb, Aartsen:2015rwa, Weaver:2015bja}. All events are required to:
\begin{itemize}
\item Pass the IceCube muon filter, which identifies track-like events. 
\item Have at least 15 DOMs triggered, with at least 6 triggered on direct light. Direct light is photons which have not significantly scattered, as determined by timing.
\item Have a reconstructed track length, based off of direct light, of at least 200~m.
\item Have a relatively smooth distribution of direct light along the track.
\item Have a reconstructed energy between 500~GeV and 9976~TeV.
\item Originate from or below the horizon: $\cos(\theta_{\rm{zenith}}) \leq 0$.
\end{itemize}

\begin{figure}
\centering
\includegraphics[width=0.75\linewidth]{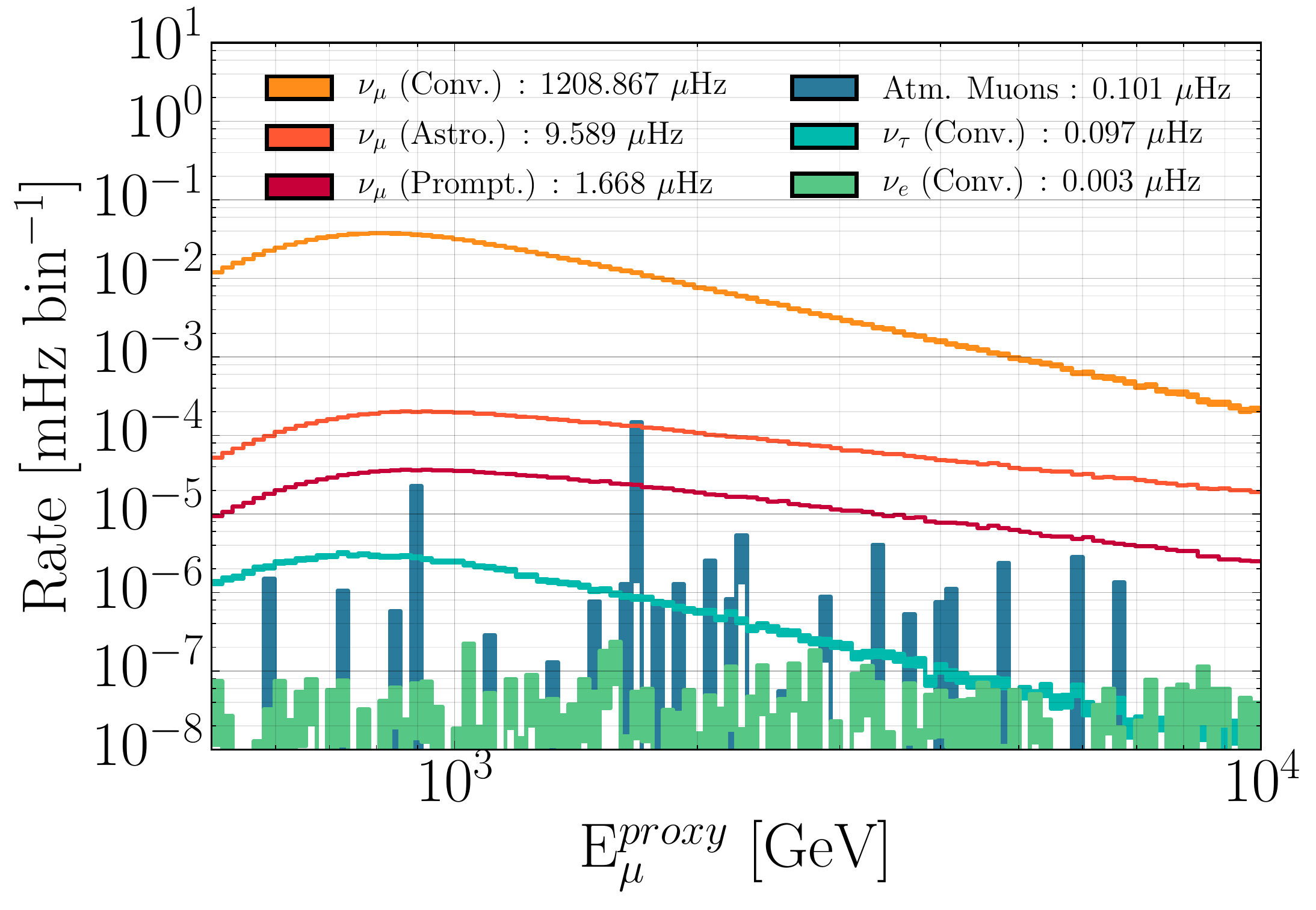}
\includegraphics[width=0.75\linewidth]{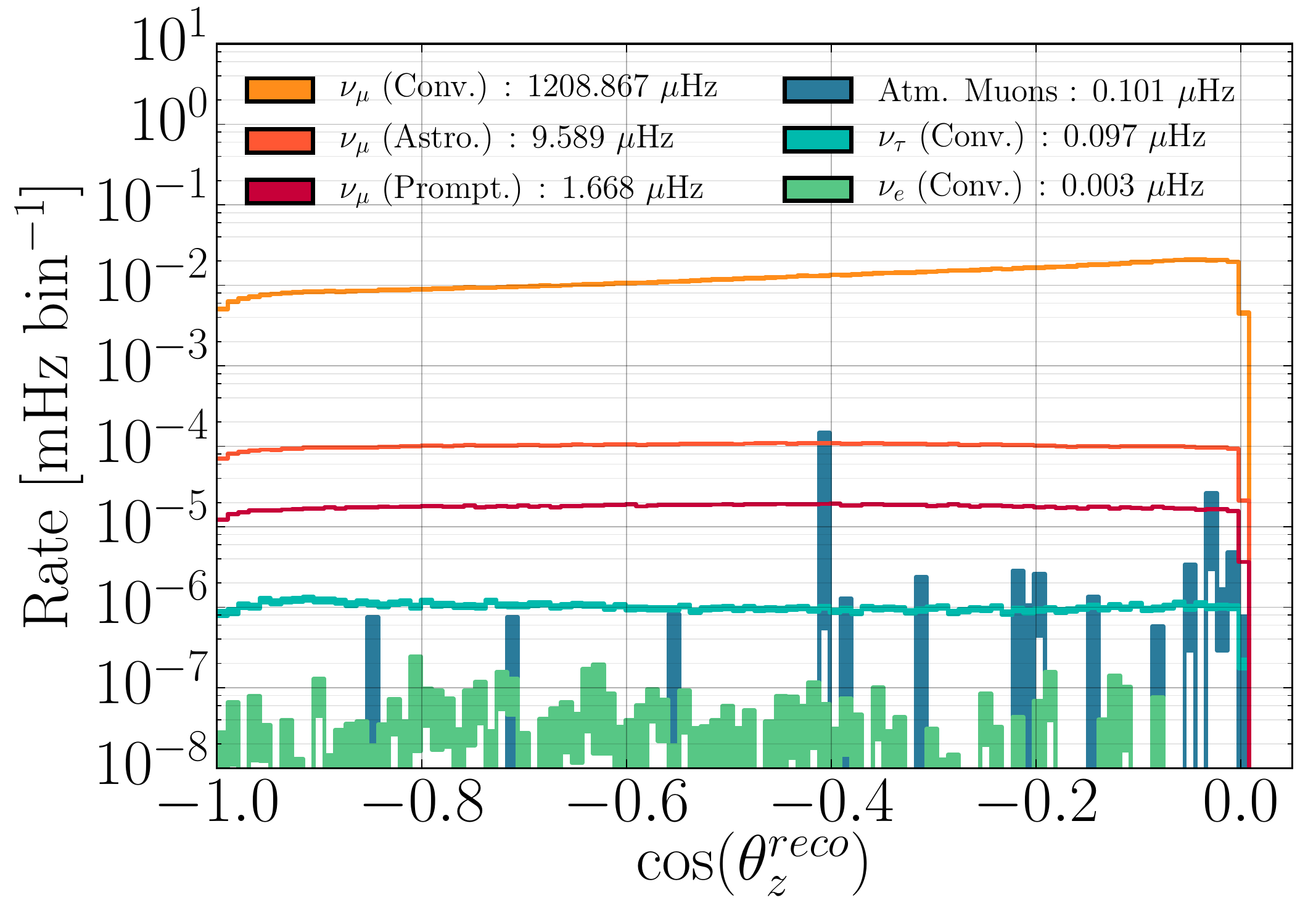}
\caption[Event selection energy and angular distribution from simulation]{(Top) Reconstructed energy distribution and (bottom) reconstructed cosine zenith distribution of the constituents of the event selection, as determined from simulation. Figures from~\cite{Axani:2020zdv, Aartsen:2020fwb}}
\label{fig:mc_event_selection_distributions}
\end{figure}

The event selection has a 99.9\% purity as determined from simulation. The expected composition of the event selection is given in Table~\ref{tab:event_composition}. The energy and cosine zenith distributions of these events is shown in~\ref{fig:mc_event_selection_distributions}.

\begin{table}[h]
\begin{center}
\begin{tabular}{l c}
\hline
\hline
Event Type & Expected Count \\
\hline
Conventional atmospheric $\nu_\mu$ & $315,214 \pm 561$ \\
Astrophysical $\nu_\mu$ & $2,350 \pm 48$\\
Prompt atmospheric $\nu_\mu$ & $481 \pm 22$\\
All $\nu_\tau$ & $23\pm5$ \\
All $\nu_e$ & $1\pm1 $\\
Atmospheric $\mu$ & $18\pm4 $\\
\hline
\end{tabular}
\end{center}
\caption[Expected contributions to the event selection]{Expected partial contributions to the event selection. Table recreated from~\cite{Axani:2020zdv, Aartsen:2020fwb}.}
\label{tab:event_composition}
\end{table}

\section{Systematic uncertainties}
\subsection{Nuisance parameters}
The treatment of systematic uncertainties in this analysis is almost identical to that of the eight-year traditional 3+1 search and more detailed descriptions can be found in~\cite{Axani:2020zdv, Aartsen:2020fwb, Aartsen:2015rwa}. Eighteen systematic uncertainties are incorporated into the analysis with nuisance parameters. These uncertainties are grouped into two broad categories: those relating to the total neutrino flux and those relating to detection. A table of these nuisance parameters, their prior central values, their prior widths, and the boundaries on them is given in Table~\ref{table:Priors}.

As described in \Cref{chap:flux}, ten nuisance parameters are used to describe the uncertainties on atmospheric neutrino flux. One is the conventional atmospheric flux normalization: $\Phi_{\textrm{Conv.}}$. Another is the uncertainty on the spectral index of the atmospheric flux: $\Delta \gamma_{\textrm{atm}}$. This shifts both the conventional and prompt atmospheric fluxes. There are six uncertainties in the production of charged kaons in cosmic ray air showers. These are the six Barr parameters WM, WP, YM, YP, ZM, and ZP, previously discussed in \Cref{chap:flux}. Another is the uncertainty in the atmospheric density, which has been shown to affect the neutrino flux~\cite{Gaisser:2013lrk}. This effect is calculated using the reported systematic uncertainty of measurements from the AIRS satellite~\cite{AIRS}. Lastly, there is an uncertainty on the cross section of kaons interacting with atmospheric nuclei, largely nitrogen and oxygen nuclei: $\sigma_{\textrm{KA}}$. This causes an uncertainty in the energy of the kaons when they decay. In the parameterization here, 1$\sigma$ represents 7.5\% of the nominal cross section.

Two nuisance parameters are used to describe the uncertainties on the astrophysical neutrino flux. They are the astrophysical flux normalization and the uncertainty on the astrophysical spectral index.
These are correlated and are derived from previous IceCube measurements~\cite{Abbasi:2020jmh, aartsen2014observation, Schneider:2019ayi, Stachurska:2019}.

The two final uncertainties that affect the flux of neutrinos at the detector are the neutrino-nucleon and antineutrino-nucleon cross sections. These cross sections enter the analysis twice: during propagation of the neutrino flux across the Earth and during interactions near the detector. The latter had previously been studied and deemed negligible~\cite{Jones:2015bya,delgado2015new}. The effects of the cross section uncertainties on the flux incident at the detector are included here.

Four nuisance parameters describe the uncertainties in the detection of neutrinos and antineutrinos. One is the effective efficiency of detecting photons, which is referred to as the DOM efficiency. The DOM efficiency reflects properties of the DOM itself, such as the photocathode, collection, and wavelength efficiencies, as well as global effects that prevents detection of photons. These include shadowing from cables, and some properties of the bulk and hole ice. The effect is determined from simulation sets where photons are down-sampled.

Another detector uncertainty is due to the effect of the hole ice, represented by the ``forward hole ice'' parameter. The refrozen ice in the drilled boreholes has different optical properties from the bulk of the ice in the detector. In particular, there is additional scattering near the DOMs. As a result, the angular acceptance of the DOMs is different from laboratory measurements. The ``forward hole ice'' parameter reflects uncertainty in the zenith-dependence of the angular acceptance of DOMs.

Finally, this analysis includes two nuisance parameter related to the bulk ice, that is, the undisturbed, natural ice between and beyond the drilled boreholes. This analysis uses the ``SnowStorm'' method to calculate the effect of the uncertainties, which are highly correlated, of the many parameters in the ice model, shown in Fig.~\ref{fig:ice_coefficients}~\cite{Aartsen:2019jcj}. The Fourier transform of the depth-dependent coefficients is analyzed. The analysis-level effect of the most significant modes is encoded into two, correlated, effective gradients. These are referred to as ``Ice Gradient 0'' and ``Ice Gradient 1''.
	
\begin{table}[h!]
\begin{center}
 \begin{tabular}{ l c c c }
 \hline
 \hline
\textbf{Parameter} & Central & Prior (Constraint) & Boundary  \\ [0.5ex]
\hline
\hline
\multicolumn{4}{c}{\textbf{Detector parameters}}\\
\hline
DOM efficiency  & 0.97 & 0.97 $\pm$ 0.10  & [0.94, 1.03]  \\
\hline
Bulk Ice Gradient 0 & 0.0 &  0  $\pm$ 1.0* & NA  \\
\hline
Bulk Ice Gradient 1 & 0.0  & 0  $\pm$ 1.0* & NA \\
\hline
Forward Hole Ice & -1.0  & -1.0  $\pm$ 10.0  & [-5, 3]    \\
\hline
\multicolumn{4}{c}{\textbf{Conventional flux parameters}}\\
\hline
Normalization ($\Phi_{\mathrm{conv.}}$)  &  1.0 & 1.0  $\pm$ 0.4 & NA  \\
\hline
Spectral shift ($\Delta\gamma_{\mathrm{conv.}}$) & 0.00 & 0.00  $\pm$ 0.03  & NA \\
\hline
Atm. Density   &  0.0 & 0.0  $\pm$ 1.0& NA \\
\hline
Barr WM       &  0.0 & 0.0  $\pm$ 0.40 & [-0.5, 0.5] \\
\hline
Barr WP       &  0.0 & 0.0  $\pm$0.40   & [-0.5, 0.5] \\
\hline
Barr YM       &  0.0 & 0.0  $\pm$ 0.30  & [-0.5, 0.5] \\
\hline
Barr YP       &  0.0 & 0.0  $\pm$0.30 & [-0.5, 0.5]  \\
\hline
Barr ZM       &  0.0 & 0.0  $\pm$ 0.12 & [-0.25, 0.5]   \\
\hline
Barr ZP       &  0.0 & 0.0  $\pm$ 0.12  & [-0.2, 0.5]  \\
\hline
\multicolumn{4}{c}{\textbf{Astrophysical flux parameters}}\\
\hline
Normalization ($\Phi_{\mathrm{astro.}}$)    &  0.787 & 0.787  $\pm$ 0.36*  & NA \\
\hline
Spectral shift ($\Delta\gamma_{\mathrm{astro.}}$)      &  0 & 0.0  $\pm$ 0.36*  & NA \\
\hline
\multicolumn{4}{c}{\textbf{Cross sections}}\\
\hline
Cross section $\sigma_{\nu_\mu}$    &  1.00 & 1.00  $\pm$ 0.03  & [0.5, 1.5]  \\
\hline
Cross section $\sigma_{\overline{\nu}_\mu}$      &  1.000 & 1.000  $\pm$ 0.075 & [0.5, 1.5]   \\
\hline
Kaon energy loss $\sigma_{KA}$   &  0.0 & 0.0  $\pm$ 1.0  &  NA \\
\hline
\hline
\end{tabular}
\caption[Nuisance parameter central values, priors and boundaries]{Table of nuisance parameters, their central values, their priors and boundaries. Correlation between two nuisance parameters is indicated with *. Table reproduced from~\cite{Aartsen:2020fwb}.}
\label{table:Priors}
\end{center}
\end{table}

\begin{figure}
\centering
\includegraphics[clip, trim = 0cm 0cm 14.8cm 0cm width=\linewidth]{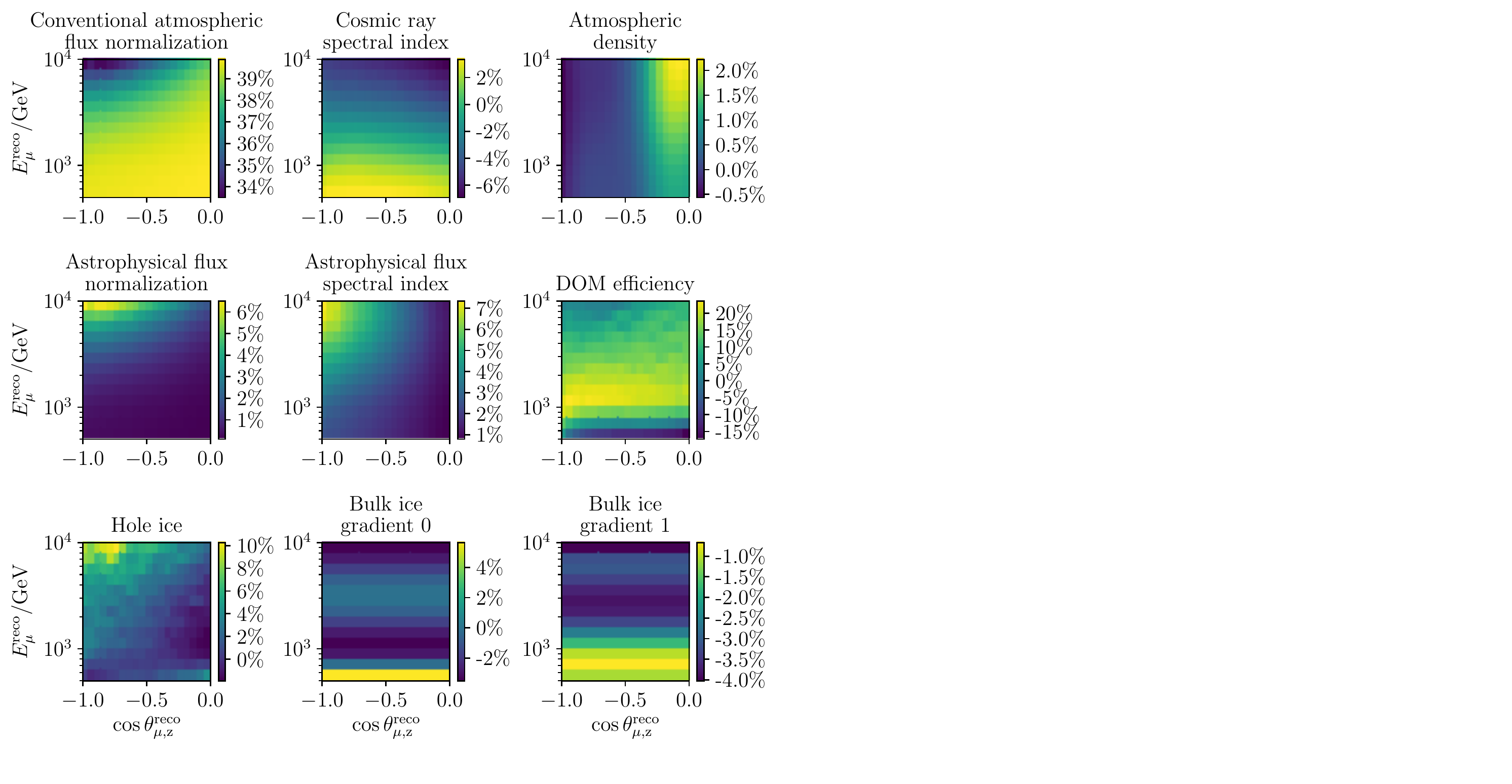}
\caption[Effect of $1\sigma$ change in nuisance parameters]{Percent difference between the expected event distribution with one nuisance parameter value increased by $1\sigma$ and the nominal expectation, shown for half the nuisance parameters. The expectations assume no sterile neurino.}
\label{fig:systematics_firsthalf}
\end{figure}

\begin{figure}
\centering
\includegraphics[clip, trim = 0cm 0cm 14.8cm 0cm width=\linewidth]{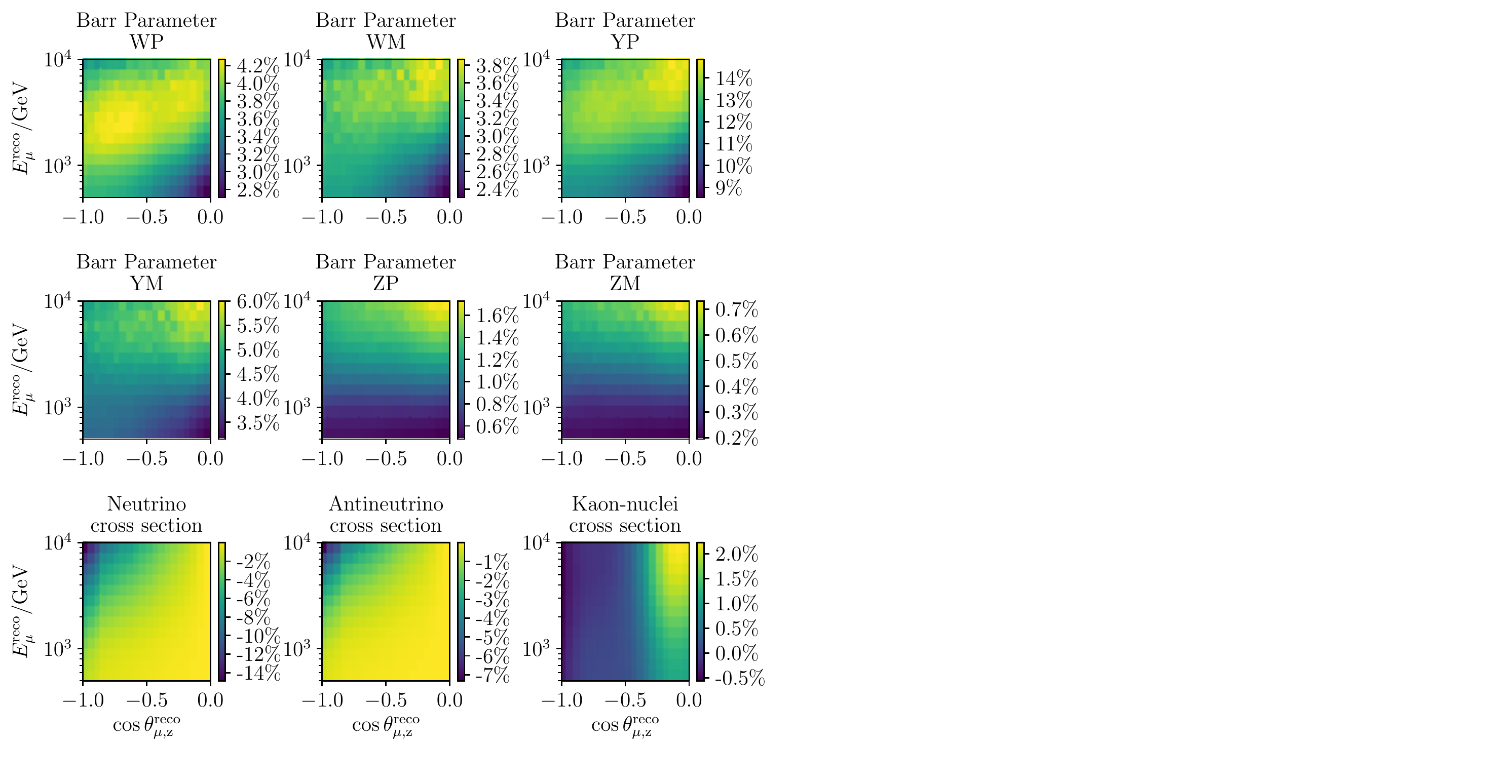}
\caption[Effect of $1\sigma$ change in nuisance parameters]{Percent difference between the expected event distribution with one nuisance parameter value increased by $1\sigma$ and the nominal expectation, shown for half the nuisance parameters. The expectations assume no sterile neurino.}
\label{fig:systematics_secondhalf}
\end{figure}

The shapes of the eighteen nuisance parameter systematics are shown in Figs.~\ref{fig:systematics_firsthalf} and~\ref{fig:systematics_secondhalf}. These figures show the fractional change that a positive $1\sigma$ variation of each of the nuisance parameters causes,
\begin{equation}
\frac{(\textrm{Nominal}+1\sigma \textrm{ variation}) - \textrm{Nominal}}{\textrm{Nominal}},
\end{equation}
assuming the null hypothesis, that is, only three neutrinos. The shapes of these systematic effects have been shown previously in~\cite{Axani:2020zdv, Aartsen:2020fwb}. One notable difference is that in~\cite{Axani:2020zdv, Aartsen:2020fwb} the overall normalization effects had been removed from all shapes except for the atmospheric and astrophysical flux normalizations. Two other differences relate to the Barr gradient and astrophysical fluxes, and are noted in~\Cref{sec:minorsysimprovements}.

\clearpage

\subsection{Minor systematic improvements from traditional 3+1 search}
\label{sec:minorsysimprovements}
Three minor improvements were made after the eight-year traditional 3+1 search:
\begin{itemize}
\item This analysis assumes a livetime of 7.634 years instead of 7.6. No additional data was added. The assumed value of the livetime was changed to be more precise.
\item The Barr gradients are calculated using atmospheric data from the AIRS satellite, instead of using a model of the atmosphere (US Standard)~\cite{AIRS, atmosphere1976national}. This makes the Barr gradient calculations consistent with that of the conventional atmospheric flux.
\item The astrophysical and prompt fluxes are calculated with a more accurate model of the Earth's density profile. This model has 3 km of ice below the surface of the Earth, instead of 30 km of ice. This change makes the calculation of these fluxes consistent with that of the conventional atmospheric flux.
\end{itemize}
This changes were made for consistency and accuracy, and are not expected to significantly affect the result.

\subsection{The Antarctic bedrock}
\label{sec:bedrock}

\begin{figure}[htb]
\centering
\includegraphics[width=4in]{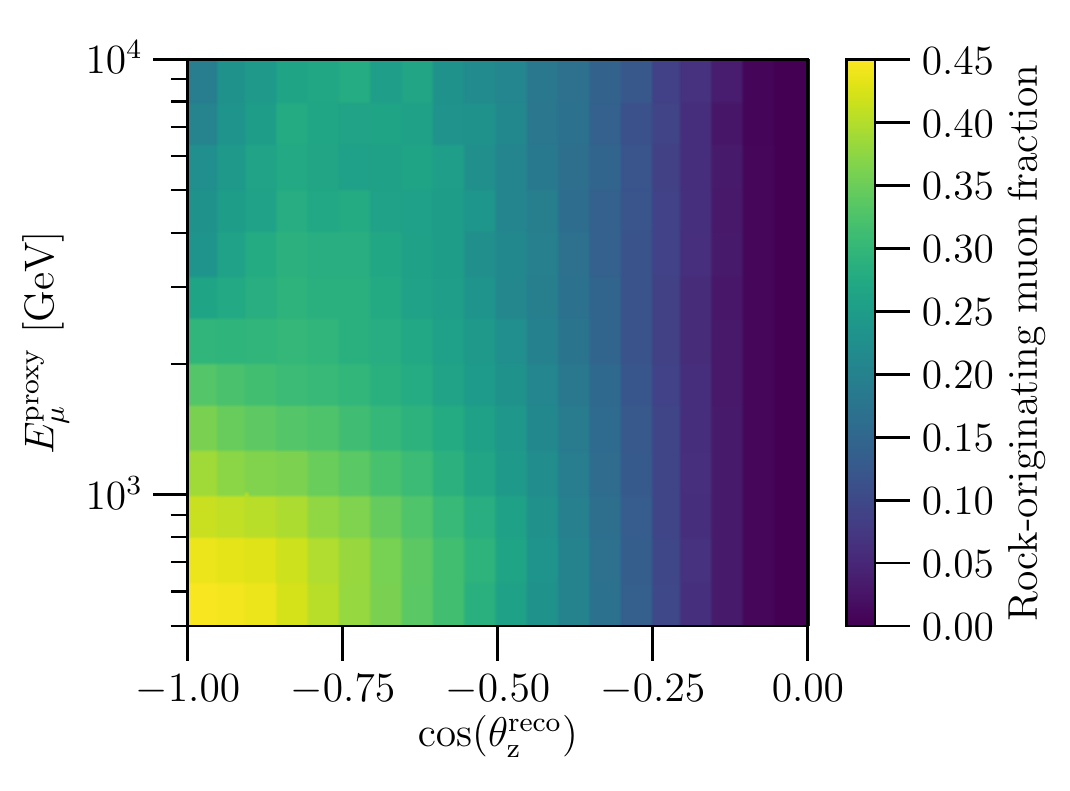}
\caption[Fraction of events that originate in the bedrock]{Fraction of events per bin that originate in the bedrock, rather than in the ice, as determined from simulation.}
\label{fig:bedrock_fraction}
\end{figure}

The Antarctic bedrock is located approximately 362~m below the bottom of the detector. Approximately 20\% of all the muons in the event selection originate in the bedrock. However, this fraction has a zenith dependence, peaking at the smallest zenith angles, as well as an energy dependence, which peaks at lower energies, shown in Fig.~\ref{fig:bedrock_fraction}. Two uncertainties associated with the bedrock are its position and its density.

\begin{figure}
\centering
\includegraphics[width=0.8\linewidth]{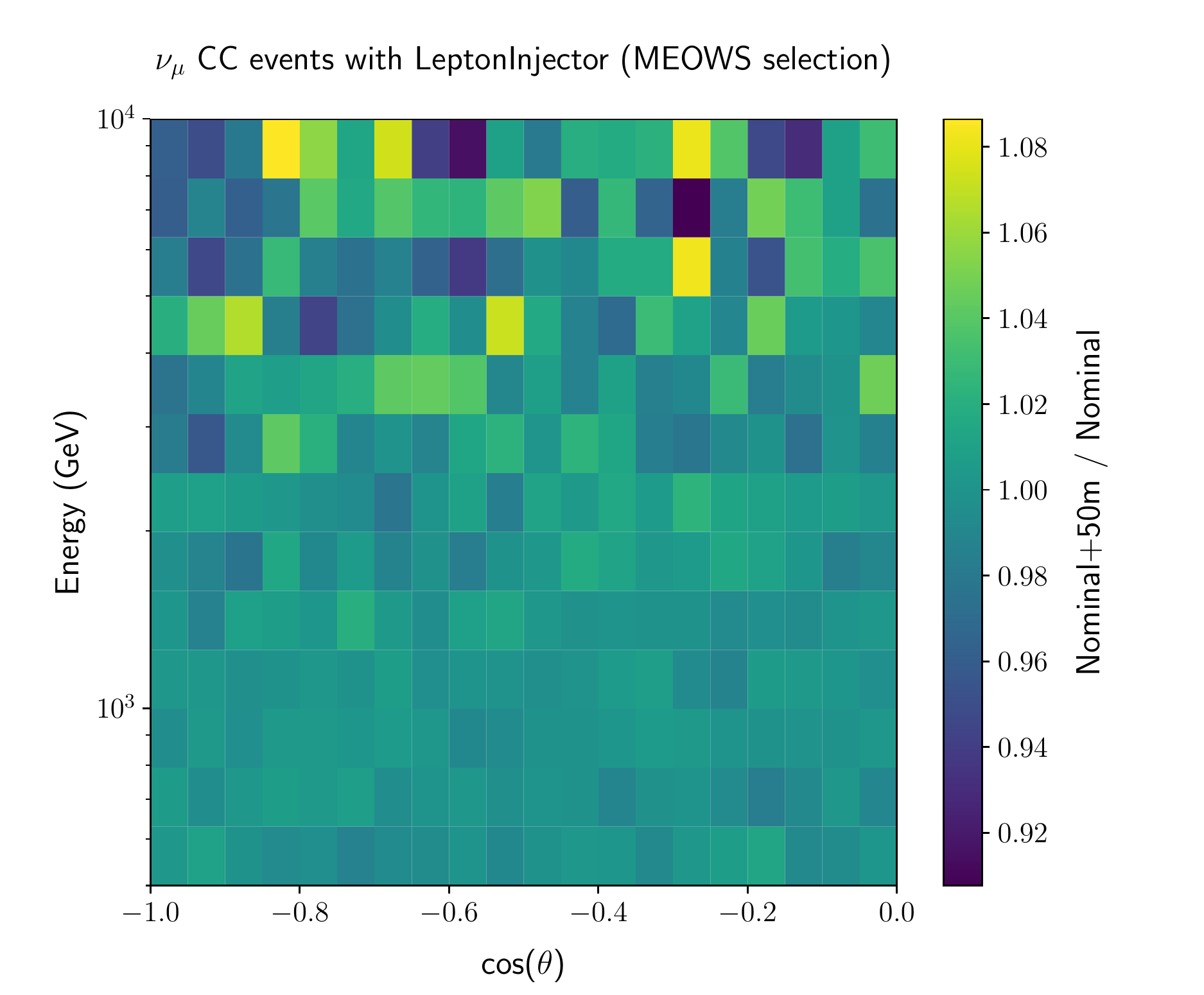}
\includegraphics[width=0.8\linewidth]{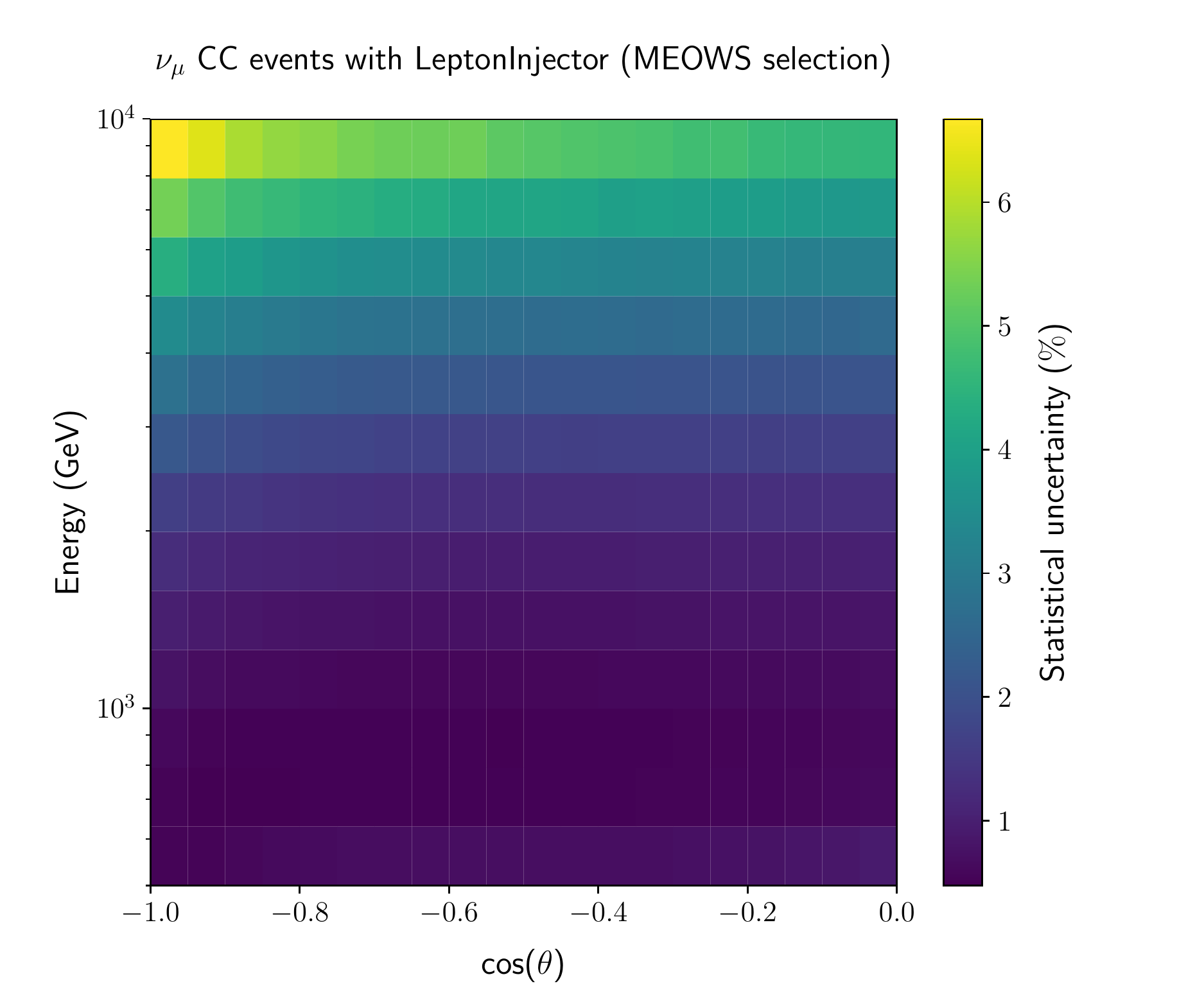}
\caption[Effect of 50 m change in bedrock depth]{(Top) The ratio of muon events per bin for a geometry with the bedrock raised 50~m to that for the nominal geometry. (Bottom) The statistical uncertainty in the simulation sets plotted above. Figures courtesy of David Vannerom.}
\label{fig:bedrock_position}
\end{figure}

Several measurements of the position of the bedrock have been made. The PolarGap campaign has made LIDAR and RADAR measurements over the South Pole~\cite{2016AGUFM.C52A..01F, forsberg_2017}. Deviations of about 45~m in the bedrock depth across the detector are indicated from the data~\cite{koskinen_2017}. Simulation studies were performed to study the effect of an uncertainty on the bedrock position. A simulation set was made where the position of the bedrock was raised 50~m with respect to the nominal position. The muon rate was found to be consistent with that of the the nominal geometry, up to the level of the statistical uncertainty, shown in Fig.~\ref{fig:bedrock_position}. The uncertainty in the position of the bedrock is thus deemed negligible for this work.

\begin{figure}[tb!]
\centering
\includegraphics[width=4in]{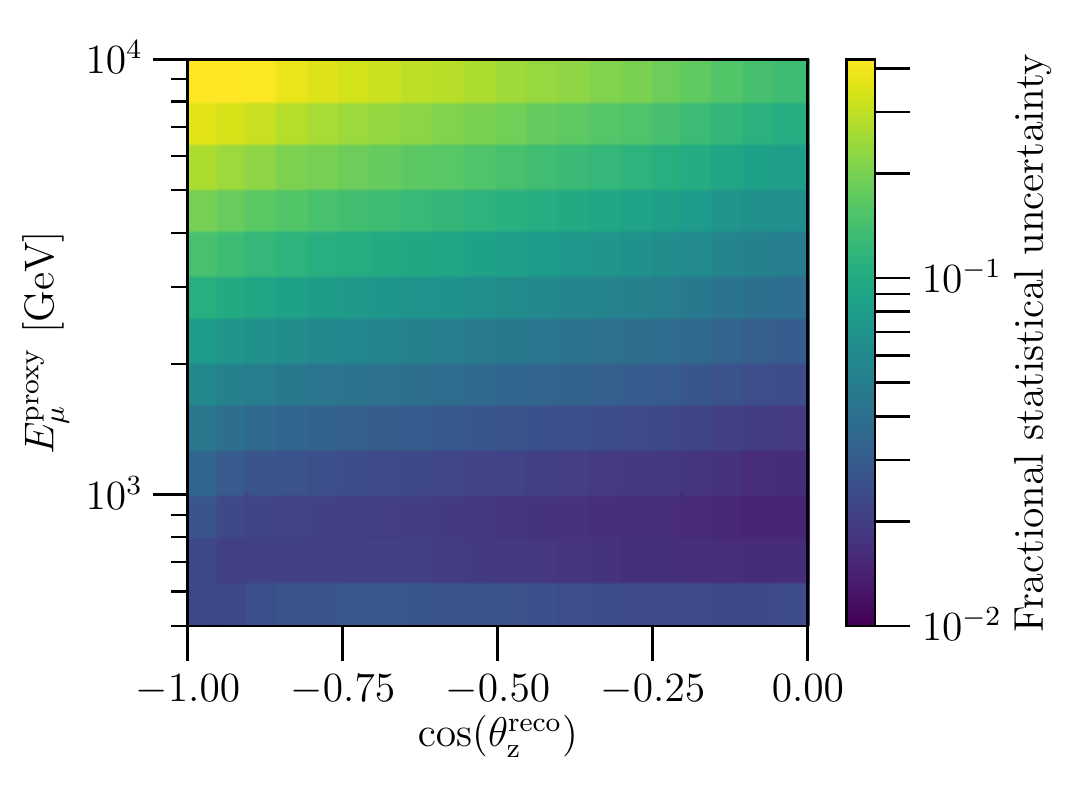}
\caption[Expected fractional statistical uncertainty]{Expected fractional statistical uncertainty per bin, assuming only three neutrinos. The asymmetrical Poisson errors are calculated using the Garwood method and averaged~\cite{10.2307/2333958, revstat2012Patil, https://doi.org/10.1002/bimj.4710350716}.}
\label{fig:expected_stat_uncertainty}
\end{figure}

The uncertainty of the density of the bedrock beneath the Antarctic ice sheet is about 7\%~\cite{hinze_2003,doi:https://doi.org/10.1002/9783527626625.ch4,tc-12-2741-2018}. An increase in density could cause two counteracting effects. A higher bedrock density would provide more targets for neutrino interactions, increasing the number of muons that are produced. However, a higher bedrock density would also provide more targets for muon interactions, causing more muons to range out before reaching the detector, or causing them to reach the detector with a lower energy. The detector's efficiency drops rapidly with energy below about 1~TeV. A precise understanding of the overall bedrock effect requires dedicated simulation studies with low statistical uncertainty. Preliminary results suggest the effect of the bedrock density uncertainty is small: less than 2\%. This is small or comparable to the expected statistical uncertainty, shown in Fig.~\ref{fig:expected_stat_uncertainty}. The simulation studies are ongoing and were not able to be fully completed and understood in time for this thesis. Therefore, the result presented in this thesis does not include this potential systematic effect. Future work will incorporate the effect of the bedrock density, if necessary, once it is understood. However, at present we believe the bedrock systematic uncertainty will be negligible.

\section{Oscillograms and expected event distributions}
\label{sec:oscillgorams_and_expected_event_distributions}
\begin{figure}[htb!]
\centering
\includegraphics[clip, trim = 0cm 0cm 0cm 1cm, width=\linewidth]{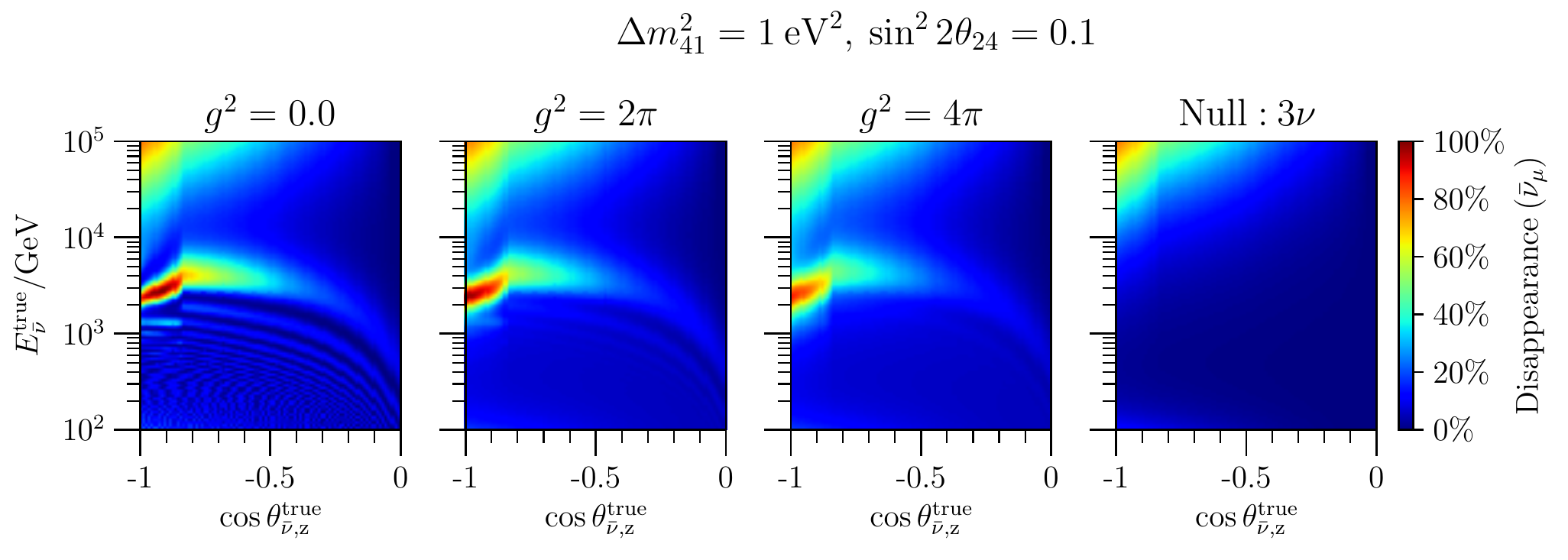}
\caption[Oscillograms for antineutrinos]{Muon antineutrino disappearance as a function of true energy and angle. The left three panels show the prediction for a 3+1 sterile neutrino model with $\Delta m^2_{41} = 1.0 \: \textrm{eV}^2$ and $\sin^2 2\theta_{24} = 0.1$ and three values of the decay-mediating coupling. The leftmost panel corresponds to a traditional 3+1 model. The rightmost panel corresponds to the scenario where there are only three neutrinos.}
\label{fig:oscillograms_anu_analysis}
\end{figure}

\begin{figure}[htb!]
\centering
\includegraphics[clip, trim = 0cm 0cm 0cm 1cm, width=\linewidth]{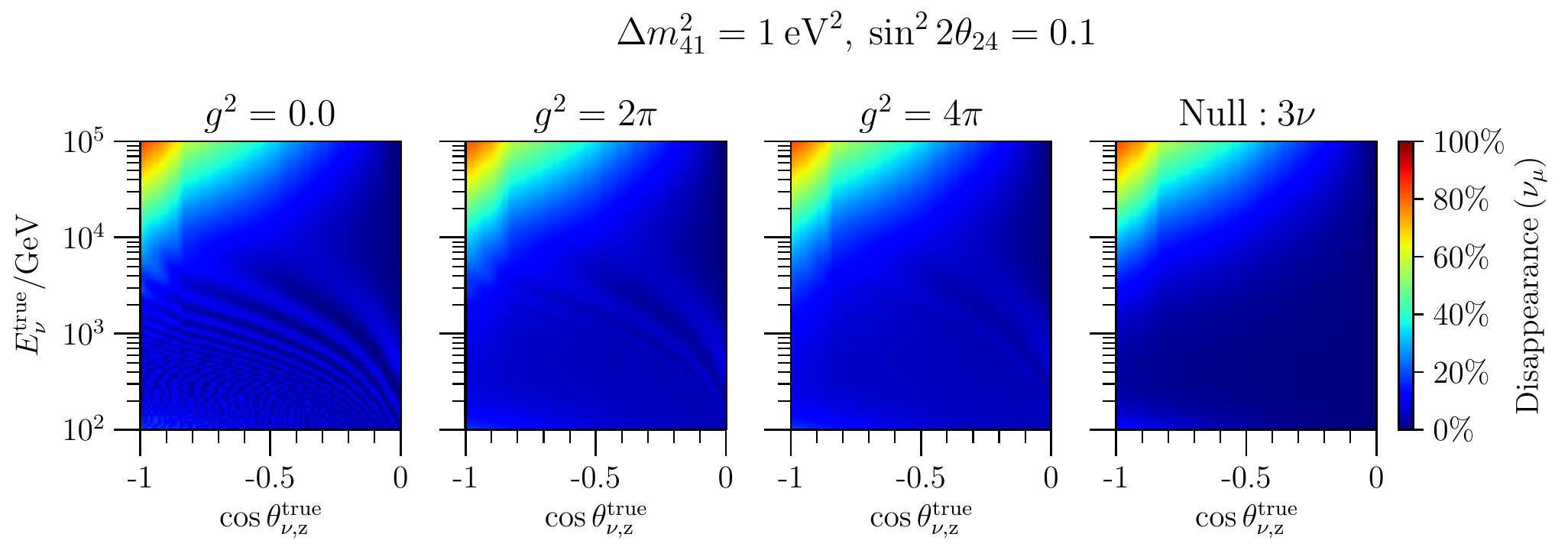}
\caption[Oscillograms for neutrinos]{Muon neutrino disappearance as a function of true energy and angle. The left three panels show the prediction for a 3+1 sterile neutrino model with $\Delta m^2_{41} = 1.0 \: \textrm{eV}^2$ and $\sin^2 2\theta_{24} = 0.1$ and three values of the decay-mediating coupling. The leftmost panel corresponds to a traditional 3+1 model. The rightmost panel corresponds to the scenario where there are only three neutrinos.}
\label{fig:oscillograms_nu_analysis}
\end{figure}

This section presents the expected event distributions assuming a sterile neutrino with parameters $\Delta m^2_{41} = 1.0 \: \textrm{eV}^2$, $\sin^2 2\theta_{24} = 0.1$, and three values for the decay-mediating coupling, $g^2 = 0,~2\pi,\textrm{ and }4\pi$. The oscillograms in Figs.~\labelcref{fig:oscillograms_anu_analysis,fig:oscillograms_nu_analysis} show the disappearance probabilities at the detector in true quantities. Figure~\ref{fig:oscillograms_anu_analysis} shows the disappearance probability of muon antineutrinos, and Fig.~\ref{fig:oscillograms_nu_analysis} shows the disappearance probability of muon neutrinos. In both these figures, the left-most panel, corresponds to a standard 3+1 model, or no $\nu_4$ neutrino decay, while the middle two panels correspond to non-zero values of the coupling, which causes $\nu_4$ decay. The rightmost panel corresponds to the three-neutrino model. The features of the oscillograms are explained in~\Cref{sec:decay_prd}.IV.B.

\clearpage

\begin{figure}[ht!]
\centering
\includegraphics[clip, trim = 0cm 0cm 0cm 1.5cm, width=\linewidth]{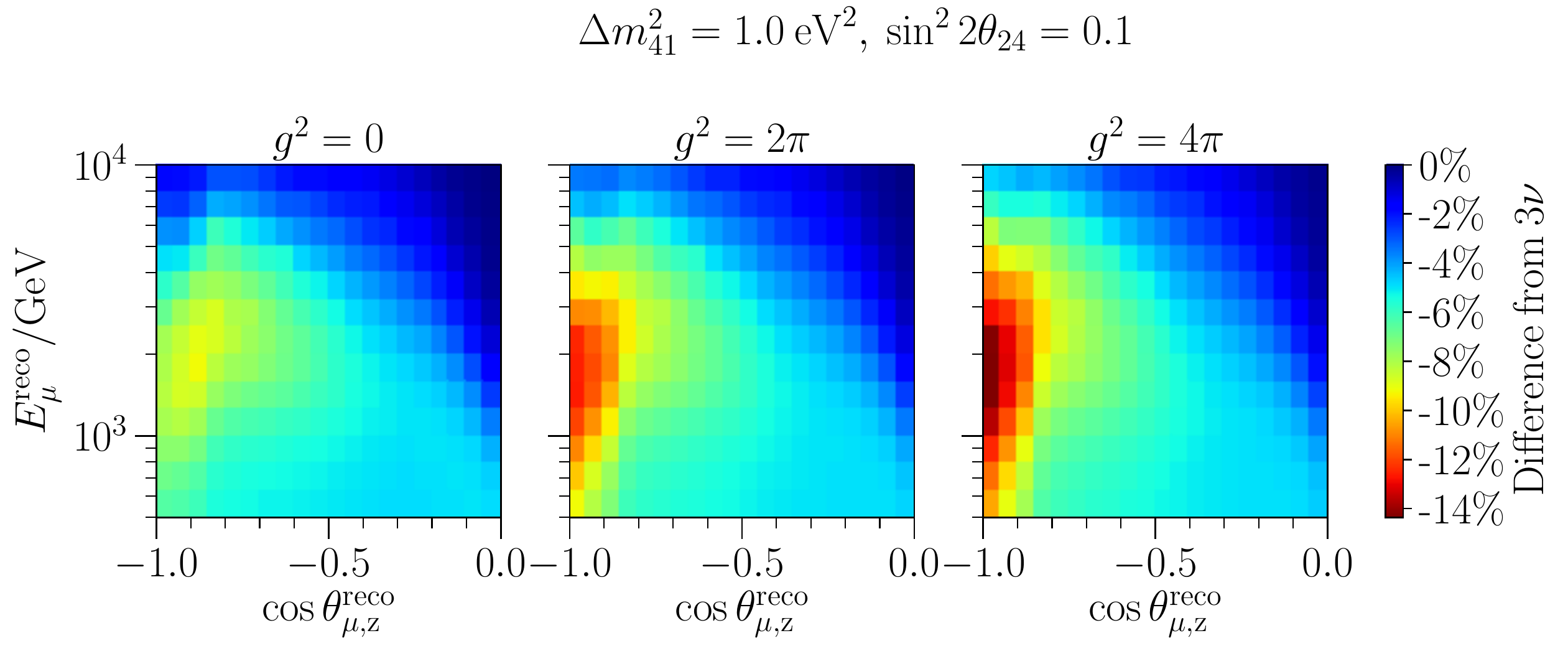}
\caption[Expected event distributions compared to three-neutrino model]{Expected event distributions, shown as a percent difference from the three-neutrino model, for  sterile neutrino parameters $\Delta m^2_{41} = 1.0 \: \textrm{eV}^2$, $\sin^2 2\theta_{24} = 0.1$, and three values of the decay-mediating coupling, $g^2 = 0,~2\pi,\textrm{ and }4\pi$.}
\label{fig:event_dist_diff_3nu}
\end{figure}

\begin{figure}[hb!]
\centering
\includegraphics[clip, trim = 0cm 0cm 0cm 1.5cm, width=0.8\linewidth]{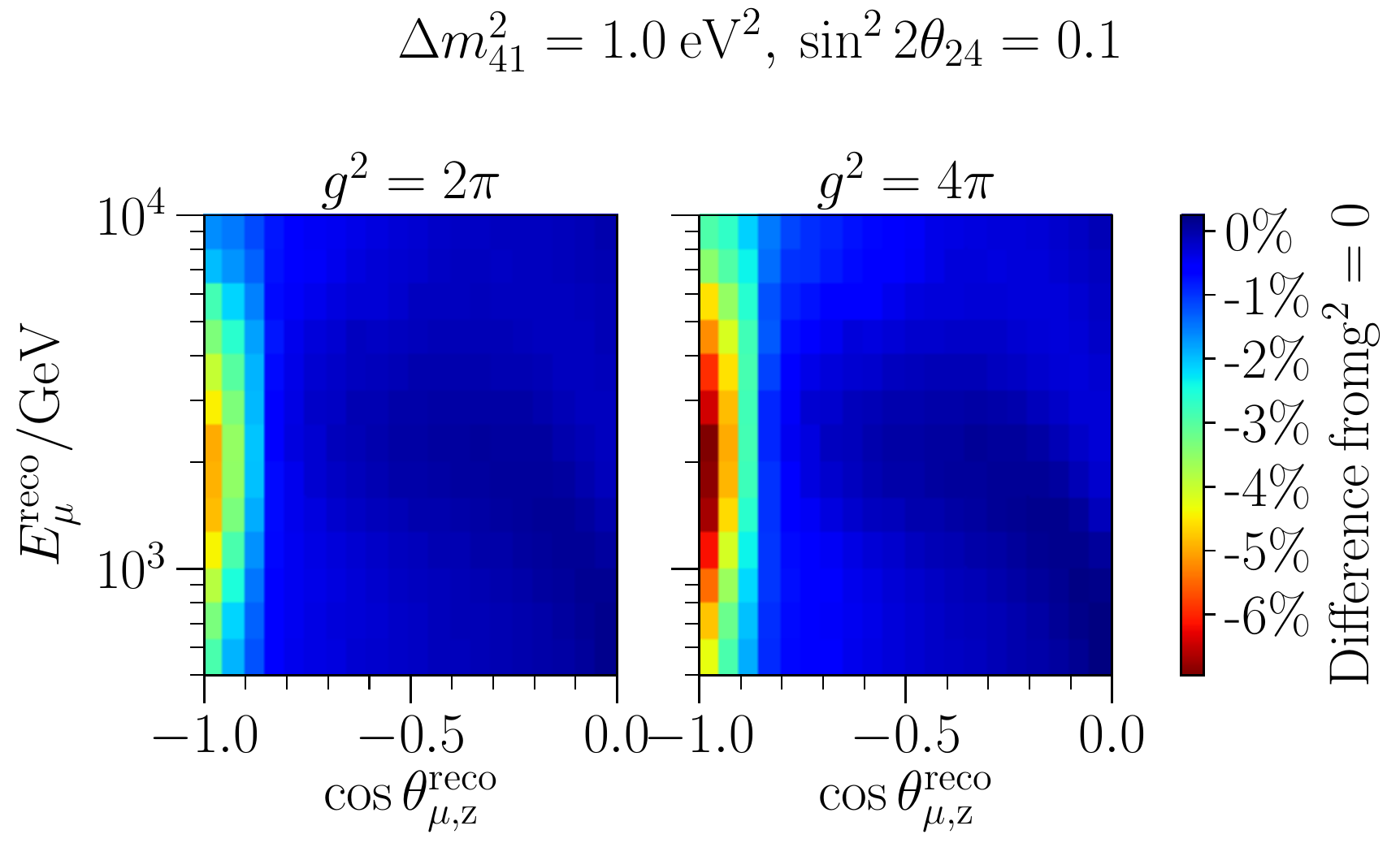}
\caption[Expected event distributions compared to traditional 3+1 model]{Expected event distributions, shown as a percent difference from the traditional 3+1 model, for  sterile neutrino parameters $\Delta m^2_{41} = 1.0 \: \textrm{eV}^2$, $\sin^2 2\theta_{24} = 0.1$, and two values of the decay-mediating coupling, $2\pi,\textrm{ and }4\pi$.}
\label{fig:event_dist_diff_nodecay}
\end{figure}

The expected event distributions for these same parameters are shown in Fig.~\ref{fig:event_dist_diff_3nu} as a percent difference from the three-neutrino scenario. For these fixed values of $\Delta m^2_{41}$ and $\sin^2 2\theta_{24}$, the traditional 3+1 model predicts a maximum 9\% per-bin deficit, which increases with non-zero $g^2$ up to 14\% and shifts in position to the minimum cosine zenith value. A comparison between the expected event distributions between zero and non-zero values of $g^2$ for fixed $\Delta m^2_{41}$ and $\sin^2 2\theta_{24}$ is shown in Fig.~\ref{fig:event_dist_diff_nodecay}. Nonzero $g^2$ causes a deficit of events at the smallest cosine zenith angles, up to about 7\% per bin.

\clearpage

\section{Likelihood function}
The Poisson likelihood function for a single bin is
\begin{equation}
\mathcal{L}(\vec{\theta}| k) =  \frac{\lambda (\vec{\theta})^k e^{- \lambda(\vec{\theta})}}{k!}
\end{equation}
where $\lambda(\vec{\theta})$  is the predicted bin count assuming hypothesis $\vec{\theta}$ and $k$ is the experimental bin count. However, finite Monte Carlo statistics do not allow for exact determination of the expected bin count, $\lambda(\vec{\theta})$. Therefore, following~\cite{Arguelles:2019izp}, we use an effective likelihood that accounts for finite Monte Carlo statistics,
\begin{equation}
\mathcal{L_{\rm{Eff}}}(\vec{\theta}| k) = \big( \frac{\mu}{\sigma^2} \big)^{\frac{\mu^2}{\sigma^2}+1} \Gamma \bigg( k +  \frac{\mu^2}{\sigma^2}+1 \bigg) \bigg[ k! \big(\frac{\mu}{\sigma^2}+1 \big)^{k + \frac{\mu^2}{\sigma^2}+1}  \Gamma \big( \frac{\mu^2}{\sigma^2}+1 \big) \bigg]^{-1} 
\label{eq:effective_likelihood}
\end{equation}
where $ \mu $ and $ \sigma^2 $ depend on $ \vec{\theta} $ and Monte Carlo weights. The parameters $\mu$ and $\sigma^2$ are given by,
\begin{equation}
\begin{split}
\begin{aligned}
\mu &\equiv \sum_{i=1}^m w_i \\
\sigma^2 &\equiv \sum_{i=1}^m w_i^2,
\end{aligned}
\end{split}
\end{equation}
where $m$ is the number of Monte Carlo events in the bin, and $w_i$ is the weight of the $i^{th}$ event, assuming $\vec{\theta}$.

To account for systematic uncertainties, nuisance parameters with Gaussian priors on their values are used. The statistical likelihood from Eq.~\ref{eq:effective_likelihood} is multiplied by:
\begin{equation}
\mathcal{L(\theta_{\vec{\eta}})}=  \prod_\eta \frac{1}{\sqrt{2 \pi \sigma_\eta^2}} e^{\frac{-(\theta_\eta - \Theta_\eta)^2}{2\sigma_\eta^2}}
\end{equation}
where $\theta_\eta$, $\Theta_\eta$, and $\sigma_\eta$ are the value, prior central value, and prior width of nuisance parameter $\eta$. The prior central values and widths are given in Table~\ref{table:Priors}. 

\section{Parameter scan}
The analysis is performed over a three-dimensional scan of the physics parameters: $\Delta m_{41}^2$, $\sin^2 2 \theta_{24}$, and $g^2$. The parameters $\Delta m_{41}^2$ and $\sin^2 2 \theta_{24}$ are sampled log-uniformly in the ranges 0.01 -- 47 eV$^2$ and 0.01 -- 1, respectively, with ten samples per decade for each parameter. These choices are driven by the region where the anomalies discussed earlier appear. The parameter $g^2$ is sampled in steps of $\pi/2$ in the range 0 -- 4$\pi$, where $g^2 = 0$ corresponds to infinite $\nu_4$ lifetime. The upper limit on scanned values of $g^2$ is chosen to preserve unitarity and perturbability, which are assumed by the calculations~\cite{Peskin:1995ev}.

\section{Frequentist analysis and Asimov sensitivity}
The frequentist analysis assumes Wilks' theorem~\cite{Wilks_1938}. This allows confidence level intervals to be drawn based on the logarithm of the profile likelihood ratio. That is 
\begin{equation}
\Delta \textrm{LLH}(\vec{\theta}) = \ln(\mathcal{L}(\hat{\vec{\theta}}))  - \ln(\mathcal{L(\vec{\theta})}),
\end{equation}
where $\hat{\vec{\theta}}$ is the set of physics and nuisance parameters that maximizes $\mathcal{L}(\vec{\theta})$, i.e. the best fit. The test statistic, TS, is
\begin{equation}
\textrm{TS}(\vec{\theta}) = - 2 \Delta \textrm{LLH}(\vec{\theta}).
\label{eq:teststatistic}
\end{equation}
Wilks' theorem states that, assuming certain criteria are met, in a likelihood space with $n$ degrees of freedom, this test statistic  has a $\chi^2$-distribution, with $n$ degrees of freedom, $\chi^2_n$. The confidence level $C$ runs through all points $\vec{\theta}$ where
\begin{equation}
\int_0^{\textrm{TS}(\vec{\theta})} dx \chi^2_n (x) = C.
\label{eq:wilks_CL}
\end{equation}
The values of the test statistic that satisfy Eq.~\ref{eq:wilks_CL} for several confidence levels, and for one and three degrees of freedom (DOF), are given in Table~\ref{tab:chi2}. If the assumptions underlying Wilks' theorem are strongly violated, the contours drawn using this technice may have improper coverage. Coverage of the contours will be checked using the Feldman Cousins method, which is always accurate but is computationally intensive, at a later date~\cite{Feldman:1997qc}. 
\begin{table}[h]
\begin{center}
\begin{tabular}{c | c c}
\hline
\hline
Confidence Level & 1 DOF & 3 DOF \\
\hline
68.27\% & 1.00 & 3.53 \\
90\% & 2.71 & 6.25 \\
95\% & 3.84 & 7.82 \\
99\% & 6.63 & 11.34 \\
\hline
\hline
\end{tabular}
\end{center}
\caption[Critical values of the $\chi^2$ test statistic]{Critical values of the $\chi^2$ test statistic for one and three degrees of freedom. Values from~\cite{Zyla:2020zbs}.}
\label{tab:chi2}
\end{table}

\begin{figure}[h!]
\centering
\includegraphics[width=\linewidth]{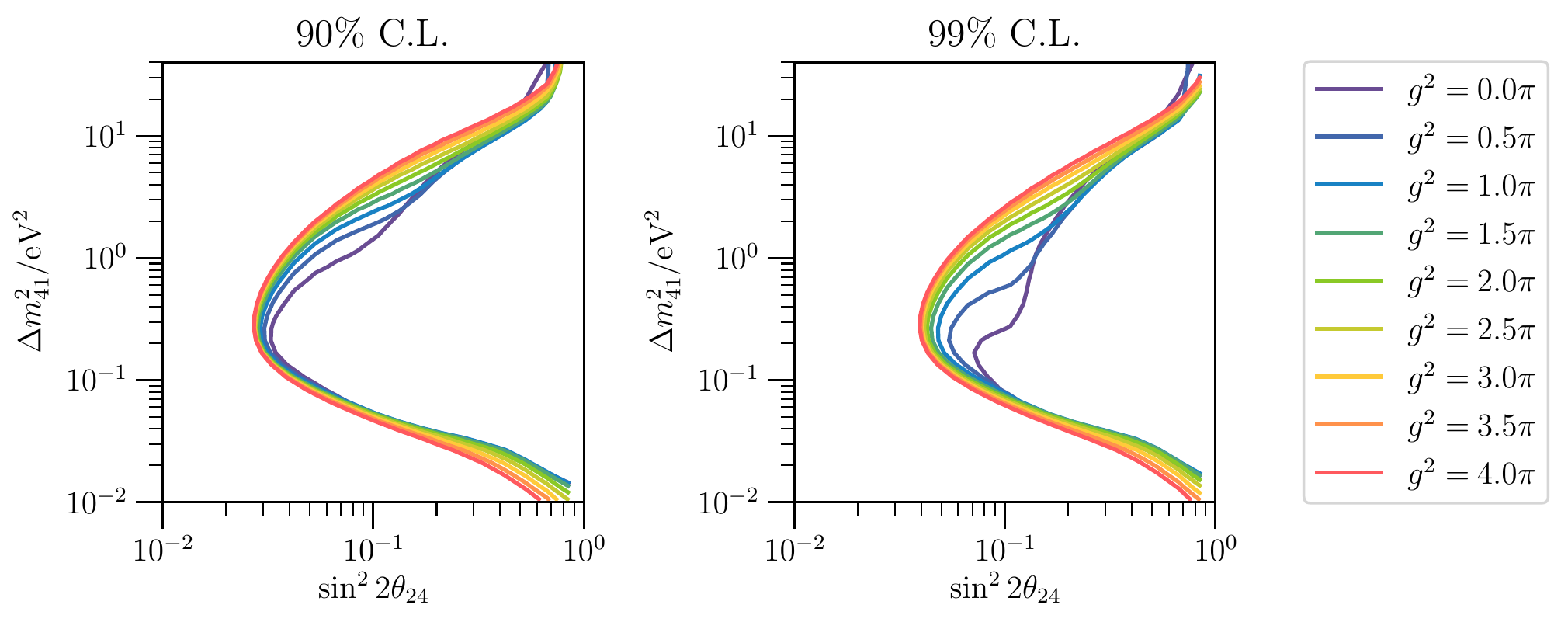}
\caption[Asimov sensitivity]{Asimov sensitivity.}
\label{fig:asimov_sensitivity}
\end{figure}

The sensitivity is approximated by the Asimov sensitivity~\cite{Cowan:2010js}. The Asimov sensitivity is calculated over a scan of the three parameters, assuming three degrees of freedom and Wilks' theorem, and is shown in Fig.~\ref{fig:asimov_sensitivity}. Increased sensitivity is observed for larger values of $g^2$ for fixed values of $\Delta m^2_{41} = 1.0 \: \textrm{eV}^2$, $\sin^2 2\theta_{24} = 0.1$ around $\Delta m^2_{41}~1~\textrm{eV}^2$. An exact sensitivity would be calculated from an ensemble simulated pseudoexperiments, using the Feldman-Cousins method~\cite{Feldman:1997qc}. This is computationally expensive and will be completed later.

\section{Bayesian analysis}

A Bayesian analysis is performed in addition to the frequentist analysis. The Bayesian evidence is computed for every sterile neutrino parameter point in the scan, as well as for the three-neutrino model. The evidence is defined as:
\begin{equation}
\mathcal{E}(\vec{\Theta}) = \int d \vec{\eta} \: \mathcal{L}(\vec{\Theta}, \vec{\eta}) \: \Pi (\vec{\eta})
\end{equation}
where $\Pi(\vec{\eta})$ are the Gaussian priors on the nuisance parameters. The integral is calculated with the MultiNest algorithm~\cite{Feroz:2013hea}. The evidence for each sterile neutrino model is compared to that of the three-neutrino model. The ratio of these evidences is the Bayes factor:
\begin{equation}
B_{ij} = \frac{\mathcal{E}_i}{\mathcal{E}_j}
\end{equation}
The Bayes factor is interpreted with Jeffreys scale, which quantifies the strength of evidence preferring one model, $i$, over the other, $j$. For negative values of the Bayes factor, model $j$ is preferred over $i$.
\begin{table}[h]
\begin{center}
\begin{tabular}{c | c}
\hline
\hline
$\log_{10}B_{ij}$ & Strength of evidence\\
\hline
$0-0.5$ & Weak \\
$0.5-1$ & Substantial \\
$1-1.5$ & Strong \\
$1.5-2$ & Very strong \\
$>2$ & Decisive \\
\hline
\hline
\end{tabular}
\end{center}
\caption[Strength of evidence determined from the Bayes factor.]{Strength of evidence determined from the Bayes factor. Adapted from~\cite{Jeffreys:1939xee}}
\label{tab:jeffreysscale}
\end{table}

\chapter{Results}
\label{chap:results}

\begin{figure}[h]
 \centering
 \includegraphics[width=\textwidth]{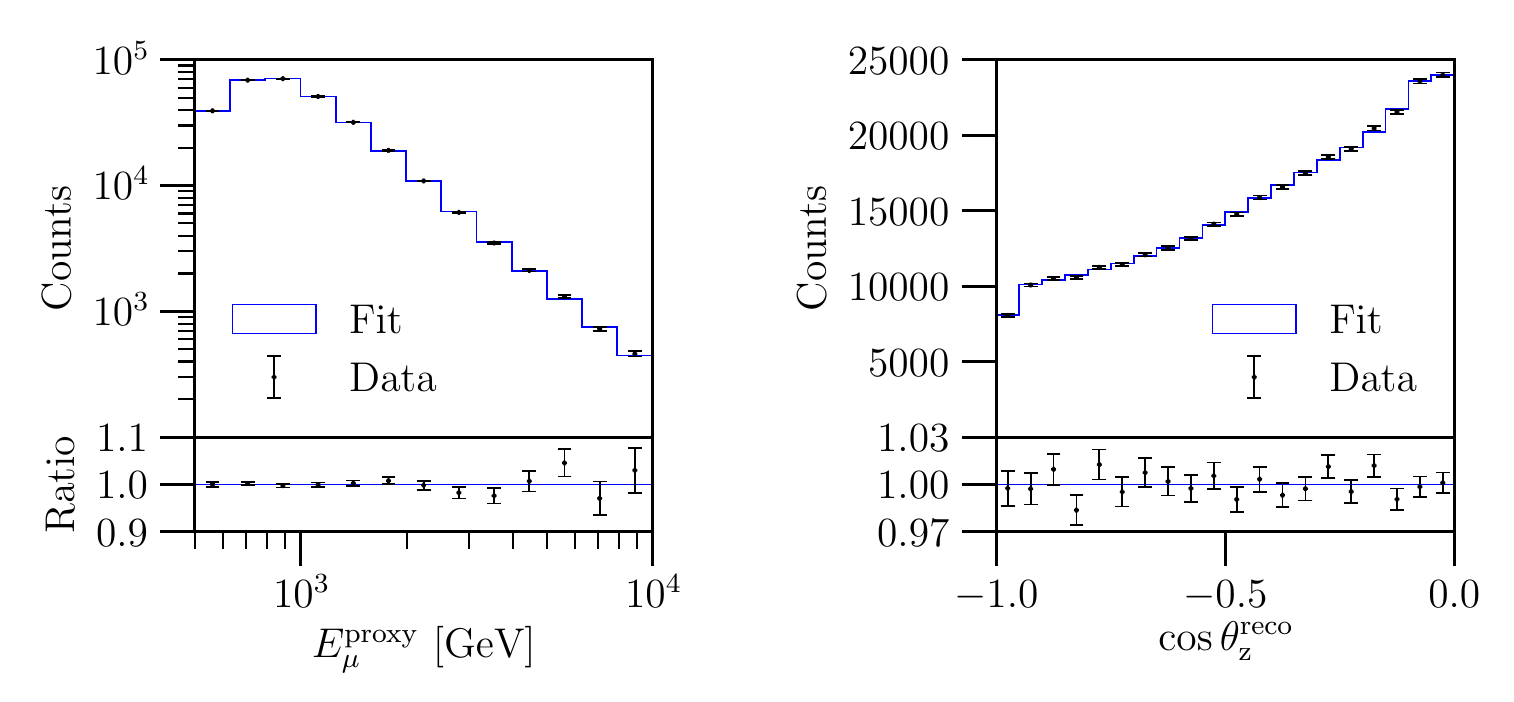}
 \caption[Projected energy and angular distribution of the data]{Projected reconstructed energy and cosine zenith distributions of the data compared to the best-fit expectation, including the best-fit systematics.}
 \label{fig:1ddist}
\end{figure}

Good agreement between data and the Monte Carlo description of the best-fit point is found. The projected distributions of reconstructed muon energy and reconstructed cosine zenith angle for both the data and the best-fit expectation are shown in Fig.~\ref{fig:1ddist}. The best-fit distribution accounts for the best-fit sterile parameters and the best-fit systematic values. Pearson $\chi^2$ tests assuming three degrees of freedom yield p-values of 46\% and 43\% for the energy proxy and cosine zenith distributions, respectively.

\clearpage

\begin{figure}[ht]
  \centering
  \includegraphics[width=4.in]{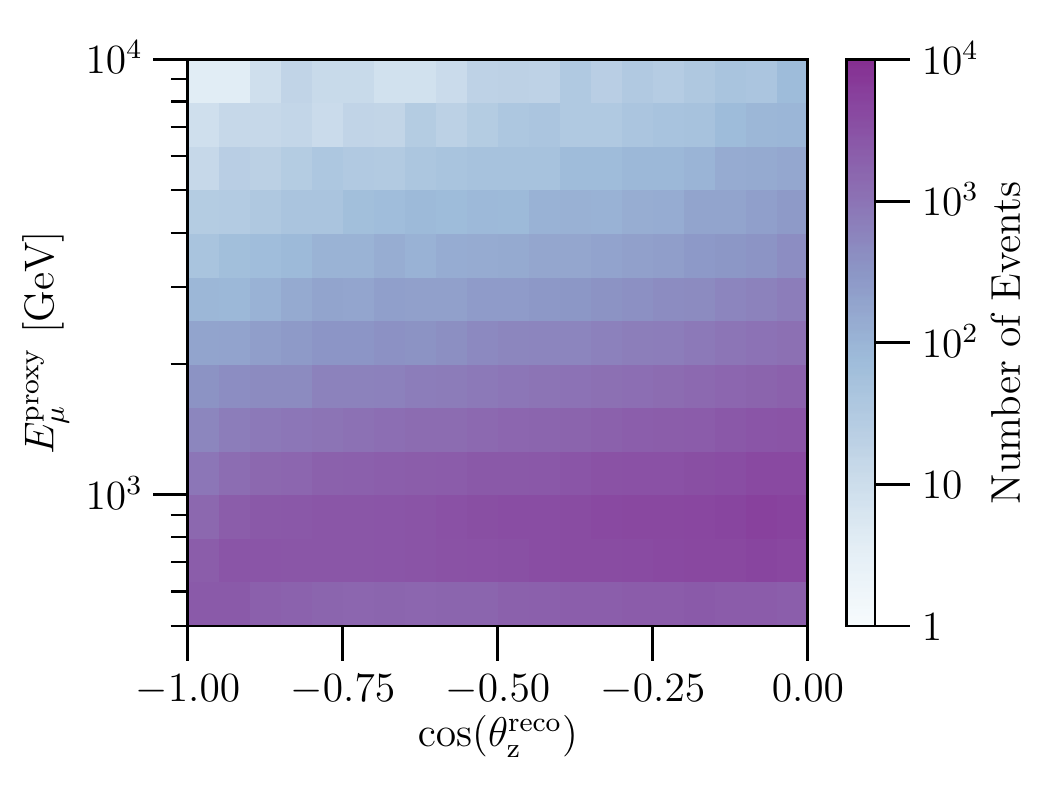}
  \caption[Data distribution]{The binned, two-dimensional distribution of the eight-year dataset. The same dataset is used in~\cite{Axani:2020zdv, Aartsen:2020fwb} and this plot is a reproduction of a plot from those references.}
  \label{fig:data}
\end{figure}
The eight-year dataset contains 305,735 reconstructed events. The two-dimensional distribution of the data is shown in Fig.~\ref{fig:data}.

\begin{figure}[h]
  \centering
  \includegraphics[width=\textwidth]{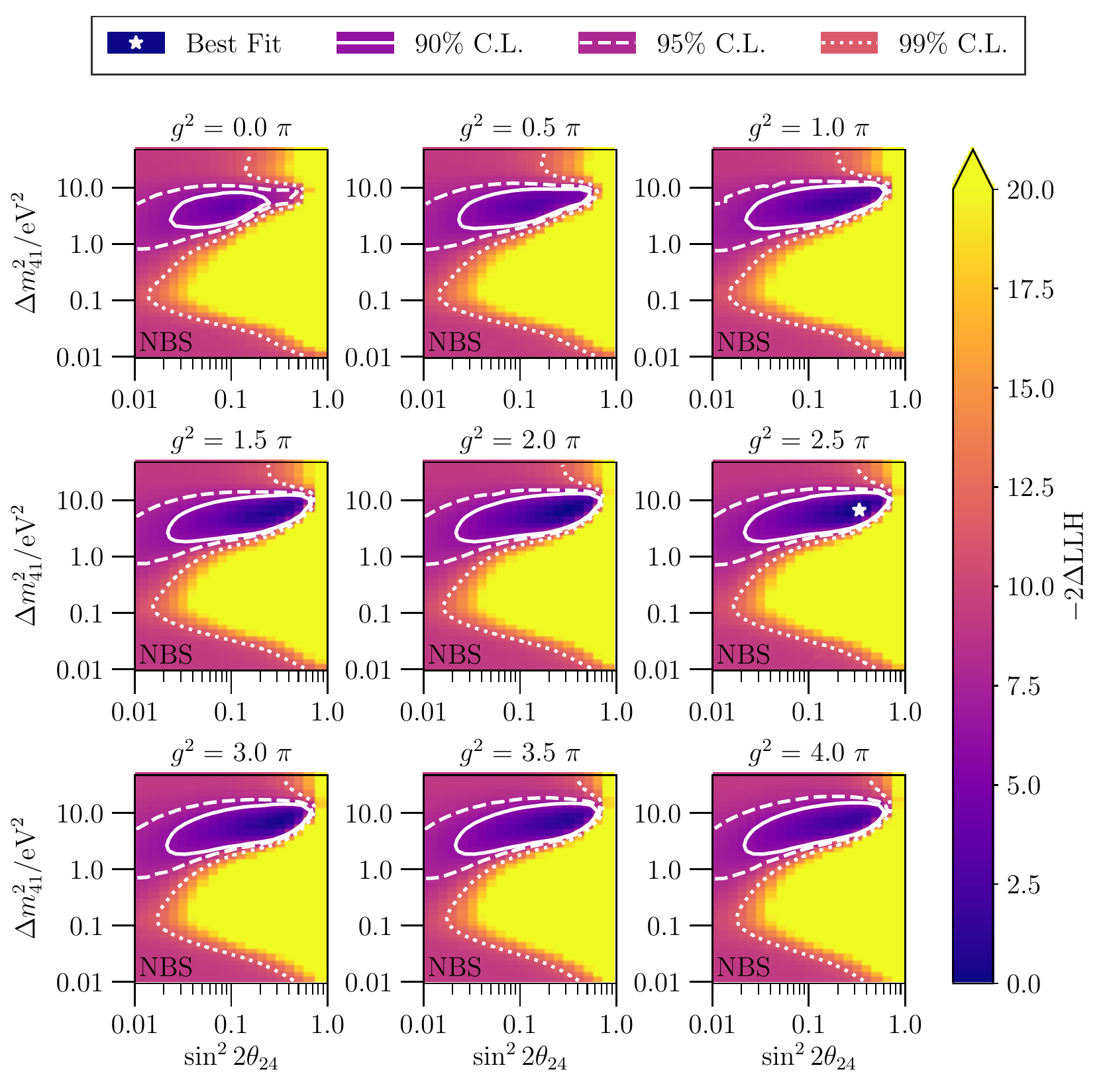}
  \caption[Result of the frequentist analysis]{Result of the frequentist analysis. ``NBS'' means that this analysis did not include a bedrock systematic uncertainty. See \Cref{sec:bedrock}.}
  \label{fig:freq_result}
\end{figure}

\begin{figure}[h]
  \centering
  \includegraphics[width=\textwidth]{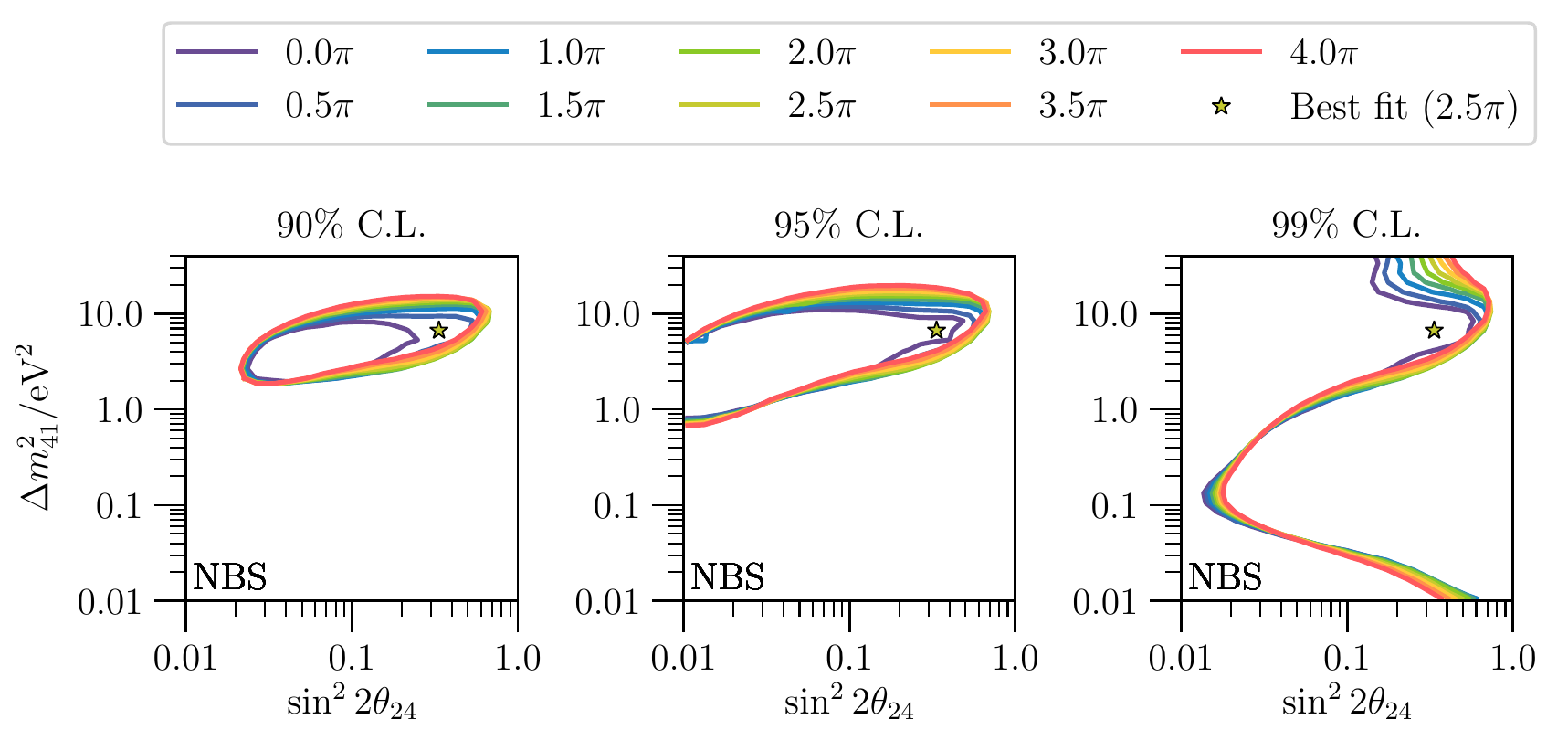}
  \caption[Result of the frequentist analysis]{Result of the frequentist analysis. ``NBS'' means that this analysis did not include a bedrock systematic uncertainty. See \Cref{sec:bedrock}.}
  \label{fig:freq_result_3panels}
\end{figure}

The best-fit point occurs at $\Delta m^2_{41}$ = 6.6 eV$^2$, $\sin^2 2\theta_{24}$ = 0.33, and $g^2 = 2.5 \pi$. The three-neutrino hypothesis is rejected with a $p$-value of 2.8\%, assuming three degrees of freedom. The best-fit physics and nuisance parameters and their uncertainties are given in 
Table~\ref{table:results}. The reported $1\sigma$ ranges come from marginalizing over all other parameters. The result of the frequentist analysis is shown in Figs.~\ref{fig:freq_result} and \ref{fig:freq_result_3panels}. In Fig.~\ref{fig:freq_result}, each of the nine panels corresponds to a fixed value of $g^2$, and the solid, dashed, and dotted white curves represent the 90\%, 95\% and 99\% C.L. regions, respectively. In Fig.~\ref{fig:freq_result_3panels}, the three panels correspond to the three confidence levels, and the different colored curves correspond to different $g^2$ values. In both figures, the best fit point is marked with a star.

\begin{table}[h!]
\begin{center}
 \begin{tabular}{ l c }
 \hline
 \hline
\textbf{Parameter} & Best fit $\pm 1 \sigma$  \\ [0.5ex]
\hline
\hline
\multicolumn{2}{c}{\textbf{Physics Parameters}}\\
\hline 
$\Delta m_{41}^2$  & $6.7^{+3.9}_{-2.5}$ eV$^{2}$\\
\hline
$\sin^2 2 \theta_{24}$ & $0.33^{+0.20}_{-0.17}$\\
\hline
$g^2$ & $2.5 \pi \pm 1.5 \pi$\\
\hline
\multicolumn{2}{c}{\textbf{Detector parameters}}\\
\hline
DOM efficiency & $0.9653 \pm 0.0004$ \\
\hline
Bulk Ice Gradient 0 & $0.008 \pm 0.015$\\
\hline
Bulk Ice Gradient 1 & $0.35 \pm 0.09$\\
\hline
Forward Hole Ice  & $-3.5 \pm 0.06$  \\
\hline
\multicolumn{2}{c}{\textbf{Conventional flux parameters}}\\
\hline
Normalization ($\Phi_{\mathrm{conv.}}$) & $1.4 \pm 0.1$  \\
\hline
Spectral shift ($\Delta\gamma_{\mathrm{conv.}}$) & $0.071 \pm 0.001$ \\
\hline
Atm. Density  & $0.005 \pm 0.035$  \\
\hline
Barr WM   & $-0.03 \pm 0.04$      \\
\hline
Barr WP   & $-0.06 \pm 0.02$     \\
\hline
Barr YM   & $0.08 \pm 0.05$     \\
\hline
Barr YP   & $-0.21 \pm 0.05$     \\
\hline
Barr ZM   & $-0.012 \pm 0.004$       \\
\hline
Barr ZP    & $-0.031 \pm 0.002$     \\
\hline
\multicolumn{2}{c}{\textbf{Astrophysical flux parameters}}\\
\hline
Normalization ($\Phi_{\mathrm{astro.}}$) & $1.1 \pm 0.05$   \\
\hline
Spectral shift ($\Delta\gamma_{\mathrm{astro.}}$) & $0.39 \pm 0.03$    \\
\hline
\multicolumn{2}{c}{\textbf{Cross sections}}\\
\hline
Cross section $\sigma_{\nu_\mu}$  & $0.998 \pm 0.002$    \\
\hline
Cross section $\sigma_{\overline{\nu}_\mu}$  & $0.998 \pm 0.003$     \\
\hline
Kaon energy loss $\sigma_{KA}$  &  $-0.38 \pm 0.04$ \\
\hline
\hline
\end{tabular}
\caption[Best-fit physics and nuisance parameter values.]{Best-fit physics and nuisance parameter values. The $1\sigma$ range are determined by marginalizng over all other parameters.}
\label{table:results}
\end{center}
\end{table}

\clearpage

\begin{figure}[ht]
  \centering
  \includegraphics[width=4.in]{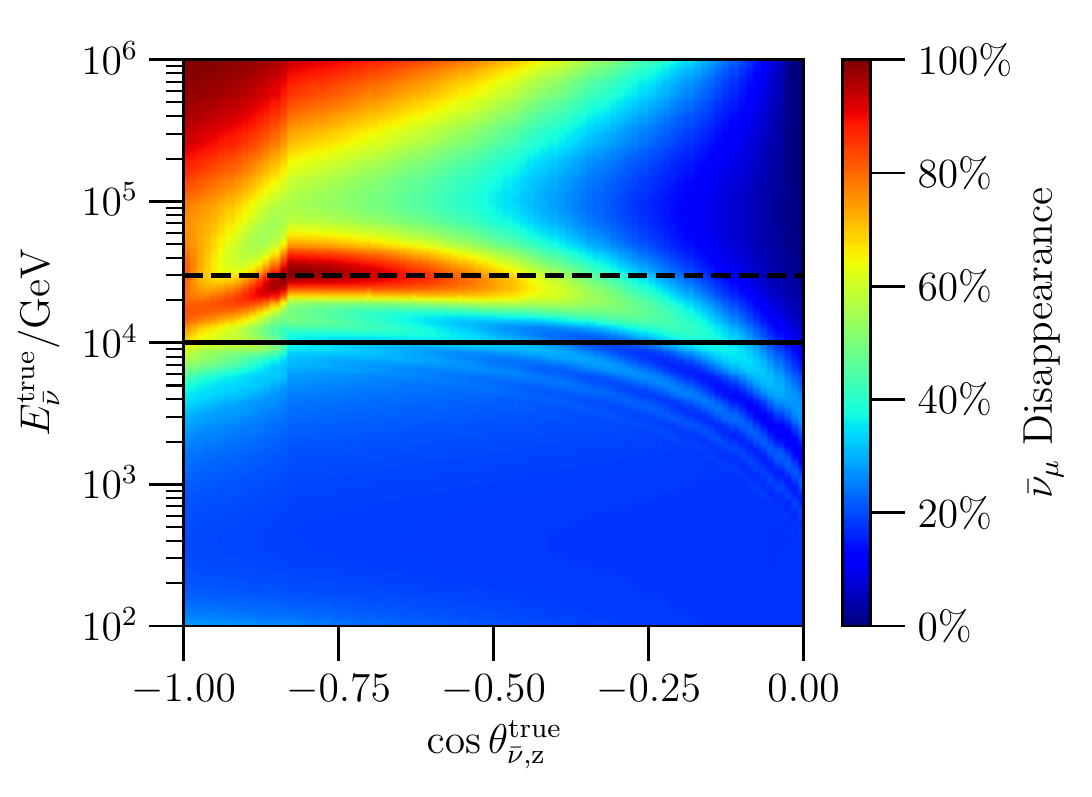}
  \caption[Oscillogram for best-fit point]{Oscillogram for the best-fit point. The lines at 10~TeV and 30~TeV indicate the true neutrino energy below which about 90\% and 99\% of the events are expected to be, respectively.}
  \label{fig:best_fit_osc}
\end{figure}

The oscillogram corresponding to the best-fit point is shown in Fig.~\ref{fig:best_fit_osc}. The horizontal line at 10~TeV indicates the true neutrino energy below which about 95\% of the events are expected to be, based off of simulation and assuming the prior values of the nuisance parameters. The horizontal line at 30~TeV indicates the true energy below which about 99\% of the events are expected to be.

\begin{figure}[htbp]
\centering
\includegraphics[width=4.in]{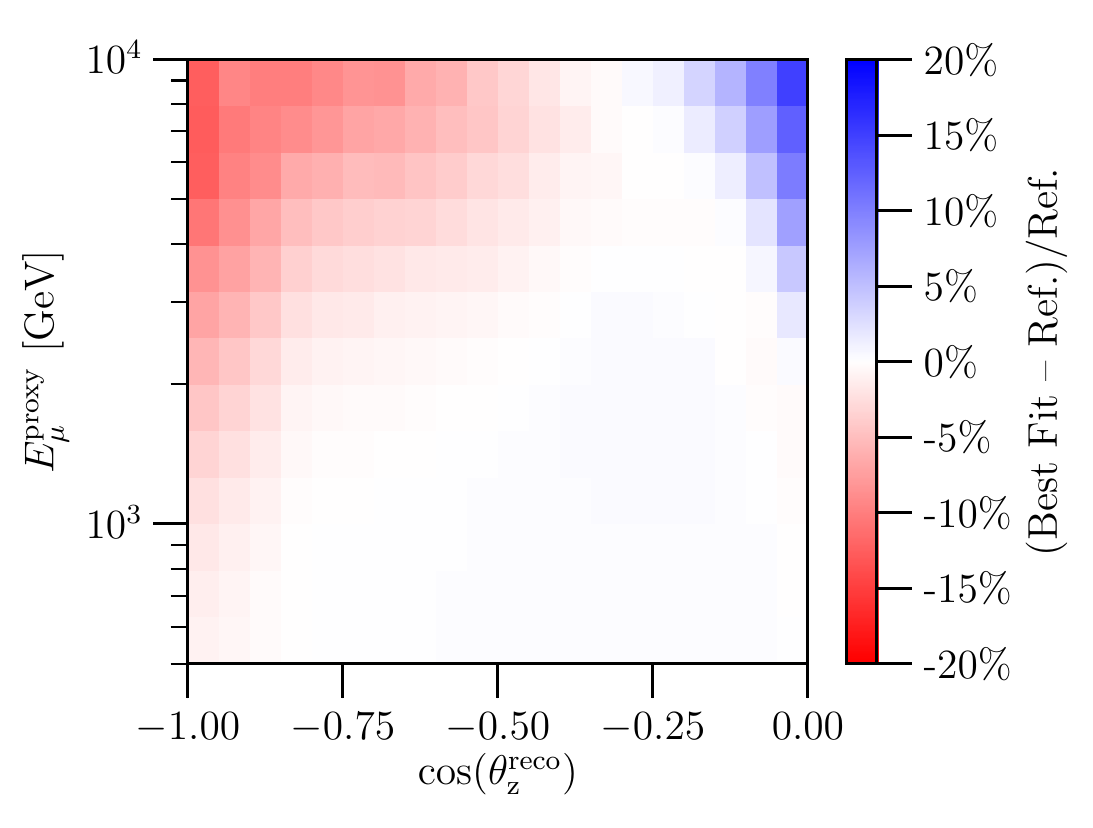}
\includegraphics[width=4.in]{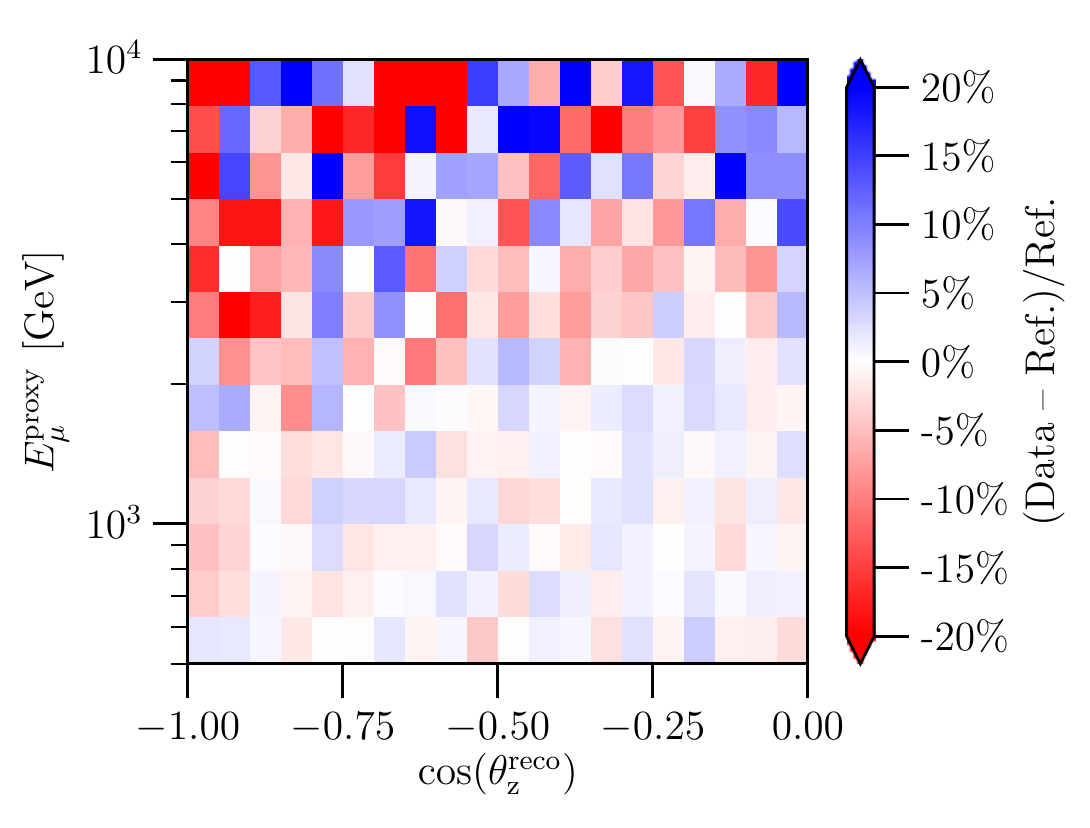}
\caption[Comparison of best-fit expectation and data to reference model]{(Top) The shape-only percent difference between the best fit expectation and a reference model. (Bottom) The shape-only percent difference between the data and a reference model. The reference model is the expectation associated with the three-neutrino model and the best-fit systematics. Overall normalization effects have been removed.}
\label{fig:quilt}
\end{figure}

The expected distribution of the best-fit point and data are each compared to a reference model in Fig.~\ref{fig:quilt}. The reference model is the expectation associated with the three neutrino hypothesis and the systematic values from the best-fit point. These comparions are shape only; overall normalization effects have been removed. Both plots show a deficit of events in the upper left corner, a relative excess of events in the upper right corner, and a modest deficit of events in the bottom left corner. Low statistics at the highest energies yield large fluctuations in the data plot.

\clearpage

\begin{figure}[h]
\centering
\includegraphics[width=4.in]{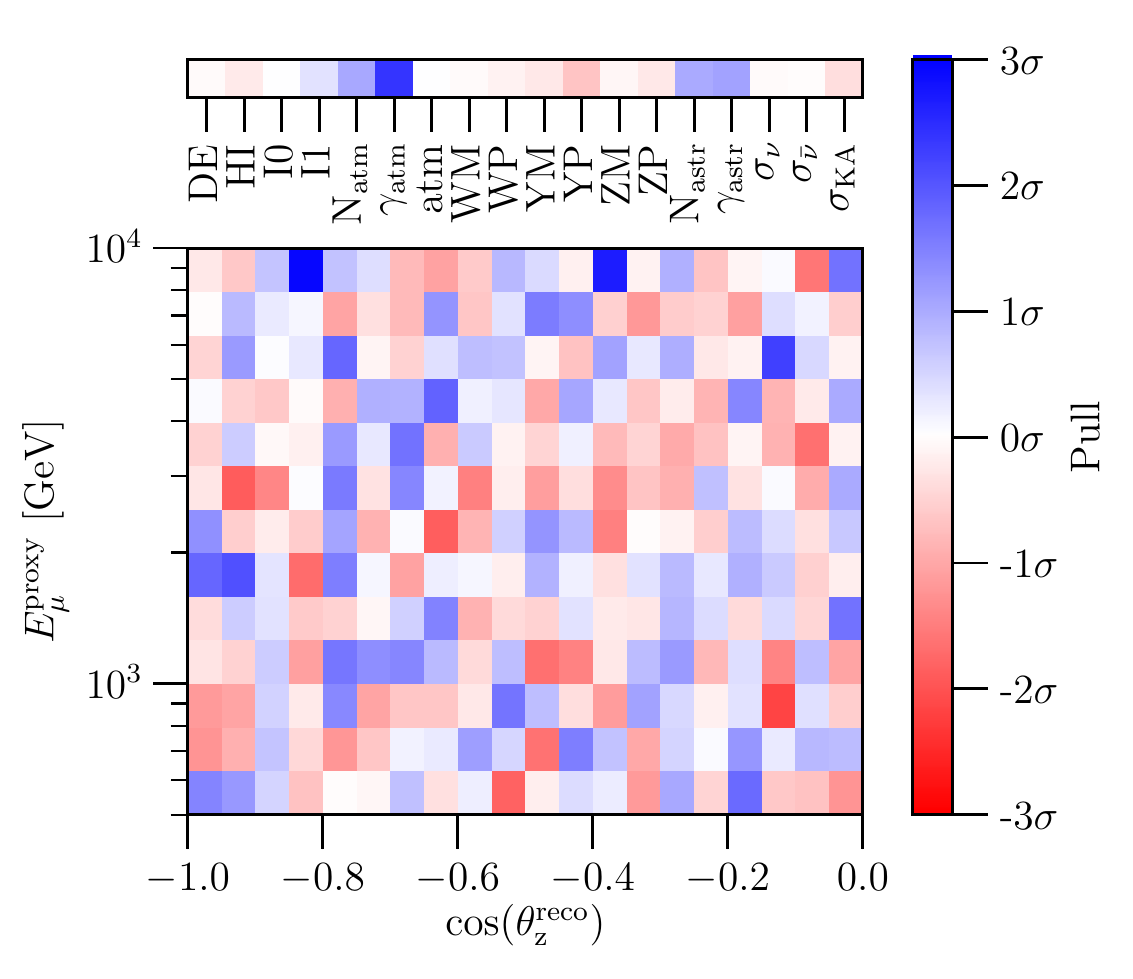}
\caption[Systematic and data pulls at best-fit point]{Systematic and data pulls at the best-fit point. The top panel shows the systematic pulls for each of the nuisance parameters. The main panel shows the data pulls.}
\label{fig:pulls_bf}
\end{figure}

Figure~\ref{fig:pulls_bf} shows the data and systematic pulls for the best-fit point. The data pull for the $i^{\rm{th}}$ bin is defined as
\begin{equation}
\rm{Pull}_{\it{i}} \equiv \frac{Data_{\it{i}} - Expectation_{\it{i}}}{\sigma_{Expectation_{\it{i}}}}, 
\end{equation}
where $\sigma_{\rm{Expectation}_i}$ is the asymmetrical Poisson error of the expectation. The Poisson error is calculated using the Garwood method~\cite{10.2307/2333958, revstat2012Patil, https://doi.org/10.1002/bimj.4710350716}. The systematic pull for the $i^{\rm{th}}$ nuisance parameter is defined as
\begin{equation}
\rm{Pull_{\it{i}}} \equiv \frac{Fit_{\it{i}} - Prior\: Center_{\it{i}}}{Prior\: Width_{\it{i}}}. 
\end{equation}
All systematics pull less than about 1$\sigma$, with the exception of the cosmic ray spectral shift, which pulls $2.4\sigma$. The data pulls appear roughly uniformly distributed. The data and systematic pulls for the fit to the three-neutrino scenario and the best-fit of the subset of points corresponding to the traditional $3+1$ model are given in~\Cref{sec:app_results_pulls}. The fit values of the nuisance parameters for each point in the scan are given in~\Cref{sec:app_results_nuisance_values}. 

\begin{figure}[ht]
  \centering
  \includegraphics[width=\textwidth]{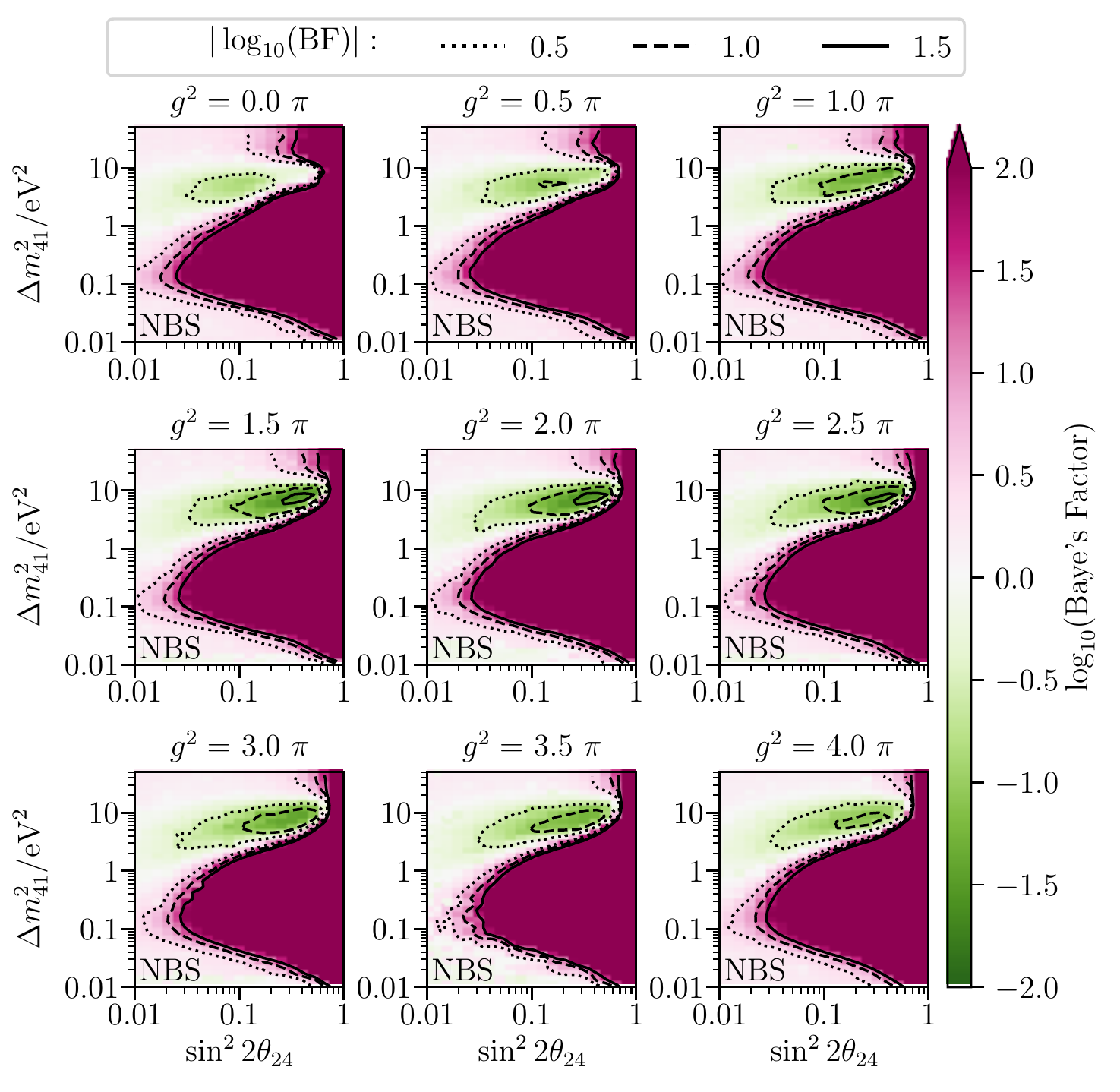}
  \caption[Result of the Bayesian analysis]{Result of the Bayesian analysis. ``NBS'' means the analysis did not include a bedrock systematic uncertainty. See \Cref{sec:bedrock}.}
  \label{fig:bayes_result}
\end{figure}

\begin{figure}[ht]
  \centering
  \includegraphics[height=8.5in]{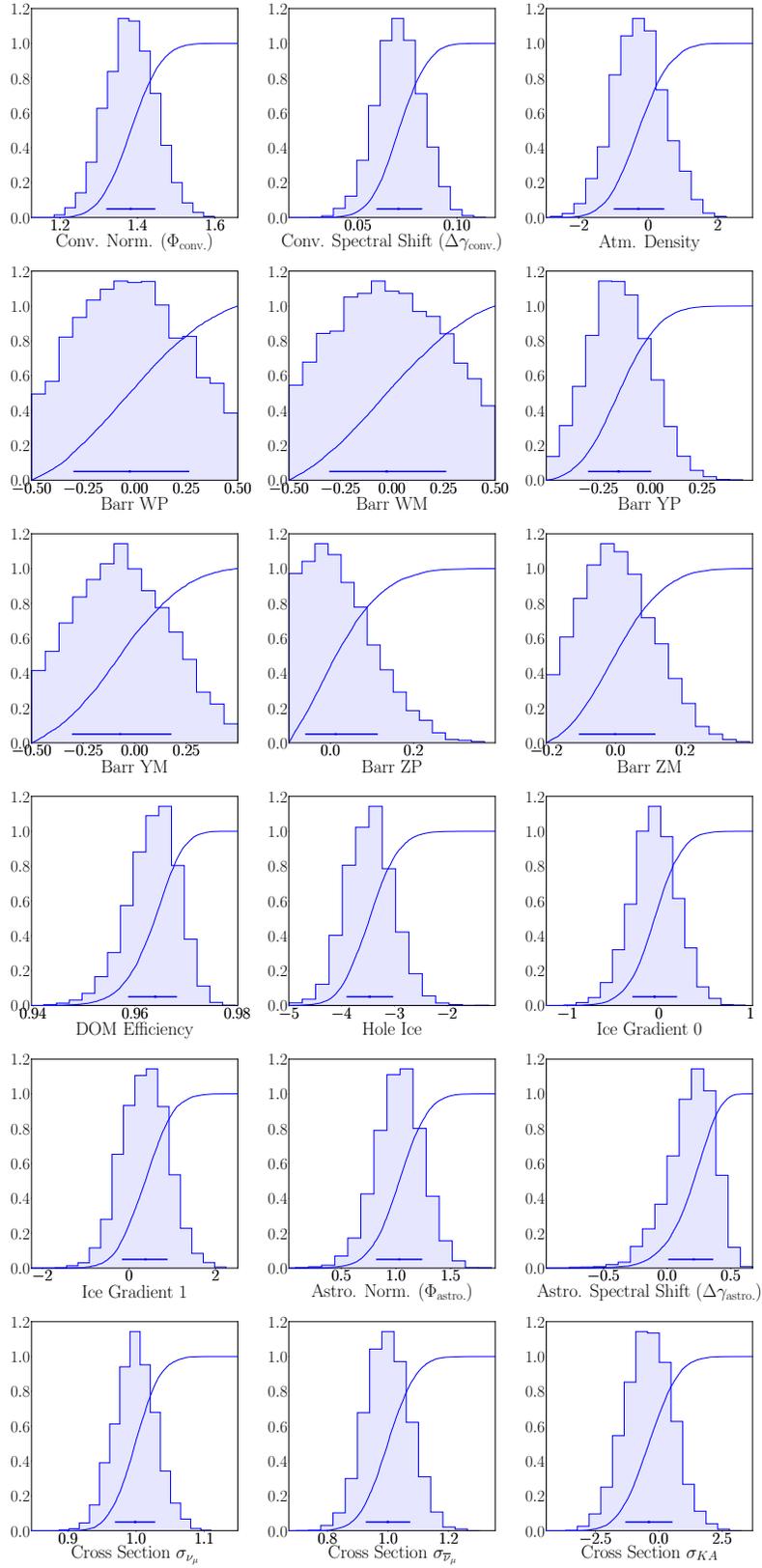}
  \caption{Posterior distributions of the nuisance parameters for the best model.}
  \label{fig:decay_posterior_alone}
\end{figure}

\begin{figure}[ht]
 \centering
 \includegraphics[width=\textwidth]{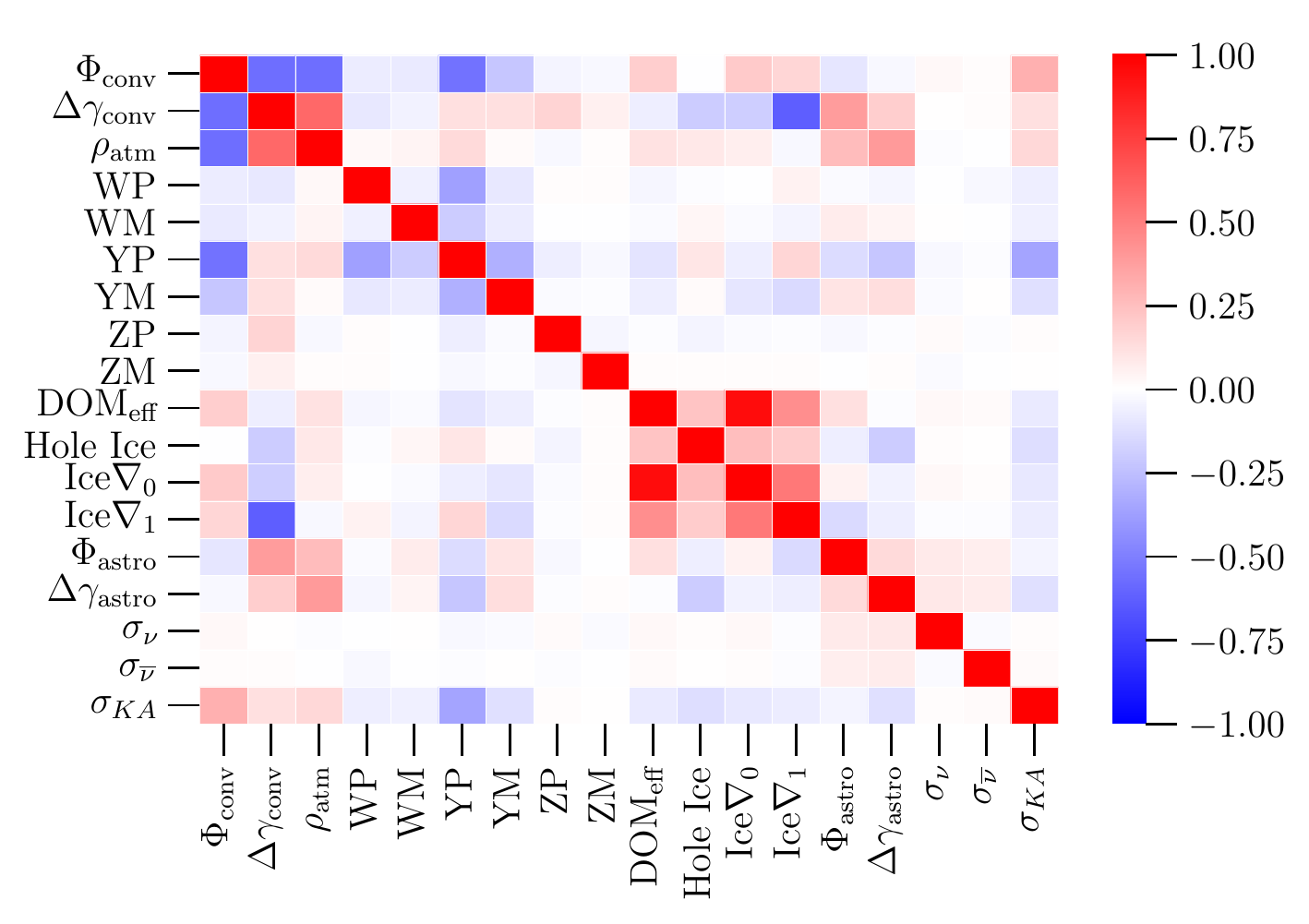}
 \caption{Correlation between the nuisance parameters.}
 \label{fig:correlation}
\end{figure}

The result of the Bayesian analysis is shown in Fig.~\ref{fig:bayes_result}. The best model has the parameters $\Delta m^2_{41}$ = 6.6 eV$^2$, $\sin^2 2\theta_{24}$ = 0.33, and $g^2 = 1.5 \pi$. The Bayes' factor is defined as the ratio of the evidence integral for a particular sterile neutrino model to the evidence integral for the three-neutrino model. The Baye's factor for the best model is 0.025, which corresponds to very strong evidence in favor of the best model. The model corresponding to the Frequentist best fit point is similarly favored over the three-neutrino model; the Bayes' factor is 0.027. The posterior distributions of each of the nuisance parameters for the best model are shown in Fig.~\ref{fig:decay_posterior_alone}. The correlation between all the nuisance parameters is shown in Fig.~\ref{fig:correlation}.

\clearpage

\begin{figure}[ht]
 \centering
 \includegraphics[width=4in]{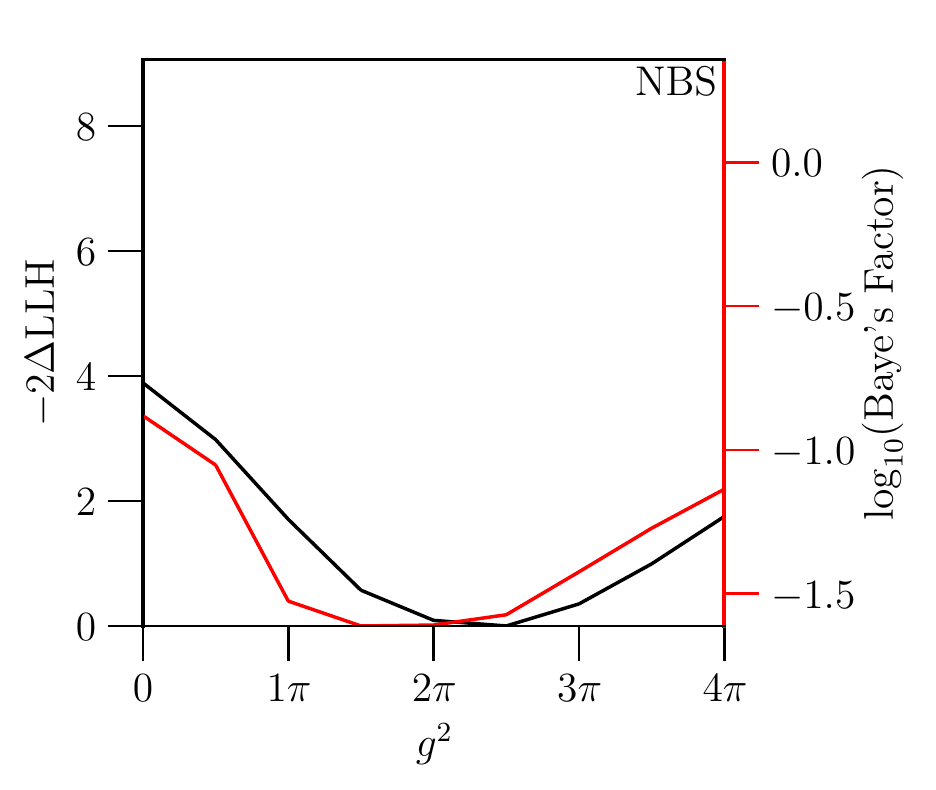}
 \caption[$-2\Delta$LLH and $\log_{10}$(Baye's factor) versus $g^2$]{Profiled -$2\Delta$LLH, plotted on the left y-axis, and $\log_{10}$(Baye's factor), plotted on the right y-axis, versus $g^2$. ``NBS'' means that this analysis did not include a bedrock systematic uncertainty.}
 \label{fig:g2_threenu}
\end{figure}

The constraining power of the frequentist and Bayesian analyses on the value of $g^2$ are compared in Fig.~\ref{fig:g2_threenu}. The black curve shows the $-2\Delta$LLH value for each scanned value of $g^2$, profiling over $\Delta m_{41}^2$ and $\sin^2 2 \theta_{24}$. This is plotted on the left y-axis. The 
red curve shows the profiled value of the $\log_{10}$(Baye's factor) for each value of $g^2$, and is plotted on the right y-axis. The two y-axes have a common scale. The value of $-2\Delta$LLH for the three-neutrino model is 9.06, the top of the figure, while on the right hand side, $\log_{10}$(Baye's factor)$=0$ corresponds to the three-neutrino model. The value of $g^2 = 0$ is rejected with a $p$-value of 4.9\%.

\clearpage

\section{Discussion}
As in the traditional 3+1 eight-year search, the only nuisance parameter that pulls greater than 2$\sigma$ is the cosmic ray spectral index shift, which pulls $2.4\sigma$. For the three-neutrino fit, this parameter pulls $2.2\sigma$. For the best fit of the subset of points with $g^2=0$, this parameter pulls $2.4\sigma$. An upgraded systematic treatment for the cosmic ray flux is warranted. In particular, recent data show a break in the cosmic ray spectral index that is not accounted for in the cosmic ray model used here~\cite{An:2019wcw,Alfaro:2017cwx}. Nevertheless, the effect of the cosmic ray spectral index shift is purely in reconstructed energy, while the signal shape is two-dimensional.

\begin{figure}[htb]
\centering
\includegraphics[width=3.7in]{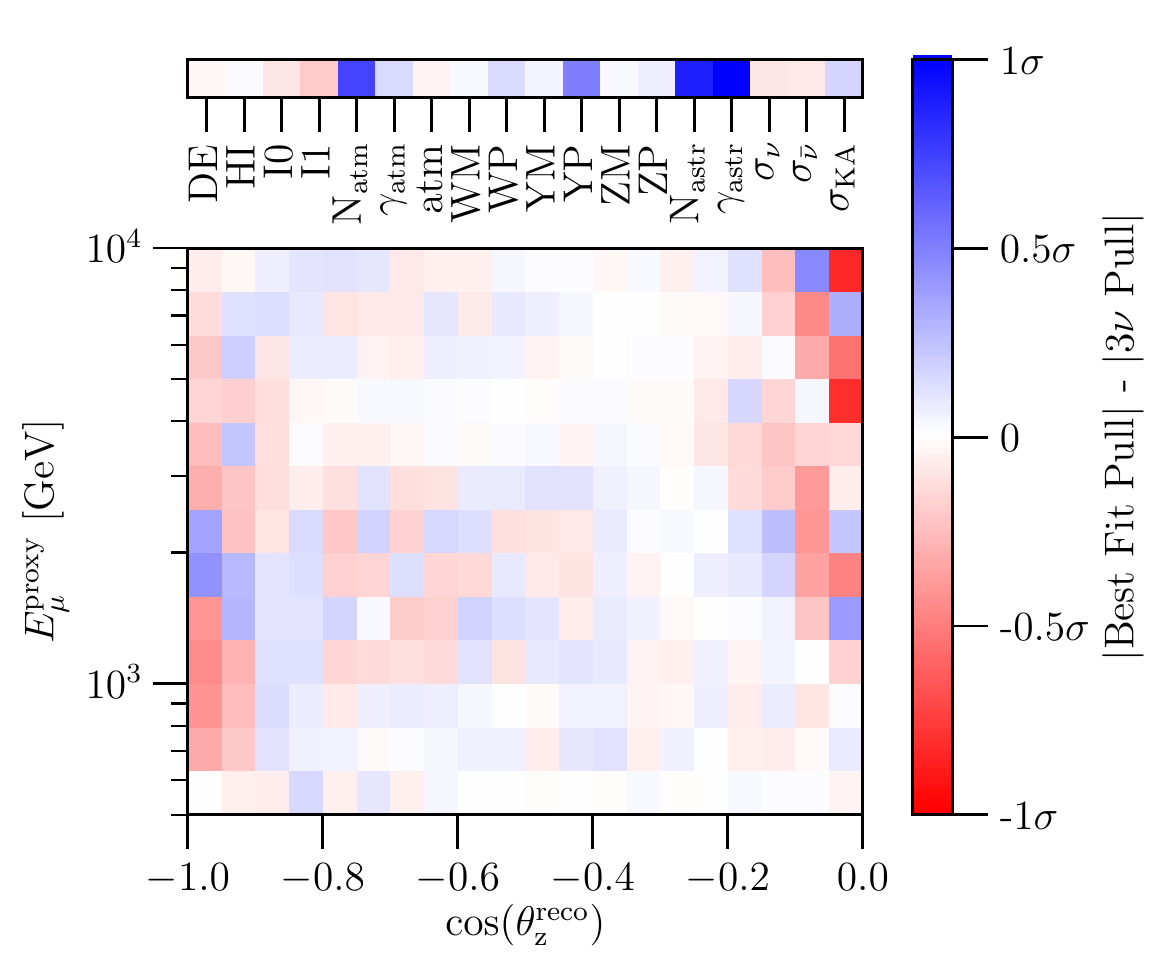}
\includegraphics[width=3.7in]{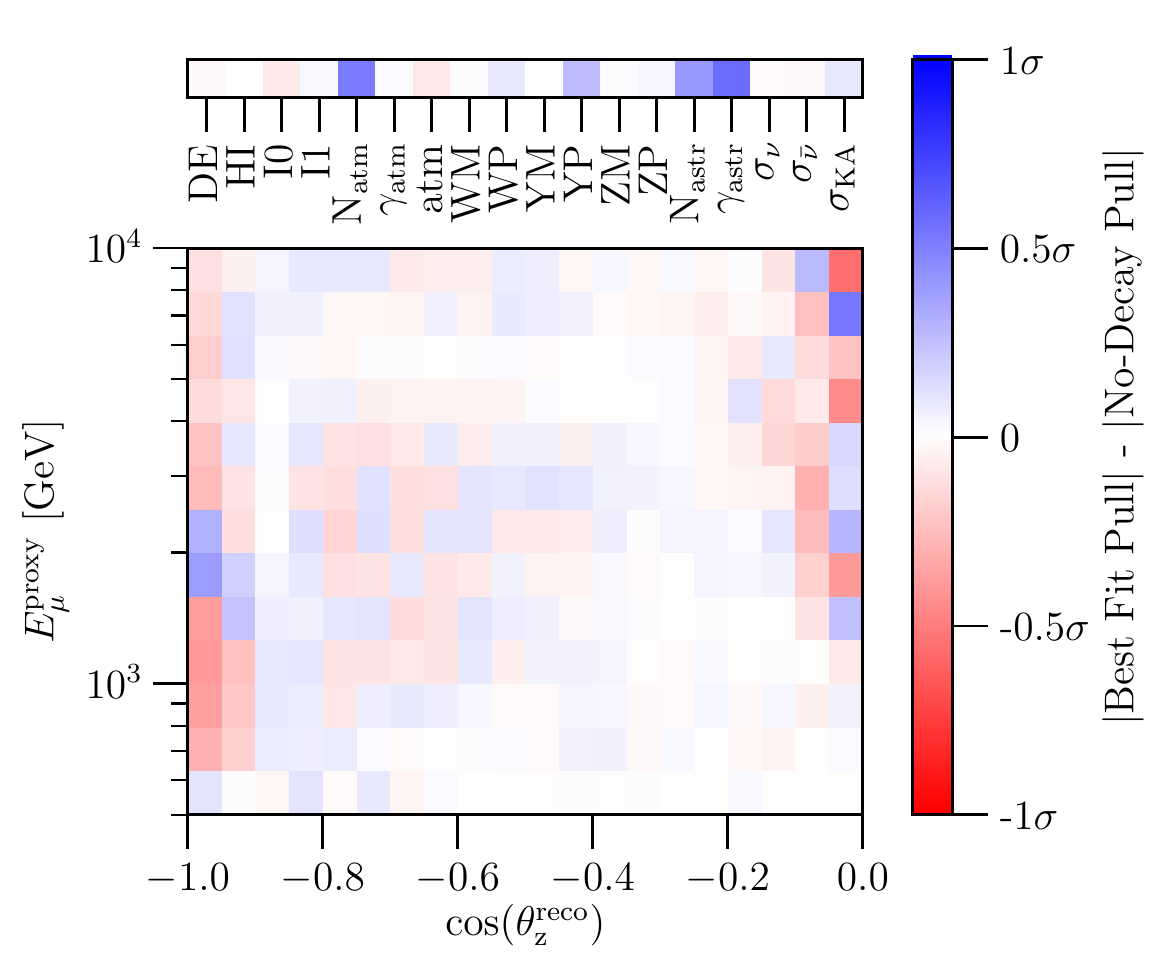}
\caption[Change in pulls by best-fit point]{(Top) Difference in the absolute value of pulls between the best-fit point and the three neutrino model. (Bottom) Difference in the absolute value of pulls between the best-fit point and the best-fit of the subset of points corresponding to no decay.}
\label{fig:delta_pull}
\end{figure}

In this analysis, the best fit of the subset of sampled points with $g^2 = 0$ is $\Delta m_{41}^2 = 4.2~\textrm{eV}^2$ and $\sin^2 2\theta_{24} = 0.11$, which is consistent with the result from the traditional 3+1 eight-year search. Figure~\ref{fig:delta_pull} show the difference in pulls between the best-fit point and the three-neutrino fit and best no-decay fit point, respectively. The red bins are where the best-fit expectation decreases the pulls. The blue bins are where the best-fit expectation increases the pulls. In comparison to both the three-neutrino fit and the traditional 3+1 best-fit, the overall best fit here reduces data pulls near the horizon at the higher energies, and for straight up-going events, especially at the lower energies. Statistical uncertainty for straight up-going events at the highest energies is large, so while the expected effect of the signal is large there, as shown in Fig.~\ref{fig:quilt} (top), data pulls are expected to be small.

\clearpage
\begin{figure}[ht]
  \centering
  \includegraphics[width=3in]{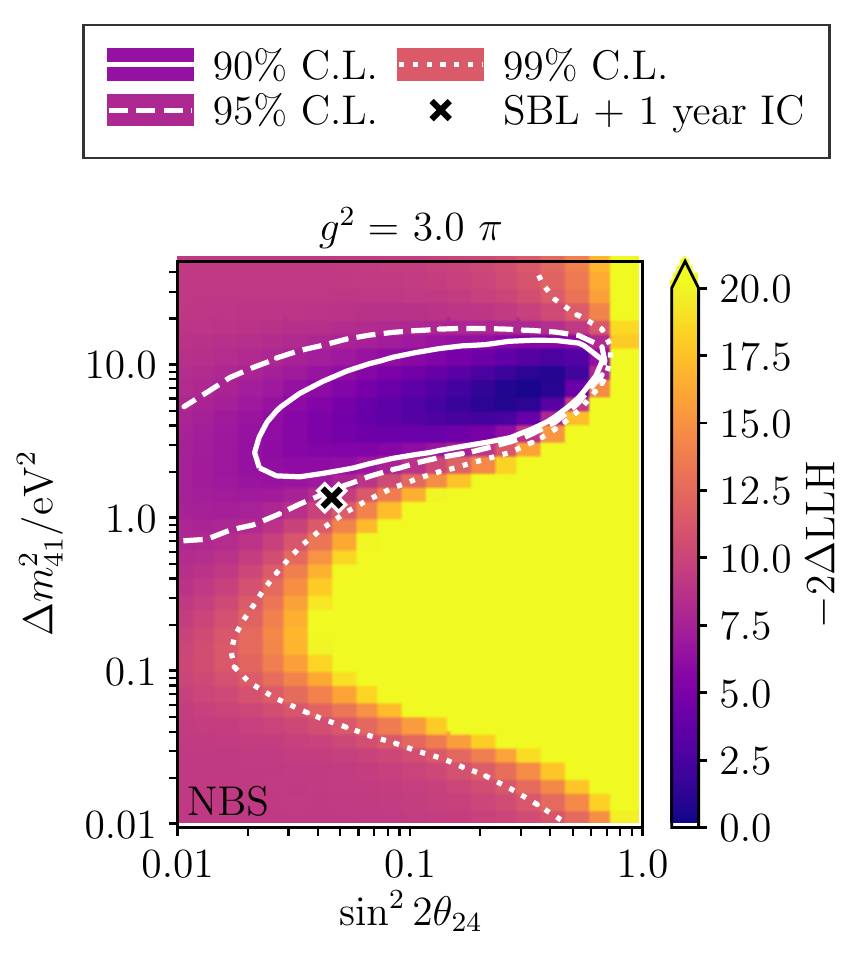}
  \caption[The $g^2=3\pi$ panel from the frequentist result.]{The $g^2=3\pi$ panel from the frequentist result. The best-fit point from combining global fits to short baseline data with one year of IceCube data is marked by an `x'. ``NBS'' means that this analysis did not include a bedrock systematic uncertainty. See \Cref{sec:bedrock}.}
  \label{fig:otherpoint}
\end{figure}

The best-fit point from combining global fits to short baseline data with one year of IceCube data, discussed in \Cref{sec:fits_prd}, is $\Delta m_{41}^2 = 1.35~\textrm{eV}^2$, $\sin^2 2\theta_{24} = 0.05$ and $g^2=3.06\pi$. Rounding 3.06 to 3, the point is marked by an `x' on the frequentist result panel for $g^2 = 3\pi$ in Fig.~\ref{fig:otherpoint}. This point sits just outside the 95\% C.L. allowed region.

\chapter{Future prospects and conclusions}

An eight-year search for light, unstable sterile neutrinos in the IceCube detector has found an anomalous result. The null hypothesis is rejected with a $p$-value of 2.8\%, which is equivalent to $2.2\sigma$. The frequentist analysis finds a best-fit point of $\Delta m^2_{41}$ = 6.7 eV$^2$, $\sin^2 2\theta_{24}$ = 0.33, and $g^2 = 2.5 \pi$. This corresponds to a $\nu_4$ lifetime of $\tau_4 / m_4 = 6 \times 10^{-16}$~s/eV. The traditional 3+1 model is rejected with a $p$-value of 4.9\%. The Bayesian analysis finds a best model with the same $\Delta m^2_{41}$ and $\sin^2 2\theta_{24}$ values as the frequentist best-fit, but the $g^2$ value is $1.5 \pi$. This model is preferred to the three-neutrino model by a factor of 41. The model associated with the frequentist best-fit is also a very good model, and is preferred to the three-neutrino model by a factor of 37.

This search scanned very coarsely in the $g^2$, and somewhat coarsely in $\Delta m_{41}^2$ and $\sin^2 2\theta_{24}$. A finer scan is likely to find a better best fit. The frequentist analysis has assumed Wilks' theorem holds. The confidence interval limits should be verified using the Feldman Cousins test. If the effective number of degrees of freedom is less than three, the frequentist result will become more significant. A precise sensitivity based off of pseudoexperiments should also be calculated.

The effect of the systematic uncertainty of the bedrock density should be finalized. If it is an important systematic effect, the best fit point and best model may change. The systematic treatment of the conventional atmospheric neutrino flux can be improved. The cosmic ray flux models used in the development of this thesis do not reflect a plethora of recent cosmic ray data~\cite{Alfaro:2017cwx, Atkin:2017vhi, An:2019wcw, IceCube:2020yct, Adriani:2019aft}. An upgraded cosmic ray flux treatment that incorporates recent data, such as a global fit, would be an important improvement~\cite{Evans:2016obt}. Similarly, the hadronic interaction uncertainties represented by the Barr parameters could be upgraded to reflect more recent measurements~\cite{Gaisser_PPPWNT2020,Aduszkiewicz:2017sei}. The hadronic interaction model used in this work, Sibyll 2.3c, has been upgraded to Sibyll 2.3d~\cite{Engel:2019dsg}.

Future searches for light, sterile neutrinos in IceCube should extend to higher energies and consider nonzero values of $\theta_{14}$ and $\theta_{34}$. They can look in an additional oscillation channel by using an independent dataset of cascade events, which correspond to electromagnetic and hadronic showers. One could simultaneously search for a disappearance of track events and an appearance of cascade events due to non-zero $\theta_{34}$.

This sterile result is the first reported anomaly in neutrino disappearance experiments. It is expected to have a significant impact on the sterile neutrino landscape. Both this result and the recently published, standard $3+1$ model result should be incorporated into global fits. These results should reduce tension between appearance and disappearance experiments, but the extent of this remains unknown. Lastly, accelerator-based neutrino experiments, such as MicroBooNE, which have completely different systematic uncertainties than IceCube, can perform follow-up searches for muon neutrino disappearance due to unstable sterile neutrinos.
\appendix
\chapter{Specific contributions}
All of the work in this thesis was done in collaboration with others. My specific contributions included:
\begin{itemize}
\item Studying and implementing systematic uncertainties and performing the analysis in~\cite{Moss:2017pur}.
\item Calculating the IceCube likelihood to combine with short-baseline fit results and analyzing the results in~\cite{Moulai:2019gpi}.
\item Studying the atmospheric neutrino flux uncertainties and implementing the Barr scheme in~\cite{Aartsen:2020iky, Aartsen:2020fwb}.
\item Performing the analysis for the IceCube search discussed in this thesis.
\item Building portions of, developing calibration methods for, and testing new gasses in a time projection chamber discussed in~\Cref{ch:MITPC}.
\end{itemize}
\chapter{Supplementary figures}
\label{ch:app_results}
\subsection{Data and systematic pulls}
\label{sec:app_results_pulls}
\begin{figure}[h]
\centering
\includegraphics[width=4.in]{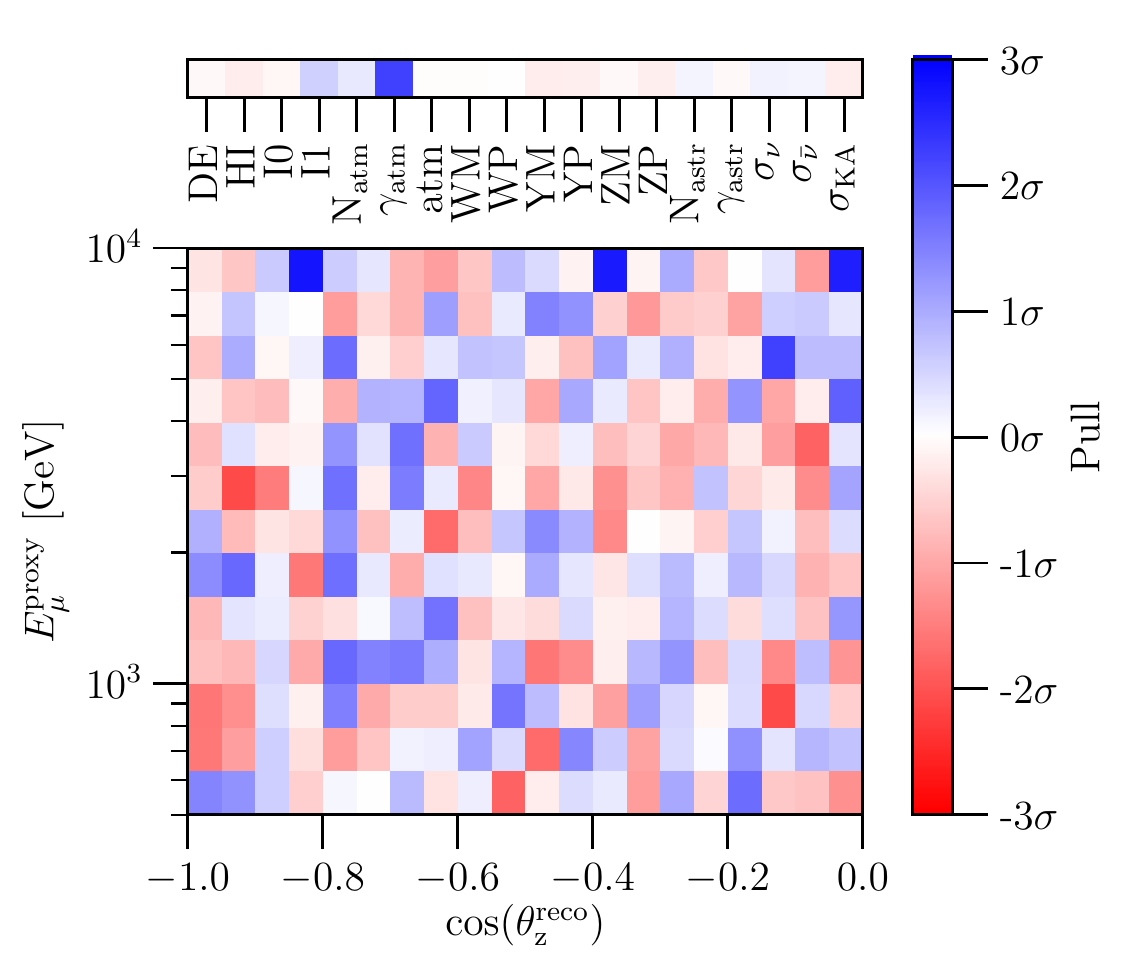}
\caption[Systematic and data pulls assuming three neutrinos]{Systematic and data pulls for the fit assuming three neutrinos. The top panel shows the systematic pulls for each of the nuisance parameters. The large, main panel shows the data pulls for each bin.}
\label{fig:pulls_null}
\end{figure}

\begin{figure}[h!]
\centering
\includegraphics[width=4.in]{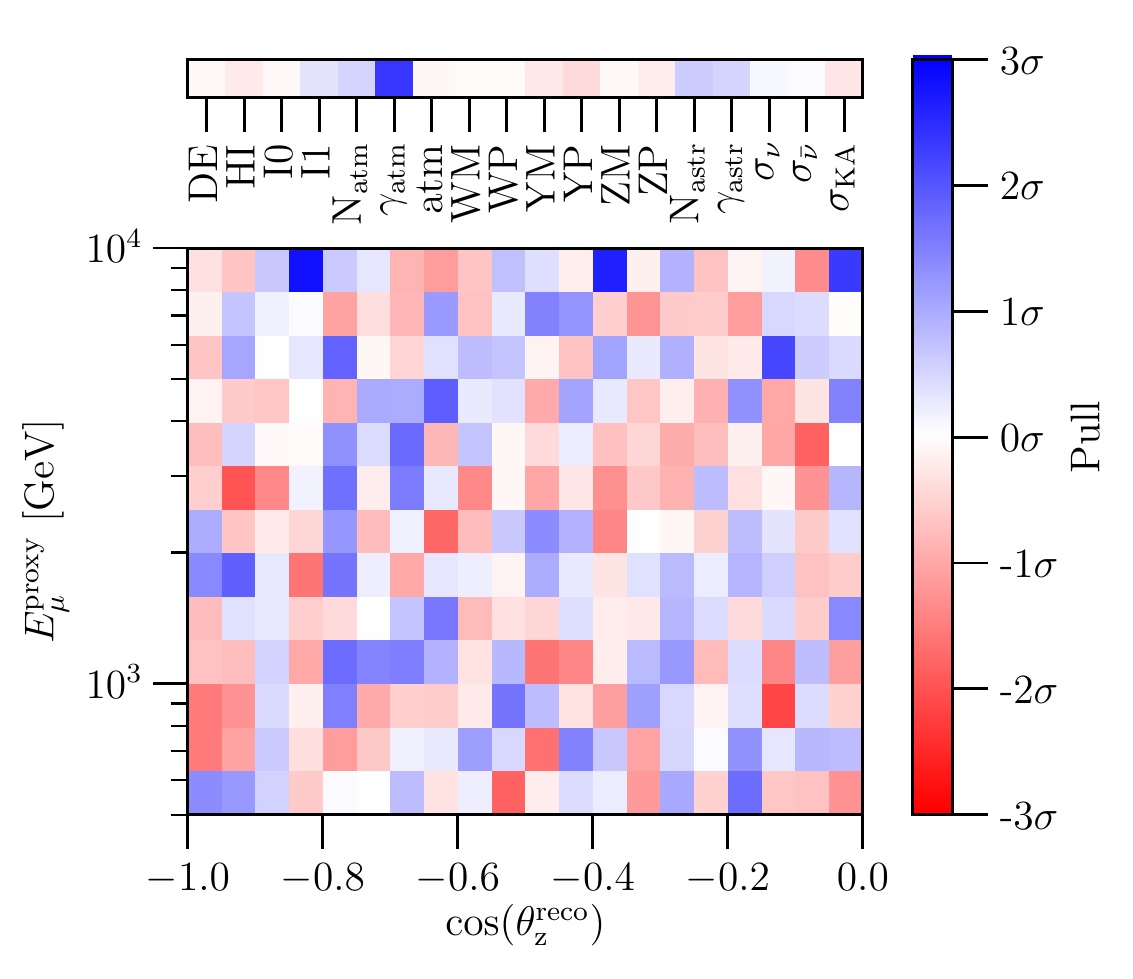}
\caption[Systematic and data pulls for the best traditional 3+1 point]{Systematic and data pulls for the best fit of the subset of points associated with the traditional $3+1$ model, i.e. $g^2 = 0$. The top panel shows the systematic pulls for each of the nuisance parameters. The large, main panel shows the data pulls for each bin.}
\label{fig:pulls_nodecay}
\end{figure}

Figure~\ref{fig:pulls_null} shows the data and systematic pulls, as defined in~\Cref{chap:results}, for the frequentist fit assuming three neutrinos. Figure~\ref{fig:pulls_nodecay} shows the data and systematic pulls for the best fit of the subset of points corresponding to $g^2 = 0$, or the traditional 3+1 model.

\FloatBarrier
\subsection{Fit systematic values}
\label{sec:app_results_nuisance_values}

\begin{figure}[h]
 \centering
 \includegraphics[width=\textwidth]{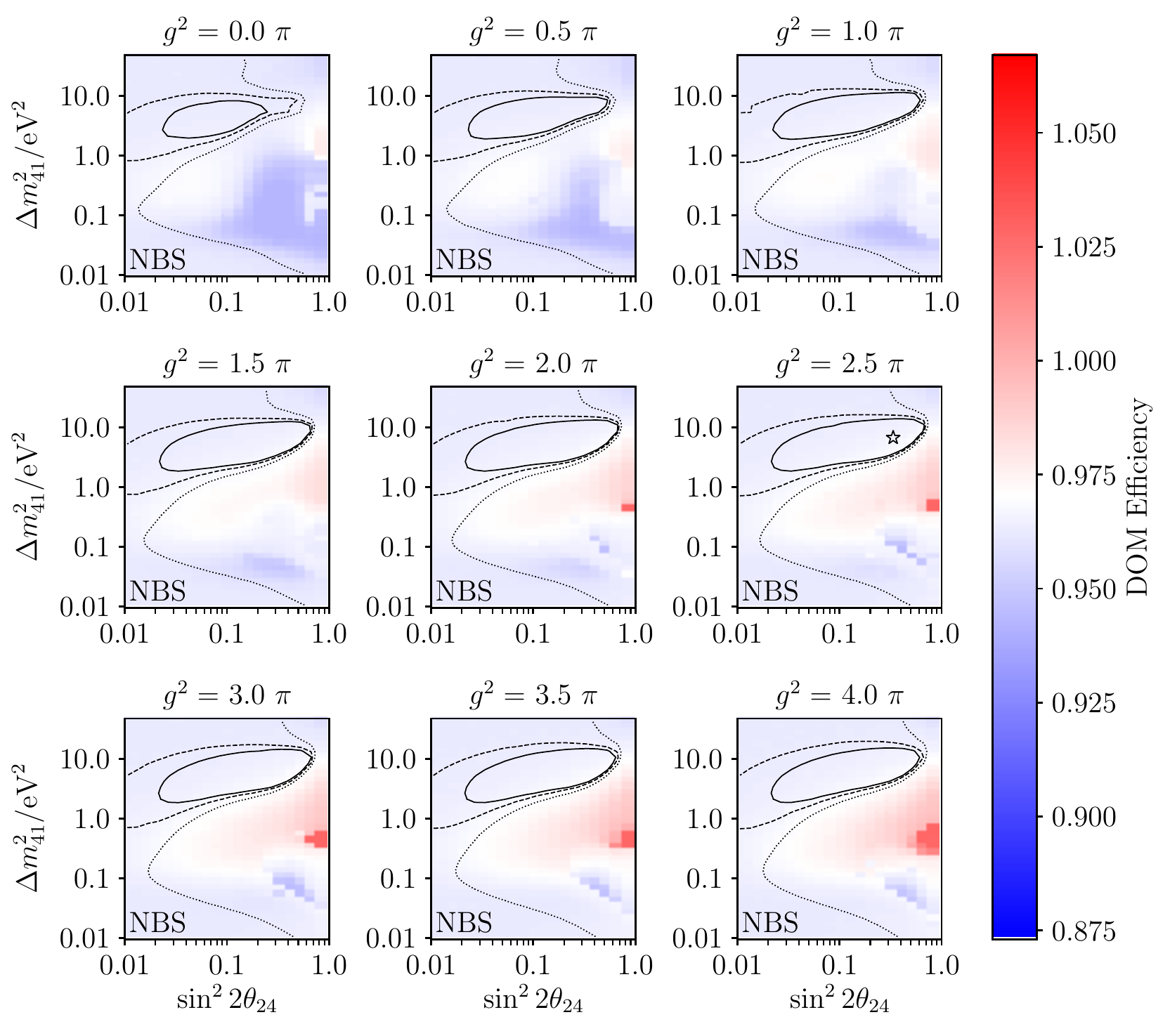}
 \caption[The fit values of the DOM efficiency]{The fit values of the DOM efficiency nuisance parameter from the frequentist analysis across the entire physics space. The color scale limits correspond to $\pm 1 \sigma$. ``NBS" means that no bedrock systematic uncertainty was included in this analysis.}
 \label{fig:nuis_DE}
\end{figure}

\begin{figure}[h]
 \centering
 \includegraphics[width=\textwidth]{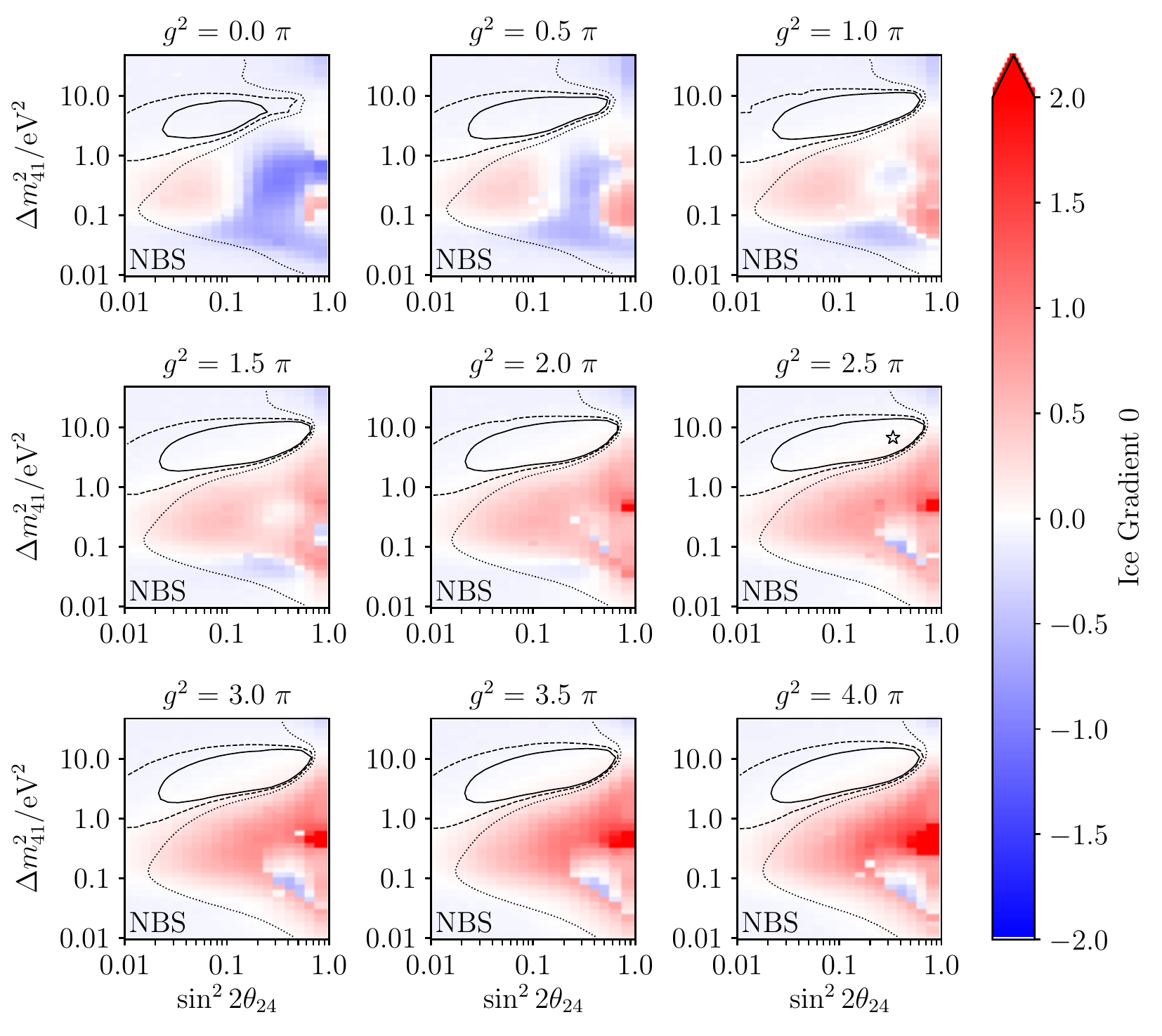}
 \caption[The fit values of Ice Gradient 0]{The fit values of the Ice Gradient 0 nuisance parameter from the frequentist analysis across the entire physics space. The color scale limits correspond to $\pm 2 \sigma$. ``NBS" means that no bedrock systematic uncertainty was included in this analysis.}
 \label{fig:nuis_I0}
\end{figure}

\begin{figure}[h]
 \centering
 \includegraphics[width=\textwidth]{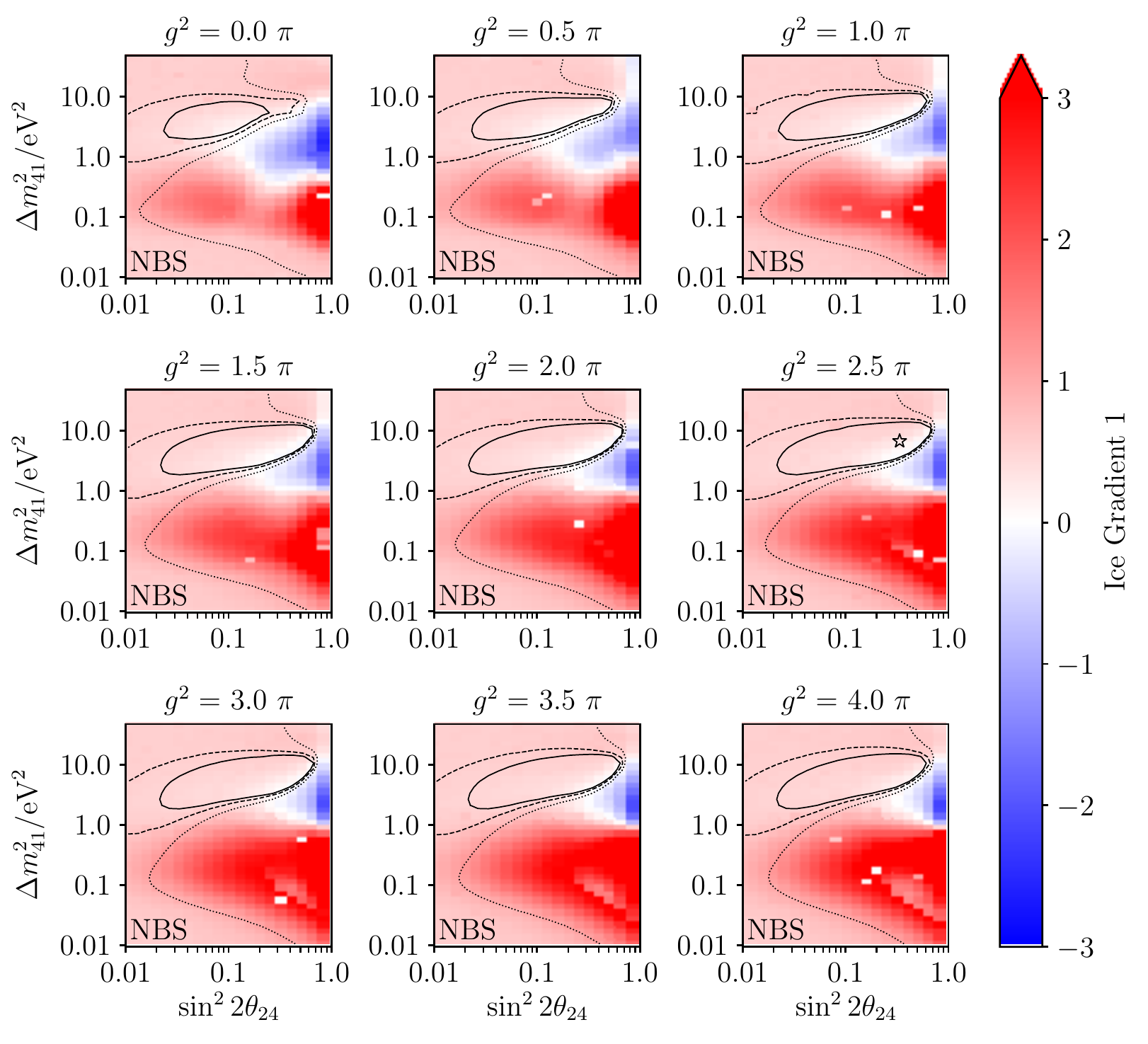}
 \caption[The fit values of Ice Gradient 1]{The fit values of the Ice Gradient 1 nuisance parameter from the frequentist analysis across the entire physics space. The color scale limits correspond to $\pm 3 \sigma$. ``NBS" means that no bedrock systematic uncertainty was included in this analysis.}
 \label{fig:nuis_I1}
\end{figure}

\begin{figure}[h]
 \centering
 \includegraphics[width=\textwidth]{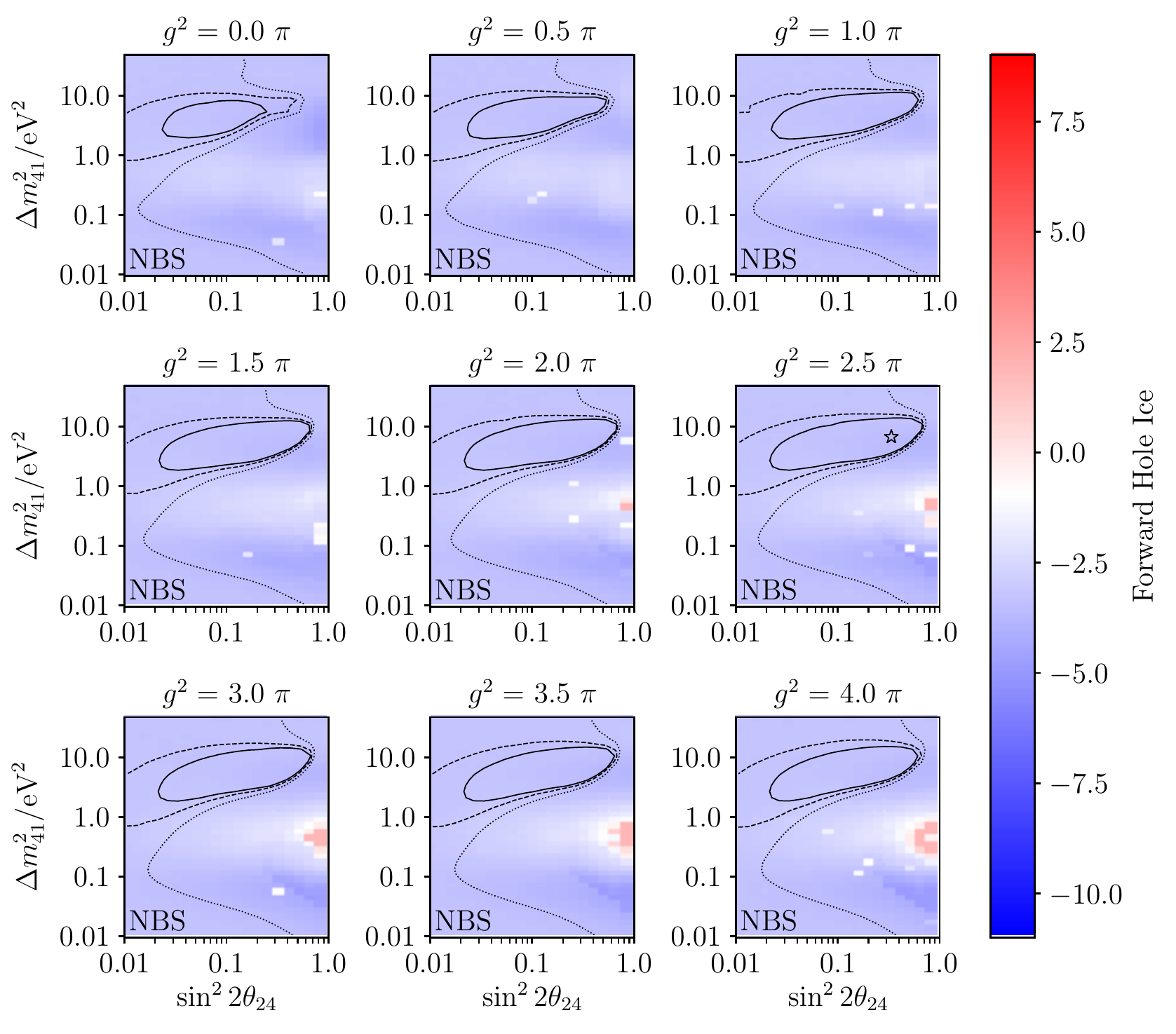}
 \caption[The fit values of Hole Ice]{The fit values of the forward hole ice nuisance parameter from the frequentist analysis across the entire physics space. The color scale limits correspond to $\pm 1 \sigma$. ``NBS" means that no bedrock systematic uncertainty was included in this analysis.}
 \label{fig:nuis_HI}
\end{figure}

\begin{figure}[h]
 \centering
 \includegraphics[width=\textwidth]{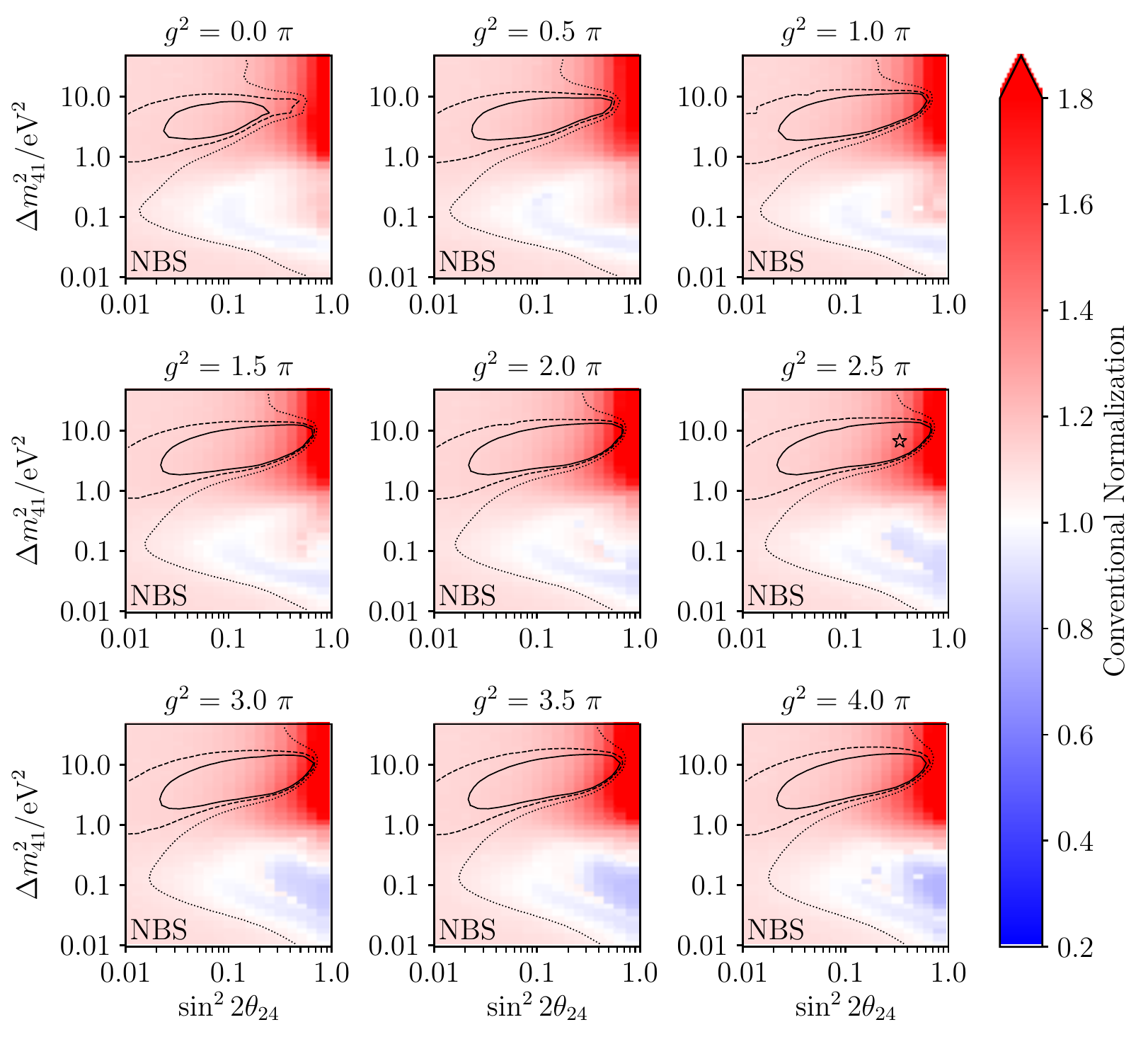}
 \caption[The fit values of conventional normalization]{The fit values of the conventional atmospheric normalization nuisance parameter from the frequentist analysis across the entire physics space. The color scale limits correspond to $\pm 2 \sigma$. ``NBS" means that no bedrock systematic uncertainty was included in this analysis. ``NBS" means that no bedrock systematic uncertainty was included in this analysis.}
 \label{fig:nuis_norm}
\end{figure}

\begin{figure}[h]
 \centering
 \includegraphics[width=\textwidth]{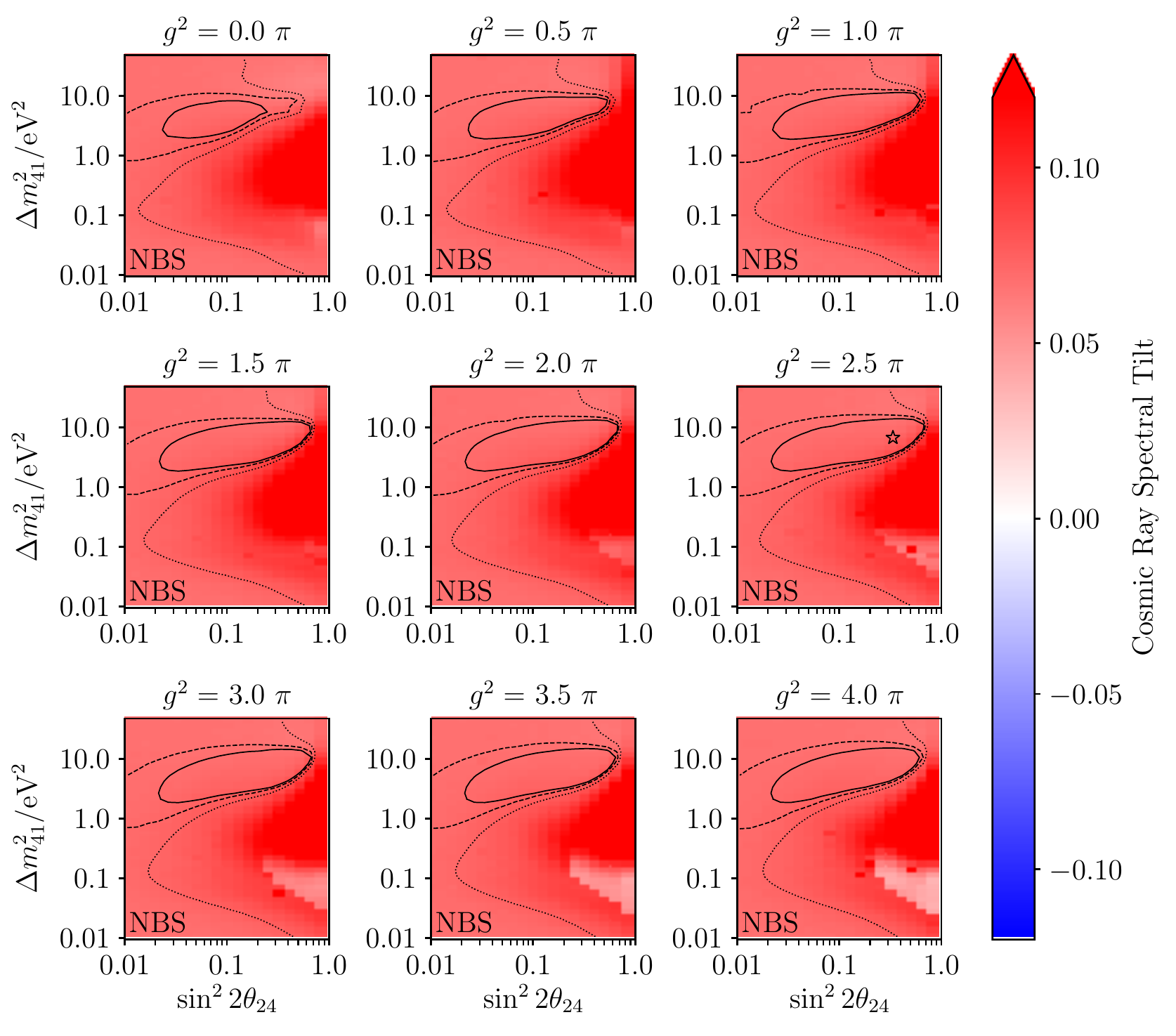}
 \caption[The fit values of the cosmic ray spectral tilt]{The fit values of the cosmic ray spectral tilt nuisance parameter from the frequentist analysis across the entire physics space. The color scale limits correspond to $\pm 3\sigma$. ``NBS" means that no bedrock systematic uncertainty was included in this analysis.}
 \label{fig:nuis_gamma}
\end{figure}

\begin{figure}[h]
 \centering
 \includegraphics[width=\textwidth]{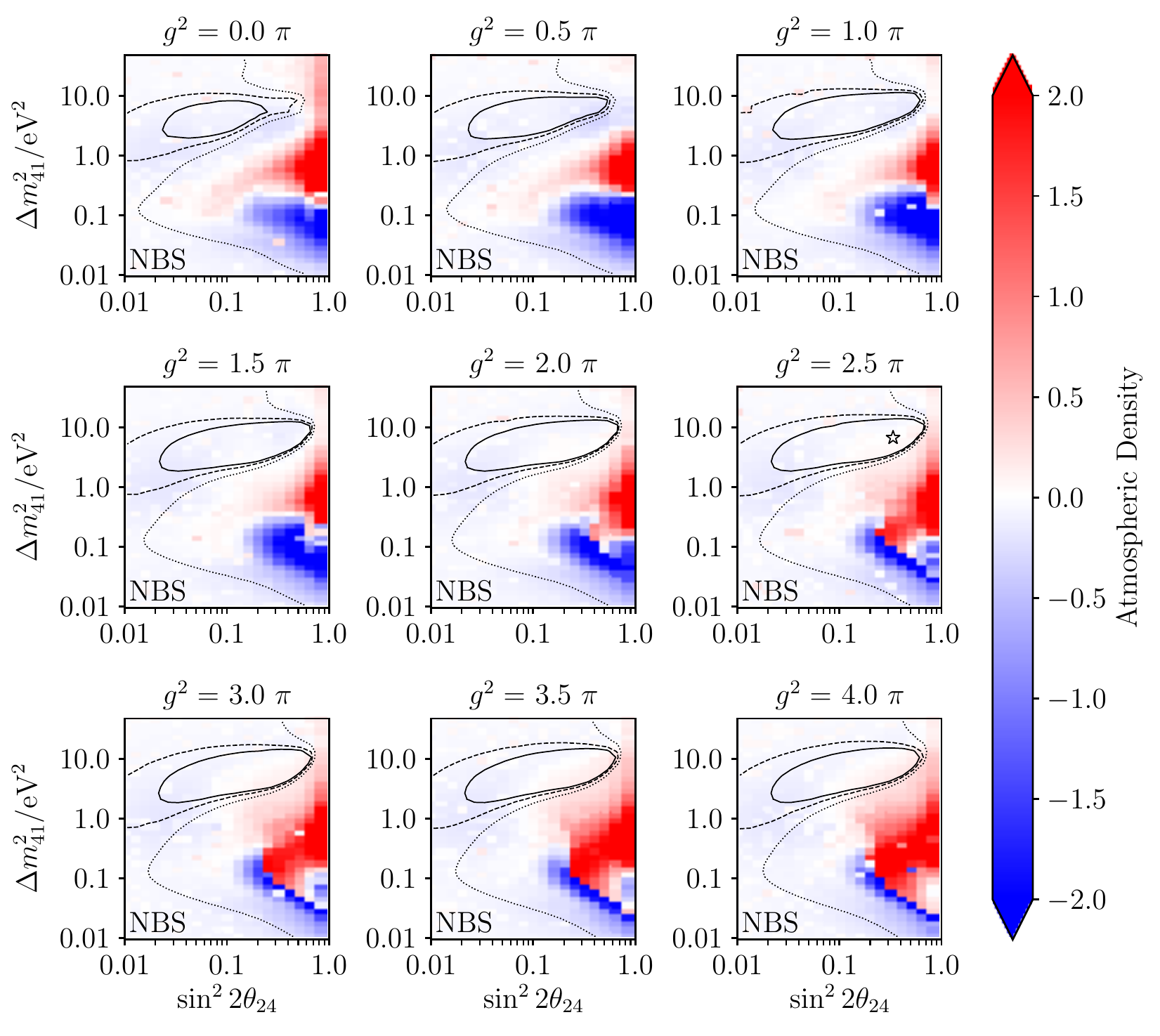}
 \caption[The fit values of the atmospheric density shift]{The fit values of the atmospheric density nuisance parameter from the frequentist analysis across the entire physics space. The color scale limits correspond to $\pm 2\sigma$. ``NBS" means that no bedrock systematic uncertainty was included in this analysis.}
 \label{fig:nuis_atm}
\end{figure}

\begin{figure}[h]
 \centering
 \includegraphics[width=\textwidth]{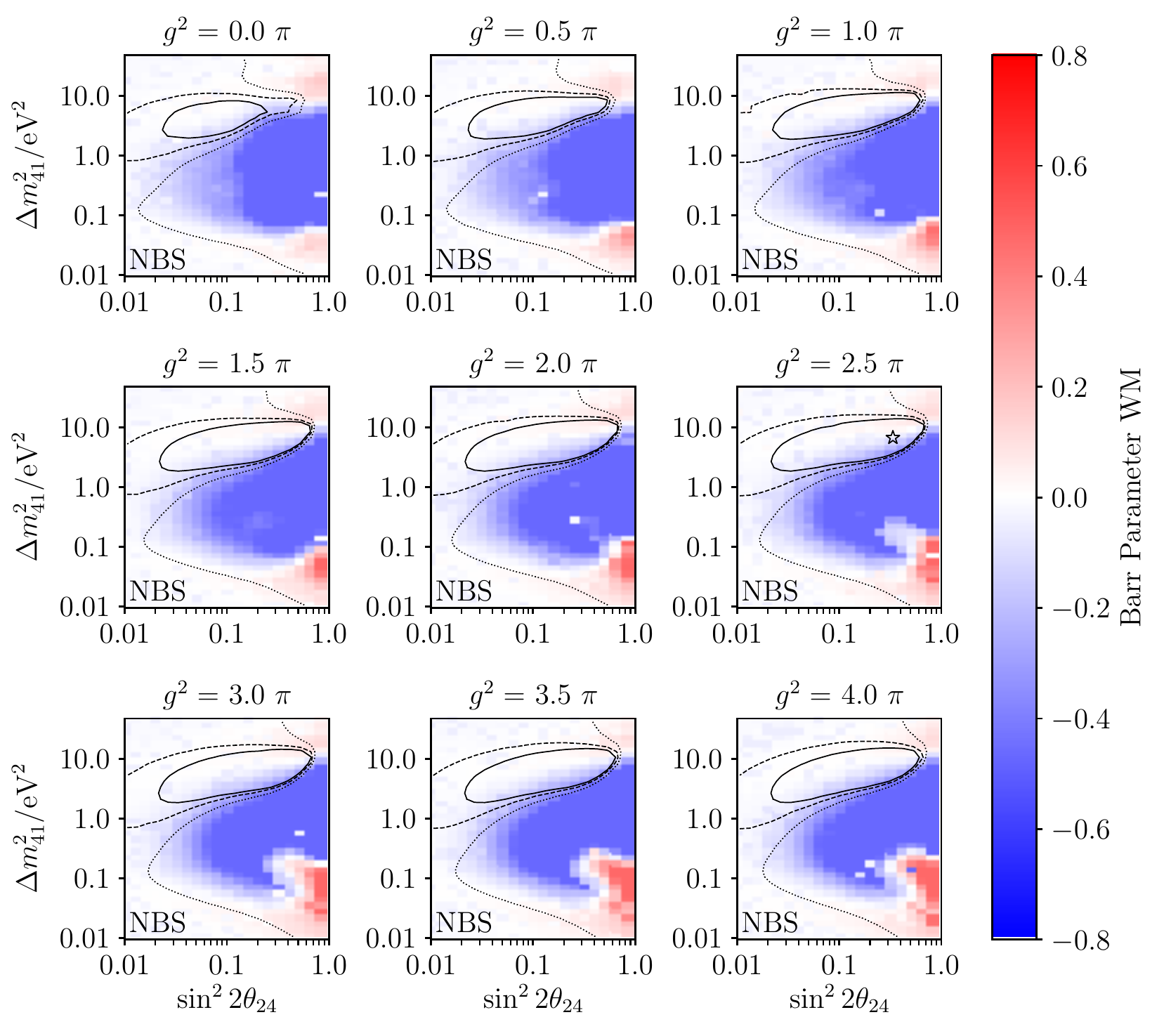}
 \caption[The fit values of Barr WM]{The fit values of the Barr WM nuisance parameter from the frequentist analysis across the entire physics space. The color scale limits correspond to $\pm 2 \sigma$. ``NBS" means that no bedrock systematic uncertainty was included in this analysis.}
 \label{fig:nuis_WM}
\end{figure}

\begin{figure}[h]
 \centering
 \includegraphics[width=\textwidth]{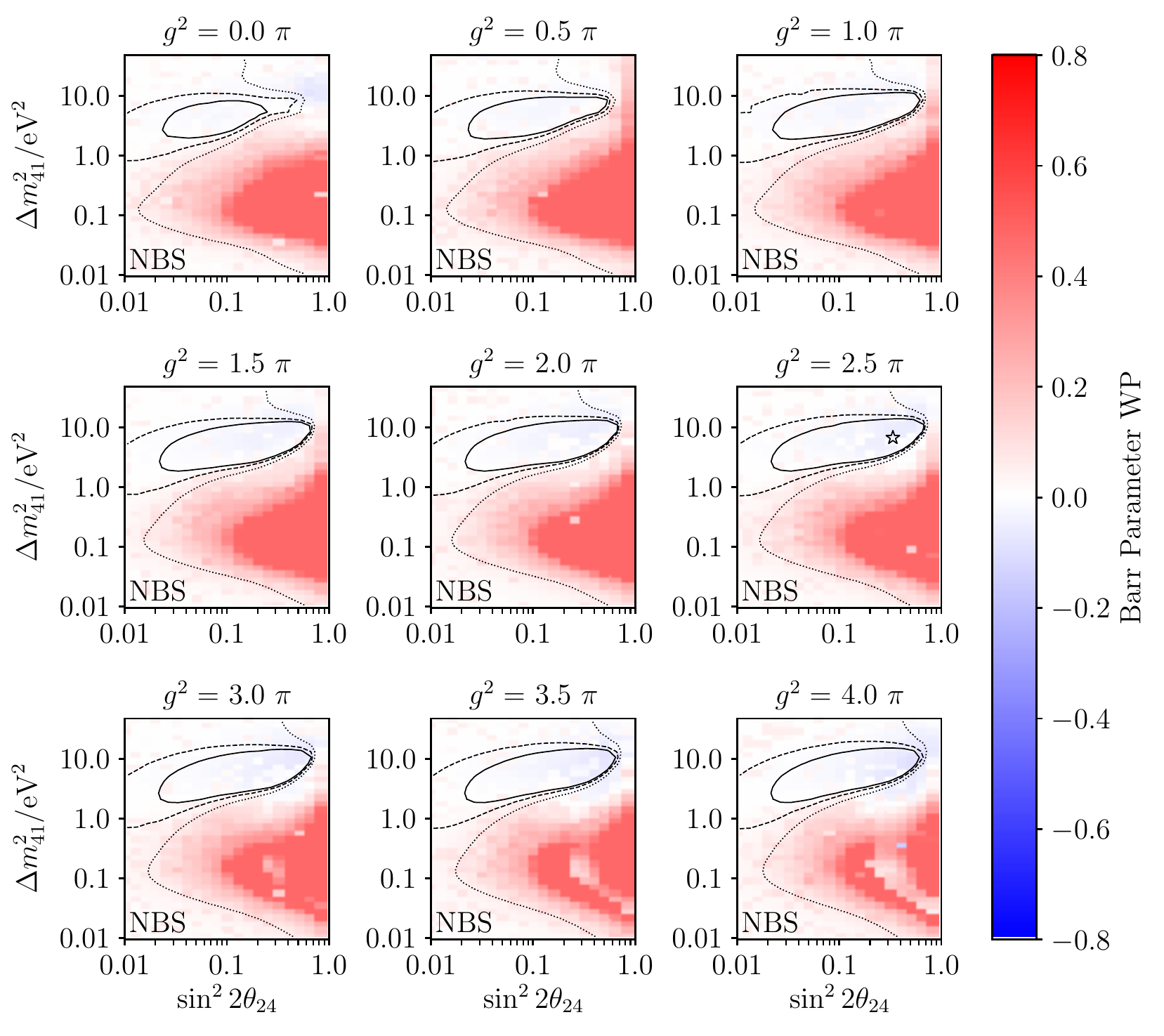}
 \caption[The fit values of Barr WP]{The fit values of the Barr WP nuisance parameter from the frequentist analysis across the entire physics space. The color scale limits correspond to $\pm 2 \sigma$. ``NBS" means that no bedrock systematic uncertainty was included in this analysis.}
 \label{fig:nuis_WP}
\end{figure}

\begin{figure}[h]
 \centering
 \includegraphics[width=\textwidth]{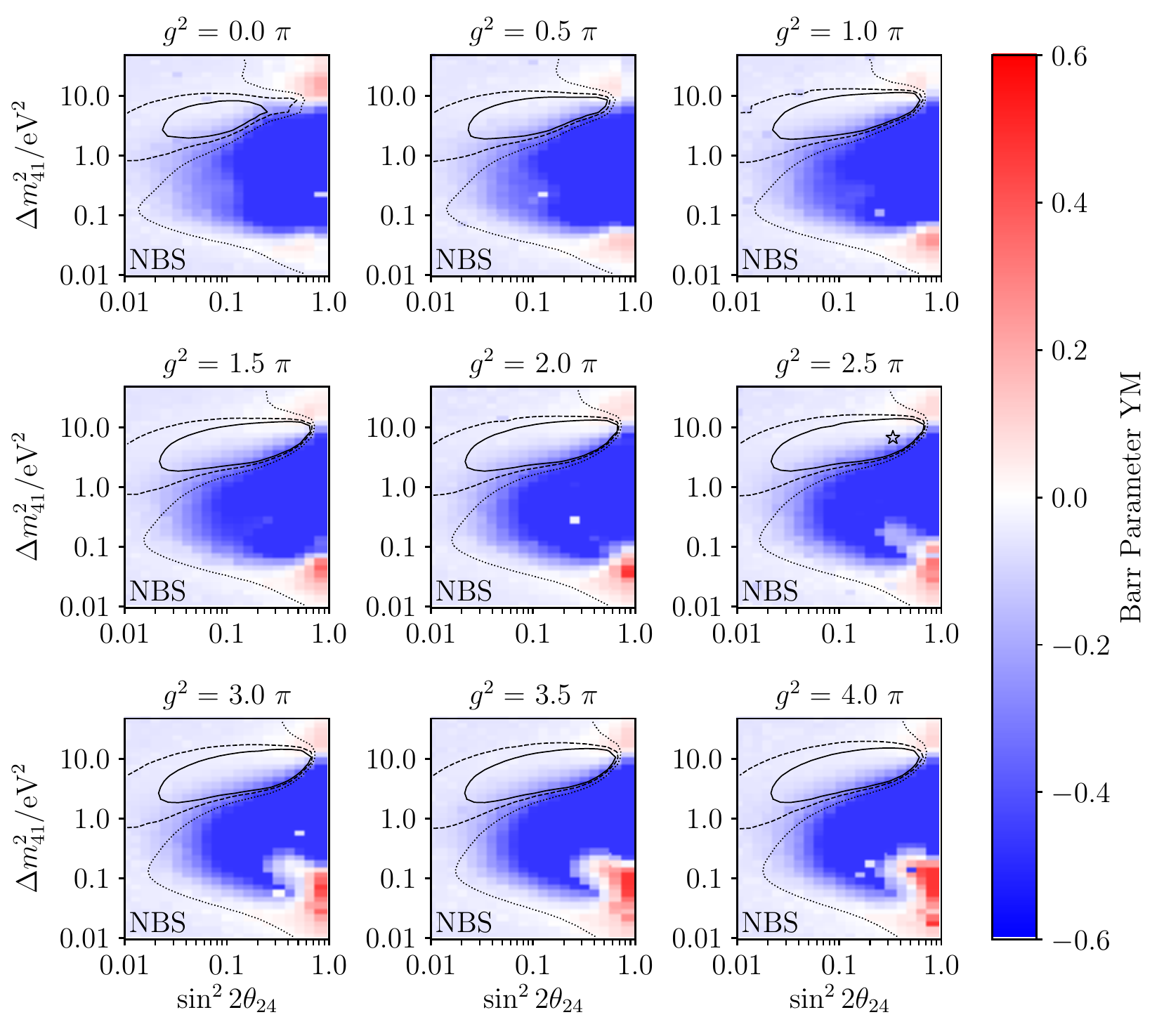}
 \caption[The fit values of Barr YM]{The fit values of the Barr YM nuisance parameter from the frequentist analysis across the entire physics space. The color scale limits correspond to $\pm 2 \sigma$. ``NBS" means that no bedrock systematic uncertainty was included in this analysis.}
 \label{fig:nuis_YM}
\end{figure}

\begin{figure}[h]
 \centering
 \includegraphics[width=\textwidth]{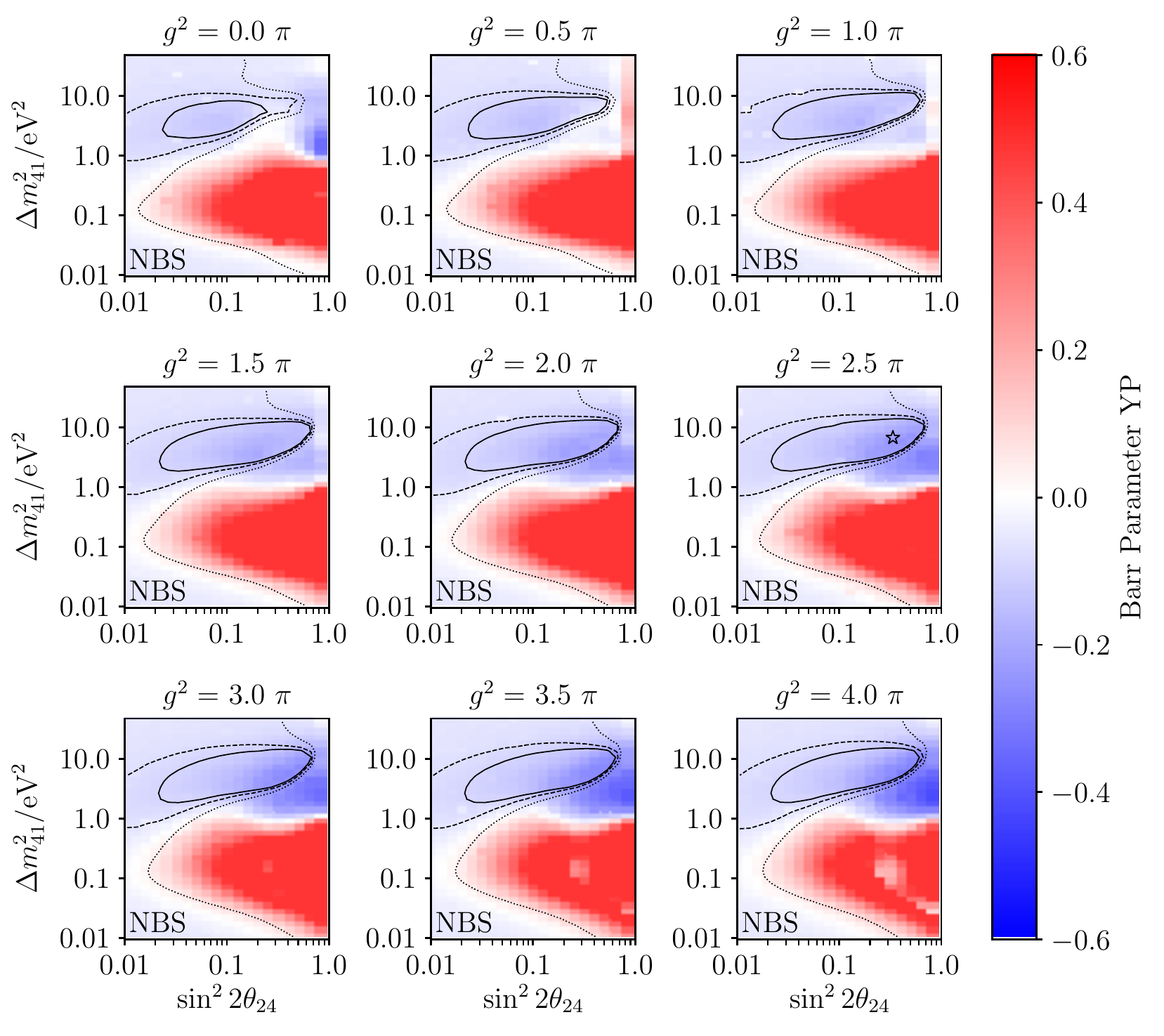}
 \caption[The fit values of Barr YP]{The fit values of the Barr YP nuisance parameter from the frequentist analysis across the entire physics space. The color scale limits correspond to $\pm 2 \sigma$. ``NBS" means that no bedrock systematic uncertainty was included in this analysis.}
 \label{fig:nuis_YP}
\end{figure}

\begin{figure}[h]
 \centering
 \includegraphics[width=\textwidth]{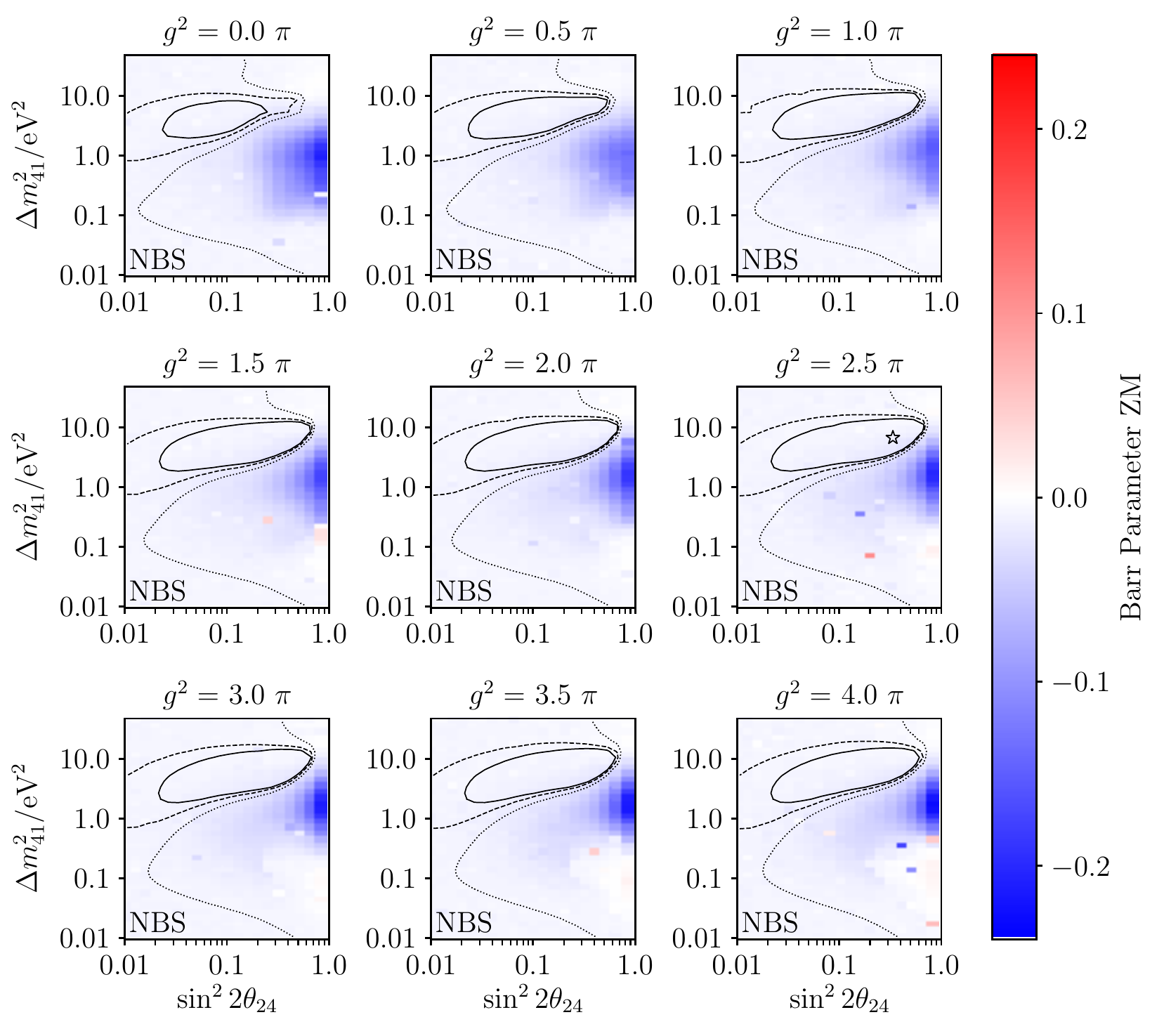}
 \caption[The fit values of Barr ZM]{The fit values of the Barr ZM nuisance parameter from the frequentist analysis across the entire physics space. The color scale limits correspond to $\pm 2 \sigma$. ``NBS" means that no bedrock systematic uncertainty was included in this analysis.}
 \label{fig:nuis_ZM}
\end{figure}

\begin{figure}[h]
 \centering
 \includegraphics[width=\textwidth]{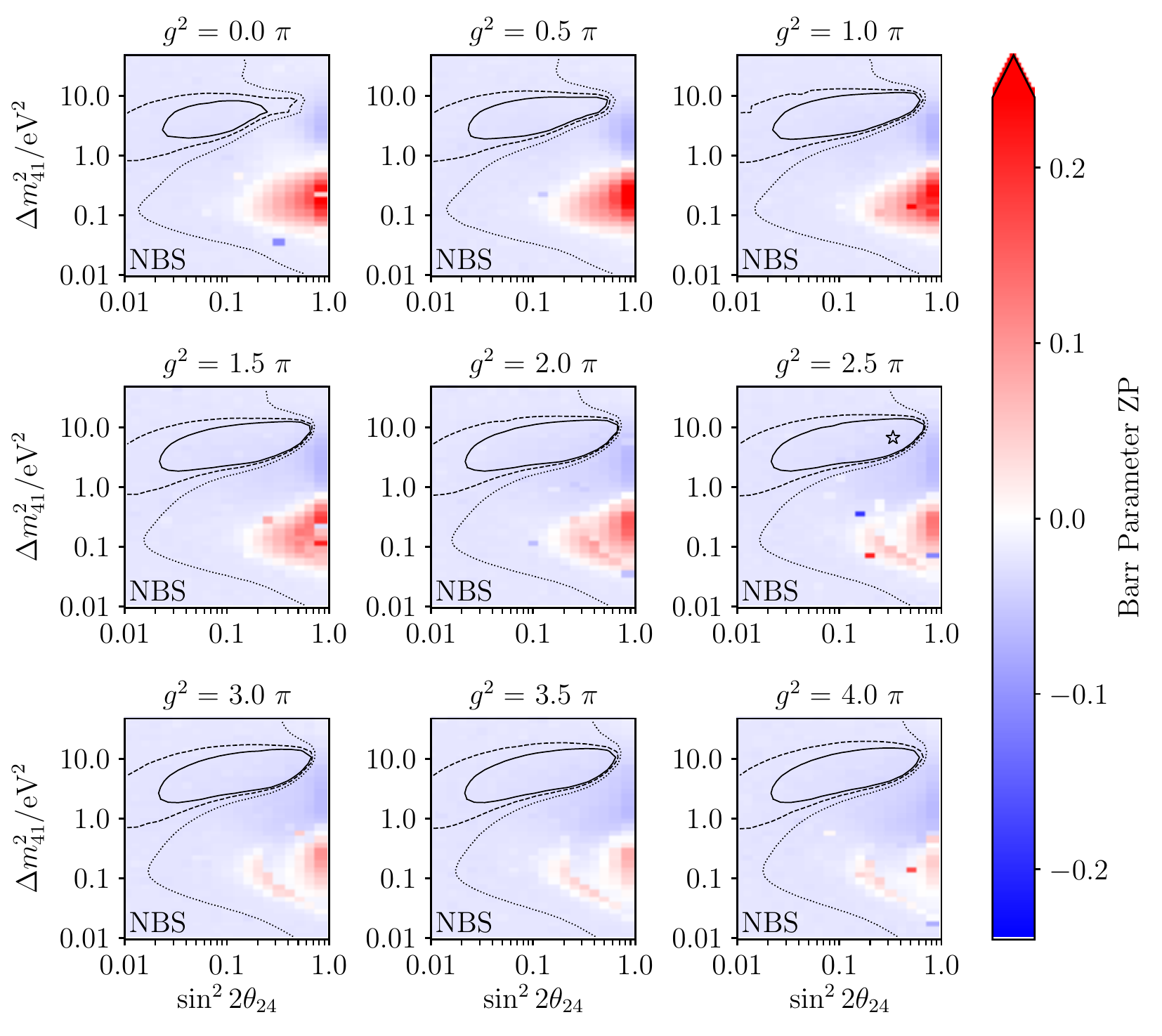}
 \caption[The fit values of Barr ZP]{The fit values of the Barr ZP nuisance parameter from the frequentist analysis across the entire physics space. The color scale limits correspond to $\pm 2 \sigma$. ``NBS" means that no bedrock systematic uncertainty was included in this analysis.}
 \label{fig:nuis_ZP}
\end{figure}

\begin{figure}[h]
 \centering
 \includegraphics[width=\textwidth]{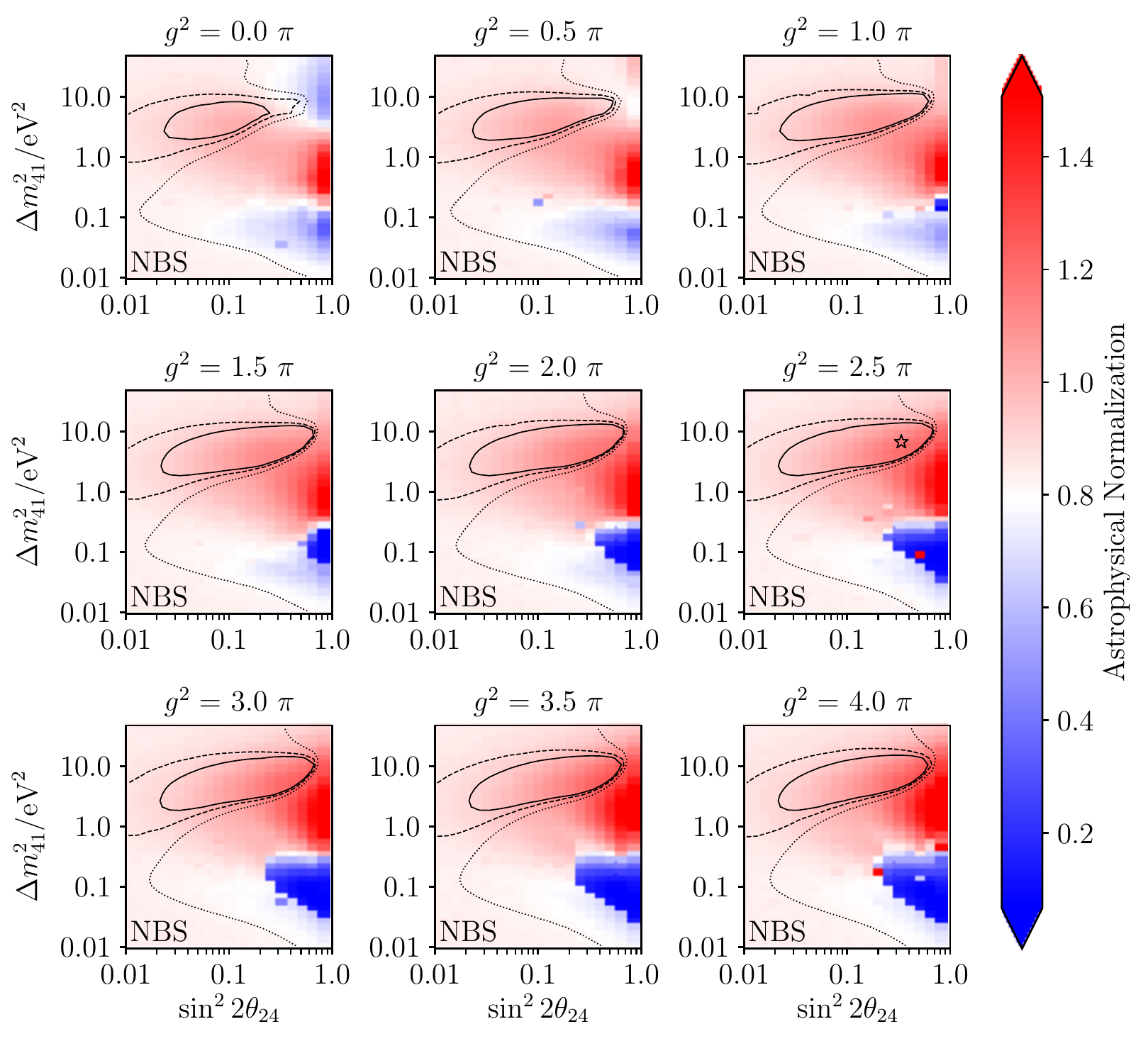}
 \caption[The fit values of the astrophysical normalization]{The fit values of the astrophysical normalization nuisance parameter from the frequentist analysis across the entire physics space. The color scale limits correspond to $\pm 2 \sigma$. ``NBS" means that no bedrock systematic uncertainty was included in this analysis.}
 \label{fig:nuis_AN}
\end{figure}

\begin{figure}[h]
 \centering
 \includegraphics[width=\textwidth]{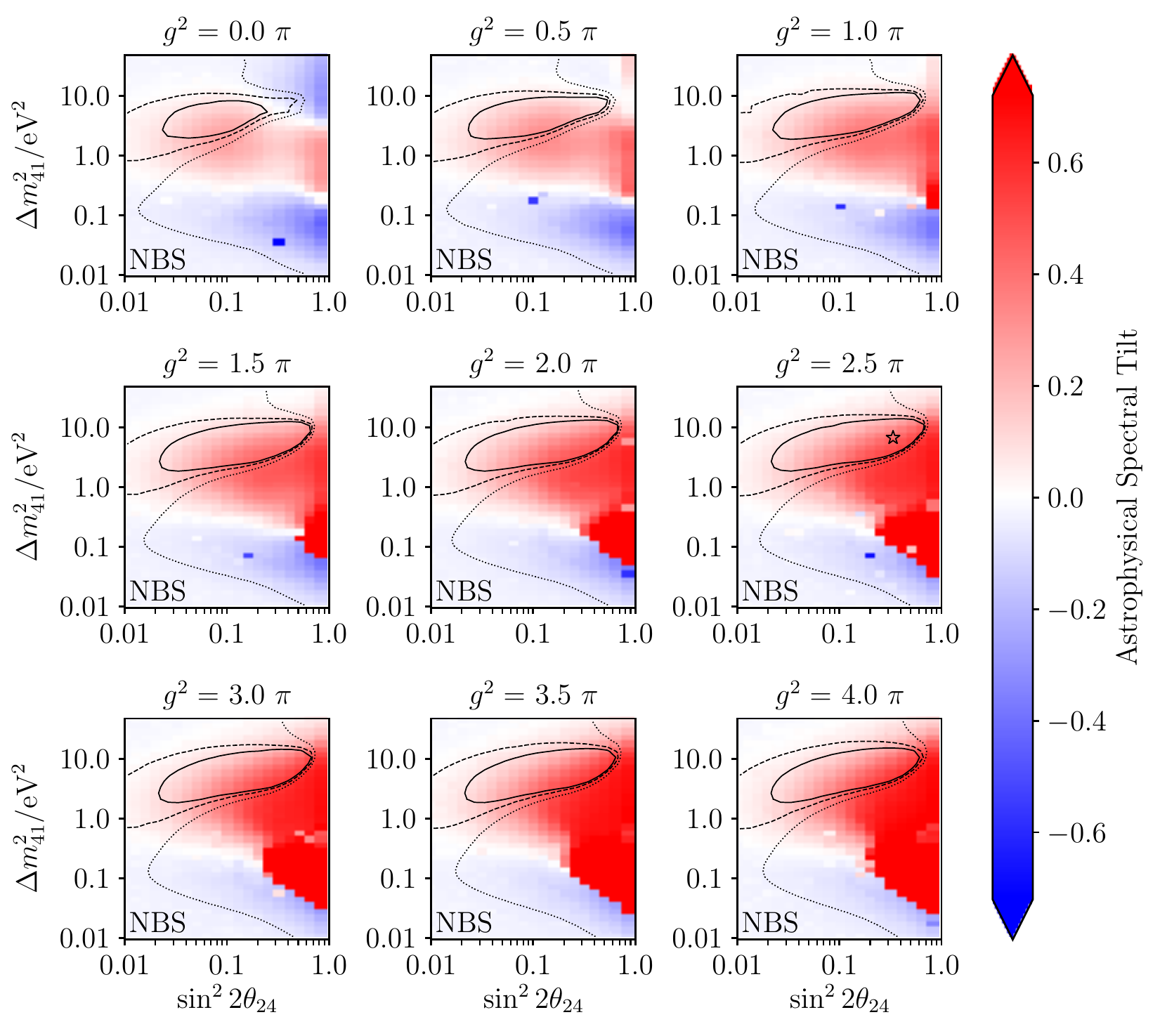}
 \caption[The fit values of the astrophysical spectral tilt]{The fit values of the astrophysical spectral tilt nuisance parameter from the frequentist analysis across the entire physics space. The color scale limits correspond to $\pm 2 \sigma$. ``NBS" means that no bedrock systematic uncertainty was included in this analysis.}
 \label{fig:nuis_ADG}
\end{figure}

\begin{figure}[h]
 \centering
 \includegraphics[width=\textwidth]{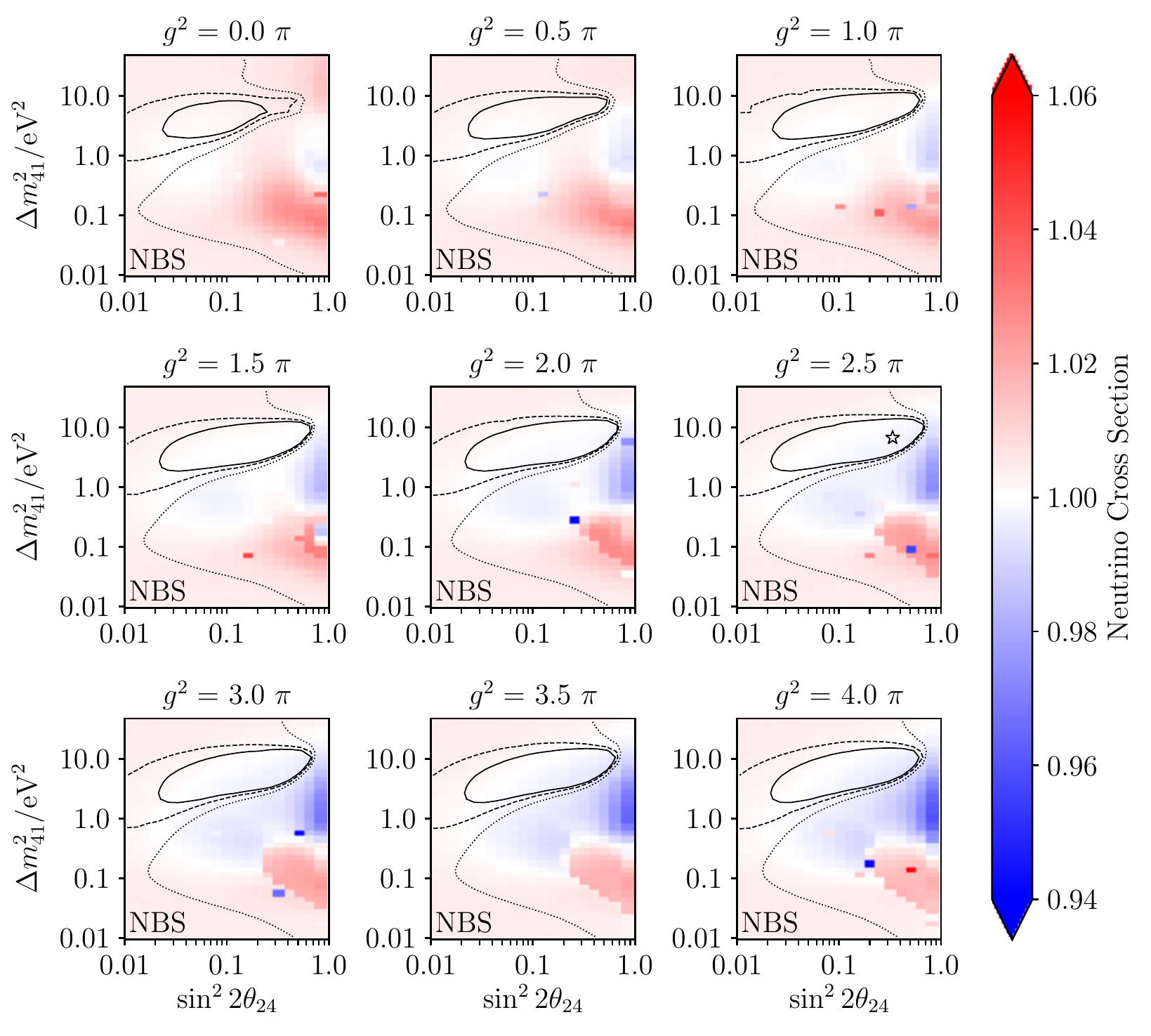}
 \caption[The fit values of the neutrino cross section]{The fit values of the neutrino cross section nuisance parameter from the frequentist analysis across the entire physics space. The color scale limits correspond to $\pm 1 \sigma$. ``NBS" means that no bedrock systematic uncertainty was included in this analysis.}
 \label{fig:nuis_nuxs}
\end{figure}

\begin{figure}[h]
 \centering
 \includegraphics[width=\textwidth]{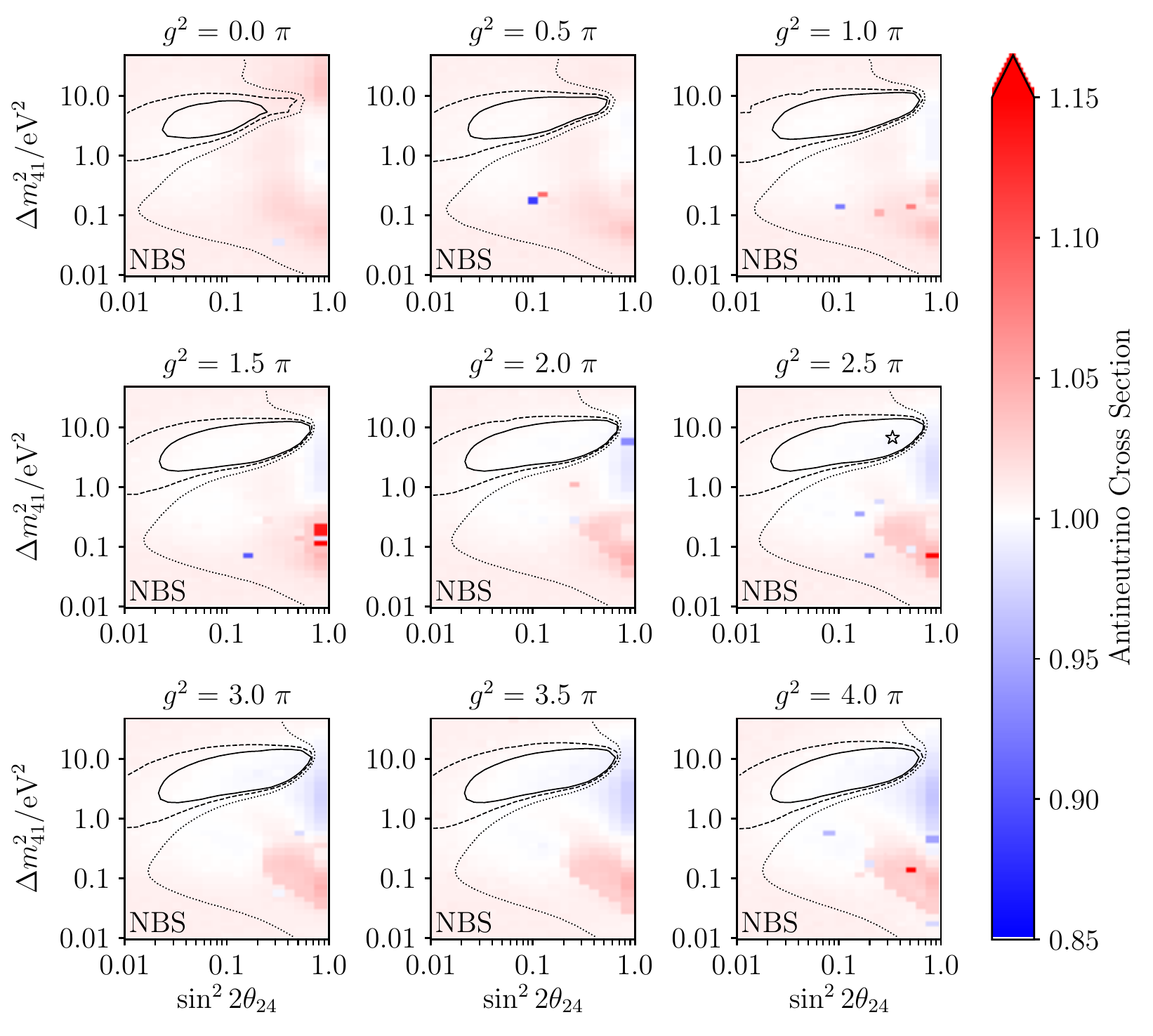}
 \caption[The fit values of the antineutrino cross section]{The fit values of the antineutrino cross section nuisance parameter from the frequentist analysis across the entire physics space. The color scale limits correspond to $\pm 1 \sigma$. ``NBS" means that no bedrock systematic uncertainty was included in this analysis.}
 \label{fig:nuis_anuxs}
\end{figure}

\begin{figure}[h]
 \centering
 \includegraphics[width=\textwidth]{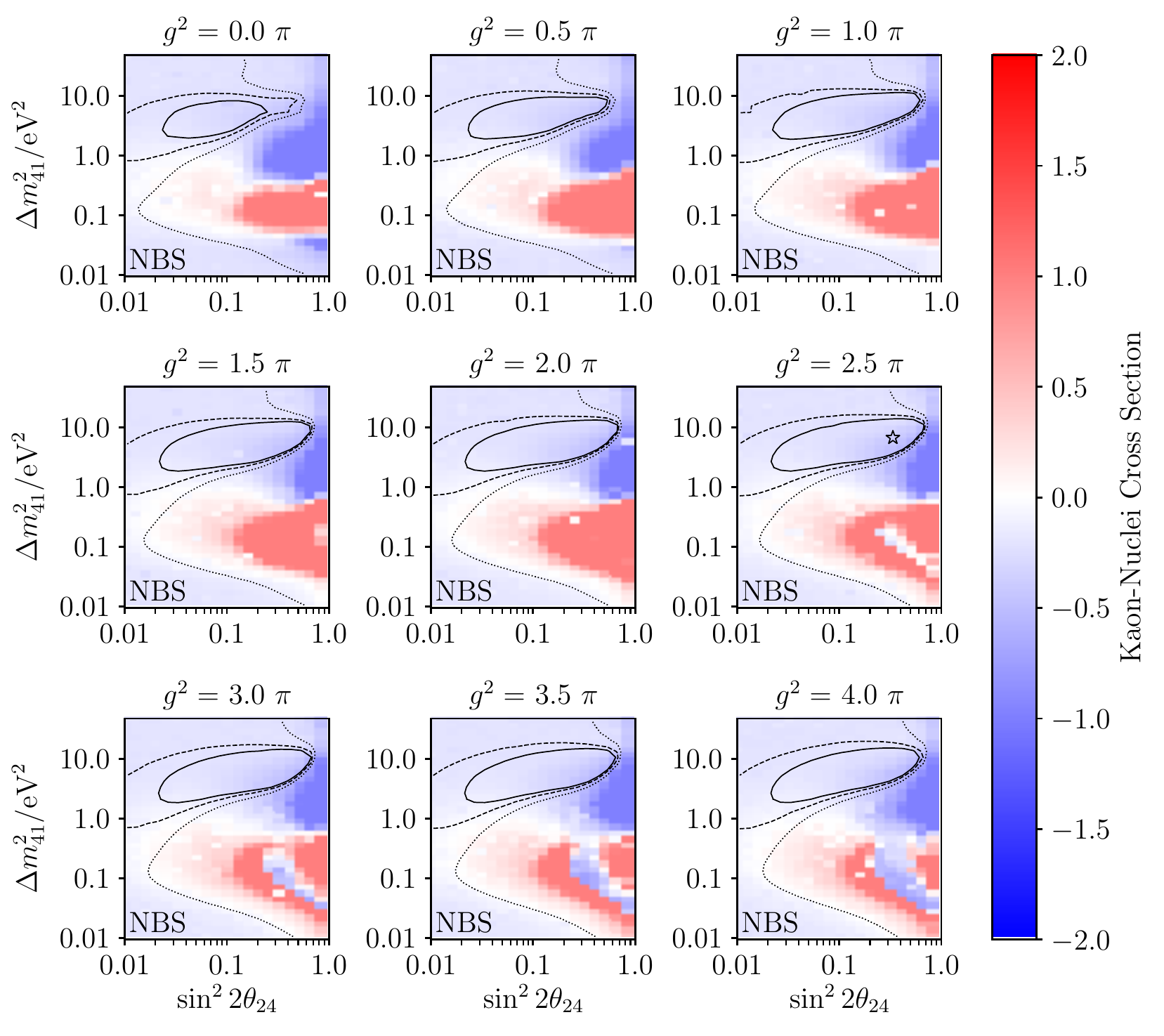}
 \caption[The fit values of the kaon-nucleus cross section]{The fit values of the kaon-nucleus cross section nuisance parameter from the frequentist analysis across the entire physics space. The color scale limits correspond to $\pm 1 \sigma$. ``NBS" means that no bedrock systematic uncertainty was included in this analysis.}
 \label{fig:nuis_KL}
\end{figure}

Figures~\labelcref{fig:nuis_DE,fig:nuis_I0,fig:nuis_I1,fig:nuis_HI,fig:nuis_norm,fig:nuis_gamma,fig:nuis_atm,fig:nuis_WM,fig:nuis_WP,fig:nuis_YM,fig:nuis_YP,fig:nuis_ZM,fig:nuis_ZP,fig:nuis_AN,fig:nuis_ADG,fig:nuis_nuxs,fig:nuis_anuxs,fig:nuis_KL} show the fit values of each of the systematic parameters across the entire scanned physics space for the frequentist analysis.

\chapter{MITPC}
\label{ch:MITPC}

For completeness, this appendix describes work performed early in my doctoral studies on developing a novel neutron detector: the MIT/UMichigan Time Projection Chamber (MITPC). A helium and carbon tetrafluoride gaseous time projection chamber (TPC) with a CCD camera was developed and deployed at the Double Chooz experiment to measure the fast neutron background~\cite{Hourlier:2016dqe}. It was brought to Fermi National Accelerator Laboratory to measure the beam-induced neutron background on the Booster Neutrino Beamline; it ran in the SciBooNE enclosure. It was necessary to demonstrate that the detector could contain and reconstruct neutron-induced nuclear recoils at higher energies. This was done by replacing the neutron target, helium, with neon. 

\addcontentsline{toc}{subsection}{\textit{Publication: Demonstrating a directional detector based on neon for characterizing high energy neutrons}}
\includepdf[pages={-}]{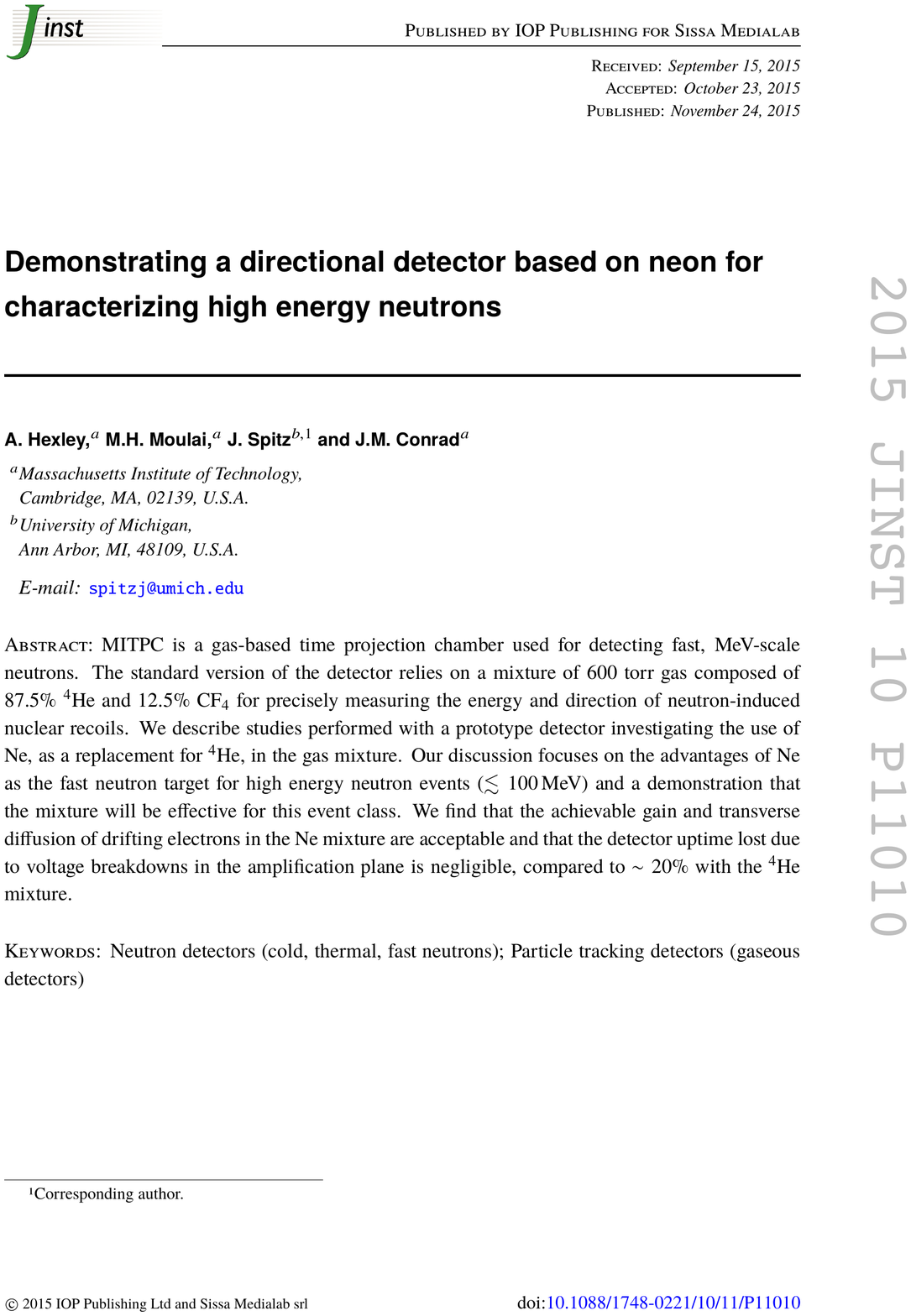}

\null
\vspace*{\fill}
\begin{figure}[h!]
  \centering
  \includegraphics[width=0.3\textwidth]{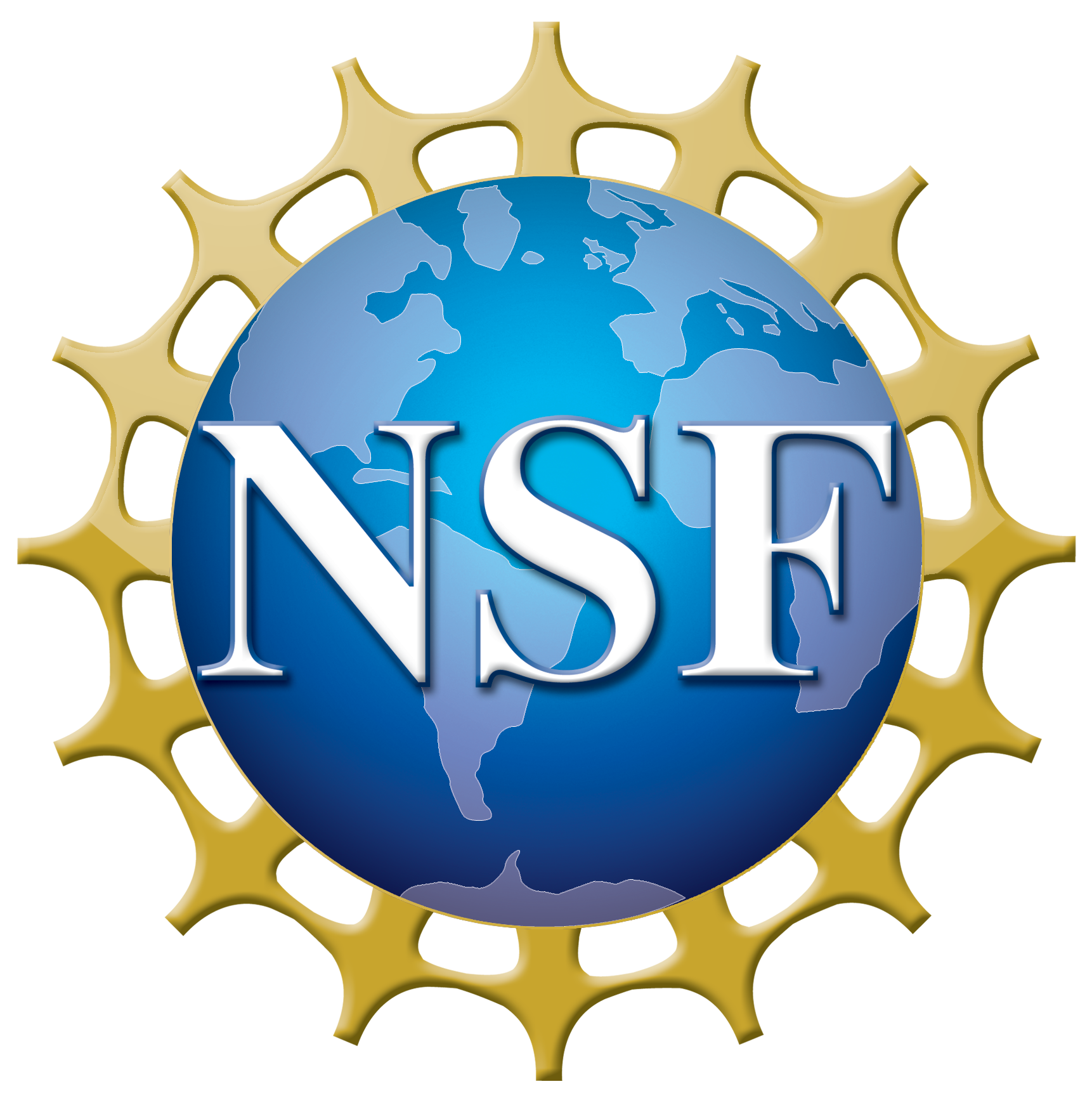}
\end{figure}
The work in this thesis was supported by the National Science Foundation.
\vspace*{\fill} 
\begin{singlespace}
\bibliography{main}
\addcontentsline{toc}{chapter}{Bibliography}
\bibliographystyle{ieeetr}
\end{singlespace}

\end{document}